**A copula-based measure for quantifying asymmetry in dependence and associations**

Robert R. Junker[1]*, Florian Griessenberger[2], Wolfgang Trutschnig[2]

[1]University of Salzburg, Department of Biosciences, Hellbrunnerstrasse 34, 5020 Salzburg

[2]University of Salzburg, Department of Mathematics, Hellbrunnerstrasse 34, 5020 Salzburg

*Corresponding author: robert.junker@sbg.ac.at, telephone number: +43/662/8044-5512

**Abstract**

Asymmetry is an inherent property of bivariate associations and therefore must not be ignored. The currently applicable dependence measures mask the potential asymmetry of the underlying dependence structure by implicitly assuming that quantity Y is equally dependent on quantity X, and *vice versa*, which is generally not true. We introduce the copula-based dependence measure *qad* that quantifies asymmetry. Specifically, *qad* is applicable in general situations, is sensitive to noise in data, detects asymmetry in dependence and reliably quantifies the information gain/predictability of quantity Y given knowledge of quantity X, and *vice versa*. Using real-world data sets, we demonstrate the relevance of asymmetry in associations. Asymmetry in dependence is a novel category of information that provides substantial information gain in analyses of bivariate associations.

**Introduction**

Despite the prevalence of asymmetry in many systems, it is ignored by measures quantifying the statistical dependence between variables. In the two-dimensional setting, both the classical Pearson's correlation and Spearman rank correlation are symmetric, i.e., $\rho(X,Y) = \rho(Y,X)$ holds; therefore, neither measure is capable of detecting asymmetries. The same is true for approaches that do not rely on specific forms of dependence and thus identify nonlinear and non-monotonic relationships (1). According to Renyi's (2) axioms, every dependence measure $d$ should be symmetric or mutual in the sense that $d(X,Y) = d(Y,X)$. However, considering, for instance, a two-dimensional sample in the form of a parabola (Fig. 1) or a sinusoidal curve (Fig. 2g), the dependence structure is clearly asymmetric. In such cases, knowing *X* strongly improves the predictability of *Y*, whereas the opposite is not true to the same extent. The dependence of random variables is asymmetric by nature, and ignoring asymmetries means ignoring valuable information. To provide a statistical tool to quantify asymmetry in dependence, we introduce the quantification of asymmetric dependence (*qad*) that induces the following related measures: $q(X, Y)$ denoting the dependence of Y on X (or, equivalently, the influence of X on Y), $q(Y, X)$ the dependence of X on Y, $\frac{1}{2}(q(X, Y) + q(Y, X))$ the average dependence, and $a = q(X, Y) - q(Y, X)$ the asymmetry of the dependence. The four measures are associated with *p*-values related to the significance of the dependence and asymmetry.

Big data and small data sets require robust statistical analyses to extract information regarding the relationships among variables. In this context, asymmetric or directional dependence adds additional and important information to the dependence structure of two variables X and Y. Both inductive and deductive approaches strongly benefit from the quantification of asymmetric dependence; notably, either more precise hypotheses and predictions will emerge from large data sets, or more precise hypotheses can be tested, i.e., quantifying the dependence of Y on X (or *vice versa*), instead of the overall dependence between X and Y. Treating dependence structures as symmetric by default goes hand in hand with ignoring key factors and relevant information. For example, if the causal relationship between two variables is known, the dependence of the affected variable on the causing variable is desired information that should be unaffected by the statistical dependence of the causing variable on the affected variable. In cases where no *a priori* knowledge of the causal relationship is available, directional dependence is a useful measure for exploring and estimating the association between variables in a more detailed and realistic way than classical (symmetric) dependence measures. Furthermore, the asymmetry $a = q(X, Y) - q(Y, X)$ provides additional information on bivariate associations. Although directional dependence is (by construction) not the same as causality, asymmetry in dependence may help to identify causal relationships among variables. In the era of big data, asymmetric directional dependence may be of particular importance for exploring pairwise relationships, identifying pairs of variables that feature either symmetric or asymmetric relationships, or pinpointing variables that affect a number of variables but are (mostly) unaffected by others and, therefore, may be key factor in their given system.

As a copula-based measure, *qad* captures all scale-invariant dependence in general situations that extend beyond standard parametric families or linear relationships and thus requires no assumptions regarding the underlying distribution of the data. These properties enable an almost universal applicability to data sets of various disciplines. We chose real-world data sets to illustrate the usefulness of *qad* and its potential in identifying important variables in large data sets.

*The copula-based asymmetric dependence measure* qad

The asymmetric, scale-invariant dependence of pairs of variables can be modeled via copulas (3-5). Copulas link multivariate distribution functions and their one-dimensional marginals and thereby capture all scale-invariant dependence (6). Moreover, *qad* can be used to detect a) the dependencies of any functional type, which can be seen as generalized asymmetric versions of the coefficient of determination $R^2$, and b) the asymmetry of the dependence between two random variables. For a concise mathematical description of the applied methodology, details regarding the motivation for this approach, and explicitly calculated examples, please refer to Supporting information 1. In short, given a two-dimensional sample $(x_1, y_1), \ldots, (x_n, y_n)$ from random vector $(X, Y)$, the *qad* method involves calculating the so-called empirical copula $\hat{A}_n$, aggregating/smoothing the copula to obtain an empirical checkerboard $\hat{C}_n$ (the unit square $[0,1]^2$ is partitioned into $N^2$ squares of edge length $\frac{1}{N}$, the mass per square is aggregated, and N is referred to as the resolution) and then calculating the normalized $D_1$-distance of $\hat{C}_n$ and the product copula $\Pi$ (modeling independence). The normalization constant 3 assures that the range of the dependence measure is within the interval of [0,1] (3). The resulting value is denoted by $\hat{q}_n(X, Y)$. After computing the quantities $\hat{q}_n(X, Y)$ and $\hat{q}_n(Y, X)$, the asymmetry $a$ of the sample $(x_1, y_1), \ldots, (x_n, y_n)$ is calculated as $a = q(X, Y) - q(Y, X)$. A permutation test (based on permutations of the conditional distributions of the empirical checkerboard) is then used to establish a *p*-value for testing the hypothesis $H_0 : q(X, Y) = q(Y, X)$ of symmetric dependence. The empirical checkerboard copula underlying *qad* together with the empirical marginals can also be exploited to predict Y given X, and *vice versa*. This process is facilitated by plots of the probability of Y for a given X, and *vice versa* (Fig. 1). Note that prediction is possible only within the range of measured X- and Y-values. Moreover, *qad* is calculated independent of a parametric regression function; therefore, no extrapolation is possible. To allow for the cross-platform application of *qad*, we wrote the R package *qad*, which is freely available from CRAN (7). The *qad* method is a non-parametric approach and thus involves no assumptions regarding the underlying distribution of the data. In addition, the approach is robust to outliers and returns reliable results. However, because the *qad* method builds upon copulas and copulas are (according to Sklar's Theorem) only unique for continuous random variables, discrete variables (i.e., ties in samples) must be handled with care. Although our simulations indicate adequate and reliable performance for data sets with ties (indicating point masses of the random variables $X, Y$), the number of unique values should be sufficiently large in comparison with the sample size. The resolution of the empirical checkerboard copula is proportional to the square root of the sample size; thus, as in any statistical method, the *qad* results become more reliable as the sample size increases. We recommend a sample size of no smaller than $n$ = 16 (i.e., resolution = 4 x 4).

**Results and Discussion**

*Properties of qad*. Dependence measures such as *qad* that capture the dependence in general situations should assign similar scores of dependence to equally noisy data in a manner independent of the concrete functional relationship (1). Accordingly, the dependence *q* between two random variables decreases with increasing noise irrespective of the functional relationship between X and Y (Fig. 2). In a linear setting, *qad* returned dependence values *q* closer to Pearson's *r* and Spearman's *rho* than the maximum information coefficient MIC (1), a symmetric dependence measure defined for general situations (Fig. 2a). Unlike Pearson's *r*, Spearman's *rho* and MIC, which are symmetric measures by construction, *qad* indicated asymmetry in settings in which (on average) more information on Y could be obtained by knowing the value of X than *vice versa*, i.e., $q(X, Y) > q(Y, X)$ (Fig. 2c, d, f, g, i). Recently, the use of Pearson's product moment correlation as a standard method for quantifying dependence was criticized for underestimating dependence (8) and for being a predictor of limited applicability in complex systems (9). In order to specifically test the relation between Pearson's *r* and *qad* we utilized the R-function plot_r() provided by Vanhove (10), which generates 16 datasets with different distributions (functional relations with noise component) all having the same Pearson correlation coefficient *r*. For *n* = 200 predefined *r*-values in the interval [-0.999, 0.999], we generated these 16 datasets and then quantified dependence using *qad*, Spearman's *rho*, and the maximum information coefficient MIC (Supporting information 4). In datasets with linear relationships and errors following a normal or a uniform distribution, *qad* produced values similar to *r*, just as Spearman's *rho* and MIC (Fig. 3 a-h) did. In datasets with non-linear underlying dependence structures, however, Pearson's *r* and Spearman's *rho* underestimated the dependence of the variables (Fig. 3 i, j, m, n) whereas *qad* and MIC detected a higher dependency in these cases. Furthermore, the graphics illustrate the well-known fact that that Pearson's *r* is particularly sensitive to (single) outliers (Fig. 3 k, l), whereas *qad* and the other rank-based quantities are not strongly affected. All dependence measures returned similar values for bimodal distributions (Fig. 3 o). For coarse data, qad, Pearson's *r* and Spearman's returned similar values, MIC failed to detect the underlying dependence structure (Fig. 3 p). Finally, *qad* was the only dependence measure capable of detecting asymmetries (Fig. 3 i, j, m, n) and thus provides additional valuable information on the underlying dependence structure.

To demonstrate the applicability and added value of *qad*, we used data sets from various disciplines to explore the asymmetry in the corresponding dependence structures.

*Cause–Effect Pairs*

We quantified the asymmetric dependence of 74 *Cause–Effect Pairs* (11, 12) to evaluate the causal discovery rate of *qad*. The *Cause–Effect Pair* benchmark data constitute a collection of data sets with known causalities for the evaluation of bivariate causal discovery methods (11). In 63.5% of all the tested pairs, the affected variable was more dependent on the causing variable than *vice versa*; therefore, the causal relationships were correctly discovered in these cases (see Supporting information SI2). The results suggest that the causal discovery rate of *qad* is as good as the best performing method specifically designed for causal discovery tested based on the same data set (11, 13). Moreover, the findings confirm that statistical dependence is not the same concept as causality; nevertheless, *qad* may help to infer causality.

*Global climate*

In the context of global warming and changes in precipitation regimes, information regarding past, present, and future climate at a local scale is required to assess the effects of climate on the environment. We retrieved bioclimatic variables for $n$ = 1862 locations homogenously distributed over the global landmass from CHELSA (14, 15) to test for asymmetric dependence between variables. As expected, the knowledge of one of the bioclimatic variables reduced the variability of the other variables, as reflected by the high dependence between pairs of variables (mean $q$ ± sd = 0.54 ± 0.18, see Supporting information SI2). Many of the associations between the bioclimatic variables were asymmetric (mean asymmetry $|a|$ ± sd = 0.06 ± 0.05, range = 0.00 – 0.23, see Supporting information SI2). For instance, annual precipitation can be better predicted by mean temperature ($q$ = 0.61, $p < 0.001$) than *vice versa* ($q$ = 0.54, $p < 0.001$, asymmetry $a$ = 0.08, $p < 0.001$, Fig. 4). Thus, *qad* helps to identify variables that are more predictive or predictable (on average) than others (in the context of climate, and in any other context), which is important knowledge for study design, particularly if assessments of many variables are cost and labor intensive.

*Microbiomes*

Diverse and complex microbial communities have become accessible due to high-throughput sequencing. Ecological relationships such as mutualism, competition, and commensalism between organisms shape the abundance distribution of taxa in communities. Often, correlation analysis testing for relationships between the abundance of pairs of taxa is used as basis for network inference, which facilitates the interpretation of microbiome structure. Ecological relationships between organisms may be reciprocal in the sense that taxa mutually affect each other, either positively (mutualism) or negatively (competition). They may, however, also be directed such that a given taxon is facilitating or inhibiting the growth of another taxon without being affected by itself by the other taxon (e.g. commensalism, amensalism). Conventional correlation analysis neither detects directed relationships nor discriminates between directed and mutual relationships, and is therefore of limited value for community dynamics (9). Here we used a dataset of bacteria associated with surfaces of the plant *Metrosideros polymorpha* (16) and tested for asymmetric relationships between pairs of $n$ = 93 operational taxonomic units (OTUs) that were observed in at least 75% of all samples ($n$ = 125). $q$-values (mean ± SD: 0.32 ± 0.094) were on average higher than Pearson's $r^2$ (0.07 ± 0.11) indicating that the majority of relationships between OTU pairs are not well described as a linear or isotonic interaction. Mostly, asymmetry in the dependence between OTUs was weak indicated by a relatively low mean value of asymmetry $|a|$ (0.034 ± 0.028). However, some pairwise interactions between OTUs were strongly asymmetric reaching $|a|$-values of >0.20. The quantification of asymmetric dependencies allows a novel definition of key species $S_f$, which influence the abundance of other species $S_j$ but are less influences by others. Thus, key species $S_f$ influence other species $S_j$ on average to a larger extend than they are dependent on other species $S_j$. Median influence of species $S_f$ is calculated as the median values of $I_f$ calculated as $I_f^j = q(S_f, S_j) - q(S_j, S_f), 1 \leq j \leq n, j \neq f$. If median ($I_f$) > 0 species $S_f$ influences most other species $S_j$ stronger than *vice versa*. As mentioned before, $q(S_f, S_j)$ denotes the dependence of $S_j$ on $S_f$ and $q(S_j, S_f)$ denotes the dependence in the other direction (Fig. 5a). In seven OTUs the corresponding medians of $I_f$ were significantly larger than zero indicating a stronger influence on most of the other OTUs than they are dependent on the other OTUs, i.e. these OTUs are defined as key species in this dataset (Fig. 5a; OTU 60, 8, 98, 56, 3433, 3977, 741; Supporting information 2). The four OTUs with the highest median influence-value ($I_f$) are members of class Gammaproteobacteria, three of them are within the order Pseudomonadales (genera Acinetobacter (OTU 8 and 60) and Pseudomonas (OTU 56)), one in the order Enterobacteriales (OUT 98, Supporting information 2). Interestingly, in an experimental study

Acinetobacter and Pseudomonas had been identified to play key roles in shaping abundances of bacteria associated with plant surfaces (17) suggesting that our results are not random findings but represent ecologically meaningful outcomes of *qad*. Using *q*-values in an adjacency matrix to create a weighted and directed network visualizing asymmetric dependencies *q* between OTUs confirmed the prominent role of the key-OTUs in shaping the abundances of a number of other OTUs. Most of the seven key-OTUs occupy central positions (betweenness centrality: 0 – 1882, mean ± sd = 431.7 ± 639.1; degree: 3 – 48, 16.0 ± 14.8) indicating their role as network hubs (hub-score: 0.04 – 1; 0.38 ± 0.33; Fig. 5b). In contrast, in an undirected and weighted network based on Pearson's $r^2$, key-OTUs defined by *qad* occupied peripheral positions (betweenness centrality: 0 – 17, mean ± sd = 5.1 ± 6.4; degree: 0 – 11, 6.3 ± 3.7). Accordingly, these OTUs were not identified as hubs (hub-score: 0.00 – 0.05; 0.019 ± 0.016; Fig. 5c). Furthermore, network roles of all OTUs in networks based on *qad* and *r* were independent in betweenness centrality and degree (*qad*: $q \leq 0.203$, $p \geq 0.08$; Pearson's product-moment correlation: $|t| \leq 0.103$, $df = 91$, $p \geq 0.61$). The hub-scores obtained by the OTUs in both networks showed a weak and even negative association (*qad*: $q = 0.34$, $p \leq 0.001$; Pearson's product-moment correlation: $|t| \leq -2.39$, $df = 91$, $p = 0.019$). Betweenness centrality, degree, and hub-score have previously been used to identify key-species in diverse microbial communities (18). These values can, however, vary depending on the statistical analysis performed to obtain the dependence of OTUs such as Pearson's *r*, Spearman's *rho*, MIC (18), or *qad*. *qad* provides two novel indices – the asymmetric dependence *q* and the asymmetry *a* – that may help to detect key-species in diverse and complex microbiome data. We do not recommend to exclusively relying on *qad* for network inference and the detection of key-species, but without doubt *qad* has the potential to discover ecologically important species that would remain undiscovered using symmetric approaches.

*World development indicators*

The World Bank provides data sets on national development indicators related to, for example, the economy, education, health, and infrastructure of states (World Development Indicators WDI, The World Bank, last accessed July 2017). Apart from hypothesis-driven studies, data exploration to develop hypotheses and find interesting associations in large data sets is a major challenge in the era of big data. We explored the WDI data set published for 2015 to identify indicator pairs that feature strong asymmetry in dependence and thus stand out from other indicator pairs (see Supporting information SI2). This approach differs from previously applied strategies to identify interesting pairs of variables in large data sets, where pairs with a strong dependence (linear or nonlinear) were selected (1). The birth rate and death rate indicators (*qad*: *mean* $q = 0.42$, $p < 0.001$) displayed strong asymmetry in dependence (*qad*: $a = 0.2016$, $p < 0.001$; Fig. 6) and thus stood out from the majority of pairs tested (higher asymmetry than in 99.8% of all indicator pairs). Further data exploration revealed that the gross domestic product per capita (GDP) was a potential underlying factor for this nonlinear relationship. The GDP was strongly and symmetrically associated with the birth rate (*qad*: *mean* $q = 0.67$, $p < 0.001$; $a = 0.02$, $p = 0.59$) and weakly and symmetrically related to the death rate (*qad*: *mean* $q = 0.31$, $p < 0.001$; $a = 0.03$, $p = 0.39$, see Supporting information SI2). Generally, in countries with a GDP below the mean GDP of all countries, the birth rate and death rate are positively associated; in countries with a GDP above the mean GDP, we found a negative association (Fig. 6, see Supporting information SI2). The relationship between the economy and population growth, i.e., the relationship between the birth and death rates, has been previously discussed (19, 20) and has important implications for societies. We conclude that *qad* strongly facilitates the detection of interesting associations in large data sets and thus contributes to more thorough exploration of data sets and the improved detection of meaningful patterns compared to traditional methods.

Our theoretical and real-world examples clearly demonstrate the necessity und universal applicability of the quantification of (a)symmetric dependence *qad* for extracting important information from bivariate associations. Asymmetry in dependence will facilitate the detection and extraction of patterns from big data and the testing of hypotheses. The *qad* provides information pertaining to a novel property of bivariate associations – asymmetry in dependence – which will enhance the understanding of large and small data sets.

**Acknowledgments:** The study was funded by the Austrian Science Fund (FWF, Y 1102-B29).

**Author contributions:** RRJ and WT designed the study; RRJ, FG, and WT analyzed the data; RRJ and WT drafted the manuscript; and all authors contributed to the final version.

**Supplementary Information:**

Supporting information 1 – Concise mathematical description of *qad*, the motivation behind this approach, and explicitly calculated examples

Supporting information 2 – Methods, data, and additional results regarding the theoretical and real-world data sets

Supporting information 3 – Data used in the study.

Supporting information 4 – R codes

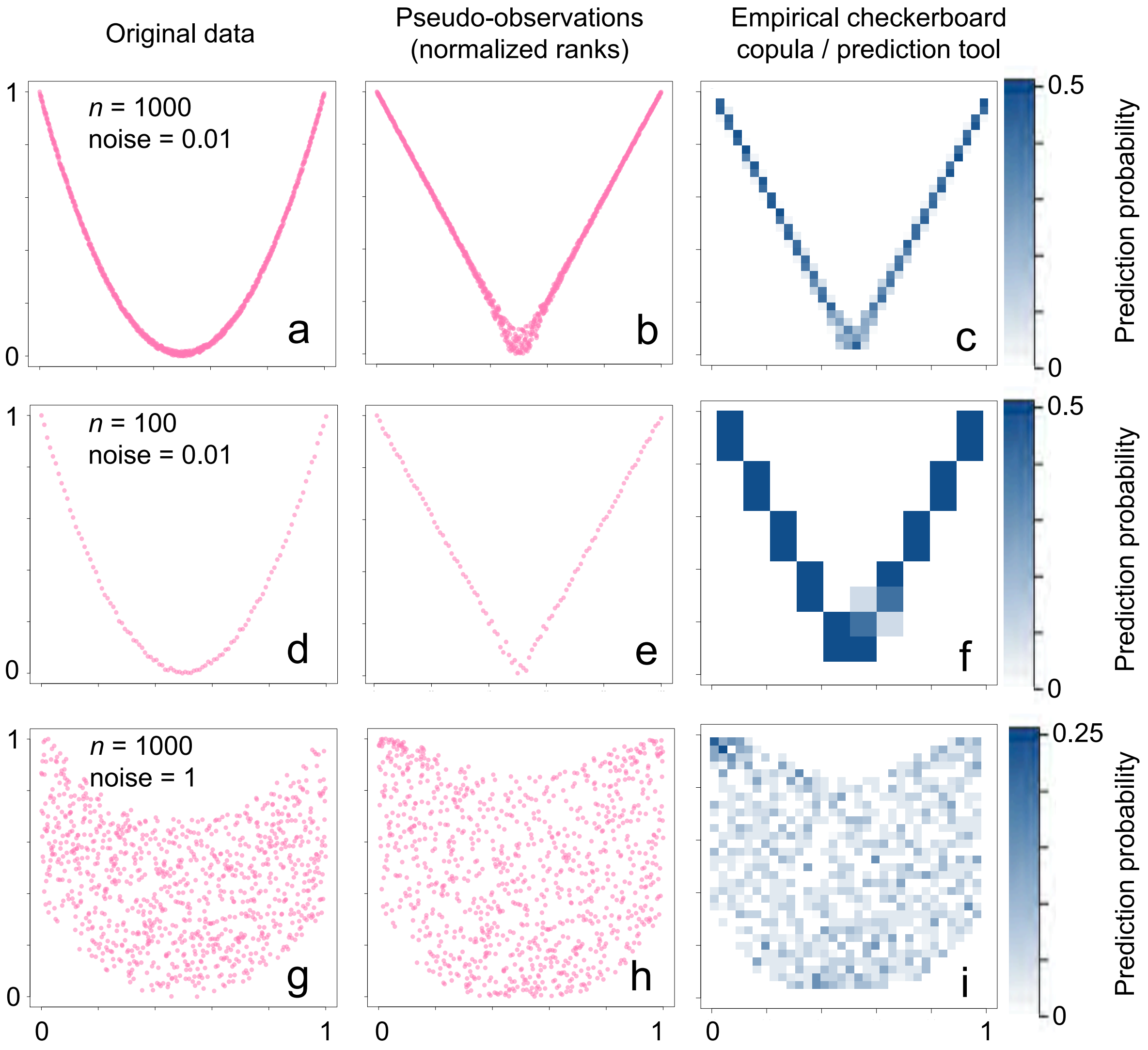

**Fig. 1** Quantification of asymmetric dependence *qad*. The quadratic function (**a**, **d**, **g**) illustrates asymmetric dependence. **a**) For each value of X, the value of Y is well predicted, whereas for each Y, two values (or regions) of X are generally possible. Accordingly, $q(X,Y)$= 0.96, $p < 0.001$, and $q(Y,X)$= 0.48, $p < 0.001$, resulting in strong and significant asymmetry $a$ = 0.48 at $p < 0.001$. Based on the pseudo-observations (normalized ranks, uniform distribution of data, **b**, **e**, **h**) the empirical checkerboard copula is calculated (**c**, **f**, **i**). The resolution of the checkerboard copula (number of stripes per dimension) depends on the sample size. For each vertical or horizontal stripe, the estimated conditional distribution is given, and the values sum to 1. Therefore, the checkerboard copula can be used to predict Y given the knowledge of X, and *vice versa*. The precision of the prediction decreases with the sample size for **d-f** ($q(X,Y)$ =

0.90, $p < 0.001$; $q(Y, X)$= 0.45, $p < 0.001$; $a = 0.45$, $p < 0.001$). For noisy and less depend data, we obtain (**g-i:** $q(X, Y)$= 0.36, $p < 0.001$; $q(Y, X)$= 0.24, $p < 0.001$; $a = 0.12$, $p < 0.001$).

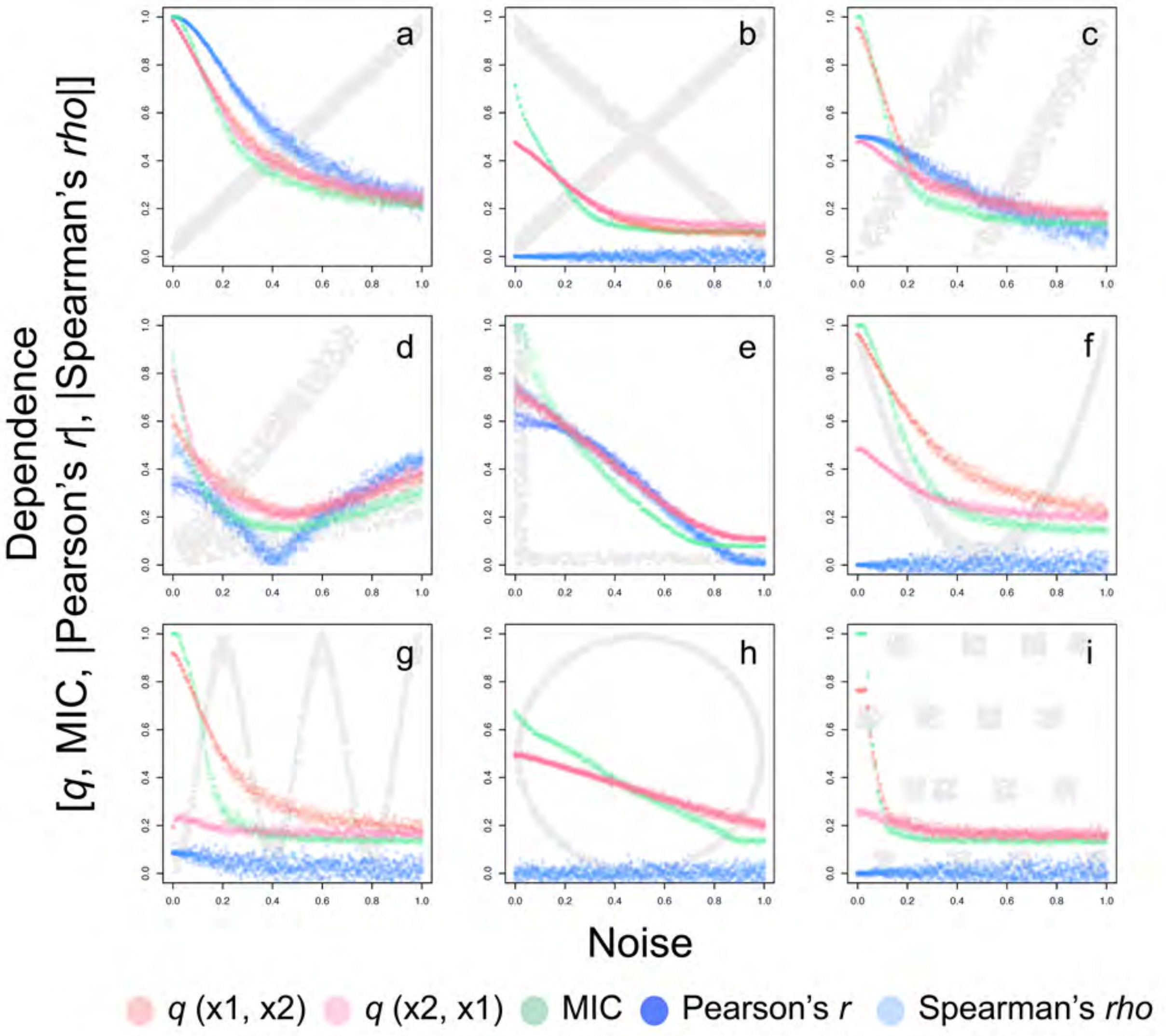

**Fig. 2** Comparison of *qad* to existing methods quantifying the dependence of two random variables. $q(X,Y), q(Y,X)$, the maximum information coefficient MIC, Pearson's *r* and Spearman's *rho* are calculated for nine different dependence structures as a function of increasing vertical noise (**a-i**). The gray points in the background depict samples from the corresponding dependence structures. Absolute values of the Pearson correlation *r* and Spearman correlation *rho* are plotted, and negative values were obtained in (**d**) and (**e**). In the linear case, the *qad* values range between Pearson's *r,* Spearman's *rho* and MIC (**a**). For the nonlinear function types, *qad* and MIC yields higher dependence values than Pearson's *r* and Spearman's *rho*. Note that the dependence structures depicted in (**c**), (**d**), (**f**), (**g**), and (**i**) are asymmetric, as reflected by the two *q*-values because we have $q(X,Y) > q(Y,X)$, particularly in situations with little noise. The other measures of dependence are not able to provide information about asymmetry in dependence.

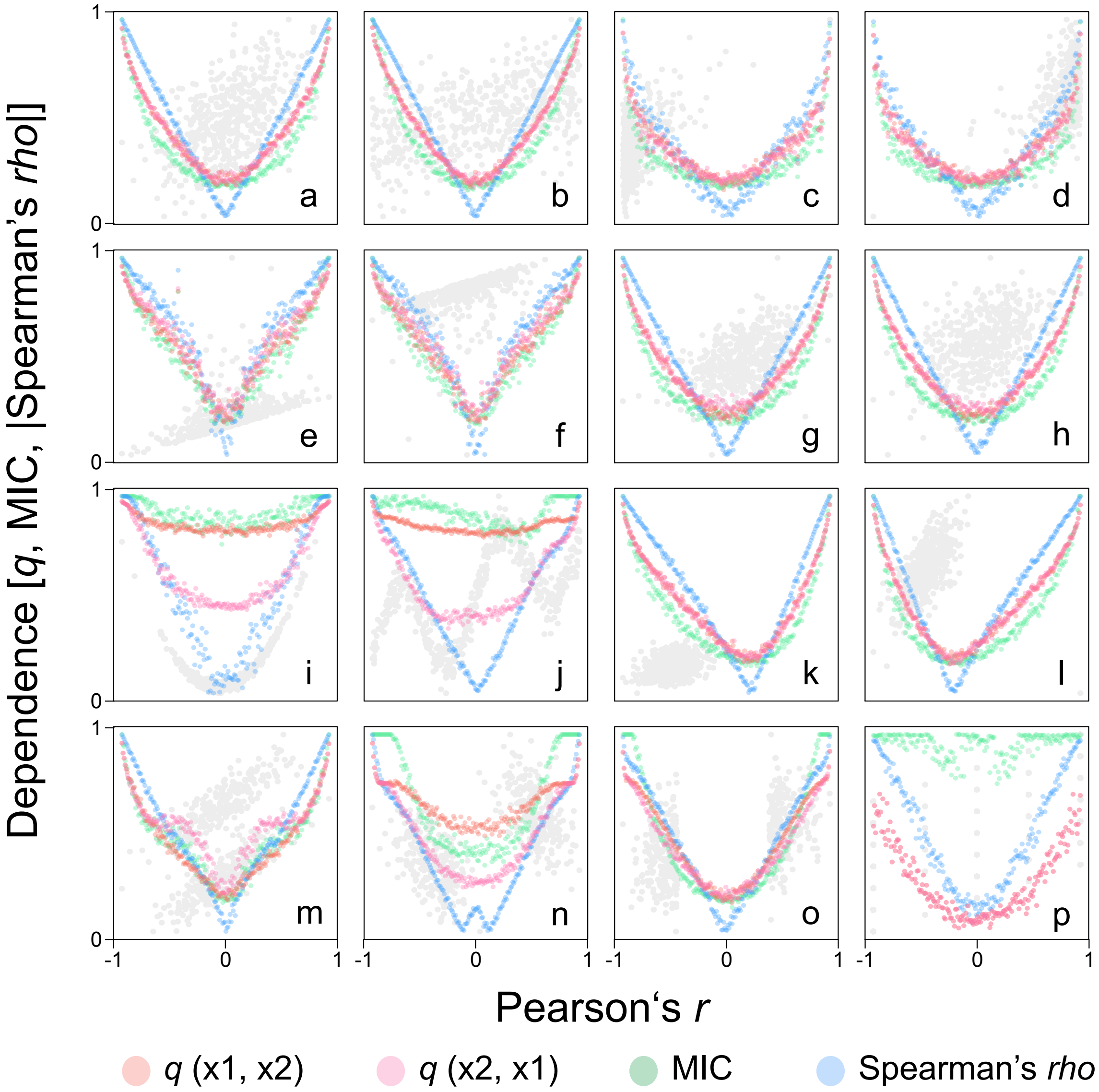

**Fig. 3** Comparison of *qad* to Pearson's correlation coefficient *r* and further existing methods quantifying the dependence of two random variables. The R-function plot_r() provided by Vanhove (10) generates 16 datasets with different distributions (functional relations with noise component) all having the same Pearson correlation coefficient *r*. For each pattern, we simulated data sets with a given *r* and quantified dependence using *qad*, Spearman's *rho*, and MIC. Absolute values for Spearman's *rho* are given. Example data with *r* = 0.5 are depicted in the background in gray in each figure. The patterns are (**a**) normal x, normal residuals; (**b**) uniform x, normal residuals; (**c**) skewed x, normal residuals; (**d**) skewed x, normal residuals; (**e**) normal x, skewed residuals; (**f**) normal x, skewed residuals; (**g**) increasing spread; (**h**) decreasing spread; (**i**) quadratic trend; (**j**) sinusoid relationship; (**k**) a single positive outlier; (**l**) a single negative outlier; (**m**) bimodal residuals; (**n**) two groups; (**o**) sampling at the extremes; (**p**)

coarse data. For further information see Vanhove (10). Note that the (dependence structures underlying the) samples depicted in (**i**), (**j**), (**m**), and (**n**) are asymmetric, which is also detected by *qad* ($q(X, Y) > q(Y, X)$), the other measures of dependence are not able to inform about asymmetry in dependence.

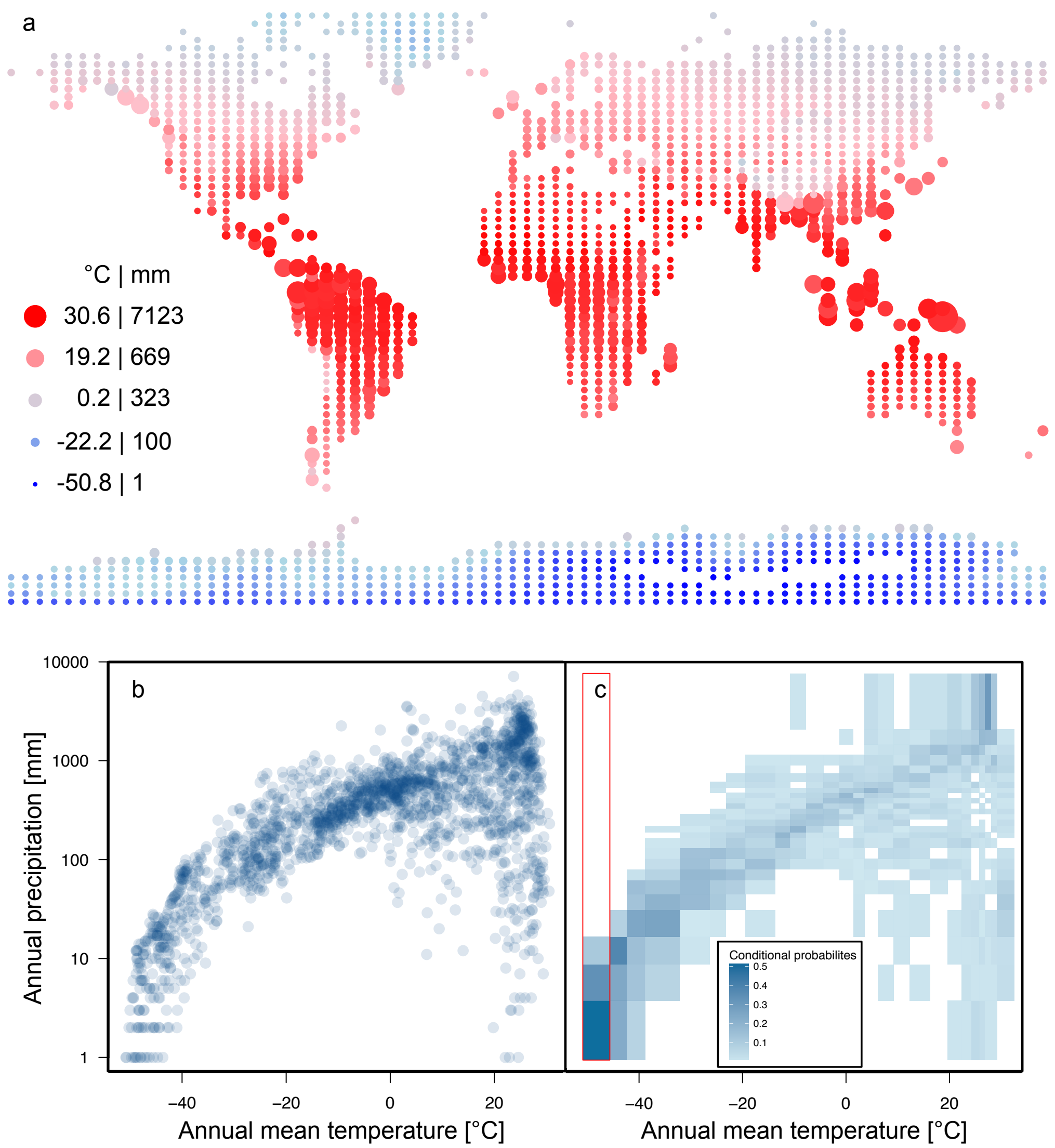

**Fig. 4** Association between the annual mean temperature AMT [°C] and annual precipitation AP [mm]. (**a**) Grid of *n* = 1862 locations homogenously distributed over the global landmass for which bioclimatic variables were retrieved. AMT is color coded, and AP is proportional to the point size. Low precipitation (small points) occurs in hot areas (red points, e.g., Sahara) and cold areas (blue, e.g., Arctic and Antarctica). High precipitation (large points) is mainly restricted to hot, tropical areas. (**b**) Accordingly, the association between AMT and AP is asymmetric. The knowledge of AMT strongly improved the predictability of AP ($q$ = 0.61, $p$ < 0.001), whereas the predictability was weaker in the opposite case ($q$ = 0.54, $p$ < 0.001; Asymmetry $a$ = 0.08, $p$ < 0.001). (**c**) The association between AMT and AP visualized as a two-dimensional empirical checkerboard distribution (retransformed empirical checkerboard copula), which can be used to predict Y-values given X-values, and *vice versa*. For instance, in an area with an AMT of -50.8

°C (first column), the estimated probability of AP totaling between 1 mm and 6 mm is 51.6%, between 6 mm and 14 mm is 35.8%, and between 14 mm and 26 mm is 12.6% (red frame).

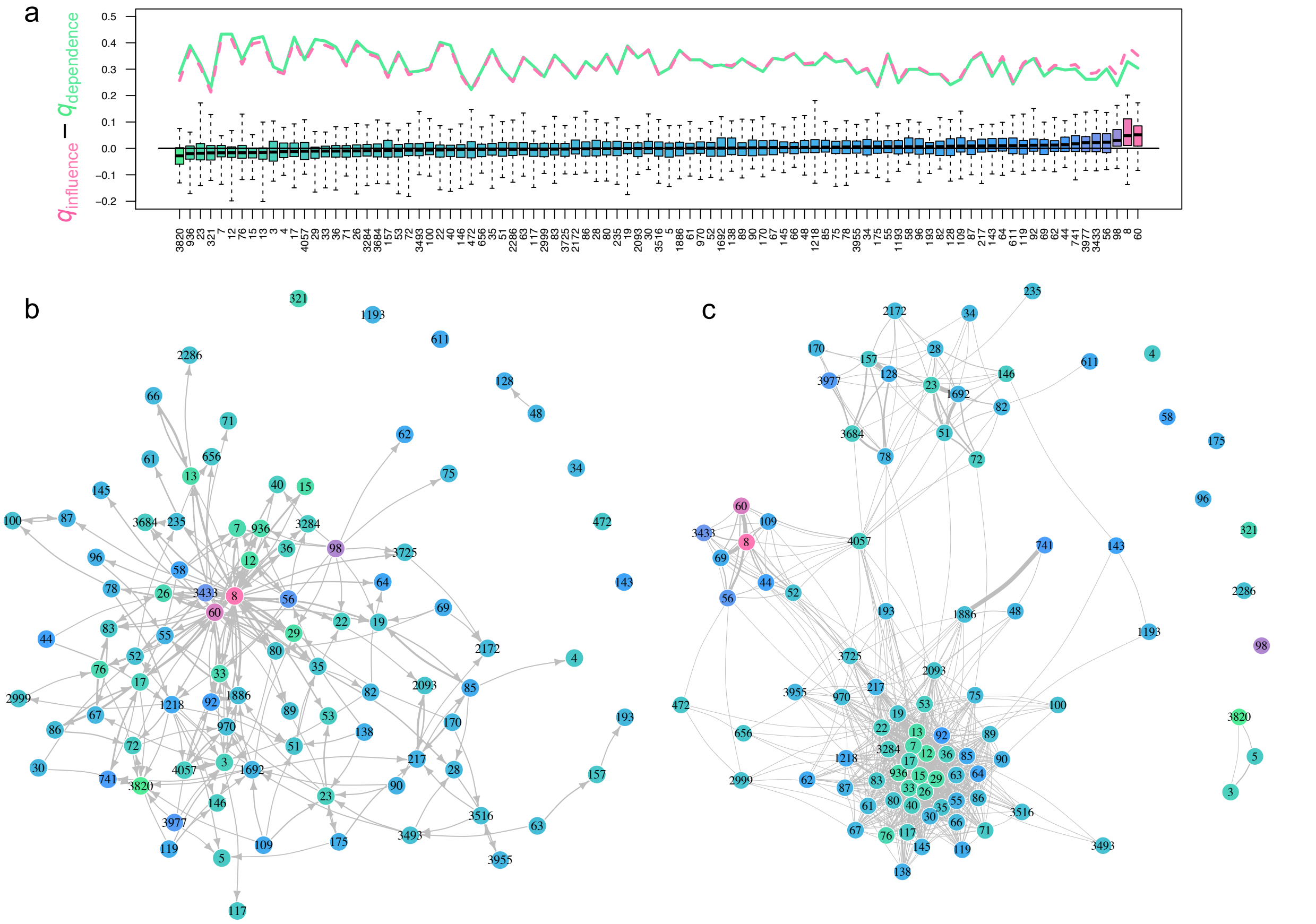

**Fig. 5** Bacterial community associated with above ground surfaces of *Metrosideros polymorpha*, a tree species endemic to the Hawaiian Islands. (a) To identify key species following the definition that key species strongly influence the abundance of (many) other species but are influenced to a much smaller extend by other species, we calculated the median of the influences $I_f^j = q(S_f, S_j) - q(S_j, S_f)$ of each species $S_f$ and every other species $S_j$. For each OTU $S_f$, the median, and quartiles of $I_f^1, \ldots, I_f^n$ is calculated. OTUs are ordered by increasing median influence *I*. In seven OTUs the corresponding medians were significantly greater than zero indicating that these OTUs have a stronger influence on most of the other OTUs than they are dependent on the other OTUs, i.e. these OTUs are defined as key species in this dataset (Fig. 5a; OTU 60, 8, 98, 56, 3433, 3977, 741; Supporting information 2). The dashed pink line above the boxplots depicts the average influence $\frac{1}{n-1}\sum_{1\leq j\leq n, j\neq f} I_f^j$ of each focal species $S_f$ on the other species $S_1, \ldots, S_n$. The green line denotes the mean influence $\frac{1}{n-1}\sum_{1\leq j\leq n, j\neq f} I_j^f$ of the other species on species $S_f$ (**b**) Weighted and directed network visualizing asymmetric dependencies between OTUs. Arrows of edges point towards the dependent species and away

from the influencing species. Width of edges is proportional to dependence *q*. Median influence *I* is lowest in nodes shown in green, and highest in nodes shown in pink; OTUs with medium influence values are plotted in blue, see scale in (a). (**c**) Undirected and weighted network based on Pearson's *r*. Edge width is proportional to dependence *r*. Colors of nodes correspond to median influence based on *q* to visualize different network roles in both networks.

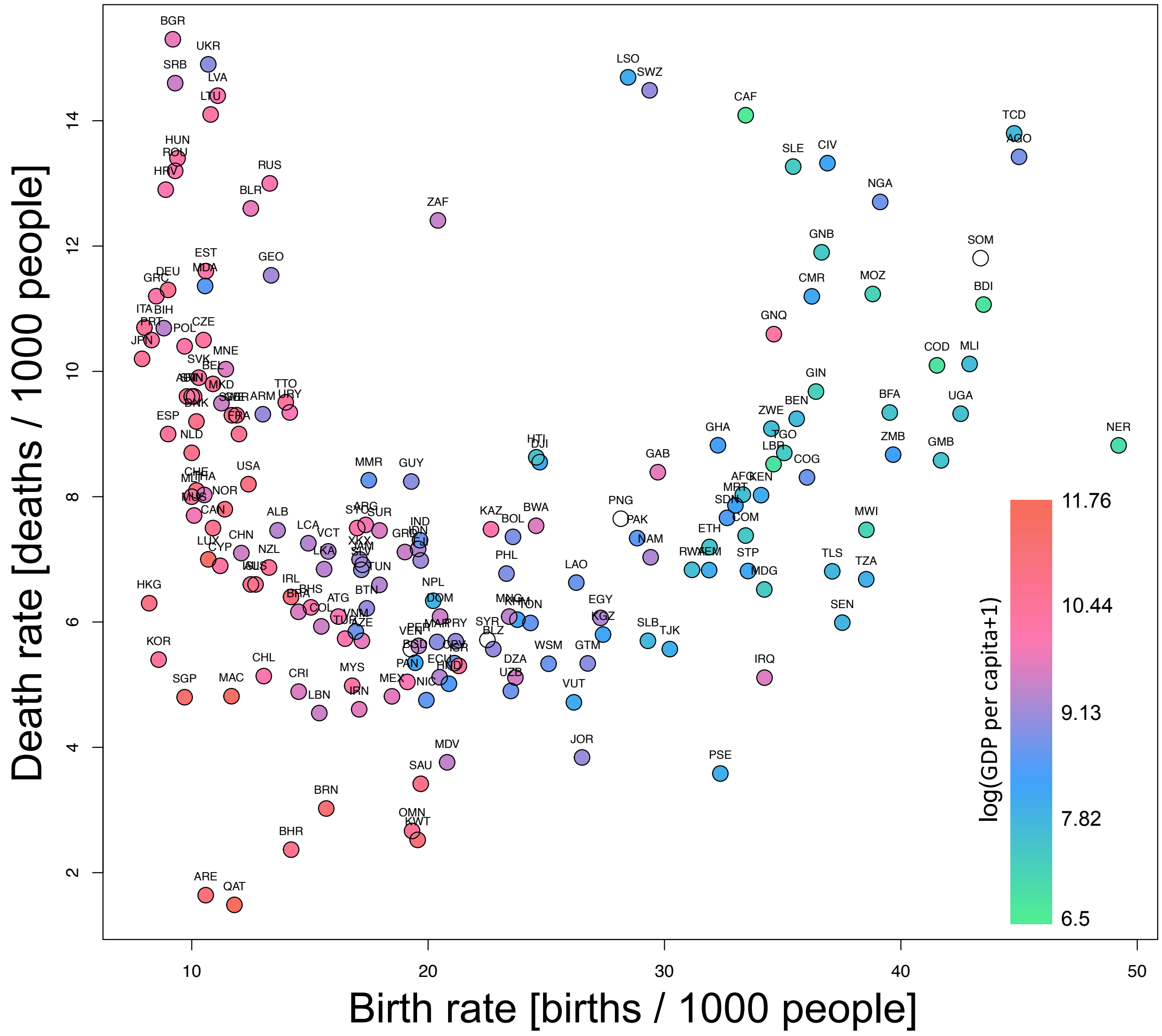

**Fig. 6** Association between the birth rate and death rate. The gross domestic product (GDP per capita) of each country is color coded. The death rate is well predicted by the birth rate (*q(birth rate, death rate)* = 0.53, *p* < 0.001). The birth rate is less well predicted by the death rate (*q(death rate, birth rate)* = 0.33, *p* < 0.001; asymmetry: *a* = 0.20, *p* < 0.001). Countries with low GDPs exhibit higher birth rates, and the corresponding death rates seem to be positively correlated with the birth rates; countries with higher GDPs display lower birth rates, and the birth and death rates seem to be negatively correlated. The detection of the relation between the birth and death rates was possible due to the strong and significant asymmetry in the relation, which is higher than that for 99.8% of all variable pairs tested. Detection exclusively based on the strength of the dependence (*qad*: mean *q* = 0.42; Pearson's *r* = 0.063) is unlikely, as the *mean q* and *r* are within the interquartile range of all variable pairs (*mean q*: 0.27 – 0.43; *r*: 0.05

– 0.28). Thus, the detection of interesting relationships and meaningful explanations is strongly facilitated by *qad*.

**A copula-based measure for quantifying asymmetry in dependence and associations**

Robert R. Junker[1*], Florian Griessenberger[2], Wolfgang Trutschnig[2]

**Supporting information 1**

## 1 Mathematical description of qad

The dependence measure $\zeta_1$, introduced in (1), is a copula-based dependence measure and hence scale-invariant. Considering that $\zeta_1$ is based on conditional distributions (Markov kernels) and that estimating conditional distribution is a difficult endeavor, it is a-priori unclear if good estimators can be derived in full generality, i.e. without any regularity assumptions on the underlying two-dimensional distribution function or the underlying copula, at all. Nevertheless we proved that the so-called empirical checkerboard estimator introduced in this paper is strongly consistent for all continuous random variables $X,Y$.

The rest of S1 is organized as follows. Section 2 gathers some preliminaries and notations that will be used throughout S1. Section 3 recalls the definition of empirical copulas which are then aggregated in Section 4 to so-called empirical checkerboard copulas ECC. Our main result saying that calculating $\zeta_1$ of the ECC and choosing an adequate resolution for the aggregation yields a strongly consistent estimator for $\zeta_1(X,Y)$ is derived and proved in Section 4 too. A simulation study for different dependence structures including some extreme cases concludes S1.

## 2 Notation and preliminaries

Throughout S1 $\mathbf{R}$ will denote the real numbers, $\mathbf{N}$ the natural numbers and $\mathscr{C}$ will denote the family of all two-dimensional copulas (for background on copulas we refer to (2, 3)). For every copula $A \in \mathscr{C}$ the corresponding doubly stochastic measure will be denoted by $\mu_A$. As usual, $d_\infty(A,B)$ will denote the uniform metric on $\mathscr{C}$, i.e.

$$d_\infty(A,B) := \max_{(x,y)\in[0,1]^2} |A(x,y) - B(x,y)|.$$

It is well known that $(\mathscr{C}, d_\infty)$ is a compact metric space. For every metric space $(\Omega, d)$ the Borel $\sigma$-field will be denoted by $\mathscr{B}(\Omega)$ and $\lambda$ will denote the Lebesgue measure on $\mathscr{B}([0,1])$. A mapping $K : \mathbf{R} \times \mathscr{B}(\mathbf{R}) \to [0,1]$ is called a Markov kernel from $\mathbf{R}$ to $\mathscr{B}(\mathbf{R})$ if $x \mapsto K(x,B)$ is measurable for every fixed $B \in \mathscr{B}(\mathbf{R})$ and $B \mapsto K(x,B)$ is a probability measure for every fixed $x \in \mathbf{R}$. A Markov kernel $K : \mathbf{R} \times \mathscr{B}(\mathbf{R}) \to [0,1]$ is called regular conditional distribution of a (real-valued) random variable $Y$ given (another random variable) $X$ if for every $B \in \mathscr{B}(\mathbf{R})$

$$K(X(\omega), B) = \mathbf{E}(\mathbf{1}_B \circ Y | X)(\omega)$$

holds $\mathbf{P}$-almost sure, whereby $\mathbf{1}_B(x)$ denotes the indicator function. It is well known that a regular conditional distribution of $Y$ given $X$ exists and is unique $\mathbf{P}^X$-almost sure. For every $A \in \mathscr{C}$ the corresponding regular conditional distribution (i.e. the regular conditional distribution of $Y$ given $X$ in the case that $(X,Y) \sim A$) will be denoted by $K_A(\cdot,\cdot)$. Note that for every $A \in \mathscr{C}$ and Borel sets $E, F \in \mathscr{B}([0,1])$ we have

$$\int_E K_A(x,F)\, d\lambda(x) = \mu_A(E \times F).$$

For more details and properties of conditional expectations and regular conditional distributions see (4, 5).

We will mainly work with the metrics $D_1$ and $D_\infty$ introduced in (1). These metrics are defined by

$$D_1(A,B) := \int_{[0,1]} \underbrace{\int_{[0,1]} |K_A(x,[0,y]) - K_B(x,[0,y])|\, d\lambda(x)}_{=:\Phi_{A,B}(y)}\, d\lambda(y)$$

and

$$D_\infty(A,B) := \sup_{y \in [0,1]} \Phi_{A,B}(y)$$

respectively. It can be shown that $\phi_{A,B}$ is Lipschitz-continuous with Lipschitz constant 2 and that both metrics generate the same topology (without being equivalent). The resulting metric space $(\mathscr{C}, D_1)$ is complete and separable and it can be shown that, firstly, $D_1(A,\Pi)$ attains only values in $[0,\frac{1}{3}]$ and that, secondly, $D_1(A,\Pi)$ is maximal if and only if $A$ is completely dependent, i.e. if a $\lambda$-preserving transformation $h : [0,1] \to [0,1]$ exists such that $K_A(x,\{h(x)\}) = 1$ for $\lambda$-a.e.

$x \in [0,1]$. In the sequel we will let $\mathscr{C}_d$ denote the family of all completely dependent copulas, and write $A_h$ and $K_h(\cdot,\cdot)$ for the completely dependent copula and the Markov kernel of the completely dependent copula induced by the $\lambda$-preserving transformation $h$ respectively. For equivalent definitions and properties of completely dependent copulas we refer to (1) and the references therein.

The asymmetric dependence measure qad calculated in the R-package coincides with $\zeta_1$ defined in (1).

**Definition 2.1** Let $X,Y$ be two continuous random variables with joint distribution function $H$ and copula $A$. Then the dependence measure $\zeta_1$ is defined by

$$\zeta_1(X,Y) = \zeta_1(A) := 3D_1(A,\Pi).$$

As a direct consequence of the properties of $D_1$ it follows immediately that for all continuous random variables $X,Y$ we have $\zeta_1(X,Y) \in [0,1]$, that $\zeta_1(X,Y) = 0$ if and only if $X$ and $Y$ are independent, and that $\zeta_1(X,Y) = 1$ if and only if the copula $A$ of $(X,Y)$ is completely dependent (or, equivalently, if there exists some Borel measurable $f : \mathbf{R} \to \mathbf{R}$ such that $Y = f \circ X$ a.s.).

## 3 Empirical copulas for general samples (possibly having ties)

Let $(X,Y)$ be a random vector with joint distribution $H$, margin distributions $F,G$ and copula $A \in \mathscr{C}$. Furthermore let $(x_1,y_1),\ldots,(x_n,y_n)$ denote a sample of $(X,Y)$, $H_n$ the bivariate empirical distribution function and $F_n,G_n$ the one-dimensional empirical distribution functions. As a consequence of Sklar's Theorem there exists a unique subcopula $A'_n : Range(F_n) \times Range(G_n) \to Range(H_n)$, which fulfills

$$H_n(x,y) = A'_n(F_n(x),G_n(y)).$$

For an exact definition of subcopulas and their properties we refer to (3). If $X,Y$ are continuous random variables then with probability one the domain of $A'_n$ is the equidistant raster $R_n = \{0,\frac{1}{n},\ldots,\frac{n-1}{n},1\}^2$, if not, then ties occur with probability greater than zero. There are uncountably many ways to extend a given (strict) subcopula to a copula - if we only consider the bilinear interpolation/extension $A_n \in \mathscr{C}$ then we can assign every sample $(x_1,y_1),\ldots,(x_n,y_n)$ a unique copula. To this (absolutely continuous checkerboard) copula we will refer to as *the empirical copula of the sample* $(x_1,y_1),\ldots,(x_n,y_n)$ in the sequel (6, 7).

More precisely we proceed as follows (we only include the description for the sake of completeness): Let $(u_i, v_i) := (F_n(x_i), G_n(y_i)), i = 1, \ldots, n$ denote the pseudo-observations and $(u'_1, v'_1), \ldots, (u'_m, v'_m)$ the distinct pairs of pseudo-observations with $m \leq n$. Set $S_1 := \{0, u_1, \ldots, u_{m_1}\}, S_2 := \{0, v_1, \ldots, v_{m_2}\}$ and define the quantities $t_i, r_i$ and $s_i$ by

$$\begin{aligned} t_i &:= \sum_{j=1}^{n} \mathbf{1}_{(u'_i, v'_i)}(u_j, v_j) \quad i = 1, \ldots, m \\ r_i &:= \sum_{j=1}^{n} \mathbf{1}_{u_i}(u_j) \quad i = 1, \ldots, m \\ s_i &:= \sum_{j=1}^{n} \mathbf{1}_{v_i}(v_j) \quad i = 1, \ldots, m. \end{aligned}$$

Define the empirical subcopula $A'_n : S_1 \times S_2 \to \{0, \frac{1}{n}, \ldots, \frac{n-1}{n}, 1\}$ by

$$A'_n(s_1, s_2) = \frac{1}{n} \sum_{i=1}^{m} t_i \cdot \mathbf{1}_{[0,s_1] \times [0,s_2]}(u'_i, v'_i) = \frac{1}{n} \sum_{i=1}^{n} \mathbf{1}_{[0,s_1] \times [0,s_2]}(u_i, v_i)$$

and extend $A'_n$ bilinearly to a copula $A_n$ by considering the product copula $\Pi \in \mathscr{C}$, defining transformations $w_i : [0,1]^2 \to [u'_i - \frac{r_i}{n}, u'_i] \times [v'_i - \frac{s_i}{n}, v'_i]$ by

$$w_i(x, y) = (u'_i - \frac{r_i}{n} + \frac{r_i}{n} x, v'_i - \frac{s_i}{n} + \frac{s_i}{n} y)$$

and setting $\mu_{A_n} := \frac{1}{n} \sum_{i=1}^{m} t_i \mu_{\Pi}^{w_i}$. In the following simple example we calculate the empirical copula of a count-data sample of size 6.

**Example 3.1.** Consider the sample $(10,10), (6,3), (5,1), (6,4), (4,1), (6,3)$ for which we obviously have $m = 5 < 6 = n$. The distinct pseudo-observations and the quantities $t_i, r_i, s_i$ are gathered in the subsequent table (Tab. S1_1). As direct consequence we get $S_1 = \{0, \frac{1}{6}, \frac{2}{6}, \frac{5}{6}, 1\}$ and $S_2 = \{0, \frac{2}{6}, \frac{4}{6}, \frac{5}{6}, 1\}$. The resulting empirical copula $A_6$ (i.e. the bilinear extension of $A'_6$) is absolutely continuous with the density depicted in Fig. S1_1.

| $i =$ | 1 | 2 | 3 | 4 | 5 |
|---|---|---|---|---|---|
| $(u_i, v_i)$ | $(1,1)$ | $(\frac{5}{6}, \frac{4}{6})$ | $(\frac{2}{6}, \frac{2}{6})$ | $(\frac{5}{6}, \frac{5}{6})$ | $(\frac{1}{6}, \frac{2}{6})$ |
| $t_i$ | 1 | 2 | 1 | 1 | 1 |
| $r_i$ | 1 | 3 | 1 | 3 | 1 |
| $s_i$ | 1 | 2 | 2 | 1 | 2 |

Tab. S1_1: Pseudo-observations and the quantities $t_i, r_i, s_i$ of example 3.1.

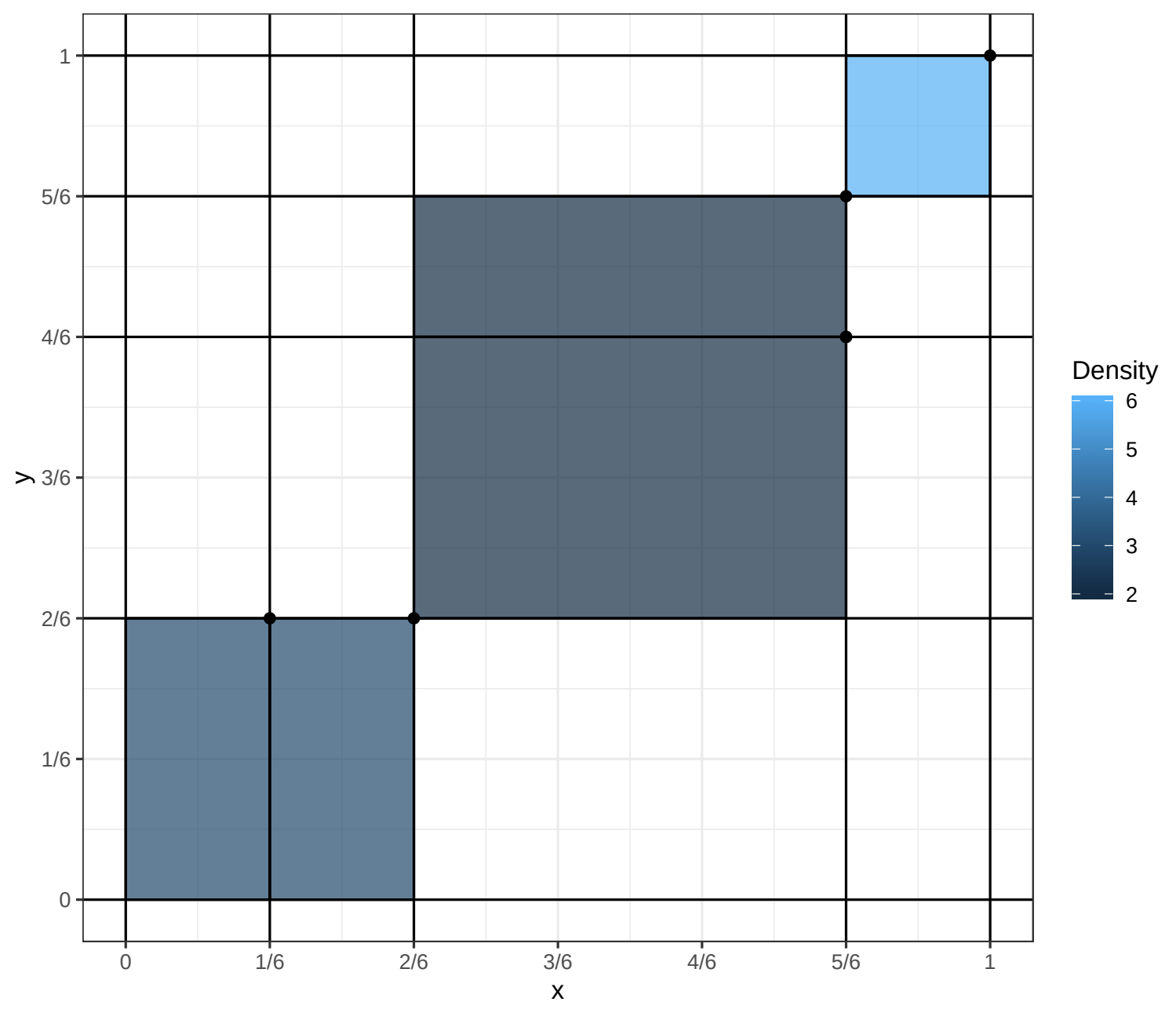

Fig. S1_1: Empirical copula for the sample considered in Example 3.1

**Remark 3.2.** The R-package calculates the empirical copula for arbitrary samples - if there are no ties the package calls the function C.n from the copula package, otherwise it calculates $A_n$ as described above.

**Remark 3.3.** The following results only consider the case of pairs $(X,Y)$ whose two-dimensional distribution function $H$ is continuous and hence the underlying copula unique. Nevertheless qad also produced very good results in the context of random variables with non-trivial discrete components. Our plan is to study the extensibility of the consistency results presented on the next pages and in Section 4 to the general setting in the near future.

The following result is key for proving consistency of the empirical checkerboard copula w.r.t.t. metric $D_1$ in the next section.

**Theorem 3.4.** (8) *Let $(X_1,Y_1),(X_2,Y_2),\ldots$ be a random sample from $(X,Y)$ and assume that $(X,Y)$ has continuous joint distribution function $H$ and copula $A$. Then the following asymptotic* *result holds for the empirical copula $A_n$ with probability $1$:*

$$d_\infty(A_n,A) = O\left(\sqrt{\frac{\log(\log(n))}{n}}\right) \tag{1}$$

The following first result shows that working with the empirical copula (without further ado) does not yield a useful estimator of $\zeta_1(A)$.

**Theorem 3.5.** *Let $(X_1,Y_1),(X_2,Y_2),\dots$ be a random sample of $(X,Y)$, where $(X,Y)$ has continuous joint distribution $H$ and copula $A=\Pi$. Then with probability $1$ we have*

$$\lim_{n\to\infty} D_1(A_n,\Pi) = \frac{1}{3}.$$

*Proof.* For every $x\in[0,1]$ there exists exactly one $i(x)\in\{0,1,\dots,n\}$ such that

$$K_{A_n}\left(x,\left[\frac{i(x)}{n},\frac{i(x)+1}{n}\right]\right)=1.$$

As direct consequence we get

$$\begin{aligned}
D_1(A_n,\Pi) &= \int_{[0,1]}\int_{[0,1]} |K_{A_n}(x,[0,y])-y|\,d\lambda(x)d\lambda(y)\\
&\geq \int_{[0,1]}\int_{\left[0,\frac{i(x)}{n}\right]\cup\left[\frac{i(x)+1}{n},1\right]} |K_{A_n}(x,[0,y])-y|\,d\lambda(y)d\lambda(x)\\
&= \int_{[0,1]}\left(\int_{\left[0,\frac{i(x)}{n}\right]} y\,d\lambda(y)+\int_{\left[\frac{i(x)+1}{n},1\right]} 1-y\,d\lambda(y)\right)d\lambda(x)\\
&= \int_{[0,1]} \frac{i(x)^2}{2n^2}+1-\frac{i(x)+1}{n}-\left(\frac{1}{2}-\frac{(i(x)+1)^2}{2n^2}\right)d\lambda(x)\\
&= \frac{1}{2}+\frac{1}{n}+\frac{1}{2n^2}+\underbrace{\int_{[0,1]}\frac{i(x)^2}{n^2}\,dx}_{=:I_1^n}-\underbrace{\int_{[0,1]}\frac{i(x)}{n}\,dx}_{=:I_2^n}+\underbrace{\int_{[0,1]}\frac{i(x)}{n^2}\,dx}_{=:I_3^n}.
\end{aligned}$$

A straightforward calculation shows $\lim_{n\to\infty} I_1^n=\frac{1}{3}$, $\lim_{n\to\infty} I_2^n=\frac{1}{2}$ and $\lim_{n\to\infty} I_3^n=0$, from which the desired result follows immediately.

One possibility to overcome this problem is to smooth or aggregate the empirical copula. Aggregation leads to so-called checkerboard copulas.

## 4 Checkerboard copulas

We follow (9) and proceed as follows. Fix $N\in\mathbf{N}$, define the squares $R_{ij}^N$ for $i,j\in\{1,\dots,N\}$ by

$$R_{ij}^N=\left[\frac{i-1}{N},\frac{i}{N}\right]\times\left[\frac{j-1}{N},\frac{j}{N}\right],$$

and let $int(R_{ij}^N)$ denote the interior of $R_{ij}^N$.

**Definition 4.1.** A copula $A_N \in \mathscr{C}$ is called $N$-checkerboard copula, if $A_N$ is absolutely continuous and (a version of) its density $k_{A_N}$ is constant on the interior of each square $R_{ij}^N$. We call $N$ the resolution of $A_N$, denote the set of all $N-$checkerboard copulas with $\mathscr{C}\mathscr{B}_N$, and set $\mathscr{C}\mathscr{B} = \bigcup_{N=1}^{\infty} \mathscr{C}\mathscr{B}_N$.

**Definition 4.2.** For $A \in \mathscr{C}$ and $N \in \mathbf{N}$ the (absolute continuous) copula $\mathscr{C}\mathscr{B}_N(A) \in \mathscr{C}\mathscr{B}_N$, defined by

$$\mathscr{C}\mathscr{B}_N(A)(x,y) := \int_0^x \int_0^y N^2 \sum_{i,j=1}^{N} \mu_A(R_{ij}^N) \mathbf{1}_{int(R_{ij}^N)}(s,t)\, d\lambda(t) d\lambda(s) \tag{2}$$

is called $N$-checkerboard approximation of $A$ or simply $N$-checkerboard of $A$.

According to (1) $\mathscr{C}\mathscr{B}$ is dense in $(\mathscr{C}, D_1)$. Furthermore, the following result going back to (9) holds (thereby $A^t$ denotes the transpose of $A$).

**Theorem 4.3.** *For every copula $A \in \mathscr{C}$ the following equality holds:*

$$\lim_{N\to\infty} D_1(\mathscr{C}\mathscr{B}_N(A), A) = 0 = \lim_{N\to\infty} D_1(\mathscr{C}\mathscr{B}_N(A)^t, A^t).$$

Out simple idea for deriving a 'good' estimator for $\zeta_1(A)$ is to plug-in the empirical copula $A_n$ in Equation (2) and consider a resolution $N$ fulfilling $N < n$. In the sequel we will refer to $\mathscr{C}\mathscr{B}_N(A_n)$ as empirical $N$-checkerboard copula.

**Example 4.4.** Suppose the vector $(X,Y)$ has distribution function $\Pi$ (in the sequel we will simply write $(X,Y) \sim \Pi$ and that $(x_1,y_1),\ldots,(x_n,y_n)$ is a sample of size $n = 100$ from $(X,Y)$. Fig. S1_2 depicts a scatterplot of the sample and Fig. S1_3 depicts the density of the corresponding empirical 5-checkerboard.

The qad packages carries out the following steps for estimating $\zeta_1(A)$ given a sample $(x_1,y_1),\ldots,(x_n,y_n)$ from $(X,Y)$ with copula $A$.

1. Calculation of the empirical copula $A_n$.

2. Selection of an appropriate resolution $N$ and calculation of the empirical $N$-checkerboard $\mathscr{C}\mathscr{B}_N(A_n)$ according to equation (2).

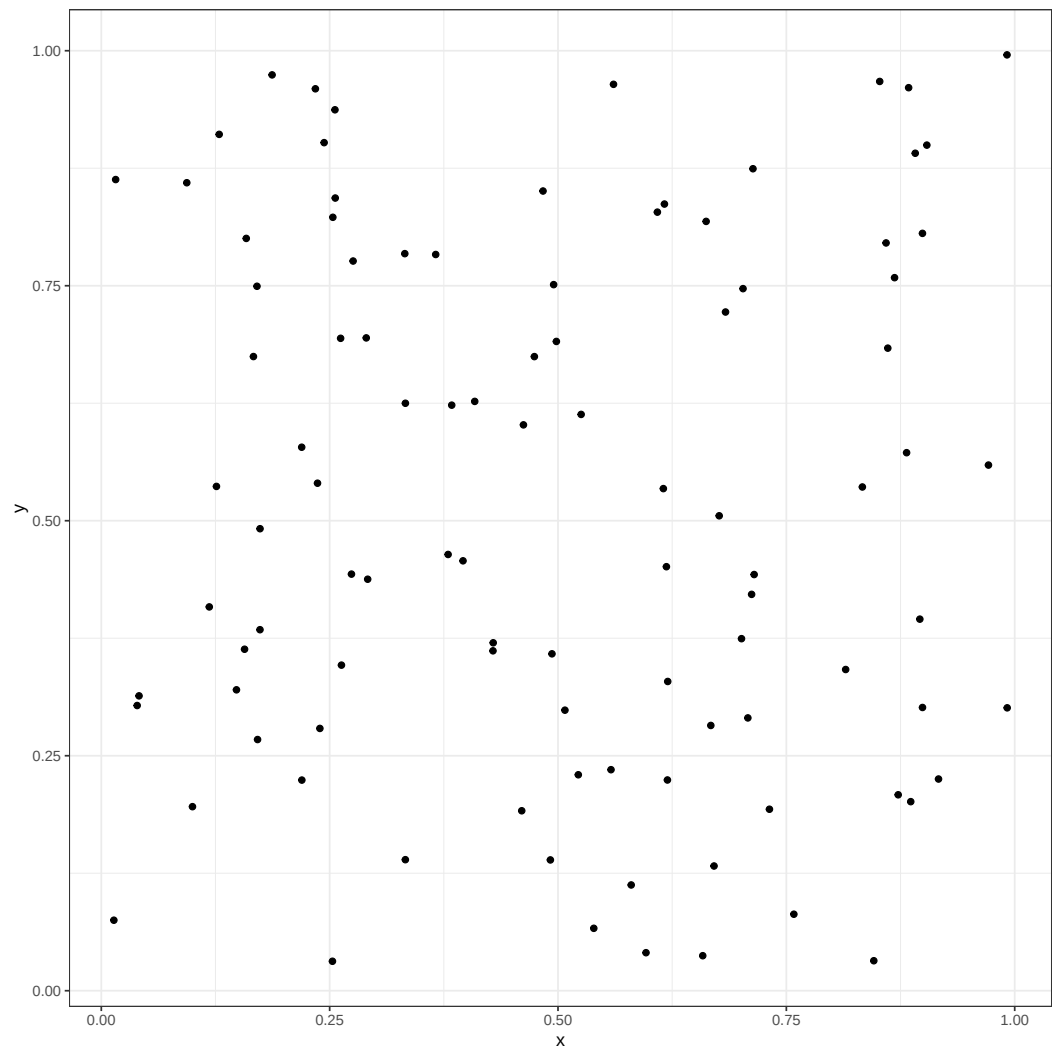

Fig. S1_2: Scatterplot of the sample $(x_1,y_1),\ldots,(x_n,y_n)$ of size $n=100$ from $(X,Y)\sim\Pi$.

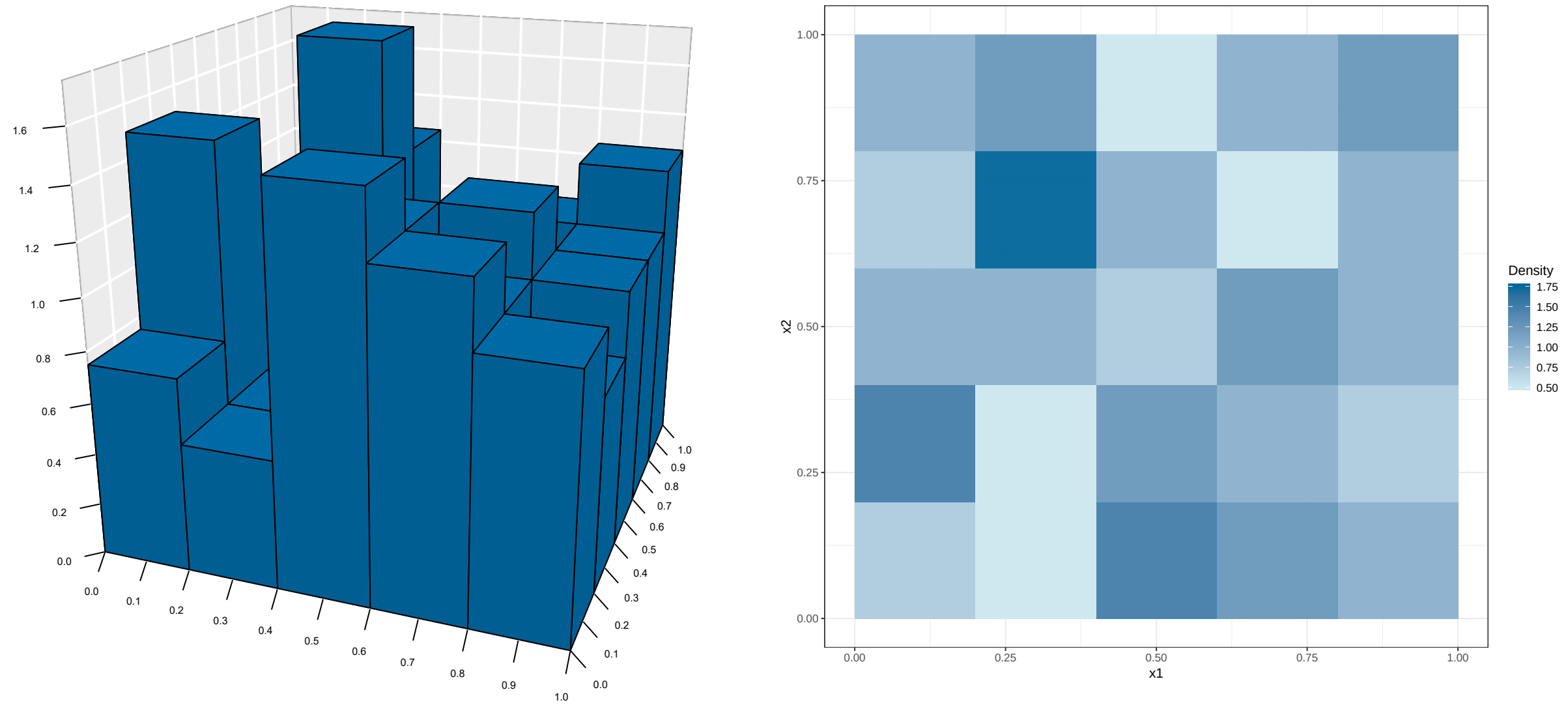

Fig. S1_3: 3d plot (left panel) and image plot (right panel) of the density of the empirical 5-checkerboard of the sample from Fig. S1_2.

3. Estimation of $\zeta_1(A)$ in terms of $\zeta_1(\mathscr{CB}_N(A_n)) = 3D_1(\mathscr{CB}_N(A_n),\Pi)$.

Choosing an adequate resolution $N=N(n)$ the estimator $\zeta_1(\mathscr{CB}_N(A_n))$ is strongly consistent - we proceed in several steps to proof this result. We start with the following lemma linking $D_\infty$ and $d_\infty$ for checkerboards.

**Lemma 4.5.** *For all $A_N, B_N \in \mathscr{CB}_N$ the following inequality holds:*

$$D_\infty(A_N, B_N) \leq 2(N-1)d_\infty(A_N, B_N).$$

*Moreover the inequality is sharp, i.e. for every $N \in \mathbf{N}$ we can find copulas $A_N, B_N \in \mathscr{CB}_N$ for which equality in (3) holds.*

*Proof.* Fix an arbitrary $i \in \{1,\dots,N\}$. Then for every $x \in \left[\frac{i-1}{N}, \frac{i}{N}\right]$ and $y \in [0,1]$ continuity of $A$, disintegration and the fact that $K_{A_N}(x,\cdot)$ is constant on each interval of the form $\left(\frac{i-1}{N}, \frac{i}{N}\right)$ yields

$$\begin{aligned} A_N\left(\frac{i}{N},y\right) - A_N\left(\frac{i-1}{N},y\right) &= \mu_{A_N}\left(\left[\frac{i-1}{N},\frac{i}{N}\right]\times[0,y]\right) \\ &= \int_{\left[\frac{i-1}{N},\frac{i}{N}\right]} K_{A_N}(s,[0,y])\,d\lambda(s) \\ &= \frac{1}{N}K_{A_N}(x,[0,y]). \end{aligned}$$

Altogether we get

$$\begin{aligned} D_\infty(A_N,B_N) &= \sup_{y\in[0,1]} \int_{[0,1]} |K_{A_N}(x,[0,y]) - K_{B_N}(x,[0,y])|\,d\lambda(x) \\ &= \sup_{y\in[0,1]} \sum_{i=1}^{N} \int_{\left[\frac{i-1}{N},\frac{i}{N}\right]} |K_{A_N}(x,[0,y]) - K_{B_N}(x,[0,y])|\,d\lambda(x) \\ &= \sup_{y\in[0,1]} \sum_{i=1}^{N} \left| A_N\left(\frac{i}{N},y\right) - A_N\left(\frac{i-1}{N},y\right) - B_N\left(\frac{i}{N},y\right) + B_N\left(\frac{i-1}{N},y\right)\right| \\ &= \sup_{y\in[0,1]} \left\{ \sum_{i=2}^{N-1} \left| A_N\left(\frac{i}{N},y\right) - A_N\left(\frac{i-1}{N},y\right) - B_N\left(\frac{i}{N},y\right) + B_N\left(\frac{i-1}{N},y\right)\right| + \right. \\ &\qquad \left. + \left|A_N(\frac{1}{N},y) - B_N(\frac{1}{N},y)\right| + \left|A_N(\frac{N-1}{N},y) - B_N(\frac{N-1}{N},y)\right| \right\} \\ &\leq \left(\sum_{i=2}^{N-1} 2d_\infty(A_N,B_N)\right) + 2d_\infty(A_N,B_N) \\ &= 2(N-1)d_\infty(A_N,B_N), \end{aligned}$$

which completes the proof of ineq. (3).

Fig. S1_4 and Fig. S1_6 depict copulas $A_N, B_N$ for which equality holds. Since the construction idea easily extends to arbitrary $N \in \mathbf{N}$ the proof is complete.

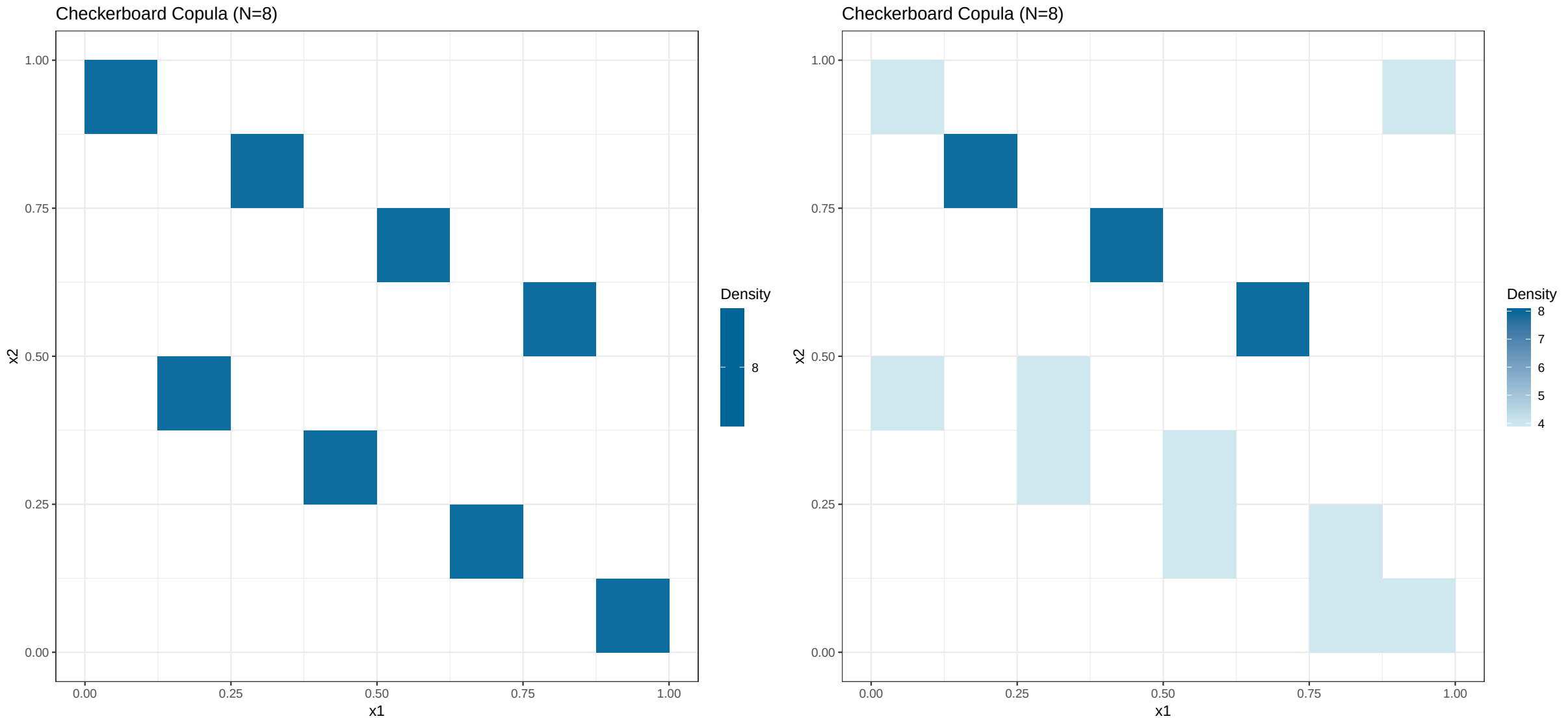

Fig. S1_4: Checkerboard copulas $A_N$ (left) and $B_N$ (right) with $N = 8$ for which equality in Lemma 4.5 holds since $D_\infty(A_N, B_N) = 7/8$ and $d_\infty(A_N, B_N) = 1/16$ (see Fig. S1_5).

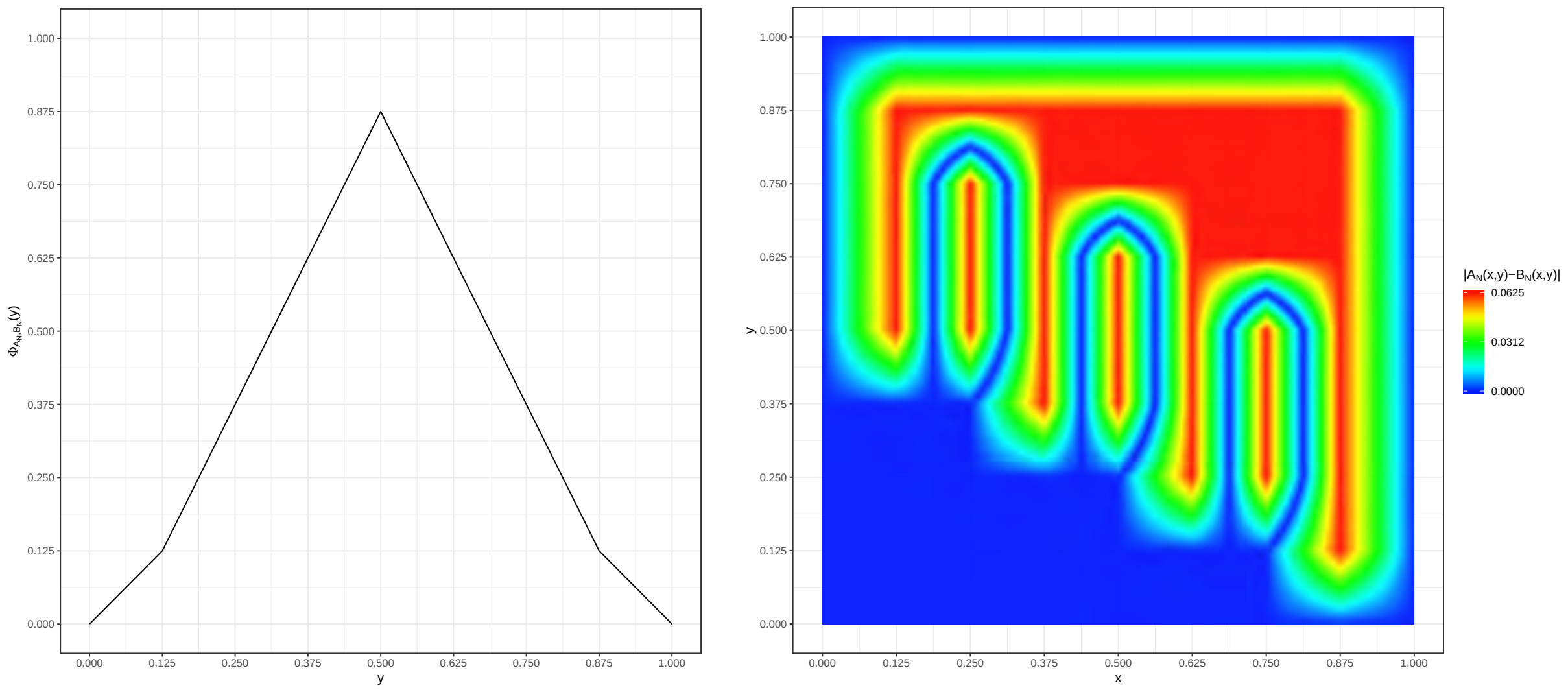

Fig. S1_5: The maps $y \mapsto \Phi_{A_N,B_N}(y)$ (left) and $(x,y) \mapsto |A_N(x,y) - B_N(x,y)|$ (right) for $A_N$ and $B_N$ defined in Fig. S1_4.

As next step we derive a slightly improved inequality linking $D_1$ and $D_\infty$ (compare with Theorem 6 in (1) and Lemma 3 in (10)):

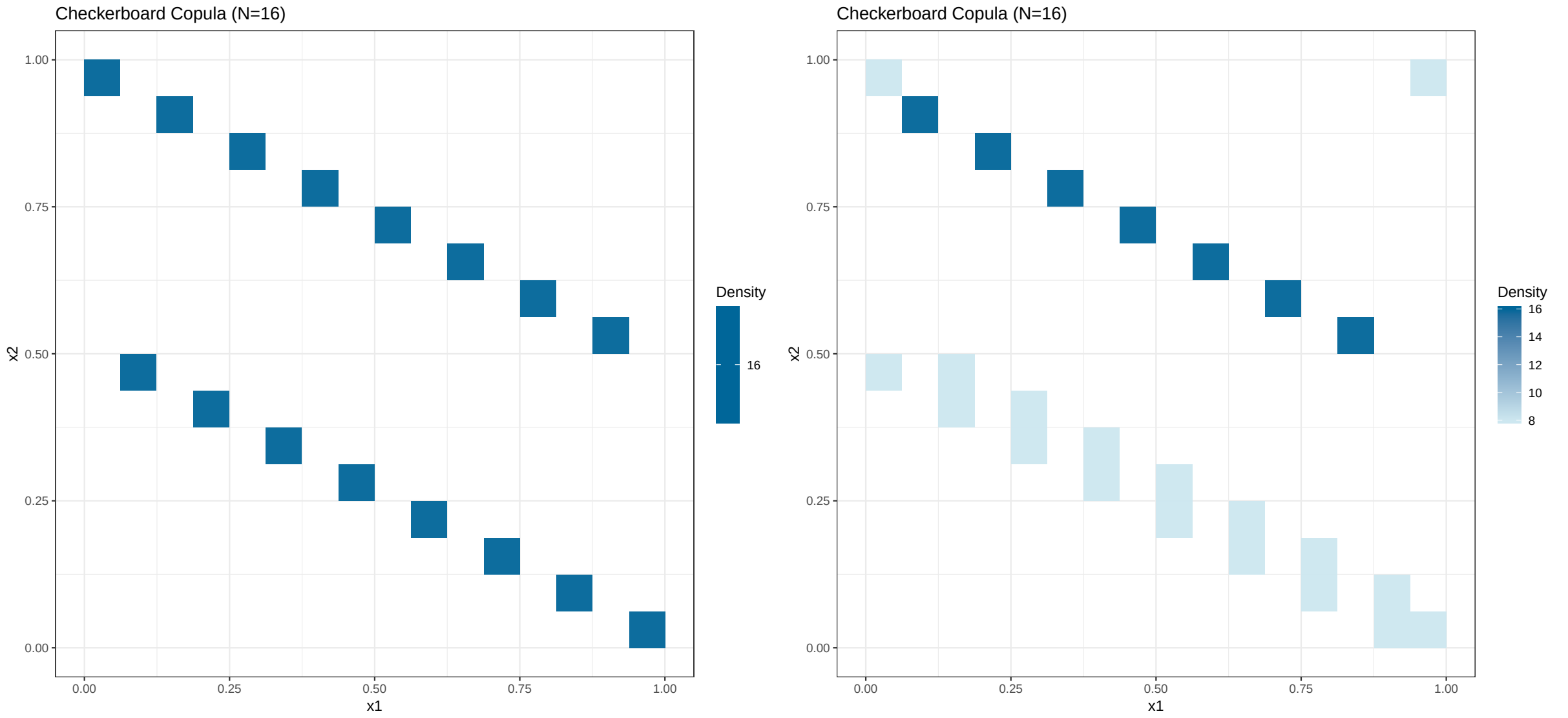

Fig. S1_6: Checkerboard copulas $A_N$ (left) and $B_N$ (right) with $N = 16$ for which equality in Lemma 4.5 holds since $D_\infty(A_N, B_N) = 15/16$ and $d_\infty(A_N, B_N) = 1/32$.

**Lemma 4.6.** *Let $N \in \mathbf{N}$ and $A_N, B_N \in \mathscr{CB}_N$. Then the following inequality holds:*

$$D_1(A_N, B_N) \leq \left(\frac{N-1}{N}\right) D_\infty(A_N, B_N). \tag{3}$$

*Inequality (3) is best possible and we have equality if there exists some $c \in [0, \frac{2}{N}]$ such that $\Phi_{A_N,B_N}(y) = c$ on $\left[\frac{1}{N}, \frac{N-1}{N}\right]$.*

*Proof.* Considering the facts that $\Phi_{A_N,B_N}(0) = \Phi_{A_N,B_N}(1) = 0$, that $y \mapsto \Phi_{A_N,B_N}(y)$ is piecewise linear and Lipschitz-continuous with Lipschitz-constant $2$ we get

$$\max_{y \in [0,1]} \Phi_{A_N,B_N}(y) = \max_{y \in \left[\frac{1}{N}, \frac{N-1}{N}\right]} \Phi_{A_N,B_N}(y),$$

from which the assertions follow.

Considering $N$-checkerboard approximations of arbitrary copulas can not increase their $d_\infty$-distance - the following lemma holds:

**Lemma 4.7.** *For arbitrary $A, B \in \mathscr{C}$ and their $N$-checkerboard approximations $\mathscr{CB}_N(A), \mathscr{CB}_N(B) \in \mathscr{CB}$ the following inequality holds:*

$$d_\infty(\mathscr{CB}_N(A), \mathscr{CB}_N(B)) \leq d_\infty(A, B). \tag{4}$$

*The inequality is sharp for every $N \in \mathbf{N}$.*

*Proof.* The inequality is easily derived via

$$\begin{aligned} d_\infty(\mathscr{CB}_N(A), \mathscr{CB}_N(B)) &= \sup_{(x,y)\in[0,1]^2} |\mathscr{CB}_N(A)(x,y) - \mathscr{CB}_N(B)(x,y)| \\ &= \sup_{(x,y)\in\{\frac{1}{N},\dots,\frac{N-1}{N}\}^2} |\mathscr{CB}_N(A)(x,y) - \mathscr{CB}_N(B)(x,y)| \\ &= \sup_{(x,y)\in\{\frac{1}{N},\dots,\frac{N-1}{N}\}^2} |A(x,y) - B(x,y)| \\ &\leq \sup_{(x,y)\in[0,1]^2} |A(x,y) - B(x,y)| \\ &= d_\infty(A,B). \end{aligned}$$

Considering the fact that for $A, B \in \mathscr{CB}_N$ we obviously have $\mathscr{CB}_N(A) = A$ and $\mathscr{CB}_N(B) = B$, obviously the inequality is best possible.

**Corollary 4.8.** *For arbitrary $A, B \in \mathscr{C}$ and their $N$-checkerboard approximations $\mathscr{CB}_N(A)$, $\mathscr{CB}_N(B) \in \mathscr{CB}$ the following inequality holds:*

$$D_1(\mathscr{CB}_N(A), \mathscr{CB}_N(B)) \leq 2\frac{(N-1)^2}{N} d_\infty(A,B). \tag{5}$$

We finally arrive at the main result saying that the empirical checkerboard is a strongly consistent estimator of the underlying copula w.r.t.t. metric $D_1$ - this result is key for finally proving consistency of $\zeta_1(\mathscr{CB}_N(A_n))$.

**Theorem 4.9.** *Let $(X_1,Y_1),(X_2,Y_2),\dots$ be a random sample from $(X,Y)$ and assume that $(X,Y)$ has continuous joint distribution function $H$ and copula $A$. Setting $N(n) := \lfloor n^s \rfloor$ for some $s \in (0, \frac{1}{2})$ the following identity holds with probability $1$:*

$$\lim_{n\to\infty} D_1(\mathscr{CB}_{N(n)}(A_n), A) = 0.$$

*Proof.* Applying Corollary 4.8 and the triangle inequality yields

$$\begin{aligned} D_1(\mathscr{CB}_{N(n)}(A_n), A) &\leq D_1(\mathscr{CB}_{N(n)}(A_n), \mathscr{CB}_{N(n)}(A)) + D_1(\mathscr{CB}_{N(n)}(A), A) \\ &\leq 2\frac{(N(n)-1)^2}{N(n)} d_\infty(A_n, A) + D_1(\mathscr{CB}_{N(n)}(A), A) \\ &\leq 2N(n)\, d_\infty(A_n, A) + D_1(\mathscr{CB}_{N(n)}(A), A) \end{aligned}$$

According to Theorem 3.4 there exists a $\Lambda \in \mathscr{A}$ with $\mathbf{P}(\Lambda) = 1$ such that for every $\omega \in \Lambda$ we can find a constant $C(\omega) > 0$ and an index $n_0 = n_0(\omega) \in \mathbf{N}$ such that for all $n \geq n_0$

$$d_\infty(A_n(\omega), A) \leq C(\omega)\sqrt{\frac{\log(\log(n))}{n}}$$

holds. Let $\varepsilon > 0$ be fixed. Theorem 4.3 implies the existence of an index $n_1 \in \mathbf{N}$ fulfilling

$$D_1(\mathscr{CB}_{N(n)}(A), A) < \frac{\varepsilon}{2}$$

for all $n \geq n_1$. For every $\omega \in \Lambda$ and $n \geq \max\{n_0, n_1\}$ we finally obtain

$$\begin{aligned} D_1(\mathscr{CB}_{N(n)}(A_n(\omega)), A) &\leq 2N(n)C(\omega) \cdot \sqrt{\frac{\log(\log(n))}{n}} + D_1(\mathscr{CB}_{N(n)}(A), A) \\ &< 2C(\omega)\lfloor n^s \rfloor n^{-\frac{1}{2}} \sqrt{\log(\log(n))} + \frac{\varepsilon}{2} \end{aligned}$$

from which the result follows since $0 < s < \frac{1}{2}$ by assumption.

The simulations in the next section insinuate that $\mathscr{CB}_{N(n)}(A_n)$ might also be a strongly consistent estimator for more flexible choices of $N(n)$, particularly for the case $N(n) := \lfloor n^s \rfloor$ and some $s \geq \frac{1}{2}$ - a clarification of this question is future work, in the R-package we considered $s = \frac{1}{2}$. We conclude this section with the main result on the estimator $\zeta_1(\mathscr{CB}_{N(n)}(A_n))$ of $\zeta_1(A)$ - this very result is the main reason for using the qad-package.

**Theorem 4.10** *Let $(X_1, Y_1), (X_2, Y_2), \ldots$ be a random sample from $(X, Y)$ and assume that $(X, Y)$ has continuous joint distribution function $H$ and copula $A$. Fix $s \in (0, \frac{1}{2})$ and set $N(n) := \lfloor n^s \rfloor$ for every $n \in \mathbf{N}$. Then with probability 1 we have*

$$\lim_{n \to \infty} \zeta_1(\mathscr{CB}_{N(n)}(A_n)) = \zeta_1(A).$$

*Proof.* Direct consequence of Theorem 4.9 and the fact that

$$|D_1(\mathscr{CB}_{N(n)}(A_n), \Pi) - D_1(A, \Pi)| \leq D_1(\mathscr{CB}_{N(n)}(A_n), A).$$

## 5 Simulations

If not specified differently throughout this section we consider $s_0 = \frac{1}{2}$ and set $N(n) = \lfloor n^{s_0} \rfloor$. In order to illustrate the performance of the estimator $\zeta_1(\mathscr{CB}_{N(n)}(A_n))$ we consider Marshall-Olkin and FGM copulas as well as a completely dependent copula $A_h$ fulfilling that $\zeta_1(A_h)$ is much bigger than $\zeta_1(A_h^t)$.

**Example 5.1.** [Marshall Olkin family] The Marshall Olkin (MO) family of copulas $(M_{\alpha,\beta})_{\alpha,\beta\in[0,1]}$ is defined by

$$M_{\alpha,\beta}(x,y) := \begin{cases} x^{1-\alpha}y & x^\alpha \geq y^\beta \\ xy^{1-\beta} & x^\alpha < y^\beta. \end{cases}$$

It contains $\Pi$ ($\alpha = 0$ or $\beta = 0$) as well as $M$ ($\alpha = \beta = 1$). It was shown in (1), that in case of $\alpha, \beta > 0$

$$\zeta_1(M_{\alpha,\beta}) = 3\alpha(1-\alpha)^z + \frac{6}{\beta}\frac{1-(1-\alpha)^z}{z} - \frac{6}{\beta}\frac{1-(1-\alpha)^{z+1}}{z+1},$$

whereby $z = \frac{1}{\alpha} + \frac{2}{\beta} - 1$.

To test the performance of qad for different dependence structures we consider MO copulas with parameter $(\alpha,\beta) \in \{(1,0),(1,1),(0.3,1),(1,0.7),(0.5,0.5)\}$, see Fig. S1_7, Fig. S1_9, Fig. S1_11, Fig. S1_13 and Fig. S1_15.

We generated samples of size $n \in \{10, 50, 100, 500, 1.000, 5.000, 10.000\}$, calculated the empirical checkerboard copula $A_n$ as well as $\zeta_1(\mathscr{CB}_{N(n)}(A_n))$. These steps were repeated $R = 500$ times and the obtained results are depicted as boxplots in Fig. S1_8, Fig. S1_10, Fig. S1_12, Fig. S1_14 and Fig. S1_16.

**Example 5.2.** [FGM family] The Farlie-Gumbel-Morgenstern family $(G_\theta)_{\theta\in[-1,1]}$ is defined by

$$G_\theta(x,y) := xy + \theta xy(1-x)(1-y).$$

According to (1) $\zeta_1(G_\theta)$ is given by

$$\zeta_1(G_\theta) = \frac{|\theta|}{4}.$$

Replacing the Marshall-Olkin copula by a Farlie-Gumbel-Morgenstern copula and proceeding analogously as before yields in the special case with $\theta \in \{-1, -0.5\}$ the results, depicted in Fig. S1_17, Fig. S1_18, Fig. S1_19 and Fig. S1_20.

**Example 5.3.** [Highly asymmetric completely dependent copulas] To test the performance of qad in a highly asymmetric situation we consider the completely dependent copula $A_{h_a}$ for $h = ax(mod 1)$ and $a = 5, 10, 50$, see Fig. S1_21, Fig. S1_23 and Fig. S1_25. Notice that in these cases we have $\zeta_1(A_{h_a}) = 1$ whereas $\zeta_1(A^t_{h_a})$ is (particularly for large $a$) close to 0. For fixed $a \in \{5, 10, 50\}$ we generated samples of size $n \in \{10, 50, 100, 500, 1.000, 5.000, 10.000\}$, calculated the empirical checkerboard copula $A_n$ as well as $\zeta_1(\mathscr{CB}_{N(n)}(A_n))$. These steps were repeated $R = 1.000$ times. The obtained results are depicted as boxplots in Fig. S1_22, Fig. S1_24 and Fig. S1_26 respectively. As expected, the parameter $a$ has a big influence on the precision of the estimator, the bigger $a$ the longer it takes qad to detect the asymmetry. Notice that, contrary to qad, classical dependence measures like Schweizer and Wolff's $\sigma$ (11) are not capable of detecting the asymmetry since both $d_\infty(A_{h_a}, \Pi)$ and $d_\infty(A_{h_a}, \Pi)$ are very small.

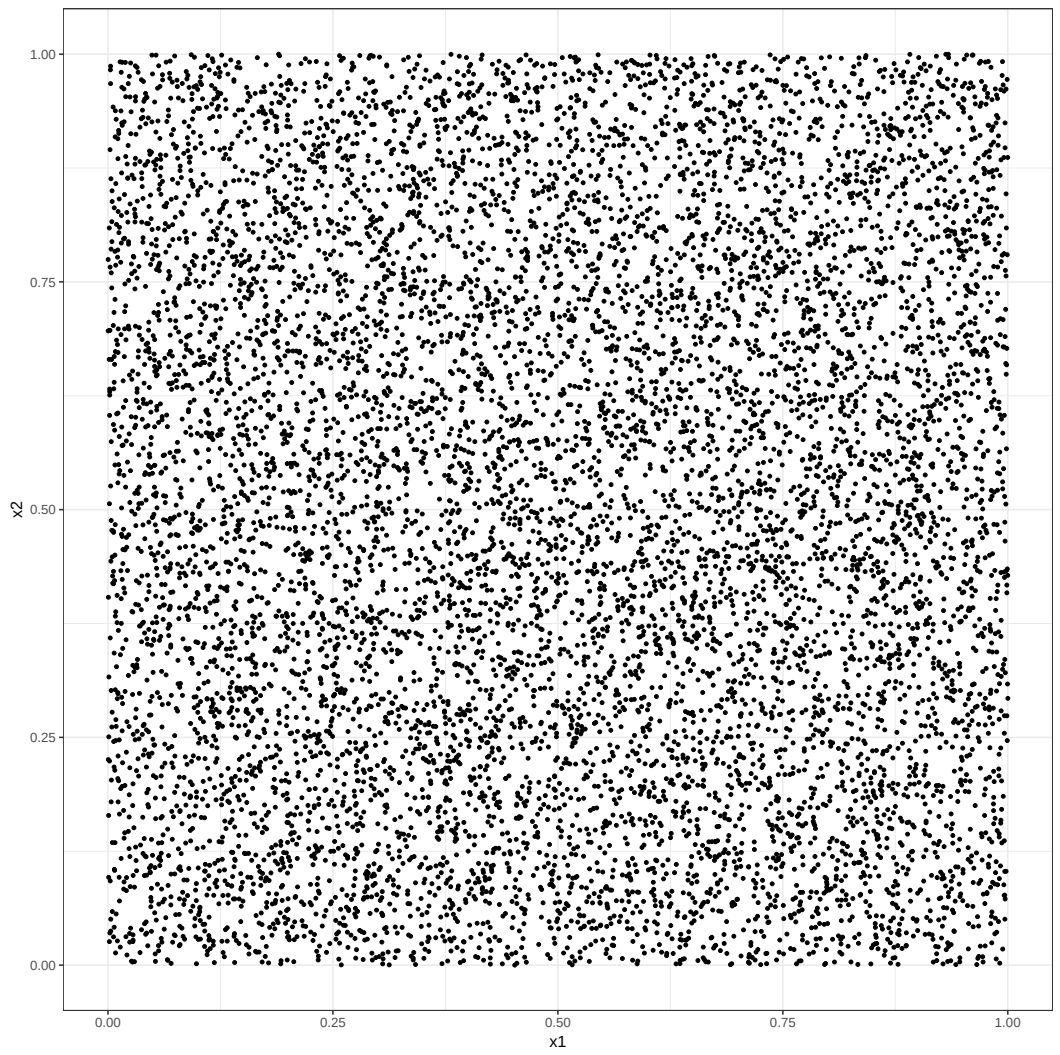

Fig. S1_7: Sample of size $10.000$ from a MO copula with parameter $\alpha = 1$ and $\beta = 0$.

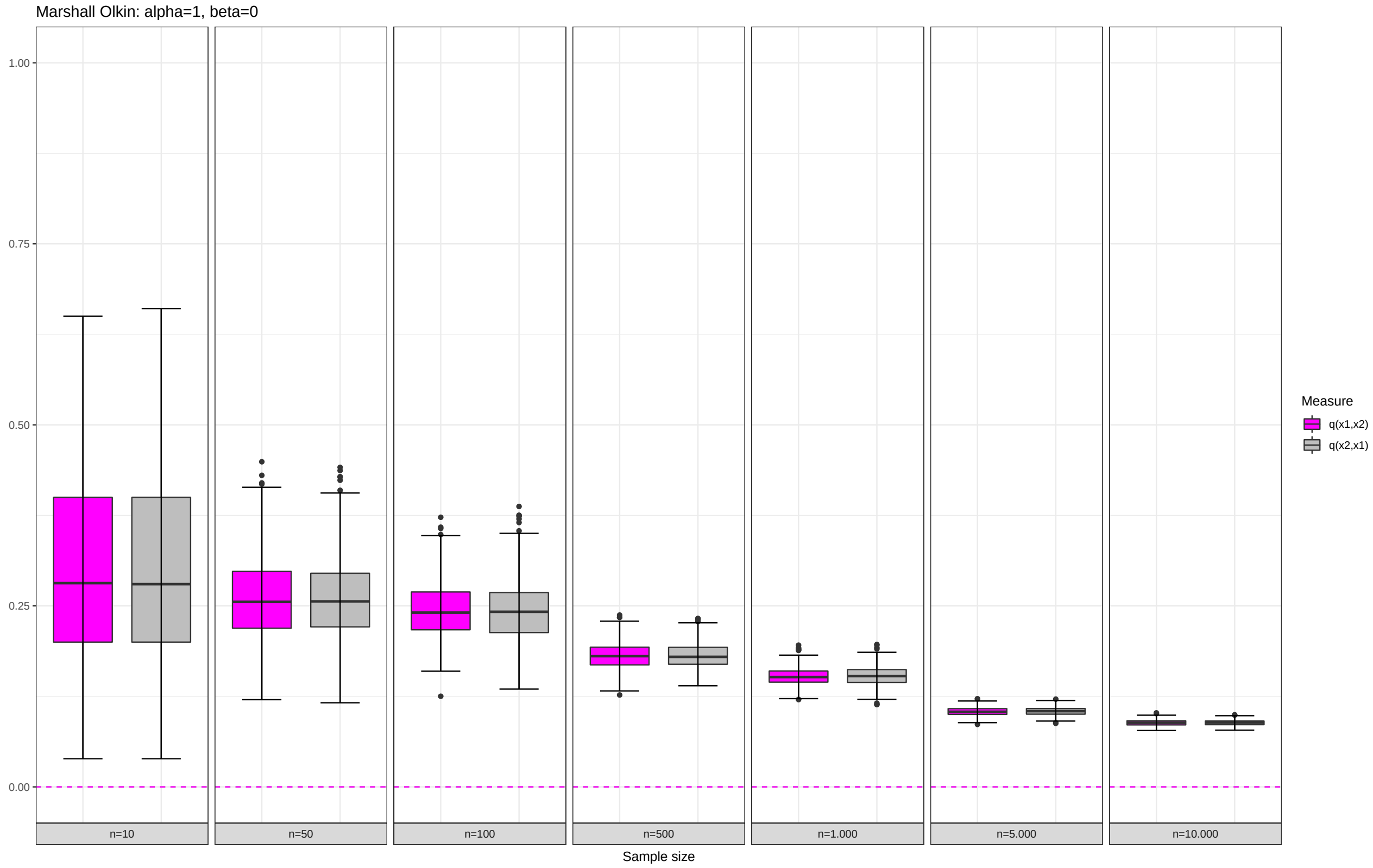

Fig. S1_8: Boxplots summarizing the 500 obtained estimates for $\zeta_1(M_{\alpha,\beta})$ (magenta) and $\zeta_1(M^t_{\alpha,\beta})$ (gray). The dashed lines depict the true dependence measure $\zeta_1(M_{\alpha,\beta})$ and $\zeta_1(M^t_{\alpha,\beta})$ for $\alpha = 1$ and $\beta = 0$.

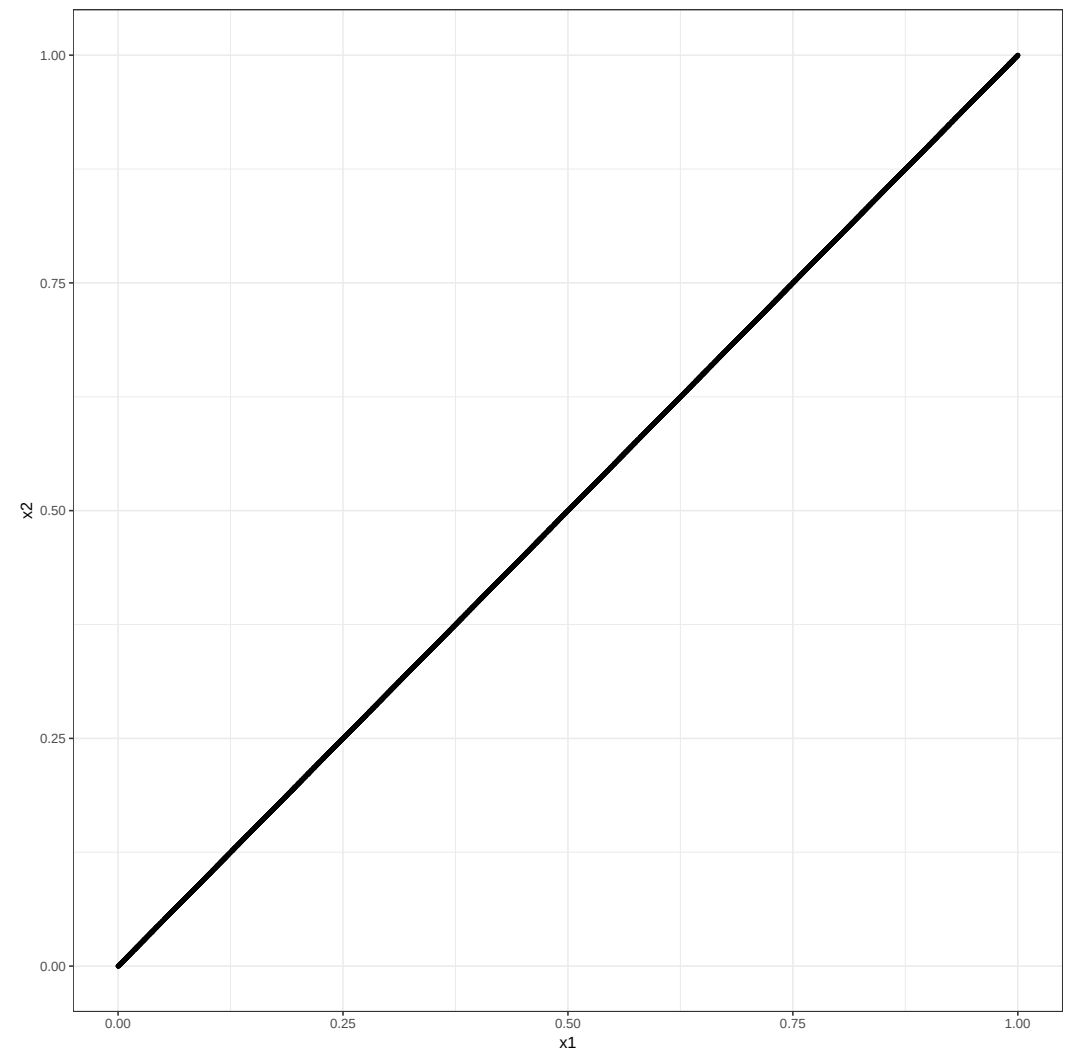

Fig. S1_9: Sample of size $10.000$ from a MO copula with parameter $\alpha = 1$ and $\beta = 1$.

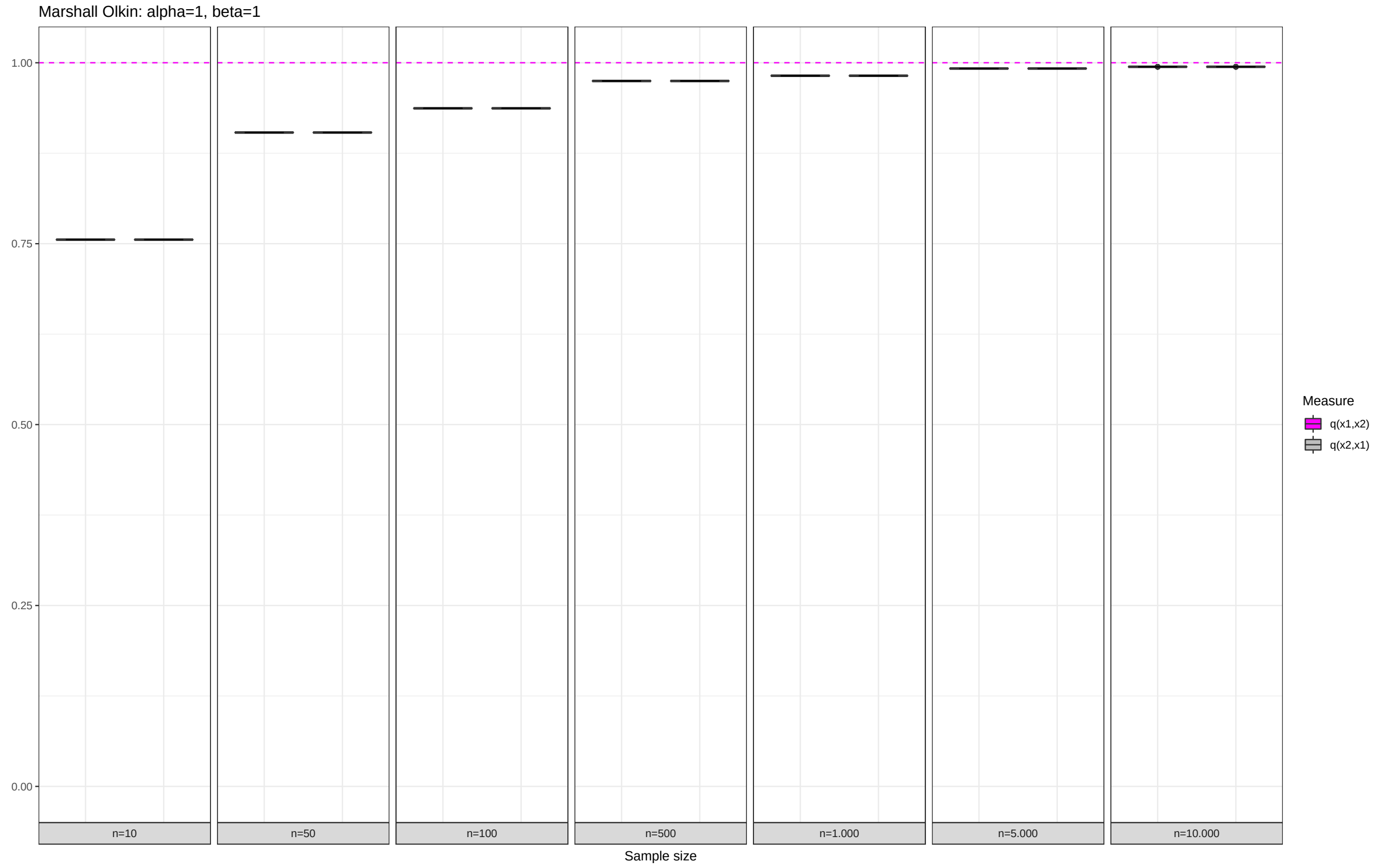

Fig. S1_10: Boxplots summarizing the 500 obtained estimates for $\zeta_1(M_{\alpha,\beta})$ (magenta) and $\zeta_1(M^t_{\alpha,\beta})$ (gray). The dashed lines depict the true dependence measure $\zeta_1(M_{\alpha,\beta})$ and $\zeta_1(M^t_{\alpha,\beta})$ for $\alpha = 1$ and $\beta = 1$.

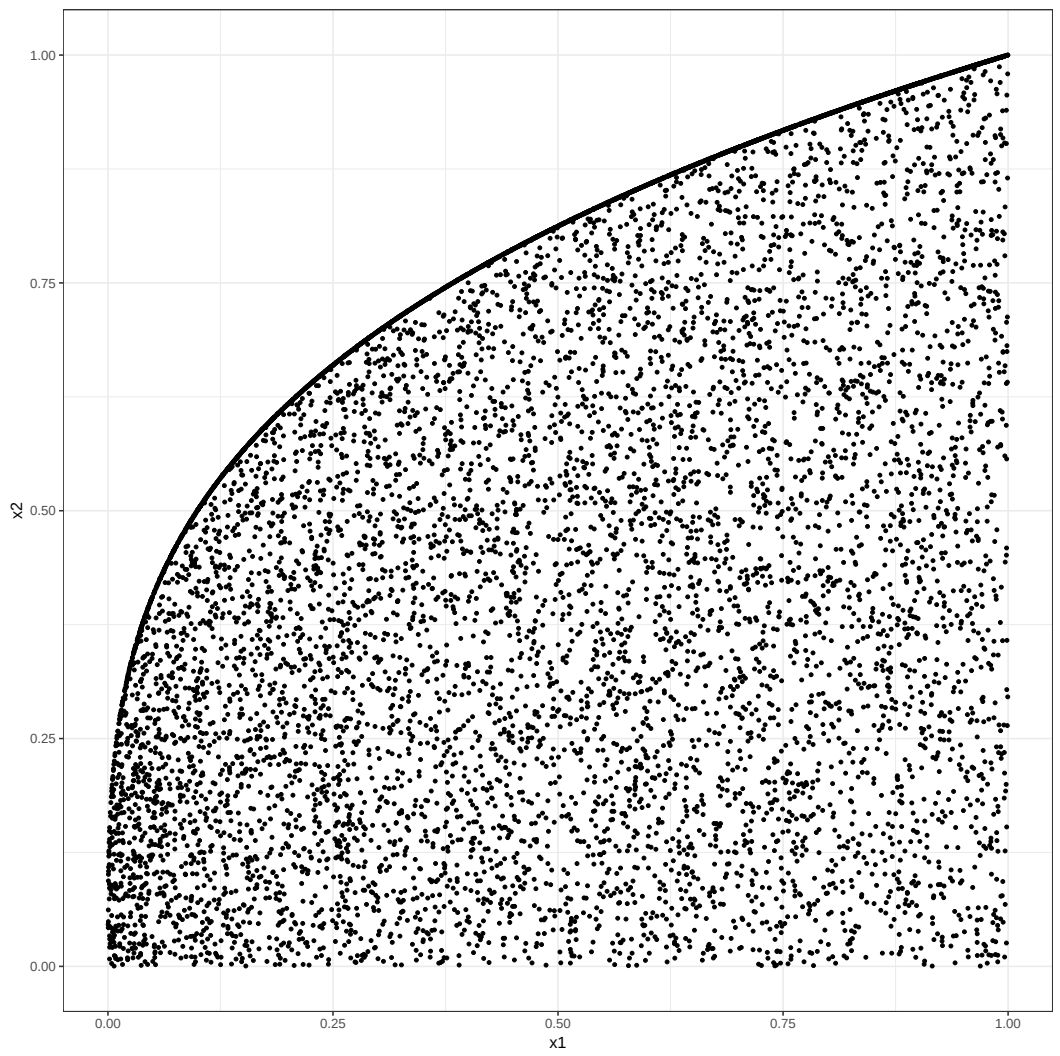

Fig. S1_11: Sample of size $10.000$ from a MO copula with parameter $\alpha = 0.3$ and $\beta = 1$.

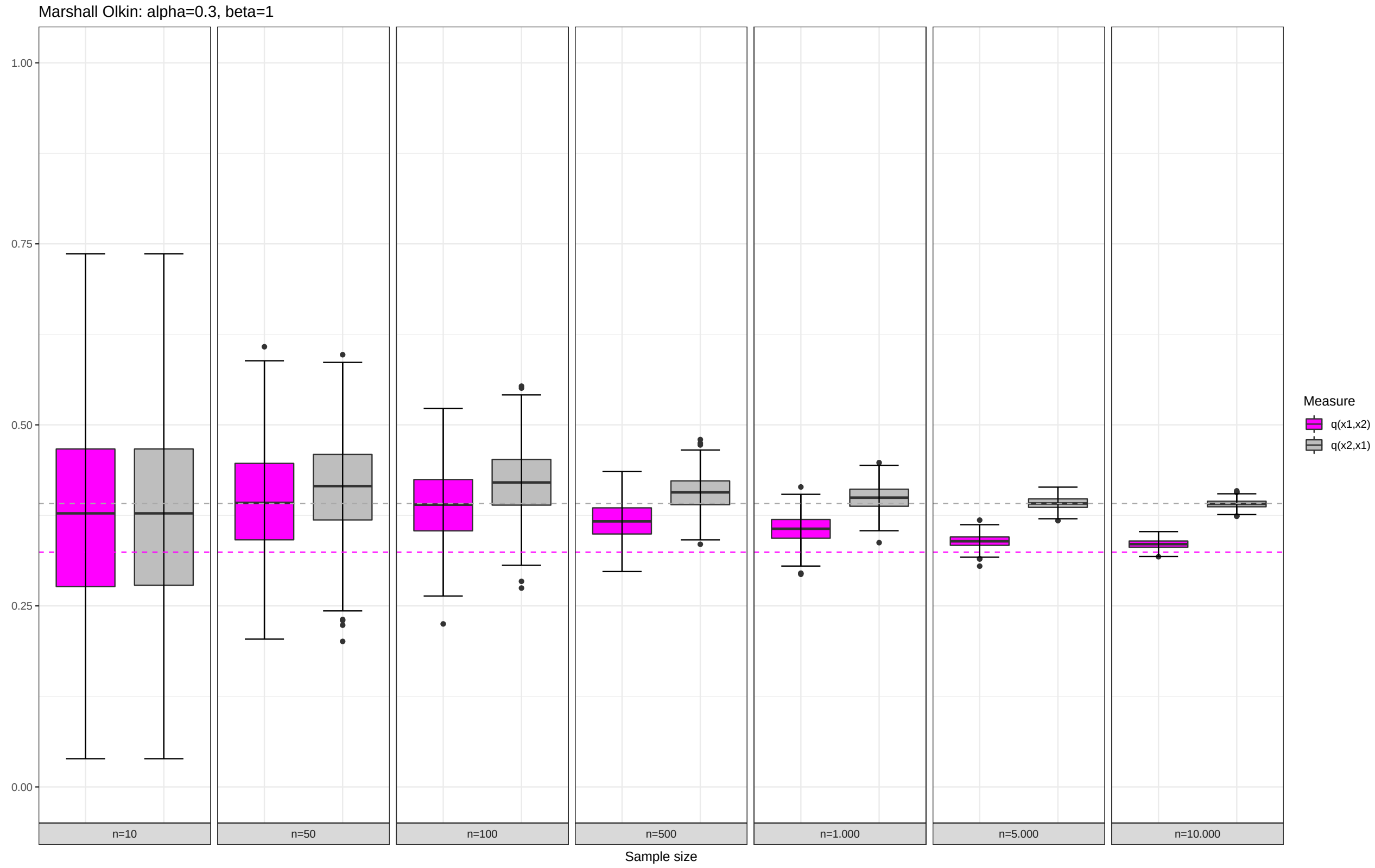

Fig. S1_12: Boxplots summarizing the 500 obtained estimates for $\zeta_1(M_{\alpha,\beta})$ (magenta) and $\zeta_1(M^t_{\alpha,\beta})$ (gray). The dashed lines depict the true dependence measure $\zeta_1(M_{\alpha,\beta})$ and $\zeta_1(M^t_{\alpha,\beta})$ for $\alpha = 0.3$ and $\beta = 1$.

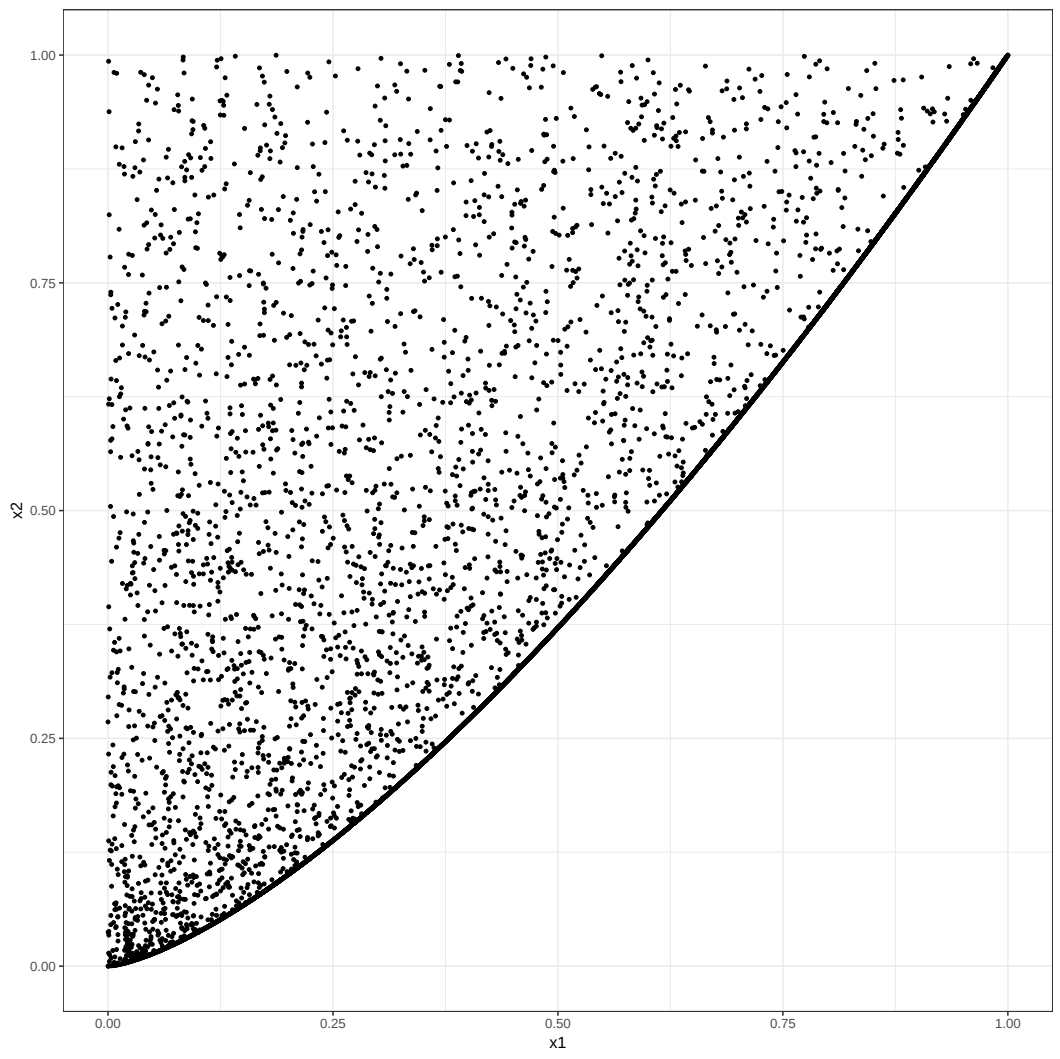

Fig. S1_13: Sample of size $10.000$ from a MO copula with parameter $\alpha = 1$ and $\beta = 0.7$.

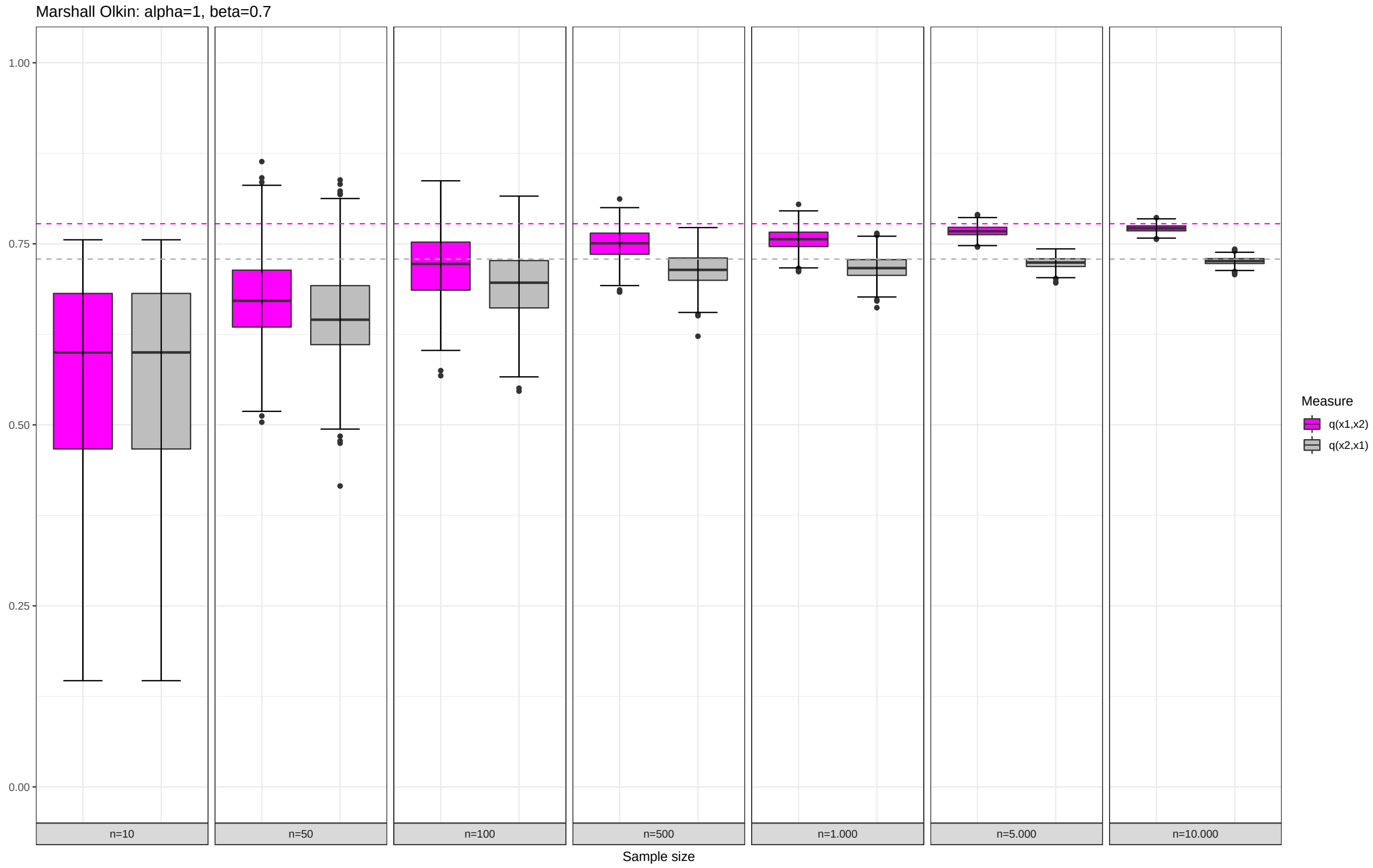

Fig. S1_14: Boxplots summarizing the 500 obtained estimates for $\zeta_1(M_{\alpha,\beta})$ (magenta) and $\zeta_1(M^t_{\alpha,\beta})$ (gray). The dashed lines depict the true dependence measure $\zeta_1(M_{\alpha,\beta})$ and $\zeta_1(M^t_{\alpha,\beta})$ for $\alpha = 1$ and $\beta = 0.7$.

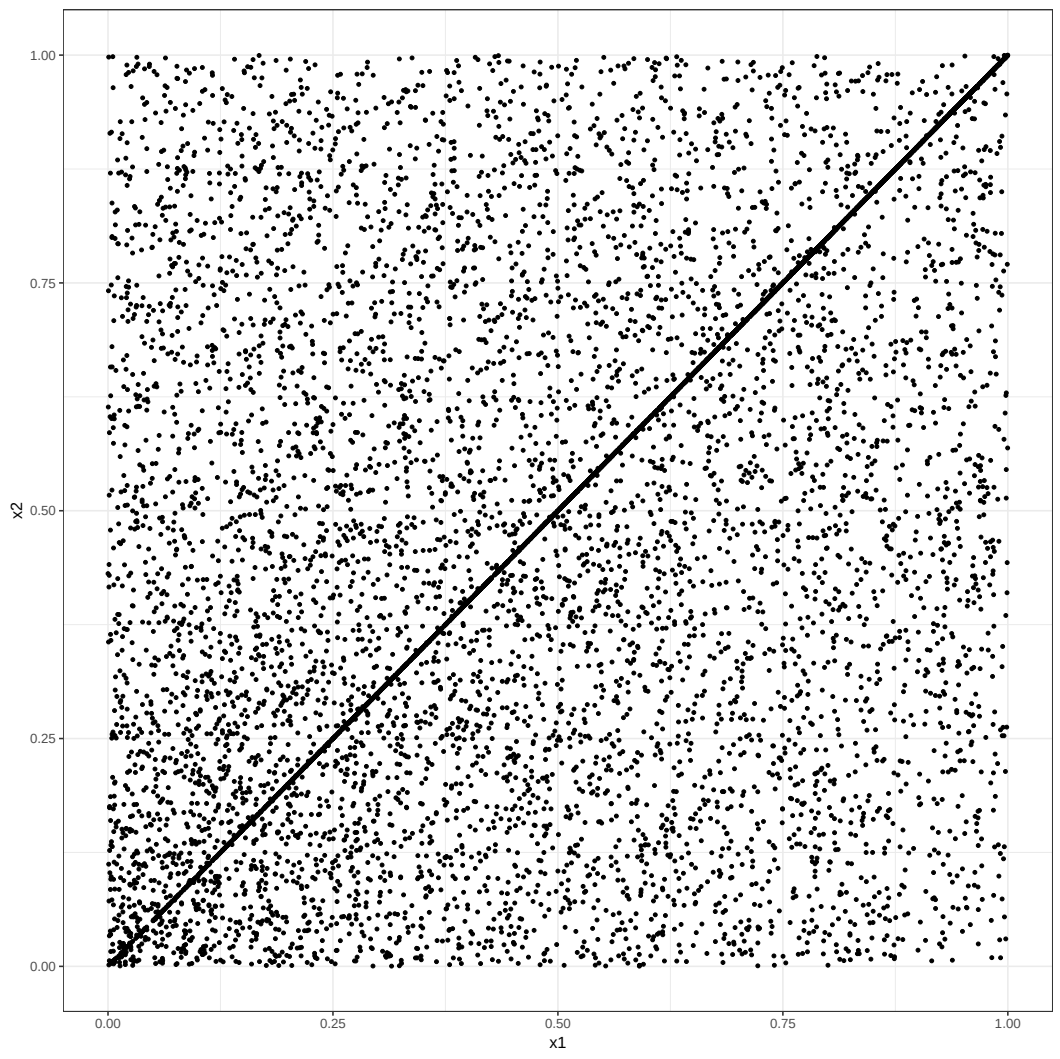

Fig. S1_15: Sample of size $10.000$ from a MO copula with parameter $\alpha = 0.5$ and $\beta = 0.5$.

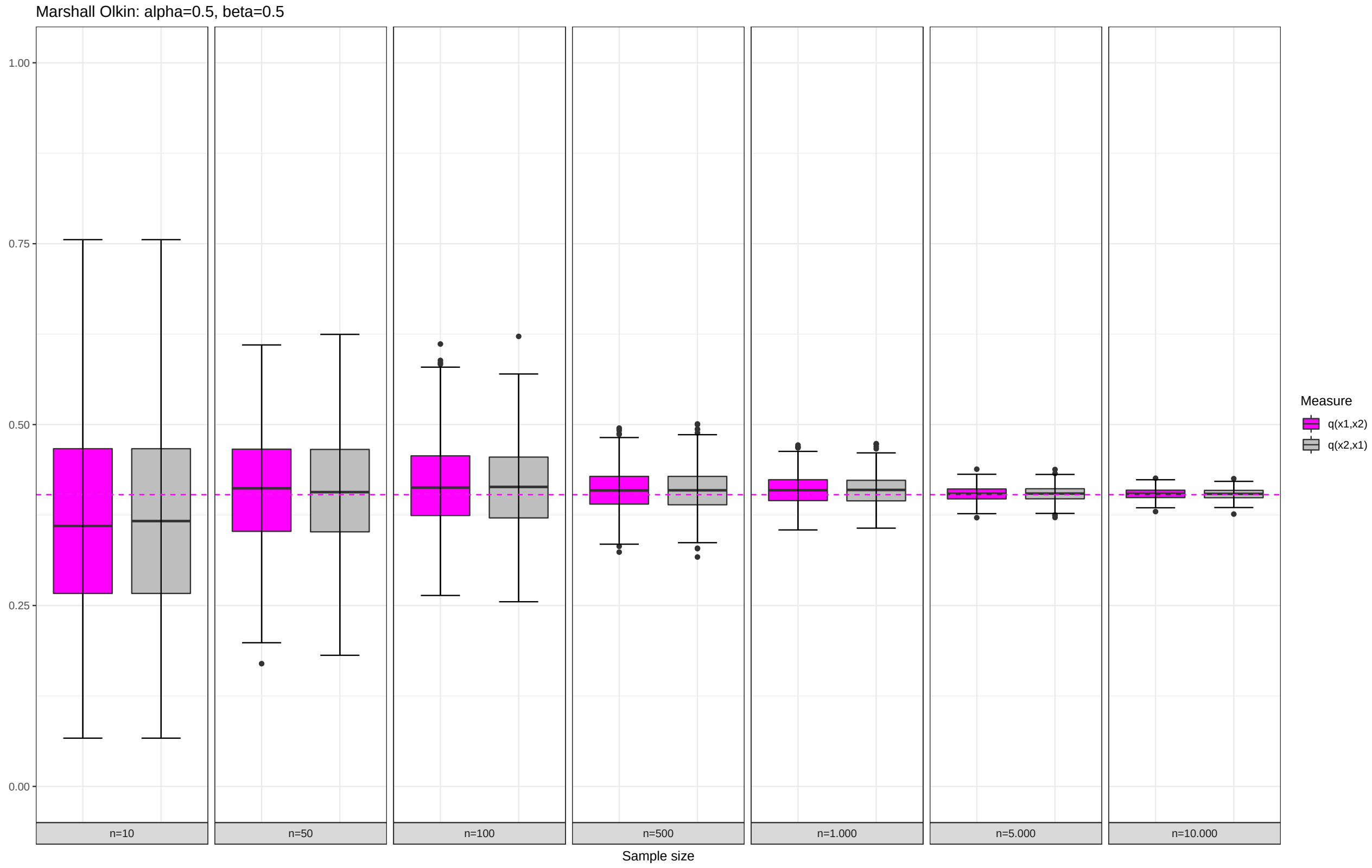

Fig. S1_16: Boxplots summarizing the 500 obtained estimates for $\zeta_1(M_{\alpha,\beta})$ (magenta) and $\zeta_1(M^t_{\alpha,\beta})$ (gray). The dashed lines depict the true dependence measure $\zeta_1(M_{\alpha,\beta})$ and $\zeta_1(M^t_{\alpha,\beta})$ for $\alpha = 0.5$ and $\beta = 0.5$.

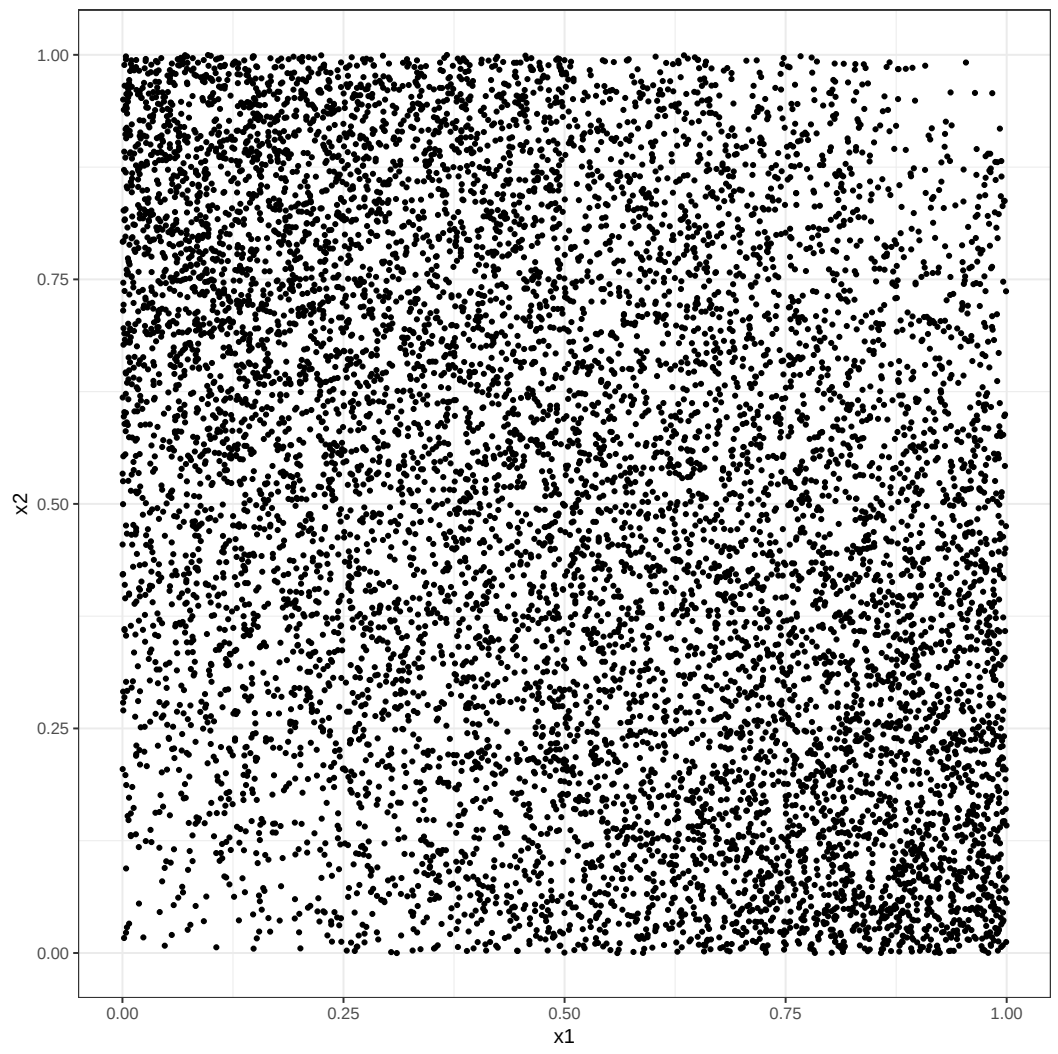

Fig. S1_17: Sample of size $10.000$ from a FGM copula with parameter $\theta = -1$.

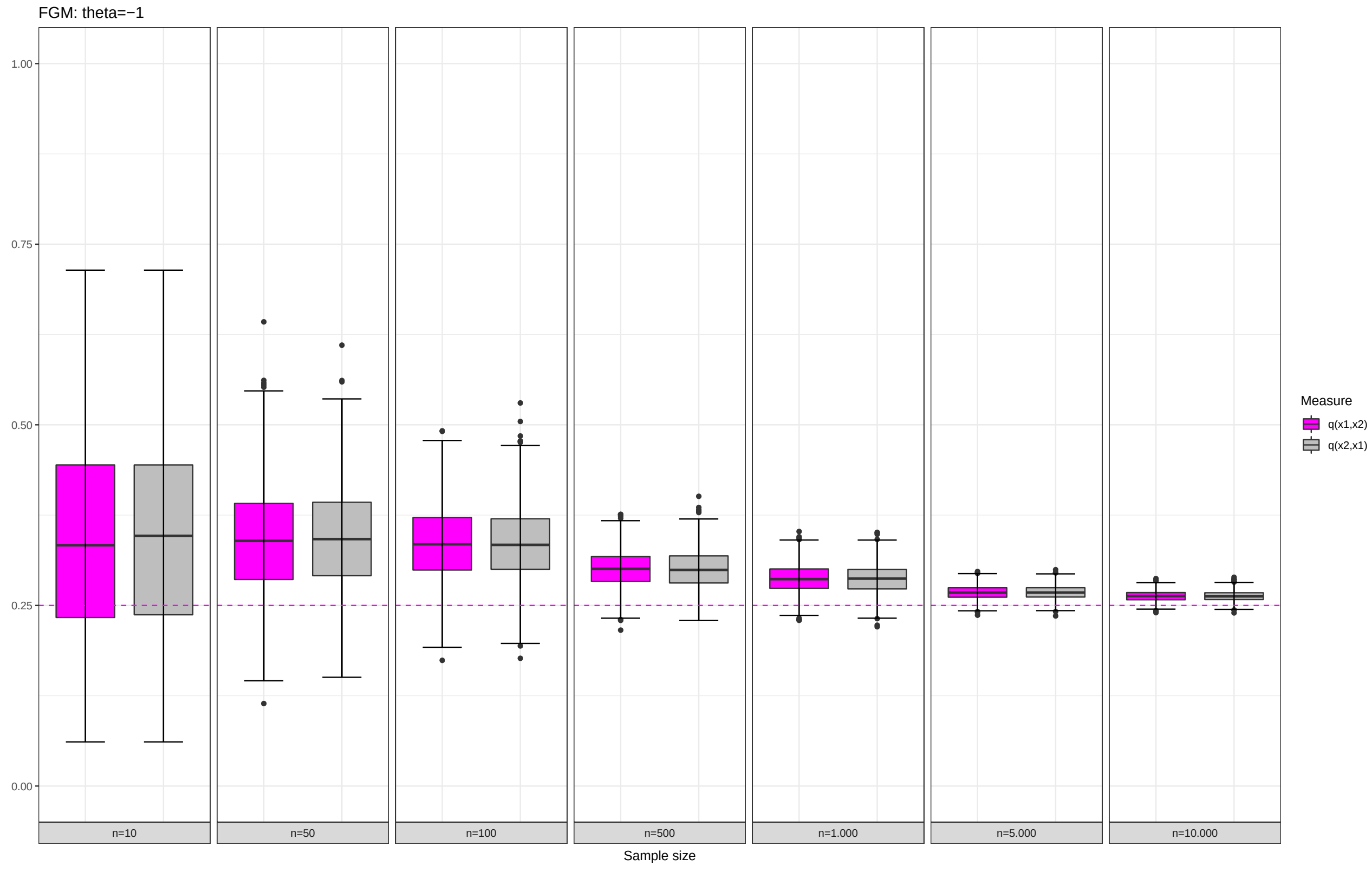

Fig. S1_18: Boxplots summarizing the 500 obtained estimates for $\zeta_1(G_\theta)$ (magenta) and $\zeta_1(G^t_\theta)$ (gray). The dashed lines depict the true dependence measure $\zeta_1(G_\theta)$ and $\zeta_1(G^t_\theta)$ for $\theta = -1$.

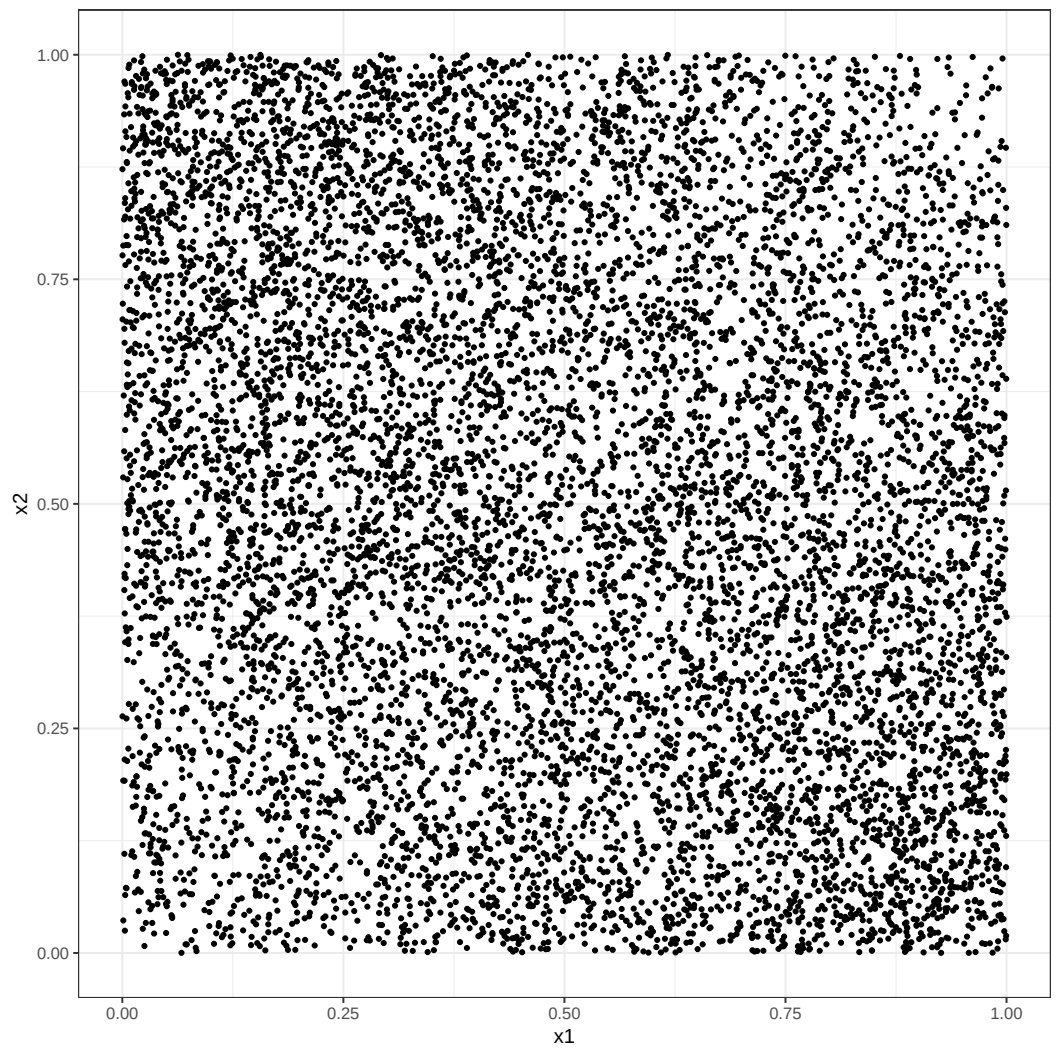

Fig. S1_19: Sample of size $10.000$ from a FGM copula with parameter $\theta = -0.5$.

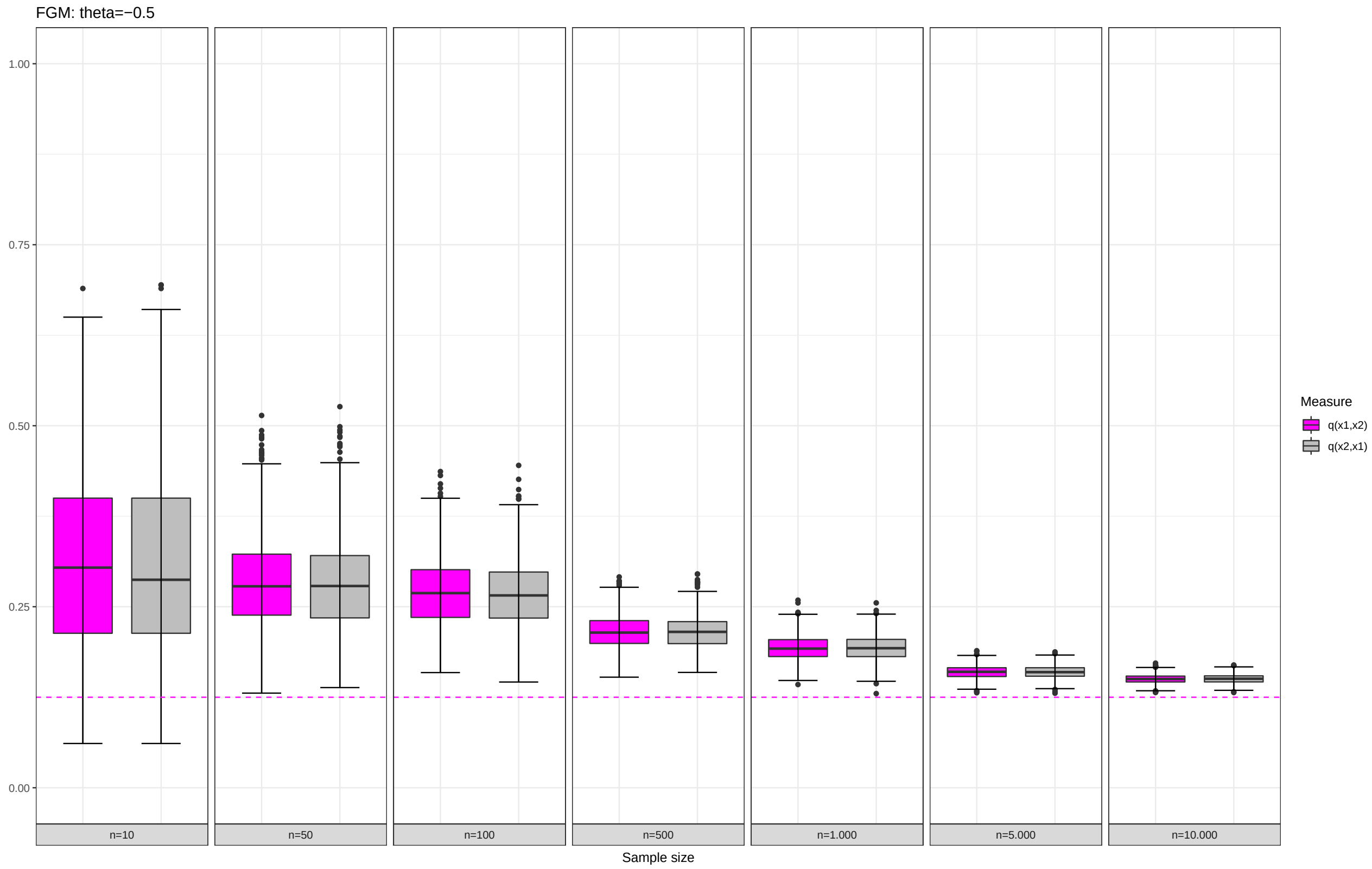

Fig. S1_20: Boxplots summarizing the 500 obtained estimates for $\zeta_1(G_\theta)$ (magenta) and $\zeta_1(G_\theta^t)$ (gray). The dashed lines depict the true dependence measure $\zeta_1(G_\theta)$ and $\zeta_1(G_\theta^t)$ for $\theta = -0.5$.

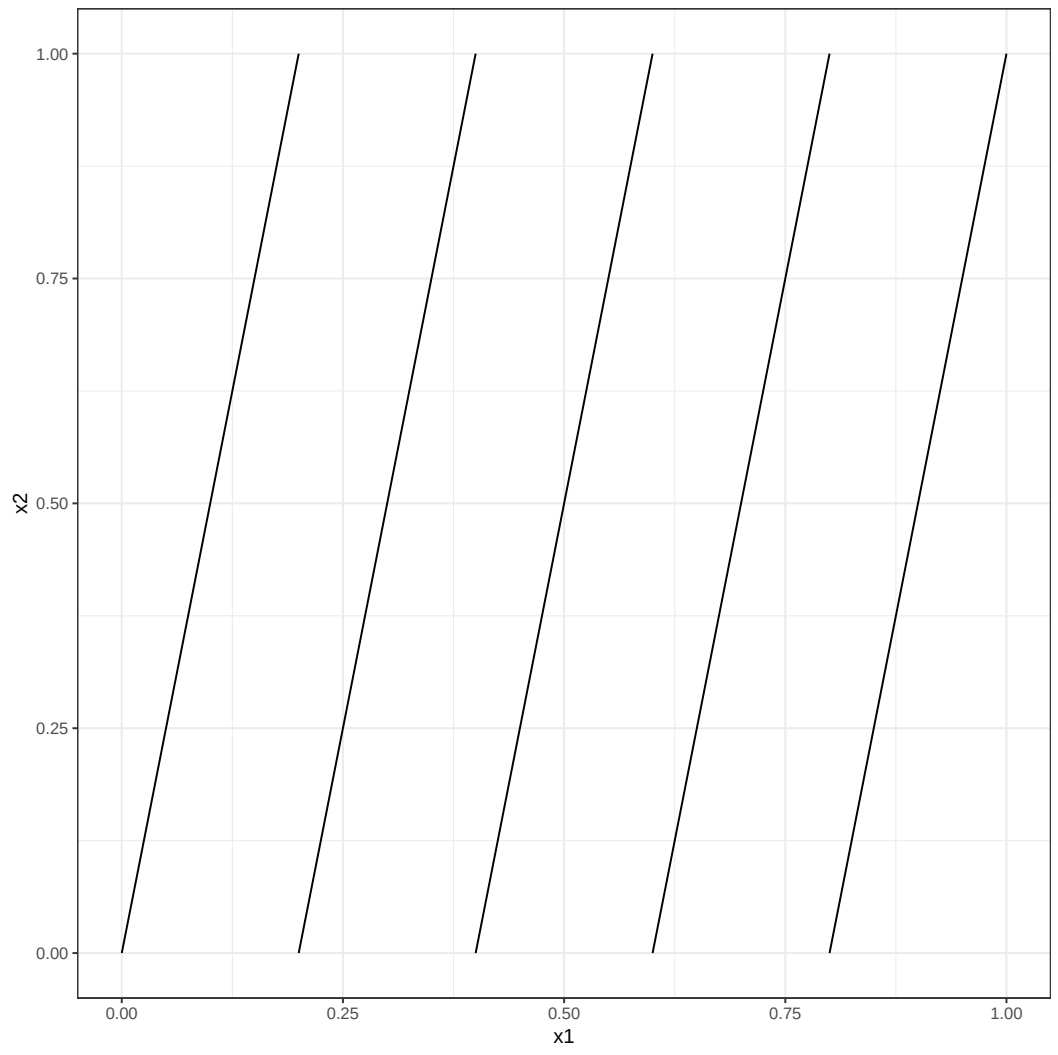

Fig. S1_21: Support of $A_{h_a}$ for $a = 5$.

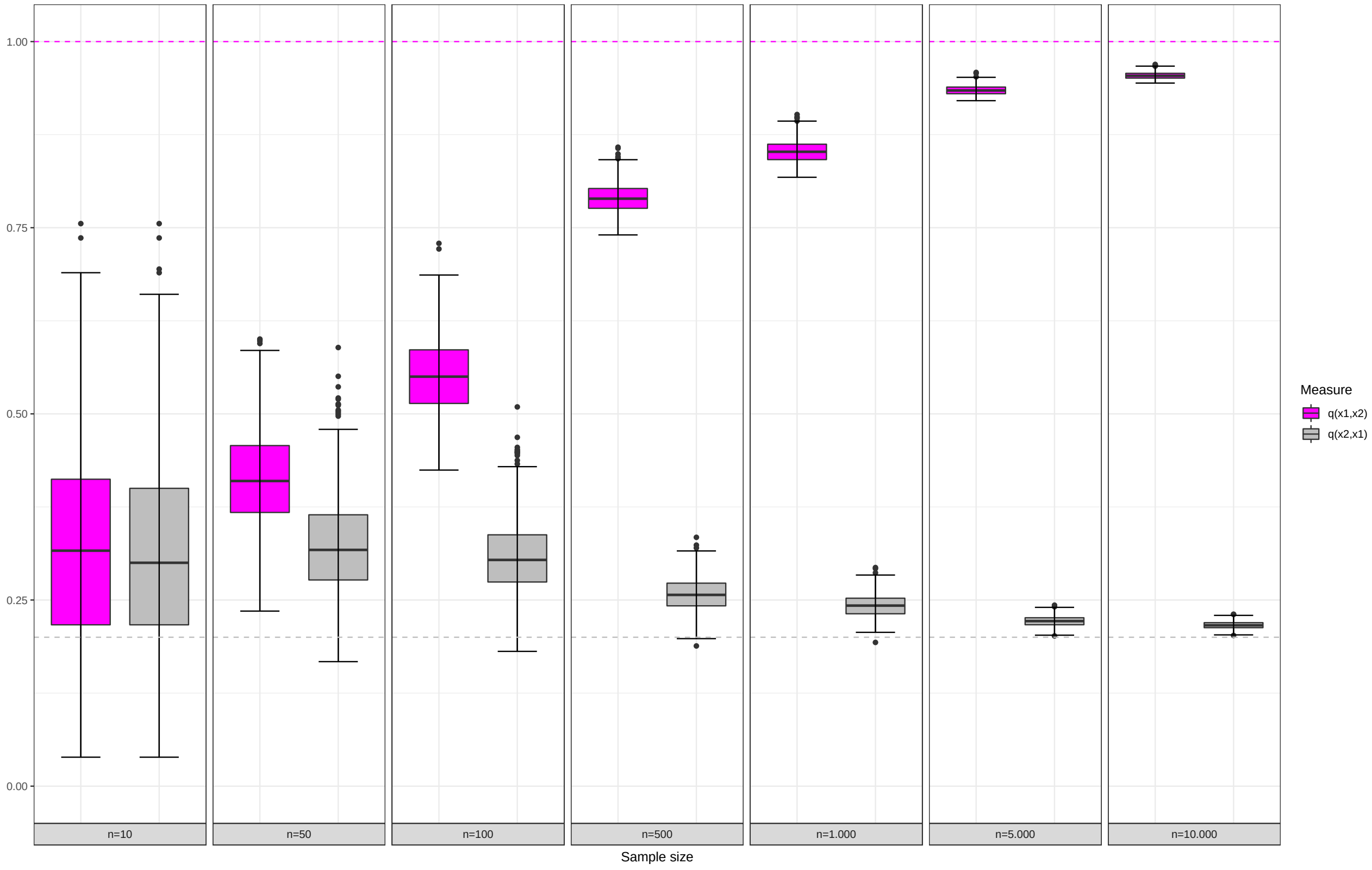

Fig. S1_22: Boxplots summarizing the 1000 obtained estimates for $\zeta_1(A_{h_a})$ (magenta) and $\zeta_1(A^t_{h_a})$ (gray) for the case $a = 5$.

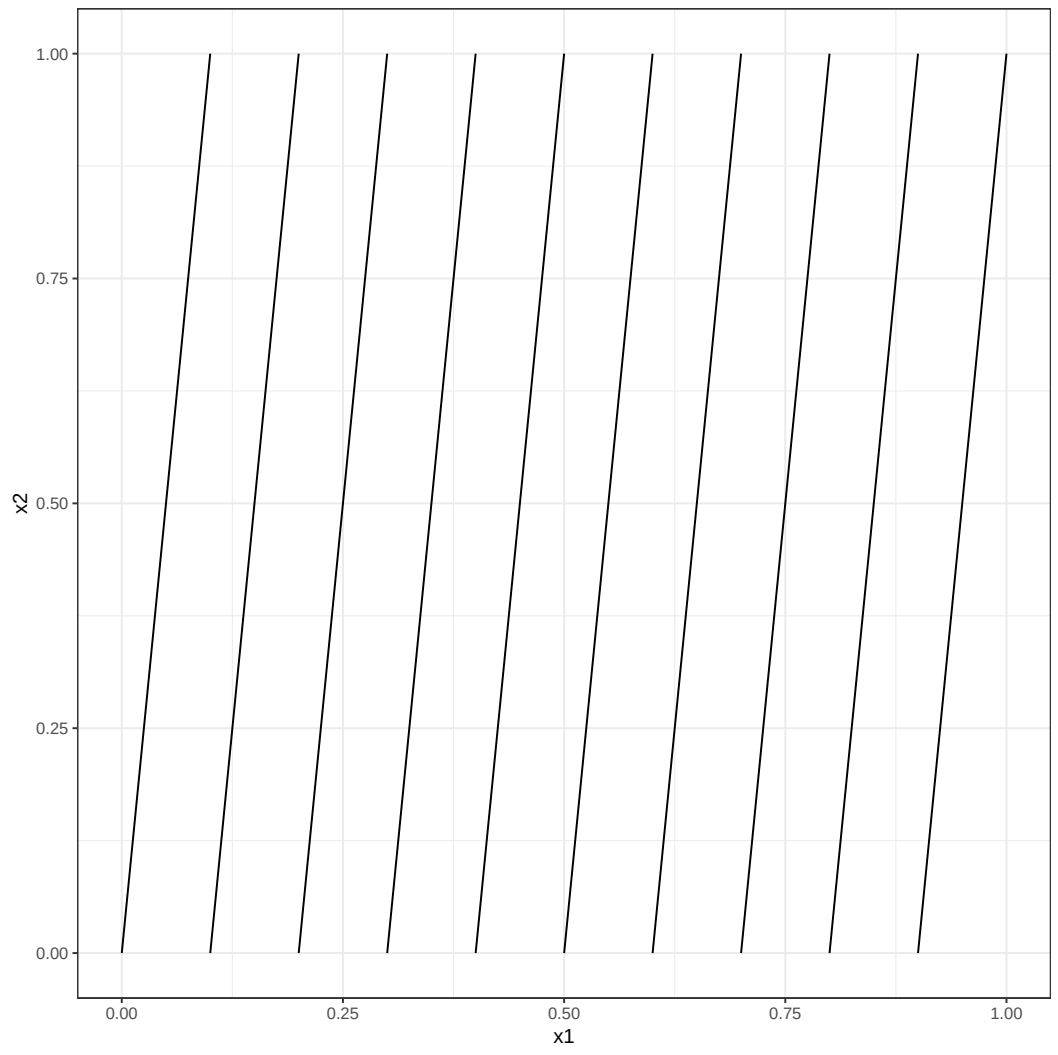

Fig. S1_23: Support of $A_{h_a}$ for $a = 10$.

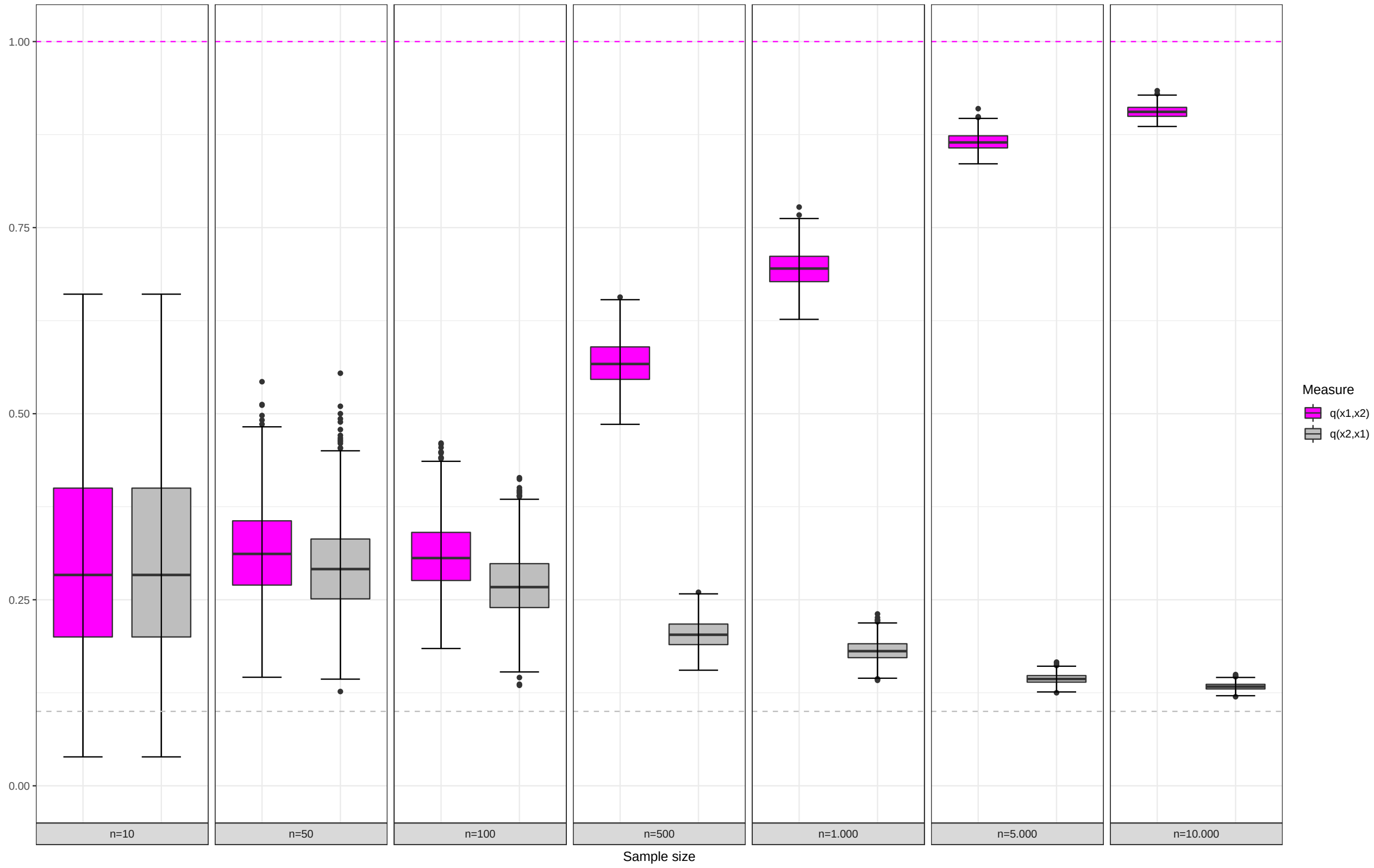

Fig. S1_24: Boxplots summarizing the 1000 obtained estimates for $\zeta_1(A_{h_a})$ (magenta) and $\zeta_1(A^t_{h_a})$ (gray) for the case $a = 10$.

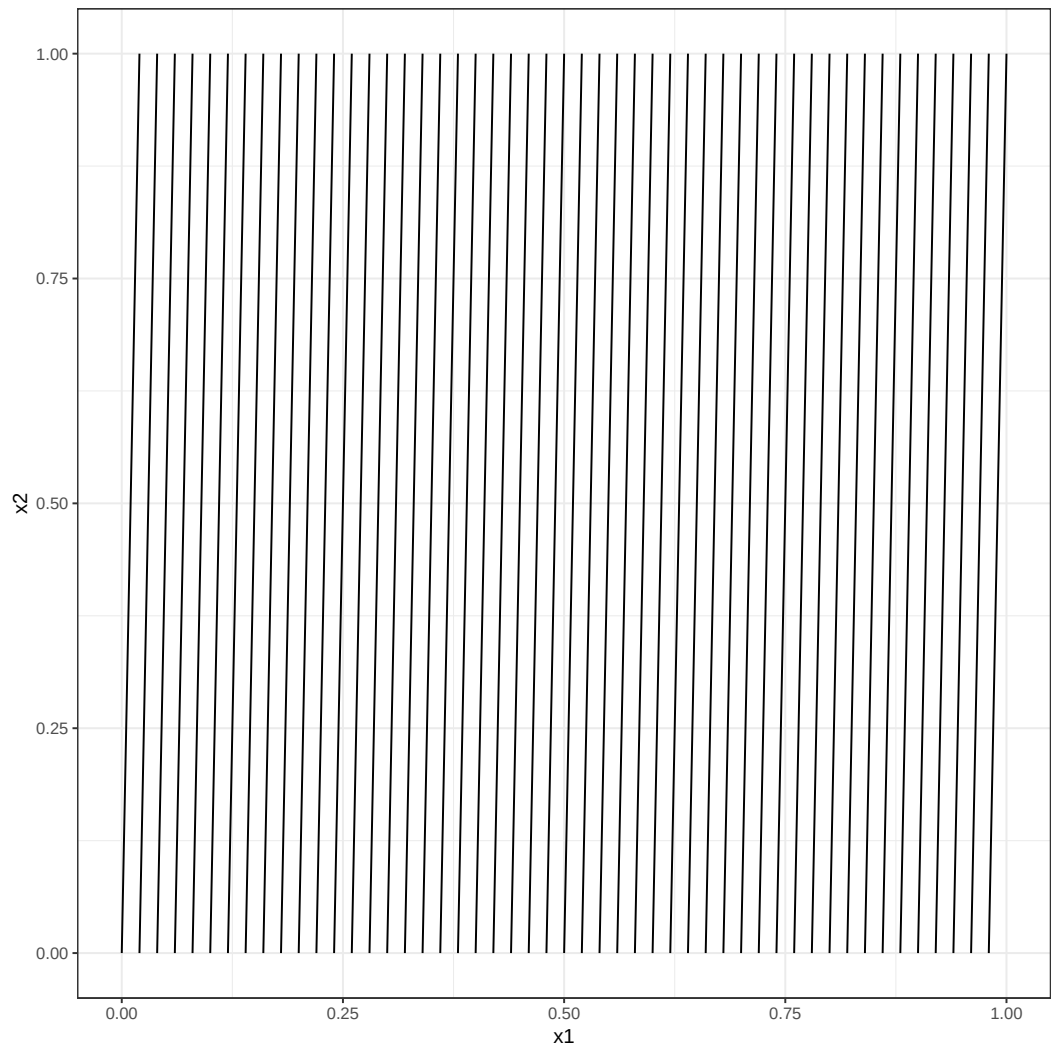

Fig. S1_25: Support of $A_{h_a}$ for $a = 50$.

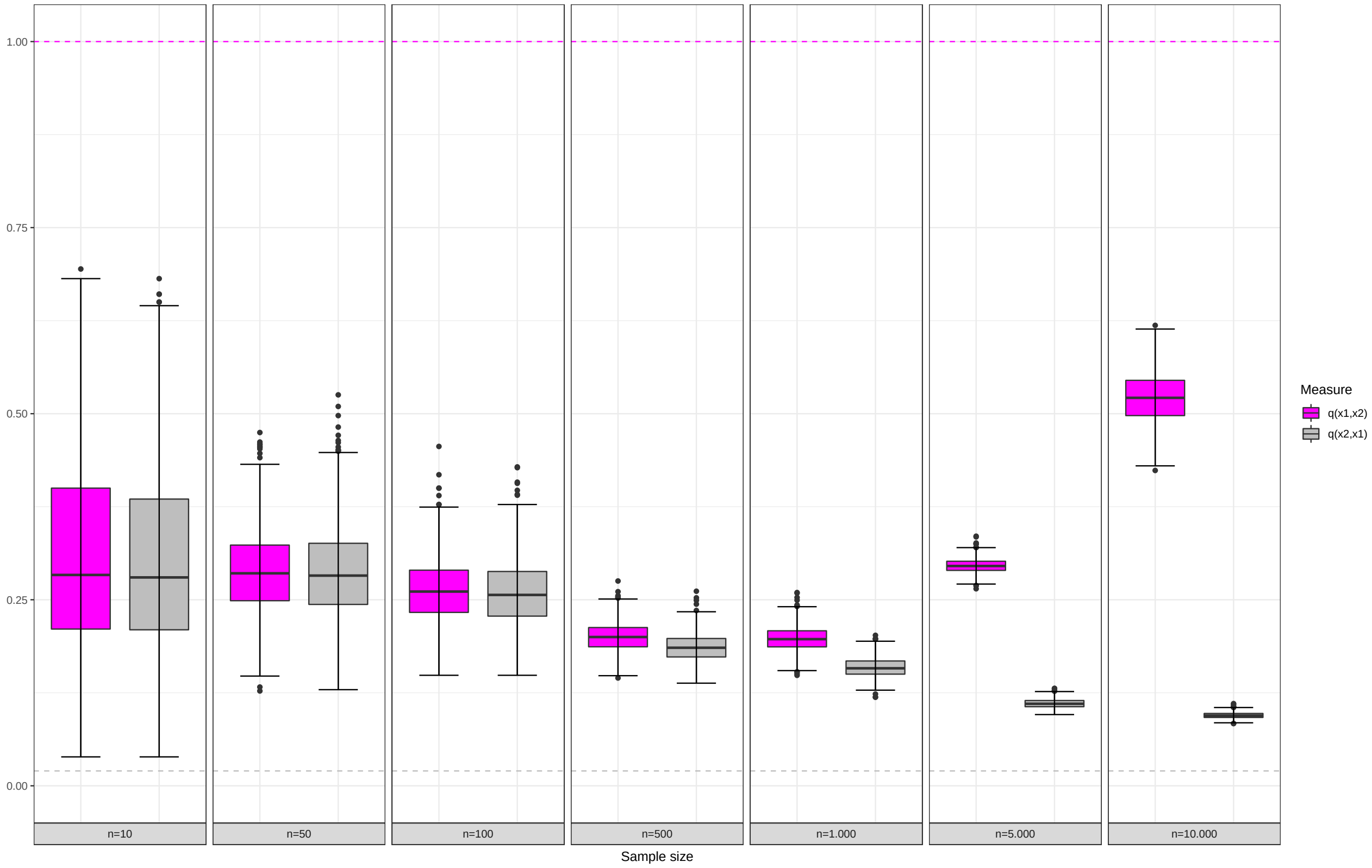

Fig. S1_26: Boxplots summarizing the 1000 obtained estimates for $\zeta_1(A_{h_a})$ (magenta) and $\zeta_1(A^t_{h_a})$ (gray) for the case $a = 50$.

**A copula-based measure for quantifying asymmetry in dependence and associations**

Robert R. Junker[1]*, Florian Griessenberger[2], Wolfgang Trutschnig[2]

## Supporting information 2

### Parabola [Fig. 1]

We generated *n* data points from a quadratic function with predefined noise ($y = x^2$ + noise; Supplementary material 4 – Fig. 1). Noise is generated by drawing random numbers from a uniform distribution on the interval $[-a, a]$. Original data and pseudo-observations (normalized ranks) were plotted. Based on pseudo-observations, the empirical checkerboard copula was calculated using the function *qad()* provided in the R-package *qad*, which also serves as a tool to predict Y given X and *vice versa*.

### Dependence between two random variables [Fig. 2]

In order to demonstrate the properties of *qad* and to compare it to other methods quantifying dependence, we generated *n* data points from nine different dependence structures with increasing vertical noises (Supplementary material 4 – Fig. 2) and quantified dependence using *qad*, Pearson's *r*, Spearman's *rho*, and the maximum information coefficient MIC (*1*). Noise was generated by drawing random numbers from a uniform distribution on the interval $[-a, a]$. For each dependence structure, we increased the noise between 0 and 1 in 1/99 increments, and for each noise value we repeated each simulation *n* = 10 times, resulting in *n* = 1,000 simulations per dependence structure.

### *CauseEffectPairs*

We quantified the asymmetric dependence of *n* = 74 *CauseEffectPairs* (*2, 3*) to evaluate the causal discovery rate of *qad*. A total of *n* = 107 *CauseEffectPairs* can be downloaded (http://webdav.tuebingen.mpg.de/cause-effect/) from which we excluded *n* = 33 with too many ties (double entries) preventing a precise estimation of the empirical checkerboard copula (see Supplementary material 1). For each of the datasets the ground truth, i.e. whether y depends on x or *vice versa*, is given. We quantified the asymmetric dependence *q(X, Y)* and *q(Y, X)* as well as the asymmetry *a = q(X, Y) – q(Y, X)* for each *CauseEffectPair* (Supplementary material 4). In the case that *q(X, Y) > q(Y, X)* and the ground truth is a dependence of Y on X, we scored a correct causal discovery (Tab. S2_1). The same is true for *q(X, Y) < q(Y, X)* if the ground truth is a dependence of X on Y. Directional dependency of *CauseEffectPairs* was mostly not significantly asymmetric (only in 5 out of 74 cases). In these cases, however, causal discovery rate was 80%. These results confirm that statistical dependence is not the same concept as causality, but *qad* may help to infer causality. Nevertheless, expert knowledge is required to correctly interpret the data and to discuss the cause of (a)symmetric dependencies.

**Tab. S2_1** Causal discovery by *qad*. For n = 74 CauseEffectPairs we quantified the asymmetric dependence *q(X, Y)* and *q(Y, X)* and asymmetry in dependence *a* between *X* and *Y* using *qad*.

Significant dependencies and asymmetries are listed in bold. Ground truth is provided for each dataset. In the case that *q(X, Y)* > *q(Y, X)* and the ground truth is a dependence of Y on X, we scored a correct causal discovery (**TRUE**), otherwise *qad* failed to discover causality (FALSE). The same is true if *q(X, Y)* < *q(Y, X)* and the ground truth is a dependence of X on Y.

| ***CauseEffectPair*** | ***n*** | ***q(X, Y)*** | ***q(Y, X)*** | ***a*** | **ground truth** | **correct causal discovery** |
|---:|---:|---:|---:|---:|:---:|:---:|
| 1 | 348 | **0.627** | **0.629** | -0.002 | X --> Y | FALSE |
| 2 | 348 | **0.588** | **0.565** | 0.023 | X --> Y | **TRUE** |
| 3 | 348 | **0.316** | **0.348** | -0.032 | X --> Y | FALSE |
| 4 | 348 | **0.255** | **0.259** | -0.004 | X --> Y | FALSE |
| 13 | 391 | **0.726** | **0.712** | 0.014 | X --> Y | **TRUE** |
| 14 | 391 | **0.686** | **0.690** | -0.005 | X --> Y | FALSE |
| 15 | 391 | **0.716** | **0.730** | -0.014 | X --> Y | FALSE |
| 16 | 391 | **0.534** | **0.518** | 0.017 | X --> Y | **TRUE** |
| 18 | 313 | **0.783** | **0.760** | 0.023 | X --> Y | **TRUE** |
| 19 | 193 | **0.704** | **0.699** | 0.006 | X --> Y | **TRUE** |
| 20 | 348 | **0.334** | **0.427** | **-0.094** | X --> Y | FALSE |
| 21 | 348 | **0.353** | **0.352** | 0.001 | X --> Y | **TRUE** |
| 22 | 449 | **0.143** | 0.086 | **0.057** | X --> Y | **TRUE** |
| 23 | 451 | **0.324** | **0.288** | 0.035 | X --> Y | **TRUE** |
| 24 | 450 | **0.139** | 0.112 | 0.027 | X --> Y | **TRUE** |
| 25 | 1029 | **0.373** | **0.358** | 0.015 | X --> Y | **TRUE** |
| 26 | 1029 | **0.239** | **0.173** | **0.066** | X --> Y | **TRUE** |
| 27 | 1029 | **0.111** | **0.111** | -0.001 | X --> Y | FALSE |
| 28 | 1029 | **0.376** | **0.296** | **0.080** | X --> Y | **TRUE** |
| 29 | 1029 | **0.276** | **0.264** | 0.013 | X --> Y | **TRUE** |
| 30 | 1029 | **0.246** | **0.204** | **0.042** | X --> Y | **TRUE** |
| 31 | 1029 | **0.220** | **0.193** | 0.027 | X --> Y | **TRUE** |
| 39 | 393 | **0.217** | **0.224** | -0.007 | X --> Y | FALSE |
| 43 | 10368 | **0.954** | **0.951** | 0.003 | X --> Y | **TRUE** |
| 44 | 10368 | **0.904** | **0.898** | 0.006 | X --> Y | **TRUE** |
| 45 | 10368 | **0.786** | **0.776** | 0.010 | X --> Y | **TRUE** |
| 48 | 167 | **0.404** | **0.445** | -0.041 | Y --> X | **TRUE** |
| 49 | 364 | **0.593** | **0.591** | 0.002 | Y --> X | FALSE |
| 50 | 364 | **0.559** | **0.514** | 0.045 | Y --> X | FALSE |
| 51 | 364 | **0.344** | **0.370** | -0.026 | Y --> X | **TRUE** |
| 52 | 10225 | **0.942** | **0.944** | -0.003 | Y --> X | **TRUE** |
| 53 | 988 | **0.873** | **0.870** | 0.003 | Y --> X | FALSE |
| 54 | 391 | **0.860** | **0.854** | 0.006 | X --> Y | **TRUE** |
| 55 | 71 | 0.222 | 0.294 | -0.073 | Y --> X | **TRUE** |
| 56 | 191 | **0.400** | **0.413** | -0.013 | Y --> X | **TRUE** |
| 57 | 191 | **0.354** | **0.390** | -0.036 | Y --> X | **TRUE** |
| 58 | 191 | **0.368** | **0.397** | -0.029 | Y --> X | **TRUE** |
| 59 | 191 | **0.385** | **0.407** | -0.022 | Y --> X | **TRUE** |
| 60 | 191 | **0.325** | **0.380** | -0.056 | Y --> X | **TRUE** |
| 61 | 191 | **0.343** | **0.373** | -0.030 | Y --> X | **TRUE** |
| 62 | 191 | **0.361** | **0.364** | -0.003 | Y --> X | **TRUE** |
| 63 | 191 | **0.348** | **0.375** | -0.026 | Y --> X | **TRUE** |
| 64 | 161 | **0.687** | **0.675** | 0.012 | X --> Y | **TRUE** |
| 65 | 1330 | **0.429** | **0.418** | 0.011 | X --> Y | **TRUE** |
| 66 | 1330 | **0.654** | **0.652** | 0.002 | X --> Y | **TRUE** |
| 67 | 1330 | **0.561** | **0.564** | -0.003 | X --> Y | FALSE |
| 68 | 497 | **0.790** | **0.775** | 0.015 | Y --> X | FALSE |
| 72 | 1631 | **0.199** | **0.189** | 0.010 | X --> Y | **TRUE** |
| 73 | 5083 | **0.865** | **0.864** | 0.001 | Y --> X | FALSE |
| 74 | 193 | **0.659** | **0.678** | -0.019 | X --> Y | FALSE |
| 75 | 204 | **0.737** | **0.743** | -0.006 | Y --> X | **TRUE** |
| 76 | 346 | **0.727** | **0.755** | -0.028 | X --> Y | FALSE |
| 78 | 720 | **0.758** | **0.737** | 0.022 | X --> Y | **TRUE** |
| 79 | 720 | **0.720** | **0.706** | 0.014 | Y --> X | FALSE |
| 80 | 720 | **0.616** | **0.626** | -0.010 | Y --> X | **TRUE** |
| 81 | 364 | **0.638** | **0.644** | -0.007 | X --> Y | FALSE |
| 82 | 364 | **0.679** | **0.662** | 0.016 | X --> Y | **TRUE** |
| 83 | 364 | **0.760** | **0.758** | 0.001 | X --> Y | **TRUE** |
| 84 | 3101 | **0.940** | **0.938** | 0.002 | Y --> X | FALSE |
| 88 | 260 | **0.512** | **0.465** | 0.048 | X --> Y | **TRUE** |
| 89 | 130 | **0.546** | **0.548** | -0.002 | Y --> X | **TRUE** |
| 90 | 125 | **0.379** | **0.395** | -0.016 | Y --> X | **TRUE** |
| 91 | 148 | **0.668** | **0.682** | -0.014 | X --> Y | FALSE |
| 92 | 149 | **0.789** | **0.787** | 0.002 | Y --> X | FALSE |
| 93 | 431 | **0.708** | **0.696** | 0.012 | X --> Y | **TRUE** |
| 97 | 201 | **0.278** | **0.286** | -0.008 | X --> Y | FALSE |
| 98 | 93 | **0.788** | **0.772** | 0.016 | X --> Y | **TRUE** |
| 100 | 208 | **0.501** | **0.522** | -0.020 | X --> Y | FALSE |
| 101 | 299 | **0.891** | **0.891** | 0.000 | X --> Y | FALSE |
| 102 | 108 | **0.475** | **0.382** | 0.094 | X --> Y | **TRUE** |
| 103 | 108 | 0.155 | 0.178 | -0.022 | X --> Y | FALSE |
| 104 | 108 | **0.278** | **0.285** | -0.007 | X --> Y | FALSE |
| 105 | 999 | **0.556** | **0.551** | 0.005 | X --> Y | **TRUE** |
| 106 | 113 | **0.756** | **0.752** | 0.003 | Y --> X | FALSE |

***Global climate [Fig. 4]***

At times of global warming and changes in precipitation regimes, information on past, present, and future climate at a local scale is demanded to assess the effect of climate on the environment. Such data can be retrieved from databases that provide climate estimates in high geographic resolution. Commonly, estimates of temperature and precipitation are used to derive 19 bioclimatic variables that inform about climate features relevant to biological processes. Since these variables are based on a limited set of data, there is a strong underlying dependence structure between them, which, in particular, prevents their independent use in statistical models. We retrieved bioclimatic variables for $n$ = 1862 locations homogenously distributed over the global landmass from CHELSA (*4, 5*) to test for asymmetric dependence between them (Fig. 4). Using these data we calculated $q(X,Y)$, $q(Y,X)$, (i.e. the mean dependence), and (i.e. the asymmetry of the dependence) and calculated the corresponding $p$-values informing about the significance of dependence and asymmetry (Supplementary material 4). Additionally, Pearson's $r^2$ was calculated. Heatmaps visualize the dependencies between the 19 bioclimatic variables based on *qad* (Fig. S2_2, asymmetry |*a*| is shown in Fig. S2_3) and Pearson's $r^2$ (Fig. S2_4).

As expected, the knowledge of one of the bioclimatic variables reduces the variability of the other variables, which is indicated by the high dependence between the pairs of variables (mean $q$ ± sd = 0.54 ± 0.18). Pearson's product moment correlation indicated weaker dependencies (mean ± sd: 0.32 ± 0.29). Many of the associations between the bioclimatic variables turned out to be asymmetric (mean asymmetry $|a|$ ± sd = 0.06 ± 0.05, range = 0.00 – 0.23). Particularly, relationships between bioclimatic variables informing about temperature and those informing about precipitation are not related linearly (**Fig. S2_2 and S2_4**). For instance, the annual precipitation can be better predicted by mean temperature ($q$ = 0.61, $p$ < 0.001) than *vice versa* ($q$ = 0.54, $p$ < 0.001, asymmetry $a$ = 0.08, $p$ < 0.001, Fig. 4). In general, precipitation related bioclimatic variables were better predictable than temperature related variables (**Fig. S2_2**).

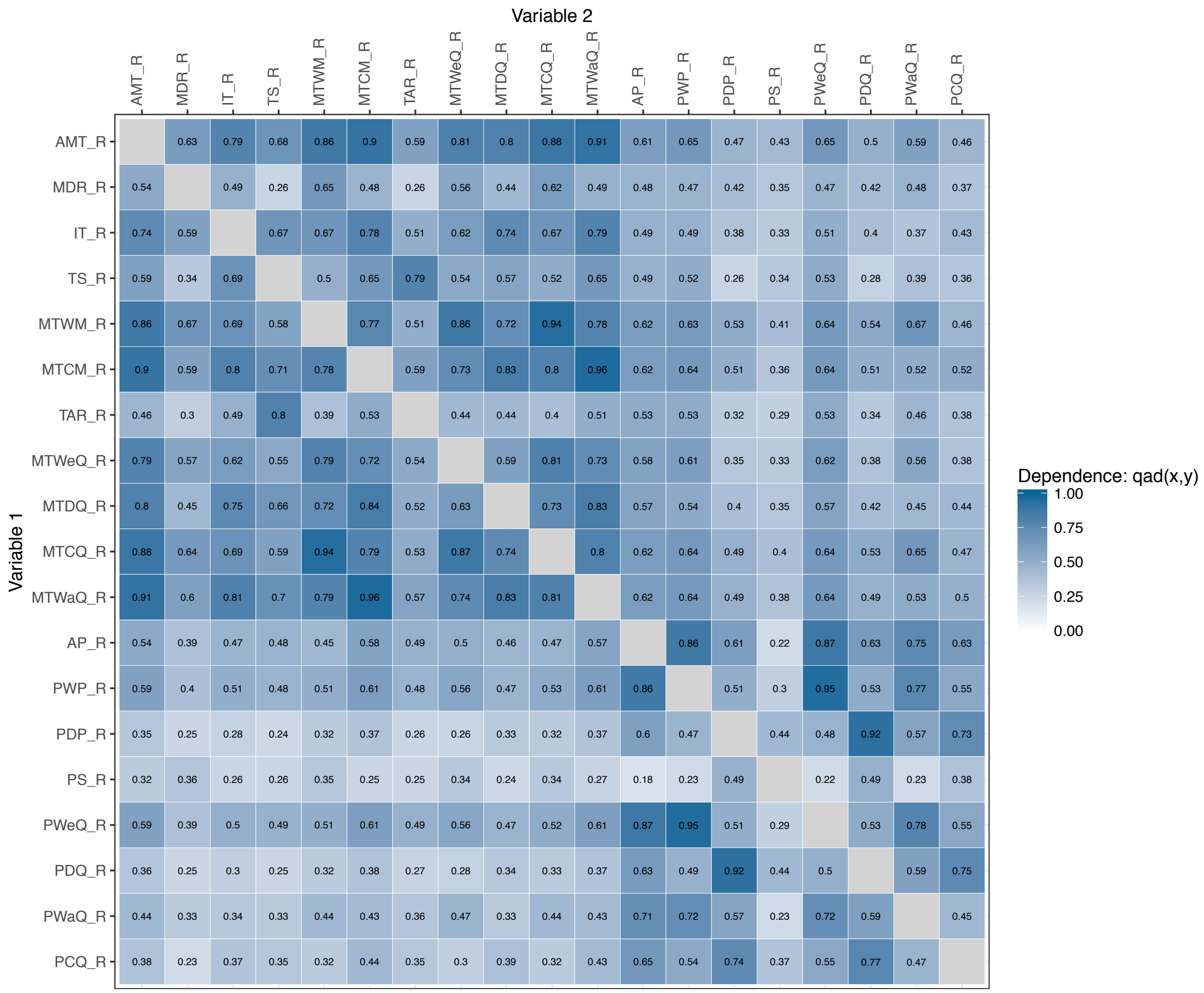

**Fig. S2_2** Asymmetric dependence *q* between 19 bioclimatic variables measured at *n* = 1862 locations homogenously distributed over the global landmass. The heatmap depicts *q(variable on the y-axis, variable on the x-axis)*, which is the dependence of the column-variable on the row-variable. Abbreviations: AMT: Annual Mean Temperature; MDR: Mean Diurnal Range; IT: Isothermality; TS: Temperature Seasonality; MTWM: Max Temperature of Warmest Month; MTCM: Min Temperature of Coldest Month; TAR: Temperature Annual Range; MTWeQ: Mean Temperature of Wettest Quarter; MTDQ: Mean Temperature of Driest Quarter; MTCQ: Mean Temperature of Coldest Quarter; MTWaQ: Mean Temperature of Warmest Quarter; AP: Annual Precipitation; PWP: Precipitation of Wettest Month; PDP: Precipitation of Driest Month; PS: Precipitation Seasonality; PWeQ: Precipitation of Wettest Quarter; PDQ: Precipitation of Driest Quarter; PWaQ: Precipitation of Warmest Quarter; PCQ: Precipitation of Coldest Quarter.

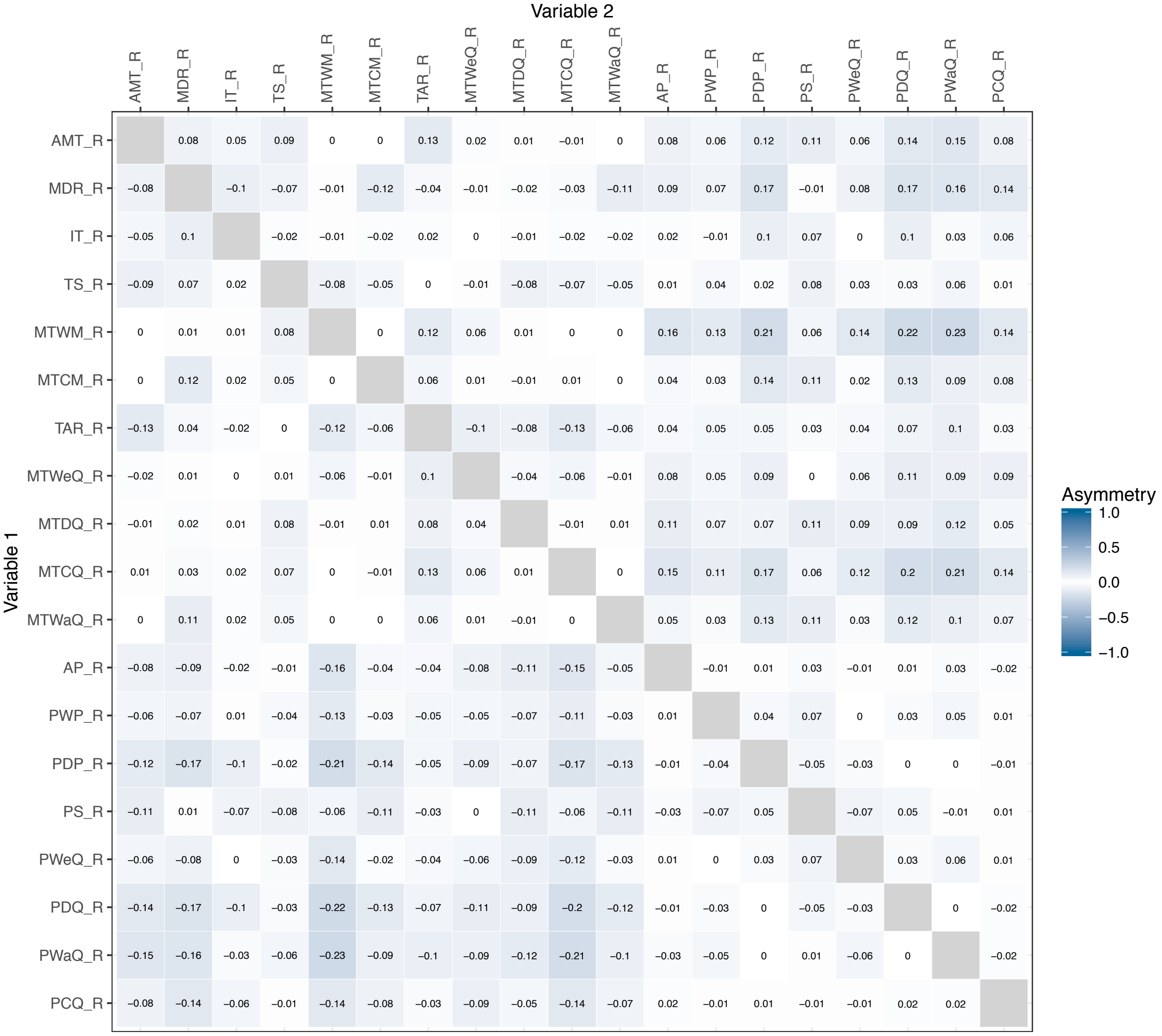

**Fig. S2_3** Asymmetry *a* in dependence *q* between 19 bioclimatic variables measured at *n* = 1862 locations homogenously distributed over the global landmass. Since |a| is symmetric, the lower triangule of the heatmap differs only in the sign from the upper triangle of the heatmap. For abbreviations see Fig. S2_2.

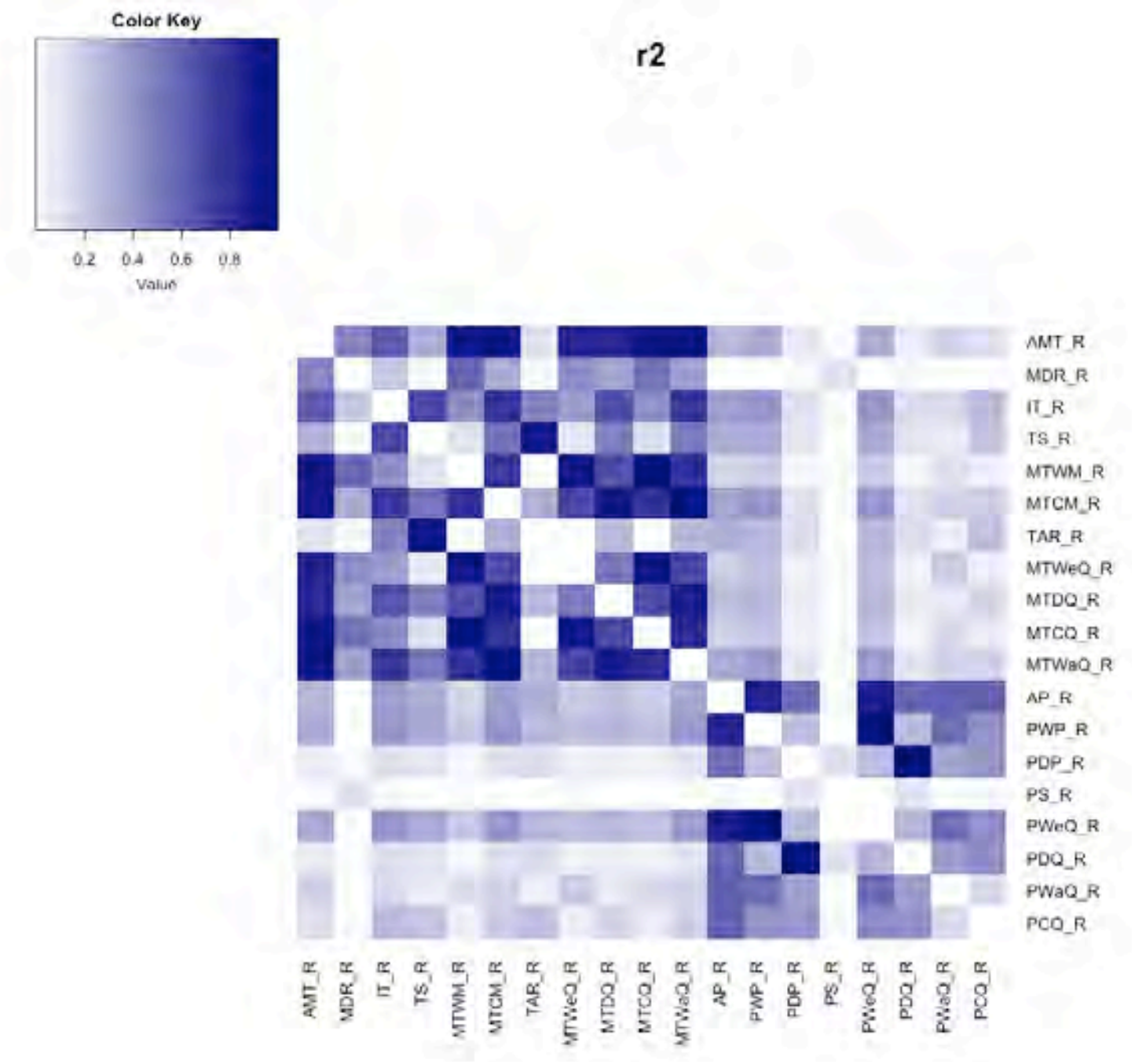

**Fig. S2_4** Dependence based on Pearson's $r^2$ between 19 bioclimatic variables measured at $n = 1862$ locations homogenously distributed over the global landmass. By construction $r^2$ is symmetric, therefore the matrix is symmetric, too. For abbreviations see Fig. S2_2.

Using mean annual temperature and annual precipitation as examples, we demonstrate the properties of *qad* to quantify asymmetric dependence of bivariate associations. The observations (Fig. S2_5a) are transformed to pseudo-observations (normalized ranks, Fig. S2_5b), which are the basis for the empirical checkerboard copula (as estimator of the true underlying dependence structure, Fig. S2_5c). The resolution $N$ of the checkerboard copula (number of stripes per dimension) depends on the sample size and is defined as number of stripes $s = r^{1/2}$, with $r$ as the number of ranks per variable (note that the number of different ranks $r$ does not necessarily coincide with the number of samples $n$). In this example, the lower number of ranks of the two variables was $r = 696$, resulting in a resolution of $696^{1/2} = N = 26$ (i.e. Fig. S2_5c displays 26 horizontal and vertical stripes). For each vertical or horizontal stripe, the mass distribution of the empirical checkerboard copula is given and sums up to 1. As a consequence the empirical checkerboard copula can be used to predict Y given the value of X and *vice versa*. In our example the annual precipitation can be better predicted by mean temperature ($q(AMT, AP) = 0.61$, $p < 0.001$) than *vice versa* ($q(AP, AMT) = 0.54$, $p < 0.001$; asymmetry $a = 0.08$, $p < 0.001$, Fig. S2_5c).

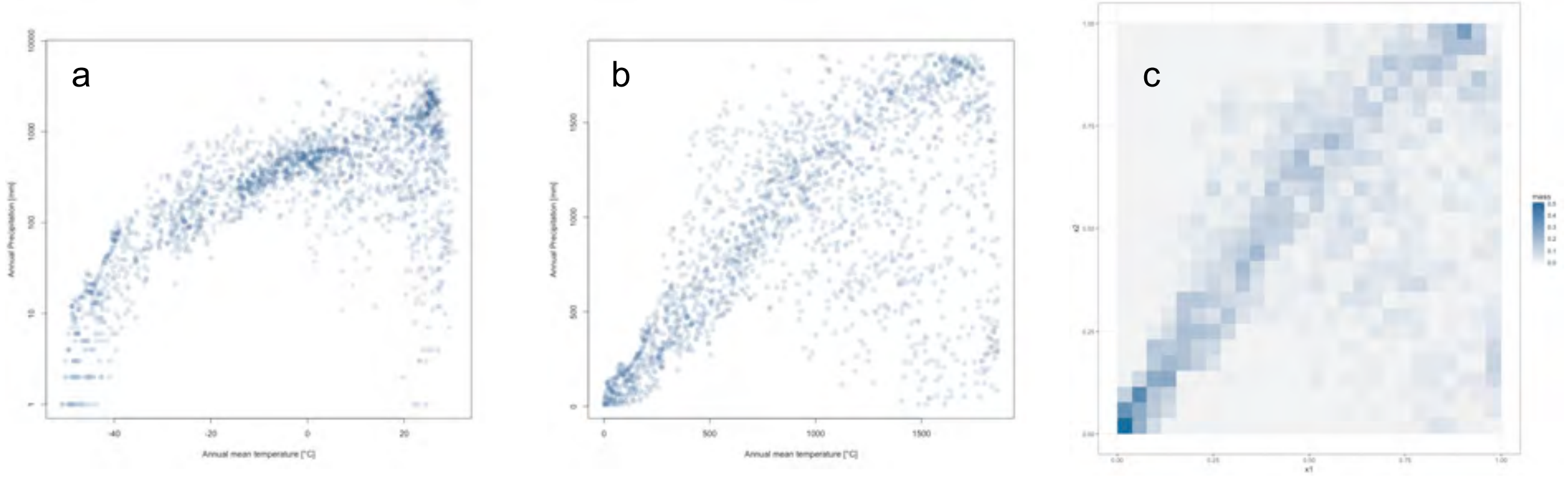

**Fig. S2_5** Relationship between mean annual temperature and annual precipitation. The observations (a) are transformed to pseudo-observations (normalized ranks, b), which are the basis for the empirical checkerboard copula (as estimator of the true underlying dependence structure, c).

***Microbiome [Fig. 5]***

We used a dataset of bacteria associated with surfaces of the plant *Metrosideros polymorpha* (*6*) (study accession number of sequences at the European Nucleotide Archive: PRJEB7828, http://www.ebi.ac.uk/ena/data/view/PRJEB7828) and tested for asymmetric relationships *qad* between pairs of $n$ = 93 operational taxonomic units (OTUs) that were observed in at least 75% of all samples ($n$ = 125) (Supporting Information 4 – Microbiomes Fig. 5). The quantification of asymmetric dependencies allows a novel definition of key species $S_f$, which influence the abundance of other species $S_j$ but are less influenced by others. Thus, key species $S_f$ influence other species $S_j$ on average to a larger extend than they are dependent on other species $S_j$. Median influence of species $S_f$ is calculated as the median of $I_f$ -values calculated as $I_f^j = q(S_f, S_j) - q(S_j, S_f), 1 \leq j \leq n, j \neq f$. If median ($I_f$) > 0 species $S_f$ influences most other species $S_j$ stronger than *vice versa*. As mentioned before, $q(S_f, S_j)$ denotes the dependence of $S_j$ on $S_f$ and $q(S_j, S_f)$ denotes the dependence in the other direction (Tab. S2_2, Supporting Information 4 – Microbiomes Fig. 5).

Networks and network indices were calculated using the R-package *igraph* (*7*). $q$-values and $r$-values below 0.325 were set to 0 in order to restrict network graphs to relatively strong dependencies only. Likewise, only significant ($p$ < 0.05) dependencies were included in the networks. Weighted and directed network visualizing asymmetric dependencies $q$ between OTUs confirmed the prominent role of the key-OTUs in shaping the abundances of a number of other OTUs. Most of the seven key-OTUs occupy central positions (Tab. S2_2) indicating their role as network hubs (Tab. S2_2). In contrast, in an undirected and weighted network based on Pearson's $r^2$, key-OTUs defined by *qad* occupied peripheral positions (Tab. S2_2). Accordingly, these OTUs were not identified as hubs (Tab. S2_2).

**Tab. S2_2** Influence and network positions of OTUs associated with surfaces of the plant *Metrosideros polymorpha*. For each OTU, the median ($I_f$), betweeness, degree, and hub score based on *q*-values is given. Betweeness, degree, and hub score are also given based on *r*-values. OTUs are ordered by increasing median $I_f$.

| OTU | Family | Genus | median ($I_f$) | betweenness_qad | betweenness_r | degree_qad | degree_r | hub_score_qad | hub_score_r |
|---|---|---|---|---|---|---|---|---|---|
| 3820 | Moraxellaceae | Acinetobacter | -0.02795 | 0 | 0 | 7 | 2 | 0.00000 | 0.00000 |
| 936 | Oxalobacteraceae | NA | -0.01985 | 0 | 47 | 5 | 42 | 0.00000 | 0.95184 |
| 23 | Bradyrhizobiaceae | Bradyrhizobium | -0.01836 | 241 | 204 | 8 | 15 | 0.07351 | 0.01049 |
| 321 | Microbacteriaceae | Curtobacterium | -0.01714 | 0 | 0 | 0 | 0 | 0.00000 | 0.00000 |
| 7 | Caulobacteraceae | NA | -0.01667 | 25 | 14 | 6 | 48 | 0.01143 | 1.00000 |
| 12 | Hyphomicrobiaceae | NA | -0.01654 | 111 | 169 | 6 | 47 | 0.01171 | 0.99004 |
| 76 | Sphingomonadaceae | Sphingomonas | -0.01640 | 29 | 5 | 4 | 27 | 0.00520 | 0.68808 |
| 15 | Comamonadaceae | NA | -0.01584 | 0 | 1 | 2 | 42 | 0.01143 | 0.95691 |
| 13 | Oxalobacteraceae | NA | -0.01534 | 189 | 124 | 6 | 48 | 0.01190 | 0.98914 |
| 3 | Weeksellaceae | Chryseobacterium | -0.01393 | 162 | 0 | 11 | 2 | 0.07348 | 0.00000 |
| 4 | Acetobacteraceae | Gluconobacter | -0.01217 | 0 | 0 | 1 | 0 | 0.00000 | 0.00000 |
| 17 | Caulobacteraceae | Phenylobacterium | -0.01156 | 240 | 14 | 10 | 47 | 0.02294 | 0.99463 |
| 4057 | Moraxellaceae | Acinetobacter | -0.01059 | 23 | 661 | 7 | 20 | 0.05168 | 0.17305 |
| 29 | Rhizobiaceae | Agrobacterium | -0.01026 | 54 | 13 | 8 | 44 | 0.01143 | 0.96599 |
| 33 | Oxalobacteraceae | Cupriavidus | -0.01012 | 147 | 19 | 6 | 41 | 0.01143 | 0.92692 |
| 36 | Aurantimonadaceae | NA | -0.01011 | 0 | 31 | 3 | 41 | 0.01143 | 0.91993 |
| 71 | Rhodobacteraceae | Paracoccus | -0.00972 | 0 | 44 | 1 | 24 | 0.00000 | 0.60429 |
| 26 | Pseudomonadaceae | NA | -0.00948 | 116 | 3 | 7 | 41 | 0.01430 | 0.94177 |
| 3284 | Oxalobacteraceae | NA | -0.00899 | 0 | 83 | 3 | 38 | 0.00000 | 0.87151 |
| 3684 | Sphingomonadaceae | Sphingomonas | -0.00884 | 0 | 25 | 3 | 7 | 0.00000 | 0.03176 |
| 157 | Oxalobacteraceae | NA | -0.00794 | 1 | 6 | 2 | 10 | 0.00000 | 0.00443 |
| 53 | Aurantimonadaceae | NA | -0.00741 | 0 | 50 | 2 | 37 | 0.00000 | 0.80187 |
| 72 | Xanthomonadaceae | Stenotrophomonas | -0.00680 | 117 | 15 | 5 | 6 | 0.02829 | 0.02250 |
| 3493 | Comamonadaceae | NA | -0.00668 | 499 | 1 | 5 | 3 | 0.00338 | 0.07761 |
| 100 | Sphingomonadaceae | Kaistobacter | -0.00636 | 0 | 7 | 3 | 6 | 0.02025 | 0.14658 |
| 22 | Oxalobacteraceae | NA | -0.00602 | 0 | 285 | 3 | 54 | 0.00000 | 0.99919 |
| 40 | Caulobacteraceae | Phenylobacterium | -0.00533 | 32 | 22 | 3 | 40 | 0.01143 | 0.92015 |
| 146 | Corynebacteriaceae | Corynebacterium | -0.00510 | 0 | 33 | 3 | 7 | 0.00000 | 0.00400 |
| 472 | Pseudomonadaceae | Pseudomonas | -0.00445 | 0 | 0 | 0 | 4 | 0.00000 | 0.03666 |
| 656 | Caulobacteraceae | NA | -0.00426 | 0 | 4 | 2 | 5 | 0.00000 | 0.08561 |
| 35 | Hyphomicrobiaceae | Devosia | -0.00402 | 342 | 11 | 7 | 36 | 0.01596 | 0.86398 |
| 51 | Nocardiaceae | Rhodococcus | -0.00396 | 123 | 145 | 6 | 11 | 0.06225 | 0.02491 |
| 2286 | Comamonadaceae | NA | -0.00389 | 0 | 0 | 1 | 0 | 0.00000 | 0.00000 |
| 63 | Caulobacteraceae | NA | -0.00350 | 0 | 19 | 2 | 30 | 0.00008 | 0.73553 |
| 117 | Caulobacteraceae | NA | -0.00329 | 0 | 26 | 1 | 30 | 0.00000 | 0.72675 |
| 2999 | Oxalobacteraceae | NA | -0.00325 | 2 | 18 | 2 | 7 | 0.03469 | 0.10711 |
| 83 | Nocardioidaceae | NA | -0.00256 | 30 | 52 | 5 | 37 | 0.00259 | 0.86209 |
| 3725 | Oxalobacteraceae | NA | -0.00235 | 54 | 308 | 3 | 21 | 0.00180 | 0.38003 |
| 2172 | Comamonadaceae | NA | -0.00163 | 0 | 8 | 3 | 6 | 0.00000 | 0.00279 |
| 86 | Rhodospirillaceae | Azospirillum | -0.00149 | 14 | 14 | 3 | 29 | 0.04164 | 0.73136 |
| 28 | Staphylococcaceae | Staphylococcus | -0.00127 | 396 | 52 | 4 | 8 | 0.00000 | 0.00685 |
| 80 | Chromatiaceae | Rheinheimera | -0.00118 | 126 | 22 | 7 | 37 | 0.01537 | 0.87546 |
| 235 | Bacillaceae | Bacillus | -0.00091 | 33 | 0 | 4 | 1 | 0.05014 | 0.00031 |
| 19 | Comamonadaceae | Limnohabitans | -0.00080 | 84 | 132 | 6 | 44 | 0.00011 | 0.90322 |
| 2093 | Moraxellaceae | Acinetobacter | -0.00048 | 14 | 305 | 3 | 26 | 0.00377 | 0.60126 |
| 30 | Caulobacteraceae | NA | -0.00037 | 0 | 11 | 2 | 38 | 0.01315 | 0.89796 |
| 3516 | Rhizobiaceae | Agrobacterium | -0.00006 | 436 | 19 | 4 | 13 | 0.00385 | 0.34912 |
| 5 | Moraxellaceae | Acinetobacter | 0.00014 | 0 | 0 | 3 | 2 | 0.00000 | 0.00000 |
| 1886 | Comamonadaceae | NA | 0.00062 | 368 | 14 | 10 | 12 | 0.05913 | 0.24324 |
| 61 | Methylobacteriaceae | NA | 0.00063 | 0 | 5 | 1 | 27 | 0.00000 | 0.65473 |
| 970 | Pseudomonadaceae | Pseudomonas | 0.00066 | 456 | 101 | 7 | 25 | 0.01462 | 0.49020 |
| 52 | Moraxellaceae | NA | 0.00099 | 215 | 13 | 5 | 14 | 0.02633 | 0.13756 |
| 1692 | Brucellaceae | Ochrobactrum | 0.00108 | 617 | 6 | 13 | 10 | 0.12783 | 0.00762 |
| 138 | Comamonadaceae | Limnobacter | 0.00133 | 0 | 21 | 3 | 21 | 0.00243 | 0.54360 |
| 89 | Planococcaceae | NA | 0.00200 | 144 | 172 | 3 | 29 | 0.00000 | 0.67224 |
| 90 | Xanthobacteraceae | NA | 0.00237 | 0 | 23 | 3 | 22 | 0.00383 | 0.56769 |
| 170 | Sphingomonadaceae | NA | 0.00255 | 0 | 0 | 2 | 5 | 0.05422 | 0.00309 |
| 67 | Erythrobacteraceae | NA | 0.00335 | 12 | 24 | 3 | 26 | 0.03920 | 0.64367 |
| 145 | Bacillaceae | NA | 0.00348 | 0 | 1 | 1 | 28 | 0.00000 | 0.70620 |
| 66 | Rhodospirillaceae | NA | 0.00356 | 0 | 23 | 2 | 33 | 0.03469 | 0.79786 |
| 48 | Weeksellaceae | Cloacibacterium | 0.00372 | 0 | 94 | 1 | 6 | 0.00000 | 0.12242 |
| 1218 | Aurantimonadaceae | NA | 0.00435 | 226 | 89 | 12 | 24 | 0.02490 | 0.55687 |
| 85 | Neisseriaceae | Vogesella | 0.00446 | 292 | 126 | 6 | 38 | 0.04868 | 0.86837 |
| 75 | Beijerinckiaceae | NA | 0.00450 | 0 | 138 | 1 | 23 | 0.00000 | 0.53894 |
| 78 | Methylobacteriaceae | Methylobacterium | 0.00514 | 50 | 164 | 4 | 10 | 0.03424 | 0.06034 |
| 3955 | Comamonadaceae | NA | 0.00577 | 409 | 46 | 2 | 10 | 0.00110 | 0.23462 |
| 34 | Moraxellaceae | Enhydrobacter | 0.00581 | 0 | 12 | 0 | 5 | 0.00000 | 0.00161 |
| 175 | Moraxellaceae | Acinetobacter | 0.00611 | 157 | 0 | 4 | 0 | 0.00605 | 0.00000 |
| 55 | Flavobacteriaceae | Flavobacterium | 0.00633 | 67 | 10 | 4 | 33 | 0.02677 | 0.81088 |
| 1193 | Comamonadaceae | Azohydromonas | 0.00637 | 0 | 0 | 0 | 2 | 0.00000 | 0.02045 |
| 58 | Streptococcaceae | Streptococcus | 0.00642 | 0 | 0 | 5 | 0 | 0.22296 | 0.00000 |
| 96 | Comamonadaceae | NA | 0.00644 | 0 | 0 | 2 | 0 | 0.02145 | 0.00000 |
| 193 | Rhodobacteraceae | Rubellimicrobium | 0.00675 | 0 | 16 | 1 | 9 | 0.00000 | 0.20472 |
| 82 | Micrococcaceae | Renibacterium | 0.00686 | 12 | 120 | 3 | 7 | 0.03035 | 0.00710 |
| 128 | Chitinophagaceae | Sediminibacterium | 0.00818 | 0 | 3 | 1 | 11 | 0.00000 | 0.00471 |
| 109 | Rhodobacteraceae | Rhodobacter | 0.00876 | 0 | 14 | 4 | 9 | 0.00889 | 0.03506 |
| 87 | Xanthomonadaceae | Lysobacter | 0.00894 | 55 | 33 | 3 | 25 | 0.00069 | 0.61503 |
| 217 | Caulobacteraceae | Mycoplana | 0.00911 | 153 | 96 | 7 | 30 | 0.00567 | 0.63203 |
| 143 | Intrasporangiaceae | NA | 0.00969 | 0 | 26 | 0 | 3 | 0.00000 | 0.02057 |
| 64 | Acetobacteraceae | NA | 0.01000 | 0 | 51 | 2 | 33 | 0.00000 | 0.79445 |
| 611 | Micrococcaceae | NA | 0.01040 | 0 | 0 | 0 | 1 | 0.00000 | 0.00021 |
| 119 | Rhodospirillaceae | Novispirillum | 0.01105 | 0 | 34 | 4 | 24 | 0.01412 | 0.61403 |
| 92 | Rhizobiaceae | Agrobacterium | 0.01187 | 287 | 1 | 9 | 36 | 0.04746 | 0.83233 |
| 69 | Aeromonadaceae | NA | 0.01219 | 0 | 19 | 3 | 9 | 0.08523 | 0.03099 |
| 62 | Rhodobacteraceae | Paracoccus | 0.01248 | 0 | 31 | 1 | 12 | 0.00000 | 0.30961 |
| 44 | Geodermatophilaceae | NA | 0.01379 | 0 | 13 | 2 | 12 | 0.06806 | 0.09644 |
| 741 | Comamonadaceae | NA | 0.01729 | 35 | 0 | 3 | 2 | 0.04004 | 0.01080 |
| 3977 | Sphingomonadaceae | NA | 0.02150 | 0 | 9 | 4 | 6 | 0.05895 | 0.00339 |
| 3433 | Oxalobacteraceae | NA | 0.02204 | 299 | 10 | 11 | 8 | 0.25974 | 0.02287 |
| 56 | Pseudomonadaceae | Pseudomonas | 0.02353 | 82 | 17 | 10 | 11 | 0.27415 | 0.04971 |
| 98 | Enterobacteriaceae | NA | 0.02986 | 0 | 0 | 10 | 0 | 0.29884 | 0.00000 |
| 8 | Moraxellaceae | Acinetobacter | 0.04838 | 1882 | 0 | 48 | 9 | 1.00000 | 0.02783 |
| 60 | Moraxellaceae | Acinetobacter | 0.05136 | 724 | 0 | 26 | 8 | 0.71298 | 0.01693 |

***World development indicators [Fig. 6]***

We retrieved World Development Indicators (WDI) for the year 2015 from the database provided by the World Bank (http://databank.worldbank.org, last accessed July 2017). The indicators were restricted to those that were available for at least 100 countries. Additionally, we excluded those WDIs with too many ties to allow for a precise estimation of the empirical checkerboard copula. This data filtering resulted in a total of $n$ = 450 WDIs of $n$ = 179 countries included in the analysis. For each indicator pair X and Y, we quantified asymmetric dependence and as well as the asymmetry in dependence (Supplementary material 4). Asymmetry in all indicator pairs ranged from $|a|$ = 0 to $|a|$ = 0.317; 99.8% of all pairs had an asymmetry below $a$ = 0.2016. The indicators birth rate and death rate that are moderately dependent on each other (*qad*: *mean q*

= 0.42, $p < 0.001$), nevertheless the dependence is significantly asymmetric (*qad*: $a = 0.2016$, $p < 0.001$; Fig. S2_6).

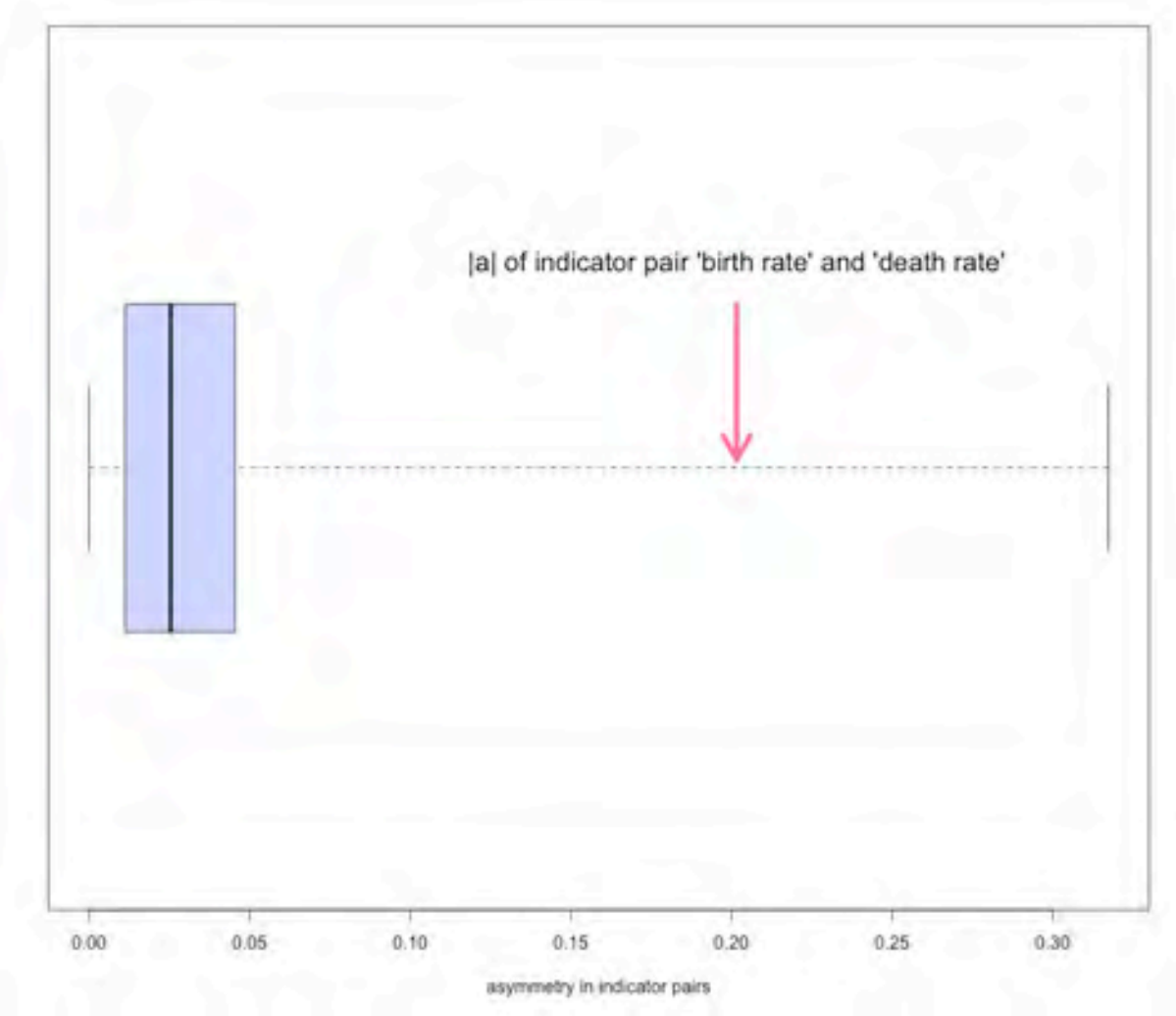

**Fig. S2_6** Boxplot of asymmetry |*a*| values quantified for all indicator pairs possible. The absolute value of the asymmetry of the indicator pair ‘birth rate’ and ‘death rate’ is indicated by the pink arrow.

Death rate is well predictable by birth rate (*q(birth rate, death rate)* = 0.53, $p < 0.001$) whereas variability in birth rate is not as strongly reduced by the knowledge of death rate (*q(death rate, birth rate)* = 0.32, $p < 0.001$; Fig. S2_7).

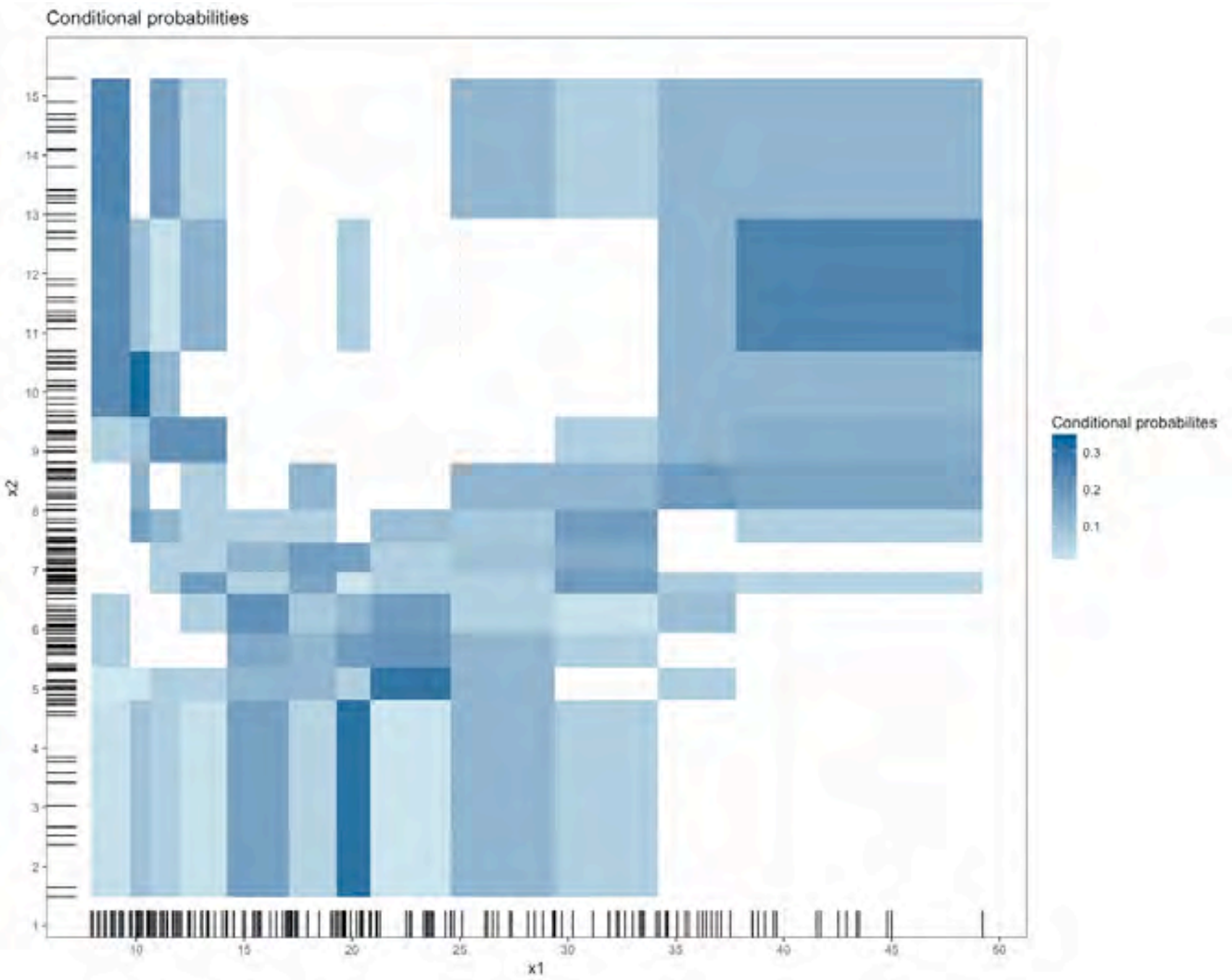

**Fig. S2_7** Empirical checkerboard copula for the indicator pair ‘birth rate’ (X1) and ‘death rate’ (X2). For each vertical or horizontal stripe, the mass distribution sums up to 1. Therefore, the checkerboard copula can be used to predict X2 given the value of X1 and *vice versa*.

The roughly ‘U’ shaped relationship between birth and death rate deserved further data exploration – we found the gross domestic product per capita (GDP) to be an underlying factor potentially explaining the non-linear relationship. GDP is strongly and symmetrically associated to birth rate (*qad*: *mean q* = 0.67, $p < 0.001$; $a = 0.02$, $p = 0.59$) and weakly and symmetrically to death rate (*qad*: *mean q* = 0.31, $p < 0.001$; $a = 0.03$, $p = 0.39$). Roughly speaking, Fig. 4 in the main text suggests that in countries with a GDP below the mean GDP of all countries, birth rate and death rate are positively associated, in countries with a GDP above the mean GDP a negative association is suggested. This pattern in supported by *qad*: In countries with a GDP below the mean GDP of all countries birth and death rate show are symmetrically associated to each other (*qad*: *mean q* = 0.48, $p < 0.001$; $a = 0.06$, $p = 0.19$; Fig. S2_8), the same is true for the same association in countries with a GDP above the mean GDP of all countries (*qad*: *mean q* = 0.53, $p < 0.001$; $a = 0.08$, $p = 0.08$; Fig. S2_9).

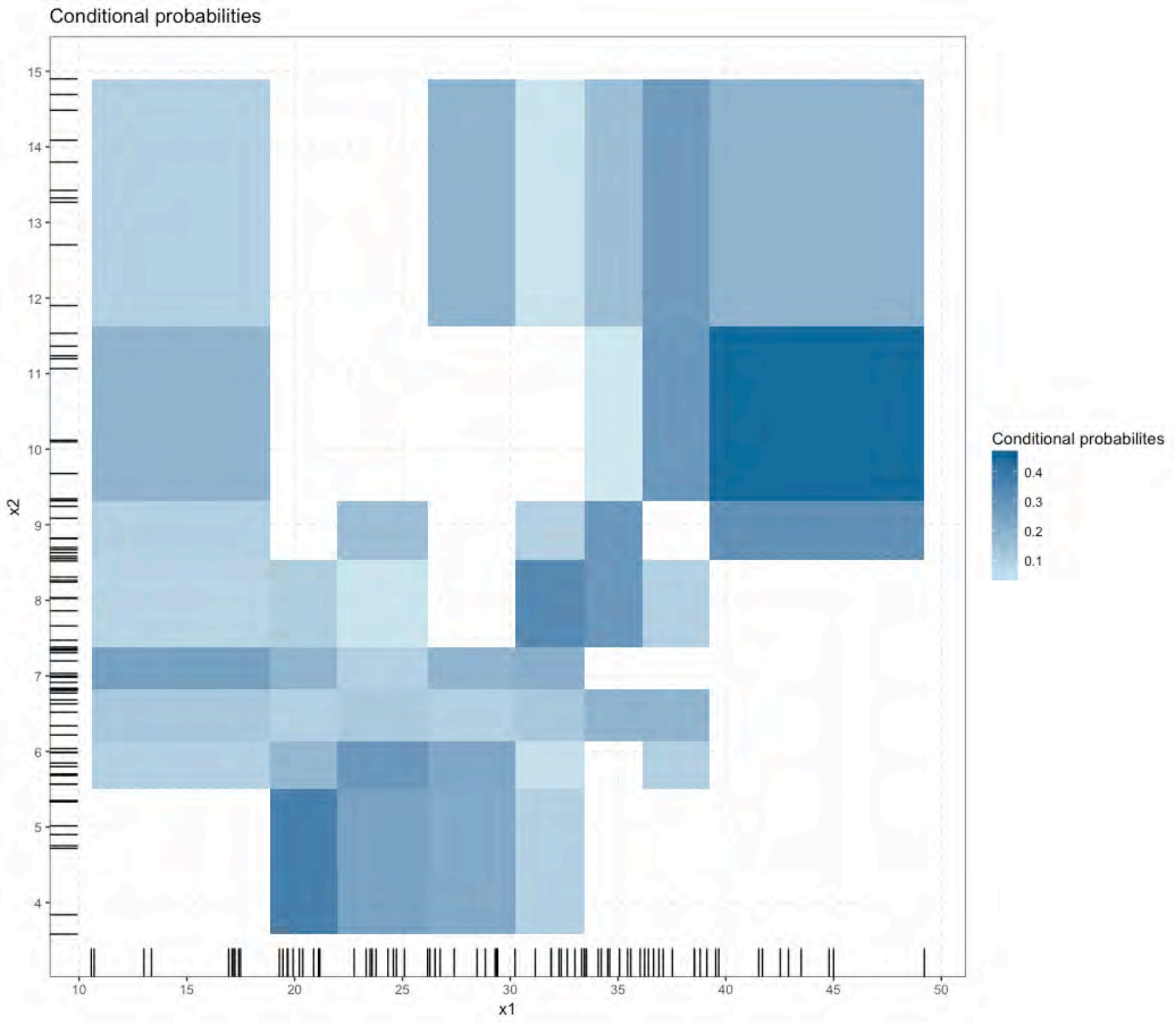

**Fig. S2_8** Empirical checkerboard copula for the indicator pair 'birth rate' (X1) and 'death rate' (X2) for countries with a GDP below the mean GDP of all countries. For each vertical or horizontal stripe, the mass distributions sums up to 1. Therefore, the checkerboard copula can be used to predict X2 given the knowledge of X1 and *vice versa*.

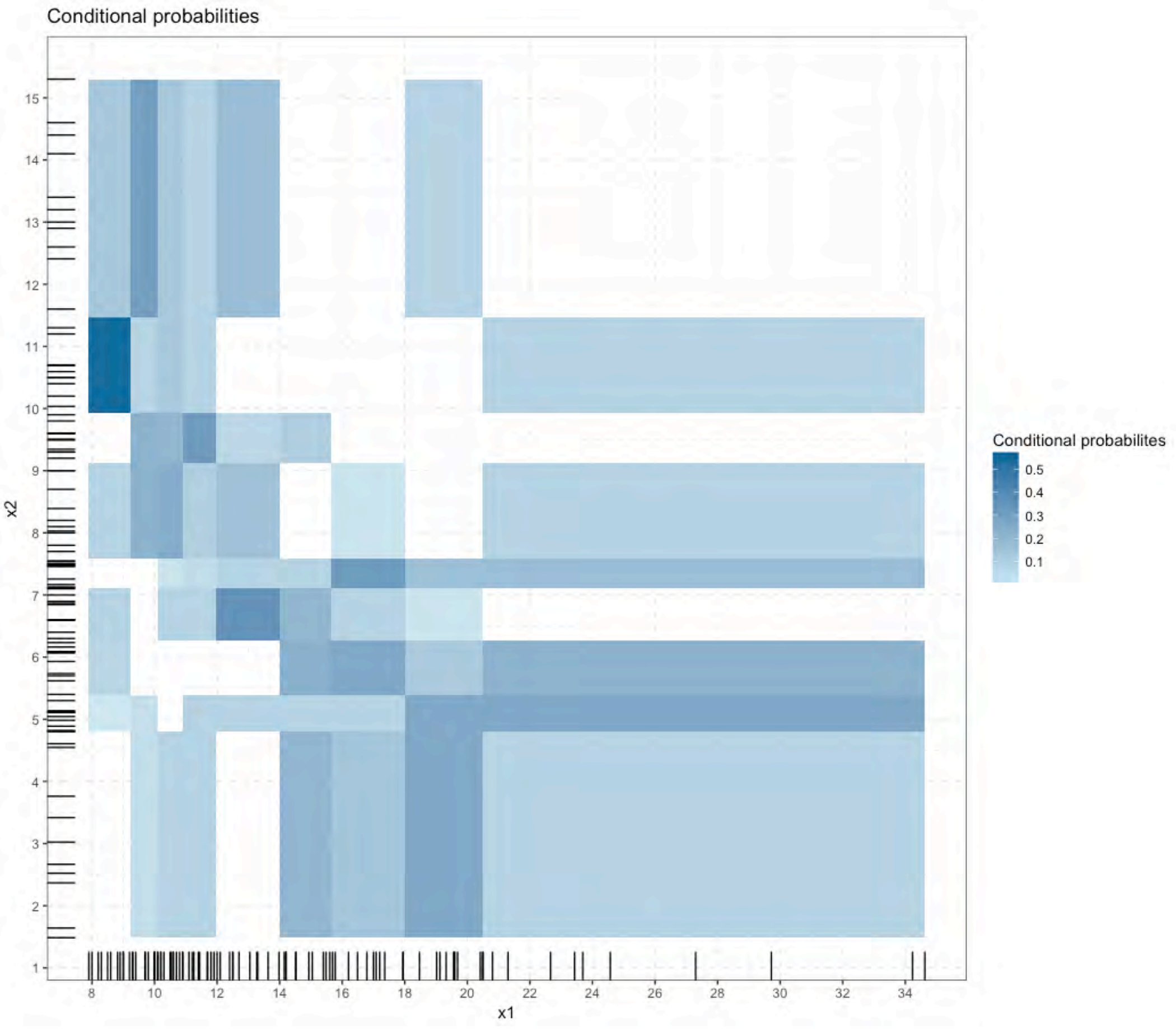

**Fig. S2_9** Empirical checkerboard copula for the indicator pair 'birth rate' (X1) and 'death rate' (X2) for countries with a GDP above the mean GDP of all countries. For each vertical or horizontal stripe, the mass distribution sums up to 1. Therefore, the checkerboard copula can be used to predict X2 given the knowledge of X1 and *vice versa*.

Linear models do not find a significant association between birth and death rate ($F_{1,170}$ = 0.44, $p$ = 0.51), however, death rate is negatively associated to GDP ($F_{1,170}$ = 8.84, $p$ = 0.003) and the interaction between birth rate and GDP significantly affects death rate ($F_{1,170}$ = 12.02, $p$ < 0.001).

| long | lat | AMT_R | MDR_ | IT_R | TS_R | MTWM_R | MTCM_ | TAR_R | MTWeQ_ | MTDQ_R | MTCQ_R | MTWa | AP_R | PWP_R | PDP_ | PS_R | PWeQ_ | PDQ_R | PWaQ_ | PCQ_ |
|---|---|---|---|---|---|---|---|---|---|---|---|---|---|---|---|---|---|---|---|---|
| -175.135274 | -87.64878754 | -411 | 29 | 107 | 8729 | -249 | -515 | 267 | -469 | -280 | -256 | -496 | 37 | 4 | 1 | 31 | 12 | 4 | 5 | 9 |
| -170.2704092 | -87.64878754 | -406 | 28 | 107 | 8636 | -245 | -509 | 264 | -441 | -276 | -253 | -490 | 53 | 6 | 2 | 30 | 18 | 7 | 8 | 13 |
| -165.4055443 | -87.64878754 | -401 | 28 | 107 | 8582 | -241 | -504 | 263 | -436 | -272 | -249 | -484 | 63 | 7 | 2 | 30 | 21 | 8 | 8 | 16 |
| -160.5406794 | -87.64878754 | -397 | 28 | 108 | 8530 | -238 | -499 | 261 | -431 | -269 | -245 | -479 | 73 | 8 | 3 | 27 | 24 | 10 | 11 | 18 |
| -155.6758146 | -87.64878754 | -394 | 28 | 109 | 8526 | -234 | -496 | 261 | -450 | -266 | -242 | -476 | 76 | 9 | 3 | 29 | 26 | 10 | 11 | 19 |
| -150.8109497 | -87.64878754 | -391 | 29 | 109 | 8544 | -231 | -493 | 262 | -447 | -262 | -239 | -473 | 80 | 9 | 3 | 28 | 26 | 11 | 11 | 21 |
| -145.9460848 | -87.64878754 | -390 | 29 | 109 | 8596 | -229 | -493 | 264 | -431 | -260 | -237 | -473 | 75 | 8 | 3 | 28 | 24 | 10 | 11 | 19 |
| -141.08122 | -87.64878754 | -391 | 29 | 109 | 8653 | -229 | -494 | 265 | -449 | -261 | -237 | -474 | 69 | 8 | 2 | 31 | 23 | 8 | 8 | 18 |
| -136.2163551 | -87.64878754 | -393 | 29 | 108 | 8719 | -230 | -497 | 267 | -436 | -262 | -238 | -477 | 64 | 7 | 2 | 29 | 21 | 8 | 8 | 16 |
| -131.3514902 | -87.64878754 | -397 | 29 | 107 | 8801 | -233 | -502 | 269 | -440 | -265 | -240 | -481 | 63 | 7 | 2 | 30 | 21 | 8 | 8 | 16 |
| -126.4866254 | -87.64878754 | -401 | 28 | 105 | 8898 | -235 | -506 | 271 | -461 | -267 | -242 | -486 | 62 | 7 | 2 | 29 | 21 | 8 | 8 | 16 |
| -121.6217605 | -87.64878754 | -403 | 28 | 103 | 9024 | -234 | -508 | 274 | -448 | -242 | -242 | -488 | 65 | 7 | 3 | 27 | 21 | 9 | 9 | 18 |
| -116.7568956 | -87.64878754 | -404 | 28 | 100 | 9163 | -233 | -510 | 277 | -451 | -266 | -241 | -491 | 67 | 7 | 3 | 25 | 21 | 10 | 11 | 18 |
| -111.8920308 | -87.64878754 | -405 | 27 | 98 | 9292 | -232 | -512 | 280 | -469 | -302 | -239 | -492 | 71 | 8 | 4 | 24 | 23 | 12 | 12 | 18 |
| -107.0271659 | -87.64878754 | -404 | 27 | 95 | 9422 | -229 | -511 | 283 | -470 | -262 | -236 | -493 | 78 | 8 | 4 | 20 | 24 | 13 | 14 | 21 |
| -102.1623011 | -87.64878754 | -401 | 26 | 92 | 9540 | -223 | -509 | 285 | -468 | -257 | -231 | -490 | 79 | 8 | 4 | 21 | 24 | 13 | 14 | 21 |
| -97.29743619 | -87.64878754 | -399 | 26 | 90 | 9629 | -220 | -507 | 287 | -467 | -293 | -227 | -489 | 75 | 8 | 4 | 22 | 23 | 13 | 14 | 20 |
| -92.43257132 | -87.64878754 | -398 | 26 | 89 | 9709 | -218 | -507 | 289 | -467 | -251 | -225 | -488 | 67 | 7 | 3 | 25 | 21 | 10 | 11 | 18 |
| -87.56770646 | -87.64878754 | -399 | 26 | 88 | 9788 | -217 | -508 | 291 | -438 | -251 | -224 | -489 | 64 | 7 | 3 | 23 | 21 | 10 | 11 | 17 |
| -82.70284159 | -87.64878754 | -400 | 25 | 86 | 9874 | -216 | -509 | 293 | -484 | -250 | -223 | -491 | 56 | 7 | 2 | 32 | 19 | 7 | 8 | 15 |
| -77.83797673 | -87.64878754 | -401 | 25 | 85 | 9980 | -216 | -511 | 295 | -441 | -250 | -223 | -493 | 53 | 6 | 2 | 28 | 18 | 7 | 8 | 15 |
| -72.97311186 | -87.64878754 | -404 | 25 | 83 | 10095 | -216 | -514 | 298 | -490 | -292 | -224 | -496 | 49 | 6 | 2 | 31 | 17 | 7 | 8 | 14 |
| -68.108247 | -87.64878754 | -407 | 25 | 81 | 10242 | -217 | -518 | 301 | -494 | -251 | -224 | -501 | 47 | 6 | 2 | 34 | 17 | 6 | 8 | 14 |
| -63.24338213 | -87.64878754 | -409 | 24 | 80 | 10399 | -217 | -522 | 305 | -498 | -252 | -224 | -505 | 45 | 5 | 2 | 31 | 15 | 6 | 8 | 14 |
| -58.37851727 | -87.64878754 | -412 | 24 | 78 | 10584 | -216 | -526 | 309 | -455 | -251 | -224 | -509 | 44 | 5 | 2 | 30 | 15 | 6 | 8 | 12 |
| -53.5136524 | -87.64878754 | -414 | 24 | 77 | 10773 | -215 | -528 | 314 | -457 | -250 | -222 | -512 | 43 | 5 | 2 | 29 | 15 | 6 | 8 | 12 |
| -48.64878754 | -87.64878754 | -415 | 24 | 76 | 10966 | -212 | -531 | 319 | -509 | -248 | -220 | -514 | 42 | 5 | 2 | 27 | 14 | 6 | 8 | 12 |
| -43.78392267 | -87.64878754 | -416 | 24 | 75 | 11123 | -210 | -533 | 323 | -511 | -247 | -219 | -517 | 42 | 5 | 2 | 27 | 14 | 6 | 8 | 12 |
| -38.91905781 | -87.64878754 | -418 | 25 | 75 | 11244 | -210 | -536 | 326 | -514 | -246 | -218 | -520 | 40 | 5 | 2 | 28 | 14 | 6 | 8 | 11 |
| -34.05419294 | -87.64878754 | -420 | 25 | 75 | 11341 | -210 | -539 | 329 | -517 | -247 | -219 | -522 | 39 | 5 | 1 | 34 | 14 | 4 | 7 | 11 |
| -29.18932808 | -87.64878754 | -423 | 25 | 75 | 11411 | -212 | -542 | 331 | -507 | -248 | -220 | -526 | 35 | 4 | 1 | 36 | 12 | 4 | 7 | 9 |
| -24.32446321 | -87.64878754 | -426 | 25 | 75 | 11474 | -214 | -546 | 333 | -525 | -251 | -222 | -529 | 31 | 4 | 1 | 37 | 11 | 3 | 7 | 9 |
| -19.45959835 | -87.64878754 | -429 | 25 | 75 | 11556 | -216 | -550 | 335 | -493 | -253 | -224 | -533 | 28 | 3 | 1 | 36 | 9 | 3 | 5 | 8 |
| -14.59473348 | -87.64878754 | -433 | 25 | 74 | 11651 | -217 | -554 | 337 | -497 | -255 | -226 | -537 | 26 | 3 | 1 | 37 | 9 | 3 | 5 | 6 |
| -9.729868619 | -87.64878754 | -436 | 25 | 73 | 11750 | -218 | -558 | 340 | -522 | -256 | -227 | -541 | 23 | 3 | 1 | 33 | 8 | 3 | 5 | 6 |
| -4.865003754 | -87.64878754 | -438 | 25 | 72 | 11832 | -219 | -561 | 342 | -541 | -257 | -229 | -544 | 22 | 3 | 1 | 30 | 8 | 3 | 5 | 6 |
| -0.000138889 | -87.64878754 | -441 | 25 | 72 | 11909 | -221 | -565 | 344 | -508 | -259 | -230 | -548 | 20 | 2 | 0 | 37 | 6 | 1 | 4 | 6 |
| 4.864725976 | -87.64878754 | -444 | 25 | 71 | 11940 | -223 | -567 | 344 | -510 | -261 | -232 | -551 | 21 | 2 | 1 | 25 | 6 | 3 | 5 | 6 |
| 9.729590841 | -87.64878754 | -447 | 25 | 71 | 11952 | -225 | -570 | 345 | -513 | -264 | -235 | -553 | 21 | 2 | 1 | 25 | 6 | 3 | 5 | 6 |
| 14.59445571 | -87.64878754 | -447 | 24 | 71 | 11928 | -226 | -570 | 344 | -514 | -316 | -236 | -554 | 17 | 2 | 0 | 45 | 6 | 2 | 5 | 3 |
| 19.45932057 | -87.64878754 | -447 | 24 | 71 | 11888 | -227 | -570 | 343 | -514 | -547 | -236 | -553 | 18 | 2 | 1 | 33 | 6 | 3 | 5 | 3 |
| 24.32418544 | -87.64878754 | -447 | 25 | 72 | 11833 | -227 | -569 | 341 | -535 | -440 | -237 | -552 | 17 | 2 | 1 | 35 | 6 | 3 | 5 | 3 |
| 29.1890503 | -87.64878754 | -448 | 25 | 72 | 11789 | -228 | -569 | 340 | -535 | -441 | -238 | -552 | 17 | 2 | 1 | 35 | 6 | 3 | 5 | 3 |
| 34.05391517 | -87.64878754 | -450 | 25 | 73 | 11770 | -231 | -571 | 340 | -538 | -443 | -241 | -554 | 17 | 2 | 1 | 35 | 6 | 3 | 5 | 3 |
| 38.91878003 | -87.64878754 | -455 | 25 | 74 | 11785 | -236 | -576 | 340 | -542 | -447 | -245 | -559 | 17 | 2 | 1 | 35 | 6 | 3 | 5 | 3 |
| 43.78364489 | -87.64878754 | -460 | 26 | 76 | 11867 | -240 | -582 | 343 | -549 | -329 | -250 | -565 | 15 | 2 | 0 | 48 | 6 | 2 | 5 | 3 |
| 48.64850976 | -87.64878754 | -466 | 27 | 77 | 12003 | -243 | -589 | 346 | -555 | -333 | -253 | -571 | 15 | 2 | 0 | 48 | 6 | 2 | 5 | 3 |
| 53.51337462 | -87.64878754 | -470 | 27 | 78 | 12170 | -245 | -596 | 351 | -255 | -396 | -255 | -577 | 11 | 2 | 0 | 54 | 5 | 1 | 5 | 3 |
| 58.37823949 | -87.64878754 | -475 | 28 | 78 | 12343 | -247 | -602 | 355 | -547 | -338 | -257 | -583 | 11 | 2 | 0 | 54 | 5 | 1 | 3 | 3 |
| 63.24310435 | -87.64878754 | -479 | 28 | 79 | 12506 | -248 | -607 | 359 | -552 | -287 | -258 | -588 | 8 | 1 | 0 | 71 | 3 | 0 | 2 | 2 |
| 68.10796922 | -87.64878754 | -482 | 29 | 79 | 12641 | -249 | -612 | 363 | -556 | -288 | -259 | -592 | 9 | 1 | 0 | 58 | 3 | 0 | 2 | 3 |
| 72.97283408 | -87.64878754 | -484 | 29 | 79 | 12751 | -250 | -615 | 365 | -559 | -289 | -260 | -596 | 8 | 1 | 0 | 71 | 3 | 0 | 2 | 2 |
| 77.83769895 | -87.64878754 | -486 | 29 | 79 | 12825 | -250 | -618 | 367 | -562 | -290 | -261 | -598 | 10 | 2 | 0 | 66 | 5 | 0 | 2 | 3 |
| 82.70256381 | -87.64878754 | -487 | 29 | 79 | 12874 | -251 | -620 | 369 | -563 | -290 | -261 | -600 | 10 | 2 | 0 | 66 | 5 | 0 | 2 | 3 |
| 87.56742868 | -87.64878754 | -488 | 29 | 79 | 12889 | -252 | -621 | 369 | -564 | -291 | -262 | -601 | 12 | 2 | 0 | 71 | 6 | 0 | 2 | 3 |
| 92.43229354 | -87.64878754 | -488 | 29 | 79 | 12857 | -253 | -622 | 369 | -563 | -291 | -263 | -601 | 12 | 2 | 0 | 71 | 6 | 0 | 2 | 3 |
| 97.29715841 | -87.64878754 | -487 | 29 | 80 | 12782 | -254 | -621 | 368 | -561 | -292 | -263 | -600 | 12 | 2 | 0 | 71 | 6 | 0 | 2 | 3 |
| 102.1620233 | -87.64878754 | -485 | 30 | 81 | 12651 | -254 | -619 | 365 | -558 | -292 | -264 | -598 | 12 | 2 | 0 | 71 | 6 | 0 | 2 | 3 |
| 107.0268881 | -87.64878754 | -483 | 30 | 83 | 12481 | -255 | -616 | 361 | -554 | -292 | -264 | -594 | 14 | 2 | 0 | 69 | 6 | 0 | 2 | 3 |
| 111.891753 | -87.64878754 | -479 | 30 | 84 | 12286 | -255 | -611 | 356 | -548 | -291 | -264 | -590 | 14 | 2 | 0 | 69 | 6 | 0 | 2 | 3 |
| 116.7566179 | -87.64878754 | -475 | 30 | 86 | 12076 | -254 | -606 | 352 | -542 | -290 | -263 | -584 | 13 | 2 | 0 | 70 | 6 | 0 | 2 | 3 |
| 121.6214827 | -87.64878754 | -470 | 30 | 88 | 11849 | -253 | -599 | 346 | -535 | -340 | -262 | -578 | 15 | 2 | 0 | 58 | 6 | 1 | 2 | 3 |
| 126.4863476 | -87.64878754 | -465 | 31 | 90 | 11603 | -253 | -593 | 340 | -528 | -288 | -262 | -571 | 16 | 2 | 0 | 47 | 6 | 1 | 2 | 3 |
| 131.3512125 | -87.64878754 | -460 | 31 | 92 | 11342 | -252 | -586 | 334 | -452 | -287 | -261 | -565 | 18 | 3 | 0 | 51 | 7 | 1 | 2 | 3 |
| 136.2160773 | -87.64878754 | -455 | 31 | 94 | 11068 | -252 | -580 | 327 | -513 | -334 | -261 | -558 | 21 | 3 | 1 | 47 | 9 | 3 | 3 | 3 |
| 141.0809422 | -87.64878754 | -450 | 31 | 97 | 10784 | -252 | -572 | 320 | -505 | -331 | -260 | -551 | 21 | 3 | 1 | 47 | 9 | 3 | 3 | 3 |
| 145.9458071 | -87.64878754 | -444 | 31 | 99 | 10494 | -251 | -565 | 313 | -497 | -329 | -260 | -543 | 21 | 3 | 1 | 47 | 9 | 3 | 3 | 3 |
| 150.8106719 | -87.64878754 | -439 | 31 | 102 | 10199 | -251 | -557 | 306 | -430 | -327 | -259 | -536 | 19 | 3 | 1 | 40 | 7 | 3 | 3 | 3 |
| 155.6755368 | -87.64878754 | -434 | 31 | 104 | 9922 | -251 | -550 | 299 | -483 | -325 | -259 | -529 | 17 | 2 | 1 | 35 | 6 | 3 | 3 | 3 |
| 160.5404017 | -87.64878754 | -430 | 31 | 106 | 9649 | -251 | -543 | 292 | -477 | -323 | -260 | -522 | 17 | 2 | 1 | 35 | 6 | 3 | 3 | 3 |
| 165.4052665 | -87.64878754 | -426 | 30 | 107 | 9412 | -252 | -537 | 285 | -472 | -322 | -260 | -517 | 17 | 2 | 1 | 35 | 6 | 3 | 3 | 3 |
| 170.2701314 | -87.64878754 | -422 | 30 | 108 | 9194 | -252 | -532 | 280 | -414 | -511 | -260 | -511 | 19 | 2 | 1 | 31 | 6 | 3 | 3 | 3 |
| 175.1349962 | -87.64878754 | -419 | 29 | 108 | 9006 | -252 | -526 | 274 | -462 | -283 | -259 | -506 | 23 | 3 | 1 | 40 | 9 | 3 | 3 | 4 |
| -175.135274 | -85.29743619 | -314 | 34 | 130 | 8212 | -159 | -418 | 259 | -368 | -190 | -171 | -397 | 94 | 11 | 5 | 25 | 31 | 15 | 15 | 20 |
| -170.2704092 | -85.29743619 | -319 | 35 | 123 | 8932 | -151 | -432 | 281 | -377 | -186 | -163 | -409 | 116 | 13 | 6 | 22 | 37 | 19 | 20 | 26 |
| -165.4055443 | -85.29743619 | -295 | 38 | 130 | 9015 | -124 | -413 | 289 | -353 | -161 | -137 | -387 | 172 | 20 | 8 | 26 | 57 | 26 | 26 | 39 |
| -160.5406794 | -85.29743619 | -231 | 37 | 130 | 8794 | -60 | -343 | 283 | -289 | -75 | -75 | -319 | 228 | 26 | 10 | 27 | 75 | 32 | 32 | 55 |
| -155.6758146 | -85.29743619 | -214 | 36 | 134 | 8315 | -52 | -323 | 271 | -268 | -67 | -67 | -299 | 221 | 26 | 9 | 30 | 75 | 27 | 27 | 55 |
| -150.8109497 | -85.29743619 | -196 | 35 | 139 | 7640 | -47 | -299 | 252 | -244 | -62 | -62 | -275 | 190 | 23 | 7 | 32 | 66 | 23 | 23 | 47 |
| -145.9460848 | -85.29743619 | -190 | 35 | 145 | 7297 | -49 | -291 | 242 | -235 | -63 | -63 | -268 | 161 | 20 | 7 | 31 | 56 | 21 | 21 | 41 |
| -141.08122 | -85.29743619 | -201 | 35 | 146 | 7147 | -64 | -301 | 237 | -245 | -94 | -76 | -277 | 159 | 20 | 7 | 30 | 56 | 22 | 23 | 38 |
| -136.2163551 | -85.29743619 | -212 | 34 | 144 | 7182 | -75 | -310 | 235 | -257 | -105 | -86 | -287 | 162 | 20 | 8 | 27 | 56 | 25 | 26 | 38 |
| -131.3514902 | -85.29743619 | -229 | 34 | 144 | 7293 | -90 | -327 | 237 | -275 | -120 | -101 | -304 | 173 | 21 | 8 | 26 | 59 | 26 | 28 | 41 |
| -126.4866254 | -85.29743619 | -249 | 34 | 145 | 7232 | -112 | -346 | 234 | -294 | -142 | -122 | -323 | 174 | 21 | 8 | 26 | 59 | 26 | 28 | 43 |
| -121.6217605 | -85.29743619 | -274 | 34 | 146 | 7096 | -140 | -371 | 230 | -318 | -170 | -149 | -348 | 172 | 20 | 9 | 23 | 57 | 28 | 29 | 42 |
| -116.7568956 | -85.29743619 | -300 | 34 | 143 | 7308 | -162 | -398 | 235 | -346 | -193 | -171 | -375 | 164 | 19 | 8 | 23 | 54 | 26 | 28 | 42 |
| -111.8920308 | -85.29743619 | -332 | 33 | 137 | 7660 | -188 | -431 | 244 | -381 | -220 | -196 | -409 | 150 | 17 | 7 | 23 | 49 | 23 | 25 | 38 |
| -107.0271659 | -85.29743619 | -354 | 34 | 131 | 8196 | -199 | -456 | 257 | -407 | -233 | -208 | -435 | 126 | 14 | 6 | 24 | 41 | 19 | 20 | 34 |
| -102.1623011 | -85.29743619 | -364 | 33 | 121 | 8825 | -198 | -470 | 273 | -423 | -233 | -207 | -450 | 106 | 12 | 5 | 24 | 35 | 16 | 17 | 30 |
| -97.29743619 | -85.29743619 | -369 | 31 | 109 | 9300 | -194 | -479 | 285 | -431 | -263 | -203 | -459 | 95 | 10 | 5 | 21 | 29 | 16 | 17 | 27 |
| -92.43257132 | -85.29743619 | -363 | 28 | 99 | 9346 | -187 | -471 | 284 | -426 | -257 | -196 | -453 | 101 | 10 | 5 | 21 | 30 | 16 | 18 | 29 |
| -87.56770646 | -85.29743619 | -329 | 26 | 91 | 9367 | -152 | -434 | 283 | -411 | -225 | -162 | -419 | 99 | 10 | 5 | 20 | 30 | 16 | 18 | 29 |
| -82.70284159 | -85.29743619 | -313 | 25 | 89 | 9200 | -138 | -416 | 278 | -393 | -211 | -149 | -401 | 95 | 10 | 5 | 23 | 30 | 15 | 15 | 28 |
| -77.83797673 | -85.29743619 | -306 | 25 | 92 | 9033 | -135 | -409 | 274 | -384 | -145 | -145 | -392 | 82 | 10 | 3 | 31 | 29 | 11 | 11 | 25 |
| -72.97311186 | -85.29743619 | -302 | 26 | 96 | 8905 | -133 | -404 | 271 | -379 | -144 | -144 | -386 | 68 | 9 | 2 | 39 | 26 | 6 | 6 | 22 |
| -68.108247 | -85.29743619 | -303 | 26 | 98 | 8809 | -136 | -405 | 269 | -338 | -147 | -147 | -387 | 63 | 8 | 2 | 38 | 24 | 6 | 6 | 20 |
| -63.24338213 | -85.29743619 | -307 | 27 | 99 | 8787 | -142 | -410 | 268 | -343 | -152 | -152 | -391 | 63 | 8 | 2 | 38 | 24 | 6 | 6 | 20 |
| -58.37851727 | -85.29743619 | -309 | 27 | 99 | 9014 | -140 | -414 | 274 | -388 | -174 | -150 | -395 | 58 | 8 | 2 | 34 | 22 | 8 | 8 | 17 |
| -53.5136524 | -85.29743619 | -322 | 27 | 98 | 9216 | -149 | -428 | 279 | -402 | -183 | -159 | -408 | 56 | 7 | 2 | 29 | 20 | 8 | 8 | 17 |
| -48.64878754 | -85.29743619 | -329 | 28 | 99 | 9397 | -153 | -437 | 284 | -367 | -188 | -163 | -418 | 51 | 6 | 2 | 29 | 18 | 7 | 8 | 14 |
| -43.78392267 | -85.29743619 | -336 | 28 | 99 | 9510 | -158 | -444 | 287 | -419 | -193 | -168 | -425 | 51 | 6 | 2 | 26 | 17 | 7 | 10 | 14 |
| -38.91905781 | -85.29743619 | -343 | 29 | 99 | 9655 | -163 | -453 | 290 | -428 | -198 | -173 | -434 | 43 | 5 | 2 | 21 | 14 | 7 | 10 | 11 |
| -34.05419294 | -85.29743619 | -354 | 29 | 98 | 9815 | -171 | -465 | 294 | -402 | -245 | -181 | -446 | 38 | 4 | 2 | 22 | 12 | 7 | 8 | 9 |
| -29.18932808 | -85.29743619 | -366 | 29 | 98 | 9967 | -180 | -477 | 297 | -415 | -216 | -190 | -460 | 33 | 3 | 2 | 16 | 9 | 6 | 8 | 9 |
| -24.32446321 | -85.29743619 | -379 | 29 | 96 | 10182 | -189 | -493 | 303 | -429 | -469 | -199 | -475 | 30 | 3 | 2 | 20 | 9 | 6 | 8 | 6 |
| -19.45959835 | -85.29743619 | -391 | 29 | 93 | 10507 | -195 | -507 | 312 | -443 | -274 | -206 | -490 | 26 | 3 | 1 | 26 | 9 | 5 | 8 | 6 |
| -14.59473348 | -85.29743619 | -402 | 29 | 90 | 10899 | -199 | -522 | 323 | -210 | -281 | -210 | -504 | 24 | 3 | 1 | 20 | 8 | 5 | 8 | 6 |
| -9.729868619 | -85.29743619 | -414 | 29 | 87 | 11321 | -204 | -538 | 334 | -215 | -346 | -215 | -520 | 21 | 3 | 1 | 34 | 7 | 3 | 7 | 6 |
| -4.865003754 | -85.29743619 | -426 | 29 | 83 | 11746 | -207 | -552 | 345 | -487 | -526 | -219 | -534 | 16 | 2 | 1 | 35 | 6 | 3 | 5 | 3 |
| -0.000138889 | -85.29743619 | -435 | 29 | 81 | 12128 | -210 | -565 | 355 | -500 | -538 | -223 | -547 | 16 | 2 | 1 | 35 | 6 | 3 | 5 | 3 |
| 4.864725976 | -85.29743619 | -442 | 28 | 78 | 12395 | -212 | -574 | 362 | -225 | -429 | -225 | -556 | 13 | 2 | 1 | 26 | 5 | 3 | 5 | 3 |
| 9.729590841 | -85.29743619 | -445 | 28 | 77 | 12528 | -212 | -577 | 365 | -226 | -306 | -226 | -560 | 12 | 2 | 0 | 41 | 5 | 2 | 5 | 3 |
| 14.59445571 | -85.29743619 | -445 | 28 | 77 | 12462 | -212 | -575 | 363 | -513 | -307 | -226 | -558 | 11 | 1 | 0 | 30 | 3 | 1 | 3 | 3 |
| 19.45932057 | -85.29743619 | -441 | 28 | 78 | 12185 | -212 | -567 | 356 | -226 | -429 | -226 | -551 | 13 | 2 | 1 | 26 | 5 | 3 | 5 | 3 |
| 24.32418544 | -85.29743619 | -436 | 27 | 79 | 11799 | -214 | -559 | 345 | -228 | -425 | -228 | -543 | 14 | 2 | 1 | 32 | 6 | 3 | 6 | 3 |
| 29.1890503 | -85.29743619 | -435 | 27 | 81 | 11470 | -219 | -555 | 336 | -233 | -538 | -233 | -539 | 13 | 2 | 0 | 46 | 6 | 1 | 6 | 3 |
| 34.05391517 | -85.29743619 | -440 | 28 | 83 | 11366 | -226 | -559 | 333 | -239 | -542 | -239 | -542 | 13 | 2 | 0 | 46 | 6 | 1 | 6 | 3 |
| 38.91878003 | -85.29743619 | -451 | 29 | 86 | 11460 | -236 | -571 | 335 | -249 | -554 | -249 | -554 | 10 | 2 | 0 | 82 | 6 | 0 | 6 | 0 |
| 43.78364489 | -85.29743619 | -462 | 30 | 88 | 11752 | -242 | -585 | 343 | -255 | -449 | -255 | -566 | 5 | 2 | 0 | 182 | 6 | 0 | 6 | 0 |
| 48.64850976 | -85.29743619 | -471 | 32 | 89 | 12074 | -245 | -598 | 352 | -259 | -458 | -259 | -578 | 2 | 1 | 0 | 224 | 3 | 0 | 3 | 0 |
| 53.51337462 | -85.29743619 | -478 | 32 | 90 | 12296 | -249 | -607 | 358 | -262 | -466 | -262 | -587 | 2 | 1 | 0 | 224 | 3 | 0 | 3 | 0 |
| 58.37823949 | -85.29743619 | -482 | 32 | 89 | 12392 | -251 | -612 | 361 | -265 | -470 | -265 | -591 | 3 | 1 | 0 | 173 | 3 | 0 | 3 | 0 |
| 63.24310435 | -85.29743619 | -484 | 32 | 89 | 12382 | -254 | -614 | 360 | -267 | -472 | -267 | -593 | 3 | 1 | 0 | 173 | 3 | 0 | 3 | 0 |
| 68.10796922 | -85.29743619 | -484 | 32 | 88 | 12331 | -255 | -613 | 358 | -554 | -473 | -268 | -593 | 2 | 1 | 0 | 224 | 2 | 0 | 2 | 0 |
| 72.97283408 | -85.29743619 | -483 | 31 | 88 | 12275 | -255 | -611 | 357 | -553 | -472 | -268 | -591 | 2 | 1 | 0 | 224 | 2 | 0 | 2 | 0 |
| 77.83769895 | -85.29743619 | -482 | 31 | 87 | 12233 | -254 | -610 | 355 | -552 | -471 | -267 | -590 | 3 | 1 | 0 | 173 | 2 | 0 | 2 | 0 |
| 82.70256381 | -85.29743619 | -481 | 30 | 85 | 12214 | -254 | -608 | 354 | -573 | -470 | -267 | -589 | 5 | 1 | 0 | 118 | 3 | 0 | 2 | 0 |
| 87.56742868 | -85.29743619 | -480 | 30 | 84 | 12207 | -252 | -606 | 354 | -549 | -582 | -265 | -588 | 6 | 1 | 0 | 100 | 3 | 0 | 2 | 0 |
| 92.43229354 | -85.29743619 | -478 | 29 | 83 | 12233 | -251 | -605 | 355 | -547 | -582 | -264 | -587 | 6 | 1 | 0 | 100 | 3 | 0 | 2 | 0 |
| 97.29715841 | -85.29743619 | -477 | 29 | 82 | 12261 | -249 | -604 | 355 | -546 | -582 | -262 | -586 | 6 | 1 | 0 | 100 | 3 | 0 | 2 | 0 |
| 102.1620233 | -85.29743619 | -475 | 29 | 82 | 12277 | -247 | -603 | 355 | -544 | -581 | -260 | -585 | 6 | 1 | 0 | 100 | 3 | 0 | 2 | 0 |
| 107.0268881 | -85.29743619 | -472 | 29 | 82 | 12235 | -245 | -600 | 354 | -540 | -578 | -258 | -582 | 6 | 1 | 0 | 100 | 3 | 0 | 2 | 0 |
| 111.891753 | -85.29743619 | -468 | 29 | 83 | 12148 | -243 | -595 | 352 | -534 | -573 | -256 | -578 | 5 | 1 | 0 | 118 | 3 | 0 | 2 | 0 |
| 116.7566179 | -85.29743619 | -461 | 30 | 85 | 11981 | -240 | -589 | 349 | -526 | -566 | -253 | -571 | 4 | 1 | 0 | 141 | 3 | 0 | 2 | 0 |
| 121.6214827 | -85.29743619 | -454 | 30 | 88 | 11724 | -238 | -581 | 343 | -516 | -557 | -250 | -562 | 4 | 1 | 0 | 141 | 3 | 0 | 2 | 0 |
| 126.4863476 | -85.29743619 | -445 | 31 | 91 | 11434 | -234 | -571 | 337 | -505 | -547 | -246 | -552 | 4 | 1 | 0 | 141 | 3 | 0 | 2 | 0 |
| 131.3512125 | -85.29743619 | -436 | 32 | 95 | 11154 | -231 | -561 | 331 | -493 | -537 | -243 | -541 | 5 | 1 | 0 | 118 | 3 | 0 | 2 | 0 |
| 136.2160773 | -85.29743619 | -426 | 32 | 100 | 10822 | -227 | -550 | 323 | -480 | -524 | -238 | -529 | 5 | 1 | 0 | 118 | 3 | 0 | 2 | 0 |
| 141.0809422 | -85.29743619 | -415 | 33 | 105 | 10405 | -223 | -536 | 313 | -466 | -509 | -234 | -515 | 5 | 1 | 0 | 118 | 3 | 0 | 2 | 0 |
| 145.9458071 | -85.29743619 | -403 | 33 | 110 | 9922 | -219 | -520 | 301 | -294 | -493 | -231 | -499 | 6 | 1 | 0 | 100 | 3 | 0 | 2 | 0 |
| 150.8106719 | -85.29743619 | -391 | 33 | 114 | 9441 | -215 | -504 | 289 | -287 | -483 | -227 | -483 | 9 | 1 | 0 | 58 | 3 | 0 | 3 | 0 |
| 155.6755368 | -85.29743619 | -377 | 32 | 114 | 9083 | -207 | -485 | 279 | -278 | -465 | -218 | -465 | 10 | 1 | 0 | 45 | 3 | 1 | 3 | 1 |
| 160.5404017 | -85.29743619 | -355 | 31 | 114 | 8826 | -188 | -460 | 272 | -349 | -440 | -200 | -440 | 9 | 1 | 0 | 58 | 3 | 1 | 1 | 1 |
| 165.4052665 | -85.29743619 | -362 | 31 | 118 | 8544 | -202 | -465 | 264 | -353 | -446 | -213 | -446 | 13 | 2 | 0 | 46 | 5 | 1 | 3 | 1 |
| 170.2701314 | -85.29743619 | -334 | 33 | 124 | 8644 | -171 | -439 | 268 | -376 | -203 | -183 | -419 | 24 | 3 | 1 | 41 | 9 | 3 | 3 | 4 |
| 175.1349962 | -85.29743619 | -348 | 36 | 136 | 8451 | -190 | -456 | 265 | -403 | -200 | -200 | -433 | 49 | 6 | 2 | 32 | 18 | 6 | 6 | 10 |
| -175.135274 | -82.94608483 | -241 | 52 | 151 | 10170 | -49 | -394 | 345 | -305 | -69 | -69 | -355 | 113 | 14 | 4 | 36 | 42 | 14 | 14 | 25 |
| -170.2704092 | -82.94608483 | -244 | 51 | 148 | 10203 | -51 | -394 | 344 | -309 | -70 | -70 | -356 | 91 | 12 | 3 | 37 | 35 | 11 | 11 | 21 |
| -165.4055443 | -82.94608483 | -252 | 50 | 143 | 10479 | -53 | -403 | 350 | -321 | -74 | -74 | -365 | 76 | 11 | 3 | 39 | 30 | 10 | 10 | 18 |
| -160.5406794 | -82.94608483 | -263 | 50 | 137 | 10962 | -55 | -417 | 362 | -336 | -98 | -76 | -380 | 62 | 9 | 2 | 40 | 24 | 8 | 9 | 15 |
| -155.6758146 | -82.94608483 | -264 | 49 | 135 | 10932 | -56 | -417 | 362 | -335 | -100 | -77 | -381 | 54 | 8 | 2 | 38 | 21 | 8 | 9 | 12 |
| -150.8109497 | -82.94608483 | -251 | 47 | 141 | 10014 | -58 | -395 | 337 | -312 | -351 | -79 | -360 | 49 | 7 | 2 | 37 | 19 | 7 | 9 | 12 |
| -145.9460848 | -82.94608483 | -243 | 48 | 146 | 9610 | -57 | -383 | 326 | -302 | -340 | -78 | -349 | 48 | 7 | 2 | 37 | 18 | 7 | 9 | 12 |
| -141.08122 | -82.94608483 | -251 | 47 | 152 | 9086 | -74 | -386 | 312 | -305 | -342 | -95 | -352 | 58 | 8 | 3 | 33 | 21 | 9 | 12 | 13 |
| -136.2163551 | -82.94608483 | -237 | 46 | 156 | 8596 | -69 | -367 | 298 | -288 | -323 | -89 | -333 | 65 | 9 | 3 | 35 | 23 | 10 | 14 | 16 |
| -131.3514902 | -82.94608483 | -225 | 45 | 159 | 8112 | -66 | -349 | 283 | -272 | -177 | -85 | -316 | 73 | 10 | 3 | 33 | 26 | 11 | 17 | 16 |
| -126.4866254 | -82.94608483 | -221 | 43 | 163 | 7603 | -72 | -337 | 265 | -265 | -177 | -89 | -306 | 91 | 12 | 4 | 32 | 32 | 14 | 20 | 20 |
| -121.6217605 | -82.94608483 | -221 | 41 | 164 | 7127 | -82 | -329 | 247 | -262 | -182 | -98 | -301 | 121 | 15 | 6 | 29 | 41 | 20 | 23 | 30 |
| -116.7568956 | -82.94608483 | -229 | 38 | 164 | 6724 | -99 | -331 | 232 | -267 | -194 | -113 | -304 | 151 | 18 | 8 | 25 | 50 | 26 | 29 | 38 |
| -111.8920308 | -82.94608483 | -251 | 37 | 163 | 6633 | -123 | -350 | 227 | -288 | -218 | -135 | -325 | 174 | 20 | 10 | 23 | 57 | 32 | 35 | 42 |
| -107.0271659 | -82.94608483 | -282 | 38 | 162 | 6899 | -150 | -384 | 233 | -322 | -248 | -162 | -358 | 155 | 17 | 9 | 20 | 49 | 29 | 32 | 37 |
| -102.1623011 | -82.94608483 | -309 | 39 | 158 | 7405 | -167 | -414 | 247 | -363 | -272 | -179 | -388 | 122 | 13 | 7 | 20 | 39 | 23 | 24 | 32 |
| -97.29743619 | -82.94608483 | -316 | 39 | 151 | 7985 | -163 | -425 | 262 | -366 | -223 | -176 | -399 | 109 | 11 | 6 | 23 | 33 | 18 | 18 | 32 |
| -92.43257132 | -82.94608483 | -311 | 36 | 136 | 8195 | -153 | -418 | 264 | -369 | -166 | -166 | -394 | 114 | 12 | 5 | 25 | 35 | 16 | 16 | 35 |
| -87.56770646 | -82.94608483 | -293 | 30 | 118 | 8052 | -138 | -393 | 255 | -368 | -201 | -150 | -375 | 109 | 12 | 5 | 25 | 36 | 16 | 18 | 34 |
| -82.70284159 | -82.94608483 | -274 | 27 | 107 | 7929 | -122 | -370 | 249 | -347 | -184 | -134 | -355 | 108 | 12 | 5 | 27 | 36 | 16 | 18 | 33 |
| -77.83797673 | -82.94608483 | -257 | 26 | 103 | 8023 | -103 | -353 | 250 | -292 | -166 | -116 | -338 | 101 | 12 | 5 | 30 | 36 | 15 | 15 | 31 |
| -72.97311186 | -82.94608483 | -247 | 26 | 102 | 8273 | -88 | -346 | 258 | -282 | -154 | -102 | -330 | 96 | 12 | 4 | 35 | 36 | 13 | 14 | 30 |
| -68.108247 | -82.94608483 | -253 | 29 | 104 | 8893 | -81 | -359 | 277 | -288 | -96 | -96 | -341 | 105 | 13 | 4 | 35 | 39 | 14 | 14 | 33 |
| -63.24338213 | -82.94608483 | -262 | 34 | 107 | 10256 | -66 | -385 | 318 | -350 | -83 | -83 | -362 | 105 | 13 | 4 | 31 | 37 | 14 | 14 | 31 |
| -58.37851727 | -82.94608483 | -268 | 38 | 114 | 10692 | -66 | -398 | 332 | -357 | -108 | -83 | -370 | 68 | 9 | 2 | 37 | 26 | 8 | 8 | 22 |
| -53.5136524 | -82.94608483 | -290 | 37 | 122 | 9751 | -106 | -410 | 304 | -373 | -179 | -121 | -384 | 58 | 7 | 3 | 24 | 20 | 10 | 11 | 17 |
| -48.64878754 | -82.94608483 | -284 | 34 | 117 | 9415 | -106 | -398 | 292 | -366 | -145 | -121 | -376 | 124 | 13 | 8 | 13 | 37 | 27 | 28 | 31 |
| -43.78392267 | -82.94608483 | -268 | 34 | 118 | 9139 | -96 | -379 | 283 | -249 | -133 | -110 | -357 | 96 | 10 | 6 | 11 | 28 | 20 | 22 | 23 |
| -38.91905781 | -82.94608483 | -291 | 33 | 120 | 8937 | -123 | -399 | 276 | -271 | -159 | -136 | -378 | 60 | 6 | 4 | 12 | 17 | 13 | 14 | 15 |
| -34.05419294 | -82.94608483 | -306 | 32 | 115 | 9172 | -134 | -415 | 281 | -287 | -170 | -147 | -395 | 52 | 6 | 3 | 17 | 16 | 11 | 13 | 12 |
| -29.18932808 | -82.94608483 | -326 | 32 | 112 | 9452 | -148 | -436 | 288 | -307 | -186 | -162 | -418 | 52 | 6 | 3 | 17 | 16 | 11 | 13 | 12 |
| -24.32446321 | -82.94608483 | -337 | 33 | 110 | 9755 | -153 | -449 | 296 | -317 | -192 | -167 | -431 | 53 | 6 | 3 | 17 | 16 | 11 | 13 | 12 |
| -19.45959835 | -82.94608483 | -341 | 33 | 107 | 10080 | -151 | -457 | 306 | -322 | -385 | -165 | -439 | 49 | 6 | 3 | 23 | 16 | 9 | 13 | 11 |
| -14.59473348 | -82.94608483 | -357 | 33 | 103 | 10497 | -159 | -477 | 318 | -338 | -404 | -174 | -459 | 40 | 5 | 2 | 22 | 13 | 7 | 11 | 9 |
| -9.729868619 | -82.94608483 | -377 | 33 | 99 | 10895 | -171 | -500 | 329 | -358 | -425 | -187 | -482 | 39 | 5 | 2 | 26 | 13 | 6 | 11 | 9 |
| -4.865003754 | -82.94608483 | -388 | 33 | 97 | 11205 | -177 | -514 | 338 | -370 | -496 | -193 | -496 | 30 | 4 | 2 | 26 | 10 | 6 | 8 | 6 |
| -0.000138889 | -82.94608483 | -400 | 33 | 96 | 11546 | -182 | -529 | 347 | -456 | -451 | -199 | -510 | 25 | 3 | 1 | 31 | 9 | 3 | 6 | 6 |
| 4.864725976 | -82.94608483 | -414 | 34 | 95 | 12025 | -187 | -547 | 361 | -474 | -517 | -205 | -528 | 17 | 2 | 1 | 35 | 6 | 3 | 6 | 3 |
| 9.729590841 | -82.94608483 | -427 | 35 | 94 | 12504 | -191 | -564 | 374 | -287 | -287 | -210 | -545 | 12 | 1 | 1 | 0 | 3 | 3 | 3 | 3 |
| 14.59445571 | -82.94608483 | -432 | 35 | 94 | 12572 | -194 | -569 | 375 | -292 | -355 | -214 | -550 | 9 | 1 | 0 | 58 | 3 | 0 | 3 | 2 |
| 19.45932057 | -82.94608483 | -425 | 34 | 94 | 12160 | -193 | -557 | 364 | -289 | -480 | -213 | -538 | 9 | 1 | 0 | 58 | 3 | 0 | 3 | 2 |
| 24.32418544 | -82.94608483 | -422 | 33 | 95 | 11721 | -198 | -549 | 351 | -290 | -531 | -217 | -531 | 8 | 1 | 0 | 71 | 3 | 0 | 3 | 0 |
| 29.1890503 | -82.94608483 | -429 | 33 | 96 | 11475 | -210 | -554 | 344 | -256 | -535 | -228 | -535 | 9 | 2 | 0 | 79 | 5 | 0 | 4 | 0 |
| 34.05391517 | -82.94608483 | -440 | 34 | 98 | 11478 | -222 | -565 | 343 | -240 | -546 | -240 | -546 | 9 | 2 | 0 | 96 | 6 | 0 | 6 | 0 |
| 38.91878003 | -82.94608483 | -457 | 35 | 100 | 11705 | -235 | -583 | 348 | -280 | -519 | -252 | -564 | 7 | 2 | 0 | 110 | 5 | 0 | 4 | 0 |
| 43.78364489 | -82.94608483 | -474 | 37 | 103 | 12099 | -245 | -604 | 358 | -335 | -539 | -263 | -583 | 3 | 1 | 0 | 173 | 3 | 0 | 3 | 0 |
| 48.64850976 | -82.94608483 | -487 | 38 | 104 | 12459 | -253 | -621 | 368 | -271 | -470 | -271 | -599 | 2 | 1 | 0 | 224 | 3 | 0 | 3 | 0 |
| 53.51337462 | -82.94608483 | -497 | 39 | 105 | 12582 | -260 | -631 | 371 | -278 | -480 | -278 | -609 | 2 | 1 | 0 | 224 | 3 | 0 | 3 | 0 |
| 58.37823949 | -82.94608483 | -502 | 39 | 106 | 12526 | -267 | -636 | 369 | -284 | -487 | -284 | -614 | 2 | 1 | 0 | 224 | 3 | 0 | 3 | 0 |
| 87.56742868 | -82.94608483 | -497 | 37 | 101 | 12346 | -266 | -631 | 365 | -283 | -482 | -283 | -609 | 1 | 1 | 0 | 332 | 2 | 0 | 2 | 0 |
| 92.43229354 | -82.94608483 | -493 | 35 | 96 | 12340 | -261 | -625 | 364 | -278 | -477 | -278 | -604 | 1 | 1 | 0 | 332 | 2 | 0 | 2 | 0 |
| 97.29715841 | -82.94608483 | -486 | 33 | 93 | 12203 | -257 | -616 | 360 | -273 | -471 | -273 | -597 | 1 | 1 | 0 | 332 | 2 | 0 | 2 | 0 |
| 102.1620233 | -82.94608483 | -479 | 32 | 91 | 12069 | -252 | -607 | 355 | -269 | -464 | -269 | -589 | 1 | 1 | 0 | 332 | 2 | 0 | 2 | 0 |
| 107.0268881 | -82.94608483 | -472 | 31 | 89 | 11927 | -247 | -598 | 351 | -264 | -458 | -264 | -581 | 1 | 1 | 0 | 332 | 2 | 0 | 2 | 0 |
| 111.891753 | -82.94608483 | -463 | 31 | 89 | 11789 | -241 | -588 | 347 | -258 | -450 | -258 | -571 | 1 | 1 | 0 | 332 | 2 | 0 | 2 | 0 |
| 116.7566179 | -82.94608483 | -455 | 30 | 88 | 11718 | -235 | -579 | 345 | -252 | -442 | -252 | -563 | 1 | 1 | 0 | 332 | 2 | 0 | 2 | 0 |
| 121.6214827 | -82.94608483 | -447 | 30 | 88 | 11612 | -229 | -571 | 342 | -245 | -435 | -245 | -554 | 1 | 1 | 0 | 332 | 2 | 0 | 2 | 0 |
| 126.4863476 | -82.94608483 | -437 | 30 | 90 | 11449 | -222 | -561 | 338 | -239 | -426 | -239 | -544 | 1 | 1 | 0 | 332 | 2 | 0 | 2 | 0 |
| 131.3512125 | -82.94608483 | -425 | 31 | 93 | 11171 | -215 | -547 | 332 | -507 | -414 | -232 | -530 | 2 | 1 | 0 | 224 | 2 | 0 | 2 | 0 |
| 136.2160773 | -82.94608483 | -410 | 32 | 96 | 10936 | -204 | -531 | 327 | -490 | -400 | -221 | -513 | 2 | 1 | 0 | 224 | 2 | 0 | 2 | 0 |
| 141.0809422 | -82.94608483 | -396 | 33 | 102 | 10647 | -194 | -515 | 321 | -451 | -491 | -212 | -496 | 4 | 1 | 0 | 141 | 3 | 0 | 2 | 0 |
| 145.9458071 | -82.94608483 | -384 | 34 | 110 | 10178 | -190 | -500 | 310 | -273 | -481 | -207 | -481 | 7 | 1 | 0 | 85 | 3 | 0 | 2 | 0 |
| 150.8106719 | -82.94608483 | -370 | 35 | 117 | 9778 | -182 | -482 | 300 | -439 | -463 | -200 | -463 | 11 | 2 | 0 | 70 | 6 | 0 | 3 | 0 |
| 155.6755368 | -82.94608483 | -333 | 36 | 122 | 9564 | -147 | -444 | 296 | -383 | -424 | -166 | -424 | 24 | 3 | 1 | 35 | 9 | 3 | 6 | 3 |
| 160.5404017 | -82.94608483 | -328 | 38 | 130 | 9318 | -149 | -444 | 295 | -392 | -188 | -166 | -421 | 61 | 8 | 3 | 35 | 24 | 9 | 9 | 12 |
| 165.4052665 | -82.94608483 | -282 | 45 | 138 | 9999 | -90 | -414 | 324 | -347 | -108 | -108 | -384 | 109 | 14 | 5 | 33 | 41 | 15 | 15 | 22 |
| 170.2701314 | -82.94608483 | -226 | 44 | 140 | 9518 | -42 | -358 | 316 | -287 | -61 | -61 | -328 | 124 | 16 | 5 | 36 | 47 | 16 | 16 | 26 |
| 175.1349962 | -82.94608483 | -244 | 51 | 148 | 10345 | -47 | -395 | 348 | -307 | -66 | -66 | -358 | 142 | 18 | 5 | 38 | 54 | 16 | 16 | 31 |
| -175.135274 | -80.59473348 | -278 | 49 | 136 | 11019 | -71 | -435 | 364 | -351 | -117 | -93 | -399 | 80 | 11 | 4 | 31 | 30 | 12 | 14 | 17 |
| -170.2704092 | -80.59473348 | -283 | 49 | 133 | 11107 | -75 | -442 | 367 | -357 | -98 | -98 | -405 | 82 | 11 | 4 | 30 | 30 | 13 | 13 | 20 |
| -165.4055443 | -80.59473348 | -285 | 48 | 130 | 11176 | -75 | -446 | 370 | -358 | -402 | -100 | -409 | 75 | 11 | 3 | 34 | 29 | 13 | 13 | 18 |
| -160.5406794 | -80.59473348 | -283 | 48 | 127 | 11402 | -69 | -448 | 379 | -354 | -117 | -94 | -411 | 70 | 10 | 3 | 33 | 26 | 11 | 12 | 18 |
| -155.6758146 | -80.59473348 | -280 | 51 | 129 | 11776 | -59 | -452 | 393 | -350 | -110 | -84 | -413 | 77 | 11 | 3 | 36 | 29 | 11 | 14 | 21 |
| -150.8109497 | -80.59473348 | -259 | 51 | 139 | 10804 | -52 | -421 | 368 | -321 | -102 | -77 | -381 | 87 | 12 | 3 | 34 | 32 | 13 | 15 | 24 |
| -145.9460848 | -80.59473348 | -244 | 51 | 147 | 9972 | -50 | -396 | 346 | -310 | -98 | -75 | -358 | 89 | 12 | 3 | 37 | 34 | 13 | 15 | 25 |
| -141.08122 | -80.59473348 | -248 | 52 | 158 | 9342 | -64 | -394 | 330 | -309 | -196 | -89 | -356 | 88 | 12 | 3 | 37 | 34 | 13 | 15 | 25 |
| -136.2163551 | -80.59473348 | -226 | 50 | 170 | 8045 | -64 | -356 | 293 | -297 | -182 | -88 | -320 | 91 | 12 | 3 | 35 | 35 | 12 | 20 | 26 |
| -131.3514902 | -80.59473348 | -223 | 47 | 176 | 7343 | -76 | -344 | 269 | -289 | -184 | -97 | -310 | 83 | 11 | 3 | 34 | 32 | 11 | 22 | 23 |
| -126.4866254 | -80.59473348 | -235 | 46 | 176 | 7260 | -90 | -353 | 263 | -283 | -197 | -110 | -320 | 73 | 9 | 3 | 29 | 25 | 10 | 20 | 20 |
| -121.6217605 | -80.59473348 | -248 | 45 | 176 | 7159 | -106 | -363 | 257 | -297 | -212 | -124 | -331 | 87 | 10 | 4 | 26 | 30 | 13 | 23 | 23 |
| -116.7568956 | -80.59473348 | -254 | 44 | 177 | 6953 | -117 | -364 | 247 | -302 | -220 | -134 | -334 | 103 | 12 | 5 | 24 | 34 | 17 | 23 | 27 |
| -111.8920308 | -80.59473348 | -268 | 43 | 181 | 6803 | -134 | -374 | 240 | -316 | -235 | -150 | -345 | 137 | 15 | 8 | 22 | 45 | 25 | 30 | 36 |
| -107.0271659 | -80.59473348 | -281 | 43 | 182 | 6763 | -147 | -384 | 236 | -319 | -249 | -163 | -356 | 140 | 16 | 8 | 22 | 46 | 26 | 27 | 36 |
| -102.1623011 | -80.59473348 | -301 | 43 | 180 | 6923 | -164 | -403 | 239 | -342 | -180 | -180 | -376 | 123 | 14 | 7 | 23 | 40 | 22 | 22 | 34 |
| -97.29743619 | -80.59473348 | -318 | 44 | 178 | 7175 | -175 | -422 | 246 | -369 | -192 | -192 | -394 | 97 | 10 | 5 | 22 | 30 | 16 | 16 | 29 |
| -92.43257132 | -80.59473348 | -307 | 41 | 171 | 7079 | -166 | -408 | 243 | -356 | -224 | -182 | -381 | 105 | 11 | 6 | 19 | 32 | 18 | 19 | 31 |
| -87.56770646 | -80.59473348 | -282 | 37 | 155 | 7165 | -140 | -382 | 241 | -353 | -198 | -156 | -358 | 127 | 13 | 7 | 18 | 39 | 22 | 24 | 37 |
| -82.70284159 | -80.59473348 | -260 | 36 | 139 | 7730 | -108 | -363 | 255 | -333 | -170 | -124 | -341 | 148 | 15 | 8 | 20 | 45 | 25 | 28 | 42 |
| -77.83797673 | -80.59473348 | -233 | 33 | 127 | 7946 | -77 | -335 | 259 | -306 | -142 | -94 | -317 | 165 | 19 | 9 | 25 | 56 | 27 | 27 | 49 |
| -72.97311186 | -80.59473348 | -238 | 36 | 124 | 8908 | -63 | -351 | 288 | -317 | -82 | -82 | -330 | 142 | 18 | 6 | 31 | 52 | 18 | 18 | 44 |
| -68.108247 | -80.59473348 | -273 | 41 | 119 | 10710 | -68 | -409 | 341 | -362 | -87 | -87 | -381 | 118 | 16 | 5 | 33 | 45 | 15 | 15 | 36 |
| -63.24338213 | -80.59473348 | -281 | 40 | 113 | 11177 | -71 | -425 | 354 | -372 | -116 | -89 | -394 | 111 | 14 | 4 | 33 | 40 | 14 | 14 | 34 |
| -58.37851727 | -80.59473348 | -282 | 41 | 112 | 11489 | -68 | -432 | 363 | -374 | -113 | -86 | -398 | 93 | 12 | 3 | 35 | 34 | 11 | 11 | 29 |
| -53.5136524 | -80.59473348 | -273 | 41 | 120 | 10666 | -74 | -415 | 341 | -359 | -117 | -91 | -381 | 86 | 11 | 3 | 34 | 31 | 10 | 11 | 25 |
| -48.64878754 | -80.59473348 | -258 | 39 | 123 | 9955 | -71 | -390 | 319 | -340 | -112 | -88 | -358 | 79 | 10 | 3 | 31 | 28 | 10 | 11 | 22 |
| -43.78392267 | -80.59473348 | -255 | 42 | 130 | 9970 | -67 | -387 | 320 | -340 | -107 | -87 | -355 | 85 | 10 | 3 | 25 | 29 | 11 | 15 | 25 |
| -38.91905781 | -80.59473348 | -244 | 39 | 128 | 9497 | -62 | -369 | 307 | -324 | -101 | -86 | -340 | 88 | 10 | 3 | 25 | 29 | 12 | 13 | 26 |
| -34.05419294 | -80.59473348 | -223 | 37 | 133 | 8652 | -54 | -335 | 281 | -297 | -93 | -77 | -310 | 88 | 9 | 3 | 23 | 27 | 13 | 13 | 26 |
| -29.18932808 | -80.59473348 | -247 | 36 | 134 | 8252 | -87 | -353 | 265 | -277 | -107 | -107 | -331 | 53 | 5 | 3 | 17 | 15 | 9 | 9 | 15 |
| -24.32446321 | -80.59473348 | -279 | 36 | 135 | 8411 | -116 | -385 | 269 | -255 | -156 | -135 | -365 | 62 | 7 | 3 | 22 | 19 | 10 | 11 | 17 |
| -19.45959835 | -80.59473348 | -288 | 37 | 136 | 8585 | -121 | -394 | 272 | -346 | -162 | -140 | -375 | 61 | 7 | 3 | 22 | 20 | 11 | 11 | 15 |
| -14.59473348 | -80.59473348 | -307 | 39 | 138 | 8862 | -135 | -417 | 282 | -282 | -347 | -154 | -396 | 55 | 7 | 3 | 26 | 19 | 10 | 11 | 12 |
| -9.729868619 | -80.59473348 | -334 | 41 | 138 | 9357 | -153 | -451 | 298 | -309 | -196 | -172 | -429 | 41 | 6 | 2 | 30 | 15 | 7 | 8 | 9 |
| -4.865003754 | -80.59473348 | -354 | 42 | 133 | 9970 | -161 | -478 | 317 | -331 | -296 | -182 | -454 | 27 | 4 | 1 | 32 | 10 | 4 | 6 | 6 |
| -0.000138889 | -80.59473348 | -370 | 41 | 127 | 10282 | -170 | -495 | 325 | -347 | -472 | -192 | -472 | 19 | 3 | 1 | 40 | 8 | 3 | 5 | 3 |
| 4.864725976 | -80.59473348 | -385 | 41 | 125 | 10384 | -183 | -511 | 328 | -362 | -476 | -205 | -488 | 16 | 3 | 1 | 47 | 7 | 3 | 5 | 3 |
| 9.729590841 | -80.59473348 | -396 | 41 | 124 | 10552 | -192 | -525 | 333 | -373 | -335 | -214 | -501 | 10 | 2 | 0 | 66 | 5 | 0 | 3 | 2 |
| 14.59445571 | -80.59473348 | -409 | 42 | 122 | 10831 | -199 | -540 | 341 | -285 | -516 | -222 | -516 | 7 | 1 | 0 | 85 | 3 | 0 | 3 | 0 |
| 19.45932057 | -80.59473348 | -424 | 41 | 118 | 11193 | -208 | -558 | 349 | -295 | -534 | -231 | -534 | 6 | 1 | 0 | 100 | 3 | 0 | 3 | 0 |
| 24.32418544 | -80.59473348 | -439 | 41 | 115 | 11443 | -220 | -574 | 354 | -308 | -551 | -242 | -551 | 6 | 1 | 0 | 100 | 3 | 0 | 3 | 0 |
| 29.1890503 | -80.59473348 | -455 | 41 | 115 | 11723 | -230 | -590 | 360 | -319 | -554 | -252 | -567 | 5 | 1 | 0 | 118 | 3 | 0 | 3 | 0 |
| 34.05391517 | -80.59473348 | -473 | 43 | 116 | 12201 | -240 | -612 | 372 | -332 | -535 | -263 | -589 | 3 | 1 | 0 | 173 | 3 | 0 | 3 | 0 |
| 38.91878003 | -80.59473348 | -490 | 45 | 117 | 12576 | -250 | -631 | 381 | -303 | -344 | -274 | -607 | 1 | 1 | 0 | 332 | 2 | 0 | 1 | 0 |
| 63.24310435 | -80.59473348 | -507 | 46 | 127 | 12002 | -275 | -638 | 363 | -326 | -370 | -298 | -615 | 1 | 1 | 0 | 332 | 2 | 0 | 1 | 0 |
| 68.10796922 | -80.59473348 | -508 | 46 | 128 | 11854 | -279 | -640 | 360 | -329 | -373 | -302 | -615 | 1 | 1 | 0 | 332 | 2 | 0 | 1 | 0 |
| 107.0268881 | -80.59473348 | -489 | 36 | 100 | 12160 | -256 | -620 | 364 | -278 | -472 | -278 | -602 | 1 | 1 | 0 | 332 | 2 | 0 | 2 | 0 |
| 111.891753 | -80.59473348 | -480 | 35 | 97 | 12015 | -250 | -609 | 358 | -272 | -464 | -272 | -591 | 1 | 1 | 0 | 332 | 2 | 0 | 2 | 0 |
| 116.7566179 | -80.59473348 | -468 | 33 | 94 | 11738 | -243 | -593 | 350 | -264 | -453 | -264 | -577 | 1 | 1 | 0 | 332 | 2 | 0 | 2 | 0 |
| 121.6214827 | -80.59473348 | -454 | 32 | 93 | 11559 | -232 | -576 | 344 | -254 | -441 | -254 | -561 | 2 | 1 | 0 | 224 | 3 | 0 | 3 | 0 |
| 126.4863476 | -80.59473348 | -444 | 31 | 93 | 11431 | -224 | -564 | 340 | -245 | -431 | -245 | -550 | 2 | 1 | 0 | 224 | 3 | 0 | 3 | 0 |
| 131.3512125 | -80.59473348 | -428 | 31 | 93 | 11158 | -213 | -546 | 333 | -234 | -416 | -234 | -532 | 2 | 1 | 0 | 224 | 3 | 0 | 3 | 0 |
| 136.2160773 | -80.59473348 | -409 | 31 | 96 | 10933 | -198 | -526 | 329 | -492 | -398 | -219 | -511 | 3 | 1 | 0 | 173 | 3 | 0 | 2 | 0 |
| 141.0809422 | -80.59473348 | -395 | 33 | 101 | 10794 | -185 | -513 | 327 | -278 | -496 | -208 | -496 | 6 | 1 | 0 | 100 | 3 | 0 | 2 | 0 |
| 145.9458071 | -80.59473348 | -374 | 35 | 109 | 10374 | -172 | -491 | 319 | -428 | -305 | -195 | -473 | 14 | 2 | 0 | 59 | 6 | 1 | 3 | 2 |
| 150.8106719 | -80.59473348 | -338 | 36 | 117 | 9988 | -142 | -453 | 312 | -389 | -434 | -165 | -434 | 22 | 3 | 1 | 44 | 9 | 3 | 5 | 3 |
| 155.6755368 | -80.59473348 | -273 | 37 | 122 | 9703 | -79 | -386 | 307 | -343 | -170 | -105 | -367 | 54 | 7 | 3 | 35 | 21 | 9 | 9 | 11 |
| 160.5404017 | -80.59473348 | -220 | 41 | 138 | 9101 | -38 | -335 | 297 | -282 | -63 | -63 | -313 | 113 | 15 | 5 | 34 | 44 | 16 | 16 | 23 |
| 165.4052665 | -80.59473348 | -238 | 48 | 142 | 10065 | -43 | -379 | 336 | -302 | -66 | -66 | -348 | 121 | 16 | 4 | 38 | 48 | 14 | 14 | 25 |
| 170.2701314 | -80.59473348 | -270 | 51 | 135 | 11483 | -54 | -435 | 381 | -342 | -77 | -77 | -398 | 95 | 13 | 4 | 37 | 38 | 13 | 13 | 20 |
| 175.1349962 | -80.59473348 | -271 | 49 | 132 | 11269 | -60 | -433 | 373 | -342 | -201 | -82 | -397 | 84 | 11 | 3 | 35 | 32 | 13 | 13 | 17 |
| -160.5406794 | -78.24338213 | -225 | 53 | 142 | 11161 | -26 | -397 | 371 | -300 | -170 | -39 | -353 | 241 | 26 | 14 | 18 | 76 | 46 | 57 | 62 |
| -155.6758146 | -78.24338213 | -211 | 53 | 163 | 9141 | -36 | -362 | 326 | -269 | -164 | -58 | -321 | 217 | 24 | 12 | 22 | 71 | 41 | 45 | 57 |
| -150.8109497 | -78.24338213 | -222 | 49 | 184 | 7100 | -82 | -349 | 267 | -266 | -141 | -105 | -311 | 322 | 39 | 19 | 24 | 112 | 59 | 60 | 85 |
| -145.9460848 | -78.24338213 | -221 | 51 | 193 | 7017 | -81 | -347 | 265 | -265 | -140 | -105 | -308 | 331 | 41 | 19 | 25 | 116 | 59 | 60 | 87 |
| -141.08122 | -78.24338213 | -211 | 53 | 205 | 6676 | -74 | -332 | 258 | -269 | -134 | -99 | -293 | 199 | 24 | 12 | 23 | 68 | 37 | 39 | 54 |
| -136.2163551 | -78.24338213 | -222 | 52 | 210 | 6406 | -88 | -336 | 248 | -263 | -148 | -113 | -300 | 182 | 21 | 11 | 22 | 60 | 34 | 36 | 49 |
| -131.3514902 | -78.24338213 | -238 | 51 | 209 | 6411 | -104 | -350 | 245 | -295 | -165 | -128 | -315 | 162 | 18 | 9 | 24 | 54 | 29 | 32 | 46 |
| -126.4866254 | -78.24338213 | -259 | 53 | 204 | 6914 | -117 | -376 | 259 | -305 | -180 | -140 | -340 | 138 | 16 | 7 | 26 | 46 | 23 | 26 | 39 |
| -121.6217605 | -78.24338213 | -264 | 49 | 191 | 7030 | -122 | -379 | 257 | -311 | -182 | -142 | -346 | 134 | 15 | 7 | 24 | 43 | 22 | 24 | 38 |
| -116.7568956 | -78.24338213 | -272 | 48 | 184 | 7265 | -129 | -389 | 259 | -311 | -187 | -148 | -356 | 228 | 25 | 12 | 20 | 71 | 38 | 42 | 62 |
| -111.8920308 | -78.24338213 | -258 | 45 | 183 | 6936 | -122 | -367 | 245 | -296 | -140 | -140 | -338 | 360 | 39 | 19 | 23 | 112 | 57 | 57 | 101 |
| -107.0271659 | -78.24338213 | -250 | 43 | 187 | 6531 | -120 | -351 | 232 | -297 | -138 | -138 | -325 | 374 | 42 | 18 | 27 | 120 | 54 | 54 | 110 |
| -102.1623011 | -78.24338213 | -256 | 43 | 190 | 6438 | -126 | -354 | 228 | -293 | -145 | -145 | -329 | 338 | 39 | 15 | 29 | 109 | 45 | 45 | 103 |
| -97.29743619 | -78.24338213 | -265 | 44 | 190 | 6551 | -131 | -363 | 232 | -304 | -152 | -152 | -338 | 288 | 33 | 13 | 29 | 92 | 39 | 39 | 89 |
| -92.43257132 | -78.24338213 | -267 | 44 | 190 | 6569 | -133 | -365 | 232 | -307 | -153 | -153 | -339 | 203 | 23 | 9 | 29 | 65 | 27 | 27 | 62 |
| -87.56770646 | -78.24338213 | -299 | 47 | 183 | 7412 | -153 | -408 | 255 | -368 | -210 | -172 | -380 | 178 | 18 | 11 | 14 | 54 | 34 | 37 | 51 |
| -82.70284159 | -78.24338213 | -228 | 41 | 145 | 8488 | -62 | -343 | 282 | -305 | -312 | -82 | -319 | 255 | 25 | 18 | 10 | 73 | 55 | 66 | 64 |
| -77.83797673 | -78.24338213 | -247 | 42 | 132 | 9750 | -59 | -378 | 318 | -332 | -344 | -80 | -351 | 265 | 27 | 19 | 11 | 78 | 58 | 64 | 67 |
| -72.97311186 | -78.24338213 | -258 | 40 | 121 | 10321 | -61 | -395 | 333 | -346 | -107 | -81 | -367 | 207 | 23 | 12 | 15 | 65 | 41 | 44 | 54 |
| -68.108247 | -78.24338213 | -260 | 41 | 122 | 10565 | -61 | -400 | 339 | -348 | -107 | -79 | -371 | 169 | 19 | 10 | 16 | 54 | 33 | 36 | 45 |
| -63.24338213 | -78.24338213 | -260 | 43 | 124 | 10804 | -58 | -405 | 347 | -346 | -103 | -76 | -372 | 156 | 18 | 9 | 17 | 50 | 29 | 33 | 39 |
| -58.37851727 | -78.24338213 | -260 | 45 | 128 | 10731 | -59 | -407 | 349 | -344 | -103 | -79 | -371 | 149 | 17 | 9 | 17 | 48 | 28 | 31 | 34 |
| -53.5136524 | -78.24338213 | -251 | 47 | 136 | 10382 | -54 | -397 | 343 | -331 | -98 | -76 | -359 | 142 | 16 | 8 | 20 | 47 | 24 | 28 | 33 |
| -48.64878754 | -78.24338213 | -243 | 47 | 143 | 10219 | -54 | -385 | 331 | -324 | -181 | -70 | -348 | 152 | 16 | 10 | 13 | 46 | 32 | 43 | 33 |
| -43.78392267 | -78.24338213 | -219 | 43 | 132 | 10297 | -32 | -356 | 324 | -300 | -72 | -45 | -323 | 216 | 24 | 15 | 13 | 69 | 46 | 53 | 51 |
| -38.91905781 | -78.24338213 | -216 | 44 | 136 | 10393 | -29 | -354 | 325 | -301 | -69 | -42 | -321 | 198 | 23 | 10 | 20 | 68 | 33 | 40 | 51 |
| -34.05419294 | -78.24338213 | -243 | 44 | 162 | 8494 | -87 | -361 | 273 | -318 | -126 | -100 | -331 | 176 | 18 | 11 | 15 | 54 | 35 | 46 | 50 |
| -29.18932808 | -78.24338213 | -236 | 50 | 183 | 7809 | -78 | -351 | 272 | -306 | -121 | -105 | -319 | 131 | 14 | 6 | 22 | 42 | 20 | 24 | 39 |
| -24.32446321 | -78.24338213 | -258 | 51 | 186 | 7927 | -97 | -372 | 276 | -331 | -142 | -124 | -342 | 109 | 12 | 4 | 28 | 36 | 14 | 16 | 34 |
| -19.45959835 | -78.24338213 | -275 | 50 | 186 | 7797 | -116 | -385 | 268 | -349 | -161 | -142 | -358 | 93 | 11 | 3 | 31 | 32 | 11 | 13 | 28 |
| -14.59473348 | -78.24338213 | -302 | 52 | 180 | 8482 | -131 | -420 | 289 | -384 | -179 | -157 | -392 | 82 | 9 | 3 | 30 | 27 | 11 | 11 | 25 |
| -9.729868619 | -78.24338213 | -329 | 50 | 165 | 9159 | -145 | -451 | 306 | -398 | -194 | -171 | -424 | 73 | 9 | 2 | 35 | 25 | 8 | 10 | 23 |
| -4.865003754 | -78.24338213 | -357 | 49 | 154 | 9732 | -162 | -482 | 320 | -430 | -247 | -189 | -457 | 50 | 7 | 2 | 44 | 19 | 6 | 6 | 17 |
| -0.000138889 | -78.24338213 | -372 | 49 | 150 | 9901 | -175 | -500 | 324 | -447 | -227 | -203 | -475 | 28 | 5 | 1 | 51 | 13 | 3 | 5 | 8 |
| 4.864725976 | -78.24338213 | -392 | 49 | 147 | 10274 | -189 | -524 | 335 | -469 | -274 | -216 | -498 | 18 | 3 | 1 | 51 | 9 | 3 | 3 | 5 |
| 9.729590841 | -78.24338213 | -413 | 49 | 141 | 10749 | -203 | -551 | 348 | -494 | -289 | -230 | -524 | 11 | 2 | 0 | 70 | 6 | 1 | 3 | 2 |
| 14.59445571 | -78.24338213 | -433 | 50 | 139 | 11146 | -217 | -575 | 358 | -518 | -548 | -244 | -548 | 6 | 2 | 0 | 129 | 5 | 0 | 3 | 0 |
| 19.45932057 | -78.24338213 | -452 | 51 | 138 | 11595 | -227 | -597 | 370 | -540 | -316 | -254 | -569 | 2 | 1 | 0 | 224 | 3 | 0 | 0 | 0 |
| 24.32418544 | -78.24338213 | -462 | 52 | 139 | 11688 | -235 | -607 | 372 | -552 | -326 | -263 | -579 | 2 | 1 | 0 | 224 | 3 | 0 | 0 | 0 |
| 29.1890503 | -78.24338213 | -475 | 52 | 140 | 11868 | -245 | -620 | 375 | -568 | -336 | -273 | -592 | 2 | 1 | 0 | 224 | 3 | 0 | 0 | 0 |
| 53.51337462 | -78.24338213 | -485 | 46 | 124 | 12103 | -247 | -615 | 368 | -347 | -466 | -276 | -596 | 9 | 1 | 0 | 58 | 3 | 1 | 3 | 3 |
| 58.37823949 | -78.24338213 | -458 | 44 | 128 | 11206 | -235 | -580 | 344 | -560 | -442 | -263 | -560 | 10 | 2 | 0 | 66 | 5 | 1 | 3 | 5 |
| 63.24310435 | -78.24338213 | -433 | 42 | 135 | 9981 | -231 | -542 | 311 | -524 | -472 | -257 | -524 | 13 | 2 | 0 | 46 | 5 | 1 | 4 | 5 |
| 68.10796922 | -78.24338213 | -427 | 39 | 130 | 9643 | -232 | -532 | 300 | -512 | -467 | -257 | -516 | 13 | 2 | 0 | 46 | 5 | 1 | 4 | 4 |
| 72.97283408 | -78.24338213 | -431 | 39 | 128 | 9854 | -233 | -538 | 306 | -511 | -319 | -258 | -522 | 16 | 2 | 1 | 35 | 6 | 3 | 4 | 4 |
| 77.83769895 | -78.24338213 | -445 | 42 | 130 | 10329 | -237 | -557 | 320 | -529 | -327 | -264 | -539 | 15 | 2 | 1 | 35 | 6 | 3 | 3 | 4 |
| 82.70256381 | -78.24338213 | -467 | 45 | 134 | 10992 | -248 | -588 | 340 | -524 | -401 | -276 | -568 | 9 | 1 | 0 | 58 | 3 | 0 | 3 | 3 |
| 87.56742868 | -78.24338213 | -494 | 50 | 137 | 11680 | -263 | -625 | 362 | -589 | -360 | -292 | -601 | 6 | 1 | 0 | 100 | 3 | 0 | 0 | 3 |
| 131.3512125 | -78.24338213 | -444 | 36 | 105 | 11418 | -220 | -565 | 346 | -246 | -430 | -246 | -550 | 2 | 1 | 0 | 224 | 3 | 0 | 3 | 0 |
| 136.2160773 | -78.24338213 | -430 | 36 | 106 | 11253 | -207 | -549 | 342 | -234 | -417 | -234 | -534 | 3 | 1 | 0 | 173 | 3 | 0 | 3 | 0 |
| 141.0809422 | -78.24338213 | -422 | 37 | 108 | 11250 | -199 | -542 | 343 | -301 | -526 | -227 | -526 | 8 | 1 | 0 | 71 | 3 | 0 | 3 | 0 |
| 145.9458071 | -78.24338213 | -415 | 38 | 111 | 11130 | -195 | -537 | 342 | -504 | -515 | -222 | -519 | 15 | 2 | 0 | 48 | 6 | 2 | 5 | 2 |
| 150.8106719 | -78.24338213 | -394 | 39 | 119 | 10452 | -187 | -514 | 327 | -476 | -489 | -213 | -493 | 21 | 3 | 0 | 53 | 9 | 2 | 8 | 2 |
| 155.6755368 | -78.24338213 | -357 | 41 | 133 | 9663 | -162 | -471 | 309 | -431 | -449 | -188 | -449 | 36 | 5 | 1 | 41 | 15 | 5 | 9 | 5 |
| 160.5404017 | -78.24338213 | -313 | 45 | 135 | 10669 | -109 | -443 | 334 | -391 | -421 | -131 | -421 | 113 | 14 | 5 | 29 | 40 | 19 | 33 | 19 |
| 165.4052665 | -78.24338213 | -216 | 41 | 128 | 9961 | -28 | -351 | 323 | -280 | -326 | -48 | -326 | 143 | 18 | 9 | 28 | 54 | 27 | 30 | 27 |
| 170.2701314 | -78.24338213 | -240 | 47 | 127 | 11229 | -29 | -397 | 368 | -311 | -366 | -52 | -366 | 166 | 22 | 10 | 28 | 61 | 30 | 36 | 30 |
| 175.1349962 | -78.24338213 | -238 | 48 | 127 | 11616 | -28 | -404 | 376 | -313 | -168 | -46 | -369 | 195 | 26 | 10 | 29 | 70 | 31 | 38 | 42 |
| -145.9460848 | -75.89203078 | -127 | 43 | 178 | 6887 | -6 | -249 | 243 | -189 | -38 | -16 | -213 | 725 | 84 | 42 | 22 | 250 | 128 | 131 | 211 |
| -141.08122 | -75.89203078 | -154 | 46 | 199 | 6182 | -35 | -266 | 232 | -210 | -78 | -51 | -231 | 593 | 68 | 34 | 22 | 203 | 103 | 110 | 174 |
| -136.2163551 | -75.89203078 | -205 | 45 | 208 | 5507 | -92 | -310 | 218 | -256 | -143 | -114 | -276 | 679 | 84 | 35 | 28 | 244 | 107 | 109 | 204 |
| -131.3514902 | -75.89203078 | -237 | 49 | 214 | 5846 | -118 | -346 | 228 | -276 | -170 | -139 | -310 | 744 | 94 | 39 | 27 | 266 | 118 | 123 | 214 |
| -126.4866254 | -75.89203078 | -250 | 45 | 208 | 5733 | -134 | -352 | 219 | -288 | -183 | -153 | -321 | 603 | 75 | 30 | 28 | 216 | 91 | 94 | 176 |
| -121.6217605 | -75.89203078 | -241 | 43 | 201 | 5668 | -126 | -338 | 213 | -278 | -173 | -144 | -309 | 653 | 76 | 33 | 26 | 225 | 100 | 102 | 184 |
| -116.7568956 | -75.89203078 | -245 | 47 | 196 | 6581 | -116 | -358 | 242 | -292 | -168 | -134 | -325 | 700 | 78 | 35 | 25 | 228 | 107 | 111 | 194 |
| -111.8920308 | -75.89203078 | -207 | 44 | 186 | 6438 | -80 | -315 | 235 | -239 | -100 | -100 | -286 | 766 | 85 | 37 | 26 | 241 | 114 | 114 | 213 |
| -107.0271659 | -75.89203078 | -185 | 42 | 199 | 5795 | -71 | -284 | 213 | -213 | -89 | -89 | -256 | 723 | 86 | 28 | 33 | 239 | 86 | 86 | 220 |
| -102.1623011 | -75.89203078 | -189 | 41 | 201 | 5592 | -77 | -281 | 204 | -216 | -95 | -95 | -255 | 782 | 97 | 26 | 37 | 267 | 79 | 79 | 251 |
| -97.29743619 | -75.89203078 | -193 | 44 | 207 | 5855 | -74 | -286 | 213 | -223 | -93 | -93 | -260 | 558 | 69 | 19 | 36 | 188 | 57 | 57 | 181 |
| -92.43257132 | -75.89203078 | -216 | 46 | 205 | 6182 | -91 | -313 | 222 | -251 | -110 | -110 | -285 | 409 | 48 | 17 | 30 | 132 | 53 | 53 | 127 |
| -87.56770646 | -75.89203078 | -244 | 44 | 194 | 6511 | -115 | -345 | 230 | -308 | -155 | -133 | -317 | 328 | 34 | 21 | 15 | 97 | 67 | 67 | 91 |
| -82.70284159 | -75.89203078 | -235 | 43 | 180 | 6912 | -100 | -340 | 240 | -116 | -303 | -116 | -312 | 436 | 43 | 30 | 11 | 127 | 92 | 127 | 101 |
| -77.83797673 | -75.89203078 | -215 | 46 | 167 | 8183 | -58 | -334 | 276 | -289 | -297 | -76 | -305 | 426 | 42 | 29 | 10 | 118 | 93 | 112 | 100 |
| -72.97311186 | -75.89203078 | -210 | 46 | 178 | 7539 | -64 | -322 | 258 | -276 | -106 | -82 | -294 | 288 | 31 | 16 | 19 | 88 | 51 | 52 | 79 |
| -68.108247 | -75.89203078 | -200 | 43 | 183 | 6906 | -64 | -300 | 236 | -81 | -271 | -81 | -275 | 187 | 24 | 11 | 26 | 70 | 35 | 70 | 36 |
| -63.24338213 | -75.89203078 | -221 | 45 | 149 | 9259 | -45 | -350 | 305 | -113 | -315 | -65 | -320 | 248 | 29 | 15 | 22 | 85 | 46 | 78 | 47 |
| -58.37851727 | -75.89203078 | -214 | 48 | 151 | 9732 | -36 | -353 | 317 | -181 | -314 | -50 | -314 | 247 | 25 | 17 | 13 | 74 | 52 | 60 | 52 |
| -24.32446321 | -75.89203078 | -202 | 42 | 181 | 6660 | -68 | -300 | 232 | -214 | -92 | -92 | -278 | 413 | 47 | 19 | 24 | 136 | 63 | 63 | 110 |
| -19.45959835 | -75.89203078 | -224 | 46 | 189 | 6861 | -84 | -326 | 242 | -264 | -112 | -112 | -302 | 468 | 55 | 21 | 26 | 162 | 71 | 71 | 126 |
| -14.59473348 | -75.89203078 | -251 | 52 | 207 | 7028 | -106 | -356 | 250 | -293 | -132 | -132 | -328 | 227 | 29 | 9 | 29 | 81 | 33 | 33 | 62 |
| -9.729868619 | -75.89203078 | -320 | 59 | 207 | 8054 | -154 | -437 | 284 | -373 | -228 | -183 | -407 | 89 | 12 | 4 | 31 | 32 | 13 | 14 | 27 |
| -4.865003754 | -75.89203078 | -355 | 60 | 194 | 8916 | -172 | -480 | 308 | -415 | -253 | -204 | -450 | 43 | 6 | 2 | 37 | 16 | 6 | 6 | 12 |
| -0.000138889 | -75.89203078 | -383 | 58 | 177 | 9642 | -188 | -513 | 325 | -449 | -271 | -220 | -484 | 33 | 5 | 1 | 42 | 13 | 4 | 5 | 9 |
| 4.864725976 | -75.89203078 | -401 | 57 | 173 | 9866 | -202 | -533 | 330 | -477 | -285 | -234 | -504 | 28 | 5 | 1 | 47 | 13 | 3 | 4 | 7 |
| 9.729590841 | -75.89203078 | -421 | 59 | 173 | 10269 | -216 | -559 | 343 | -492 | -248 | -248 | -528 | 14 | 3 | 0 | 59 | 7 | 1 | 1 | 3 |
| 14.59445571 | -75.89203078 | -443 | 60 | 167 | 10851 | -228 | -586 | 358 | -528 | -492 | -260 | -555 | 10 | 2 | 0 | 82 | 6 | 1 | 1 | 3 |
| 19.45932057 | -75.89203078 | -453 | 58 | 164 | 10904 | -237 | -593 | 356 | -534 | -324 | -269 | -564 | 5 | 2 | 0 | 154 | 4 | 0 | 1 | 1 |
| 24.32418544 | -75.89203078 | -462 | 56 | 161 | 10784 | -247 | -596 | 349 | -548 | -512 | -278 | -569 | 5 | 1 | 0 | 118 | 3 | 0 | 1 | 1 |
| 29.1890503 | -75.89203078 | -470 | 54 | 155 | 11004 | -251 | -601 | 351 | -566 | -442 | -281 | -577 | 12 | 2 | 0 | 58 | 6 | 1 | 3 | 4 |
| 34.05391517 | -75.89203078 | -471 | 53 | 153 | 10986 | -251 | -600 | 348 | -567 | -342 | -283 | -577 | 15 | 2 | 1 | 35 | 6 | 3 | 4 | 4 |
| 38.91878003 | -75.89203078 | -480 | 53 | 150 | 11209 | -256 | -610 | 354 | -349 | -521 | -288 | -588 | 9 | 1 | 0 | 58 | 3 | 0 | 3 | 3 |
| 43.78364489 | -75.89203078 | -488 | 52 | 143 | 11684 | -255 | -621 | 366 | -351 | -545 | -288 | -599 | 7 | 1 | 0 | 85 | 3 | 0 | 3 | 3 |
| 48.64850976 | -75.89203078 | -474 | 48 | 136 | 11361 | -246 | -599 | 353 | -278 | -449 | -278 | -581 | 11 | 2 | 0 | 70 | 6 | 1 | 6 | 3 |
| 53.51337462 | -75.89203078 | -439 | 42 | 130 | 10505 | -227 | -552 | 326 | -257 | -418 | -257 | -537 | 21 | 3 | 0 | 53 | 9 | 1 | 9 | 8 |
| 58.37823949 | -75.89203078 | -396 | 40 | 131 | 9698 | -199 | -501 | 302 | -486 | -378 | -227 | -486 | 34 | 5 | 1 | 38 | 14 | 4 | 10 | 14 |
| 63.24310435 | -75.89203078 | -354 | 38 | 137 | 8892 | -171 | -450 | 279 | -436 | -392 | -198 | -436 | 37 | 4 | 2 | 21 | 12 | 7 | 7 | 12 |
| 68.10796922 | -75.89203078 | -337 | 41 | 148 | 8774 | -155 | -435 | 280 | -378 | -375 | -184 | -419 | 33 | 4 | 2 | 26 | 11 | 6 | 6 | 10 |
| 72.97283408 | -75.89203078 | -348 | 45 | 162 | 8595 | -168 | -448 | 280 | -386 | -198 | -198 | -429 | 25 | 3 | 1 | 36 | 9 | 3 | 3 | 7 |
| 77.83769895 | -75.89203078 | -373 | 47 | 167 | 8499 | -194 | -473 | 279 | -438 | -410 | -225 | -453 | 20 | 3 | 1 | 37 | 8 | 3 | 3 | 6 |
| 82.70256381 | -75.89203078 | -410 | 48 | 165 | 9000 | -223 | -516 | 292 | -451 | -449 | -254 | -495 | 15 | 2 | 0 | 48 | 6 | 1 | 3 | 4 |
| 87.56742868 | -75.89203078 | -446 | 51 | 163 | 9711 | -245 | -558 | 312 | -524 | -391 | -278 | -536 | 13 | 2 | 0 | 70 | 6 | 0 | 1 | 4 |
| 92.43229354 | -75.89203078 | -477 | 52 | 157 | 10500 | -263 | -596 | 333 | -564 | -355 | -295 | -573 | 7 | 1 | 0 | 85 | 3 | 0 | 1 | 3 |
| 97.29715841 | -75.89203078 | -493 | 53 | 148 | 11370 | -265 | -622 | 357 | -591 | -362 | -298 | -598 | 4 | 1 | 0 | 141 | 3 | 0 | 1 | 3 |
| 102.1620233 | -75.89203078 | -496 | 52 | 141 | 11841 | -259 | -630 | 370 | -598 | -473 | -292 | -605 | 4 | 1 | 0 | 141 | 3 | 0 | 3 | 1 |
| 107.0268881 | -75.89203078 | -494 | 51 | 134 | 12085 | -253 | -630 | 377 | -601 | -473 | -287 | -607 | 4 | 1 | 0 | 141 | 3 | 0 | 3 | 1 |
| 111.891753 | -75.89203078 | -500 | 52 | 132 | 12706 | -248 | -642 | 395 | -613 | -478 | -283 | -620 | 3 | 1 | 0 | 173 | 3 | 0 | 2 | 1 |
| 131.3512125 | -75.89203078 | -455 | 44 | 128 | 11109 | -230 | -575 | 345 | -262 | -438 | -262 | -557 | 2 | 1 | 0 | 224 | 3 | 0 | 3 | 0 |
| 136.2160773 | -75.89203078 | -443 | 43 | 125 | 11034 | -219 | -562 | 342 | -251 | -427 | -251 | -544 | 4 | 1 | 0 | 141 | 3 | 0 | 3 | 0 |
| 141.0809422 | -75.89203078 | -434 | 43 | 123 | 11076 | -209 | -554 | 345 | -313 | -536 | -242 | -536 | 8 | 1 | 0 | 71 | 3 | 0 | 3 | 0 |
| 145.9458071 | -75.89203078 | -424 | 44 | 126 | 11019 | -200 | -546 | 346 | -305 | -351 | -233 | -527 | 11 | 1 | 0 | 30 | 3 | 1 | 3 | 3 |
| 150.8106719 | -75.89203078 | -400 | 44 | 133 | 10417 | -188 | -520 | 333 | -485 | -390 | -219 | -499 | 19 | 2 | 1 | 31 | 6 | 3 | 6 | 3 |
| 155.6755368 | -75.89203078 | -349 | 46 | 147 | 9434 | -154 | -464 | 310 | -422 | -440 | -185 | -440 | 31 | 4 | 1 | 29 | 11 | 5 | 9 | 5 |
| 160.5404017 | -75.89203078 | -277 | 44 | 154 | 8670 | -97 | -385 | 288 | -185 | -363 | -129 | -363 | 102 | 13 | 4 | 28 | 37 | 16 | 31 | 16 |
| -126.4866254 | -73.54067943 | -145 | 40 | 187 | 6001 | -42 | -258 | 216 | -206 | -52 | -52 | -224 | 1120 | 127 | 50 | 27 | 373 | 157 | 157 | 351 |
| -97.29743619 | -73.54067943 | -156 | 46 | 228 | 5309 | -53 | -253 | 200 | -223 | -71 | -71 | -223 | 671 | 78 | 26 | 33 | 222 | 78 | 78 | 222 |
| -92.43257132 | -73.54067943 | -170 | 46 | 226 | 5494 | -63 | -265 | 202 | -225 | -79 | -79 | -236 | 793 | 87 | 41 | 23 | 247 | 125 | 125 | 241 |
| -87.56770646 | -73.54067943 | -165 | 41 | 215 | 5307 | -61 | -254 | 193 | -217 | -76 | -76 | -228 | 911 | 98 | 52 | 20 | 282 | 159 | 159 | 260 |
| -77.83797673 | -73.54067943 | -136 | 45 | 203 | 6323 | -18 | -242 | 224 | -192 | -34 | -34 | -211 | 630 | 71 | 34 | 22 | 203 | 104 | 104 | 183 |
| -72.97311186 | -73.54067943 | -125 | 41 | 202 | 5671 | -21 | -225 | 203 | -173 | -36 | -36 | -195 | 663 | 78 | 36 | 22 | 221 | 110 | 110 | 193 |
| -68.108247 | -73.54067943 | -175 | 44 | 217 | 5521 | -71 | -272 | 201 | -222 | -86 | -86 | -242 | 665 | 77 | 43 | 17 | 213 | 129 | 129 | 176 |
| -63.24338213 | -73.54067943 | -218 | 49 | 201 | 7134 | -86 | -330 | 244 | -101 | -297 | -101 | -298 | 309 | 38 | 16 | 29 | 114 | 50 | 114 | 53 |
| -19.45959835 | -73.54067943 | -170 | 45 | 162 | 8812 | -18 | -297 | 280 | -225 | -31 | -31 | -268 | 453 | 53 | 21 | 26 | 150 | 67 | 67 | 137 |
| -14.59473348 | -73.54067943 | -229 | 58 | 209 | 7981 | -72 | -349 | 278 | -273 | -99 | -99 | -320 | 456 | 59 | 22 | 27 | 163 | 72 | 72 | 129 |
| -9.729868619 | -73.54067943 | -231 | 54 | 231 | 6264 | -95 | -330 | 235 | -265 | -126 | -126 | -302 | 215 | 30 | 9 | 32 | 81 | 31 | 31 | 64 |
| -4.865003754 | -73.54067943 | -300 | 57 | 211 | 7631 | -141 | -411 | 270 | -349 | -211 | -172 | -383 | 148 | 20 | 8 | 27 | 52 | 25 | 28 | 44 |
| -0.000138889 | -73.54067943 | -320 | 57 | 202 | 8194 | -150 | -434 | 283 | -384 | -223 | -182 | -407 | 141 | 18 | 8 | 26 | 48 | 24 | 26 | 44 |
| 4.864725976 | -73.54067943 | -356 | 58 | 207 | 8035 | -189 | -469 | 280 | -418 | -260 | -220 | -441 | 40 | 6 | 2 | 33 | 16 | 6 | 7 | 10 |
| 9.729590841 | -73.54067943 | -382 | 55 | 194 | 8411 | -211 | -496 | 285 | -448 | -462 | -240 | -470 | 39 | 5 | 2 | 34 | 15 | 7 | 7 | 11 |
| 14.59445571 | -73.54067943 | -384 | 52 | 184 | 8442 | -213 | -493 | 280 | -463 | -417 | -240 | -471 | 55 | 8 | 3 | 34 | 22 | 9 | 13 | 20 |
| 19.45932057 | -73.54067943 | -369 | 47 | 179 | 7985 | -206 | -471 | 265 | -443 | -405 | -232 | -451 | 76 | 11 | 4 | 30 | 30 | 14 | 16 | 27 |
| 24.32418544 | -73.54067943 | -380 | 45 | 171 | 8103 | -215 | -482 | 267 | -454 | -417 | -241 | -462 | 48 | 8 | 2 | 41 | 22 | 8 | 10 | 18 |
| 29.1890503 | -73.54067943 | -366 | 45 | 162 | 8441 | -192 | -468 | 276 | -441 | -402 | -220 | -449 | 112 | 16 | 7 | 27 | 43 | 21 | 25 | 40 |
| 34.05391517 | -73.54067943 | -374 | 49 | 171 | 8711 | -194 | -481 | 287 | -450 | -411 | -224 | -459 | 66 | 9 | 3 | 27 | 25 | 11 | 14 | 21 |
| 38.91878003 | -73.54067943 | -405 | 50 | 168 | 9180 | -215 | -515 | 300 | -484 | -442 | -247 | -493 | 38 | 5 | 2 | 28 | 14 | 6 | 10 | 11 |
| 43.78364489 | -73.54067943 | -422 | 51 | 163 | 9737 | -221 | -536 | 316 | -507 | -398 | -255 | -515 | 36 | 5 | 2 | 33 | 14 | 6 | 12 | 11 |
| 48.64850976 | -73.54067943 | -435 | 51 | 158 | 10109 | -226 | -551 | 325 | -261 | -472 | -261 | -531 | 24 | 3 | 1 | 35 | 9 | 3 | 9 | 7 |
| 53.51337462 | -73.54067943 | -419 | 46 | 144 | 9922 | -215 | -531 | 316 | -248 | -394 | -248 | -514 | 36 | 6 | 1 | 54 | 18 | 4 | 18 | 11 |
| 58.37823949 | -73.54067943 | -365 | 38 | 133 | 9093 | -178 | -465 | 287 | -207 | -438 | -207 | -450 | 53 | 7 | 2 | 39 | 21 | 7 | 21 | 17 |
| 63.24310435 | -73.54067943 | -279 | 35 | 134 | 8218 | -109 | -370 | 261 | -356 | -257 | -135 | -356 | 48 | 6 | 3 | 23 | 17 | 10 | 10 | 17 |
| 68.10796922 | -73.54067943 | -273 | 54 | 171 | 9630 | -65 | -382 | 317 | -355 | -106 | -106 | -360 | 21 | 3 | 1 | 34 | 7 | 3 | 3 | 7 |
| 72.97283408 | -73.54067943 | -268 | 45 | 189 | 7044 | -117 | -356 | 239 | -322 | -167 | -146 | -337 | 41 | 5 | 2 | 30 | 15 | 6 | 6 | 10 |
| 77.83769895 | -73.54067943 | -332 | 47 | 190 | 7270 | -175 | -422 | 247 | -386 | -206 | -206 | -403 | 30 | 4 | 1 | 35 | 12 | 4 | 4 | 7 |
| 82.70256381 | -73.54067943 | -377 | 49 | 188 | 7837 | -208 | -472 | 263 | -437 | -412 | -241 | -452 | 25 | 4 | 1 | 46 | 11 | 3 | 4 | 7 |
| 87.56742868 | -73.54067943 | -413 | 53 | 186 | 8608 | -230 | -516 | 286 | -452 | -366 | -265 | -494 | 26 | 3 | 1 | 37 | 9 | 3 | 4 | 7 |
| 92.43229354 | -73.54067943 | -446 | 55 | 177 | 9522 | -247 | -558 | 311 | -529 | -335 | -282 | -533 | 26 | 5 | 1 | 53 | 13 | 3 | 4 | 11 |
| 97.29715841 | -73.54067943 | -456 | 55 | 168 | 10157 | -245 | -575 | 330 | -544 | -396 | -281 | -547 | 28 | 5 | 1 | 47 | 13 | 3 | 6 | 11 |
| 102.1620233 | -73.54067943 | -452 | 51 | 155 | 10327 | -238 | -567 | 329 | -542 | -431 | -273 | -544 | 26 | 5 | 1 | 56 | 13 | 3 | 7 | 11 |
| 107.0268881 | -73.54067943 | -455 | 51 | 145 | 11215 | -226 | -578 | 353 | -555 | -433 | -262 | -555 | 27 | 5 | 1 | 55 | 13 | 3 | 6 | 13 |
| 111.891753 | -73.54067943 | -462 | 57 | 152 | 11737 | -223 | -600 | 377 | -569 | -390 | -261 | -571 | 23 | 4 | 0 | 54 | 11 | 2 | 6 | 5 |
| 116.7566179 | -73.54067943 | -455 | 59 | 163 | 11041 | -228 | -589 | 360 | -556 | -390 | -264 | -558 | 18 | 3 | 0 | 51 | 8 | 1 | 5 | 3 |
| 121.6214827 | -73.54067943 | -458 | 60 | 167 | 10912 | -234 | -591 | 357 | -558 | -394 | -270 | -559 | 21 | 3 | 0 | 47 | 9 | 2 | 6 | 3 |
| 126.4863476 | -73.54067943 | -457 | 57 | 165 | 10601 | -238 | -585 | 347 | -553 | -395 | -274 | -554 | 18 | 3 | 0 | 51 | 8 | 1 | 5 | 3 |
| 131.3512125 | -73.54067943 | -443 | 52 | 157 | 10165 | -231 | -561 | 329 | -533 | -384 | -266 | -535 | 17 | 2 | 0 | 45 | 6 | 1 | 6 | 6 |
| 136.2160773 | -73.54067943 | -427 | 48 | 149 | 10095 | -215 | -539 | 324 | -516 | -366 | -251 | -519 | 25 | 3 | 1 | 24 | 8 | 4 | 6 | 7 |
| 141.0809422 | -73.54067943 | -414 | 47 | 145 | 10143 | -201 | -528 | 327 | -504 | -351 | -237 | -508 | 27 | 3 | 1 | 26 | 9 | 4 | 6 | 9 |
| 145.9458071 | -73.54067943 | -404 | 50 | 150 | 10157 | -191 | -521 | 330 | -482 | -339 | -227 | -499 | 32 | 4 | 1 | 28 | 11 | 4 | 8 | 9 |
| 150.8106719 | -73.54067943 | -389 | 50 | 157 | 9750 | -184 | -506 | 321 | -282 | -328 | -219 | -482 | 39 | 4 | 2 | 22 | 12 | 6 | 11 | 9 |
| 155.6755368 | -73.54067943 | -373 | 51 | 161 | 9504 | -174 | -489 | 315 | -451 | -313 | -207 | -463 | 54 | 7 | 3 | 25 | 18 | 9 | 16 | 12 |
| 160.5404017 | -73.54067943 | -351 | 51 | 168 | 9104 | -160 | -467 | 307 | -425 | -378 | -194 | -440 | 74 | 9 | 4 | 22 | 24 | 14 | 22 | 15 |
| 165.4052665 | -73.54067943 | -286 | 52 | 173 | 8908 | -113 | -412 | 298 | -178 | -380 | -134 | -380 | 293 | 43 | 15 | 34 | 119 | 49 | 115 | 49 |
| -72.97311186 | -71.18932808 | -93 | 46 | 224 | 5736 | 6 | -202 | 207 | -145 | -10 | -10 | -168 | 548 | 61 | 30 | 21 | 172 | 91 | 91 | 155 |
| -68.108247 | -71.18932808 | -84 | 45 | 235 | 5089 | 10 | -182 | 192 | -106 | -7 | -7 | -150 | 457 | 48 | 28 | 16 | 132 | 86 | 86 | 110 |
| -63.24338213 | -71.18932808 | -205 | 44 | 219 | 5597 | -100 | -300 | 200 | -116 | -270 | -116 | -272 | 407 | 62 | 18 | 44 | 182 | 55 | 182 | 58 |
| -9.729868619 | -71.18932808 | -173 | 51 | 203 | 7414 | -34 | -286 | 251 | -210 | -54 | -54 | -257 | 541 | 67 | 31 | 23 | 186 | 93 | 93 | 160 |
| -4.865003754 | -71.18932808 | -169 | 62 | 225 | 7961 | -11 | -288 | 278 | -211 | -70 | -36 | -258 | 566 | 71 | 32 | 23 | 192 | 97 | 105 | 167 |
| -0.000138889 | -71.18932808 | -155 | 58 | 229 | 6908 | -11 | -266 | 255 | -189 | -41 | -41 | -235 | 324 | 43 | 14 | 33 | 112 | 47 | 47 | 111 |
| 4.864725976 | -71.18932808 | -162 | 49 | 216 | 6521 | -35 | -261 | 226 | -226 | -127 | -55 | -236 | 310 | 39 | 17 | 29 | 110 | 54 | 63 | 102 |
| 9.729590841 | -71.18932808 | -181 | 45 | 208 | 6213 | -58 | -274 | 216 | -241 | -77 | -77 | -251 | 229 | 30 | 12 | 34 | 87 | 41 | 41 | 80 |
| 14.59445571 | -71.18932808 | -181 | 41 | 202 | 5774 | -64 | -268 | 205 | -237 | -86 | -86 | -247 | 230 | 31 | 10 | 35 | 89 | 34 | 34 | 83 |
| 19.45932057 | -71.18932808 | -158 | 44 | 196 | 6441 | -30 | -252 | 222 | -218 | -51 | -51 | -230 | 300 | 40 | 15 | 30 | 112 | 48 | 48 | 108 |
| 24.32418544 | -71.18932808 | -159 | 45 | 198 | 6580 | -27 | -255 | 229 | -220 | -50 | -50 | -232 | 320 | 44 | 12 | 35 | 122 | 39 | 39 | 122 |
| 29.1890503 | -71.18932808 | -181 | 43 | 195 | 6240 | -51 | -273 | 221 | -251 | -78 | -78 | -251 | 273 | 37 | 8 | 37 | 103 | 27 | 27 | 103 |
| 34.05391517 | -71.18932808 | -228 | 41 | 187 | 6278 | -98 | -319 | 222 | -241 | -124 | -124 | -297 | 159 | 20 | 4 | 39 | 56 | 14 | 14 | 55 |
| 38.91878003 | -71.18932808 | -268 | 45 | 179 | 7196 | -117 | -366 | 249 | -288 | -146 | -146 | -344 | 258 | 34 | 9 | 34 | 93 | 29 | 29 | 82 |
| 43.78364489 | -71.18932808 | -324 | 55 | 194 | 8246 | -149 | -434 | 285 | -355 | -184 | -184 | -407 | 122 | 17 | 5 | 35 | 46 | 15 | 15 | 36 |
| 48.64850976 | -71.18932808 | -366 | 59 | 196 | 8781 | -179 | -479 | 300 | -440 | -399 | -217 | -452 | 77 | 9 | 5 | 23 | 25 | 15 | 17 | 23 |
| 53.51337462 | -71.18932808 | -396 | 59 | 189 | 9204 | -202 | -512 | 310 | -239 | -370 | -239 | -485 | 42 | 5 | 2 | 25 | 14 | 7 | 14 | 11 |
| 58.37823949 | -71.18932808 | -374 | 47 | 163 | 8653 | -193 | -479 | 286 | -225 | -410 | -225 | -459 | 59 | 9 | 2 | 44 | 27 | 8 | 27 | 15 |
| 63.24310435 | -71.18932808 | -302 | 36 | 147 | 7597 | -146 | -393 | 247 | -194 | -361 | -170 | -377 | 76 | 11 | 3 | 38 | 31 | 10 | 31 | 20 |
| 68.10796922 | -71.18932808 | -206 | 40 | 150 | 8112 | -36 | -303 | 267 | -285 | -107 | -64 | -285 | 61 | 8 | 4 | 25 | 22 | 12 | 13 | 22 |
| 72.97283408 | -71.18932808 | -209 | 46 | 193 | 6980 | -62 | -304 | 241 | -263 | -92 | -92 | -281 | 58 | 8 | 2 | 36 | 22 | 6 | 6 | 15 |
| 77.83769895 | -71.18932808 | -278 | 47 | 208 | 6432 | -139 | -364 | 224 | -324 | -168 | -168 | -344 | 72 | 9 | 3 | 33 | 26 | 10 | 10 | 19 |
| 82.70256381 | -71.18932808 | -333 | 51 | 206 | 7173 | -177 | -424 | 247 | -388 | -209 | -209 | -403 | 61 | 7 | 3 | 25 | 20 | 10 | 10 | 17 |
| 87.56742868 | -71.18932808 | -357 | 52 | 203 | 7513 | -193 | -449 | 256 | -424 | -267 | -226 | -427 | 78 | 10 | 3 | 30 | 29 | 11 | 12 | 26 |
| 92.43229354 | -71.18932808 | -380 | 52 | 196 | 7926 | -208 | -474 | 266 | -449 | -286 | -241 | -451 | 72 | 11 | 3 | 39 | 30 | 9 | 11 | 27 |
| 97.29715841 | -71.18932808 | -389 | 51 | 183 | 8374 | -209 | -488 | 279 | -462 | -291 | -243 | -463 | 81 | 12 | 4 | 37 | 33 | 12 | 14 | 22 |
| 102.1620233 | -71.18932808 | -392 | 51 | 172 | 8917 | -203 | -497 | 294 | -470 | -287 | -236 | -472 | 77 | 12 | 3 | 42 | 33 | 10 | 14 | 19 |
| 107.0268881 | -71.18932808 | -395 | 56 | 173 | 9648 | -193 | -516 | 323 | -480 | -341 | -228 | -487 | 76 | 11 | 3 | 36 | 31 | 11 | 16 | 19 |
| 111.891753 | -71.18932808 | -380 | 58 | 186 | 9067 | -190 | -500 | 310 | -461 | -333 | -223 | -467 | 62 | 9 | 2 | 32 | 25 | 8 | 13 | 17 |
| 116.7566179 | -71.18932808 | -370 | 58 | 193 | 8728 | -186 | -485 | 299 | -451 | -327 | -219 | -455 | 85 | 11 | 3 | 28 | 31 | 12 | 18 | 26 |
| 121.6214827 | -71.18932808 | -373 | 59 | 201 | 8512 | -192 | -485 | 293 | -453 | -332 | -226 | -455 | 89 | 12 | 3 | 28 | 33 | 12 | 20 | 26 |
| 126.4863476 | -71.18932808 | -388 | 57 | 198 | 8439 | -208 | -496 | 289 | -467 | -347 | -242 | -468 | 87 | 11 | 3 | 26 | 30 | 11 | 22 | 23 |
| 131.3512125 | -71.18932808 | -395 | 54 | 186 | 8649 | -209 | -501 | 292 | -472 | -350 | -244 | -475 | 89 | 11 | 3 | 28 | 31 | 11 | 22 | 30 |
| 136.2160773 | -71.18932808 | -393 | 51 | 171 | 9061 | -199 | -499 | 300 | -475 | -343 | -234 | -475 | 99 | 11 | 3 | 25 | 31 | 12 | 27 | 31 |
| 141.0809422 | -71.18932808 | -375 | 50 | 163 | 9195 | -179 | -483 | 304 | -460 | -323 | -214 | -460 | 105 | 11 | 4 | 25 | 33 | 14 | 27 | 33 |
| 145.9458071 | -71.18932808 | -353 | 52 | 171 | 9133 | -160 | -466 | 306 | -333 | -301 | -195 | -440 | 135 | 16 | 4 | 30 | 46 | 15 | 32 | 39 |
| 150.8106719 | -71.18932808 | -325 | 56 | 193 | 8402 | -146 | -437 | 291 | -300 | -279 | -180 | -407 | 186 | 23 | 6 | 31 | 65 | 22 | 43 | 52 |
| 155.6755368 | -71.18932808 | -310 | 59 | 214 | 7728 | -145 | -421 | 276 | -284 | -200 | -178 | -387 | 156 | 19 | 6 | 31 | 55 | 20 | 35 | 36 |
| 160.5404017 | -71.18932808 | -242 | 50 | 192 | 7544 | -82 | -345 | 263 | -218 | -206 | -111 | -317 | 235 | 30 | 11 | 32 | 89 | 33 | 62 | 56 |
| 165.4052665 | -71.18932808 | -223 | 56 | 195 | 8341 | -64 | -353 | 289 | -174 | -188 | -87 | -316 | 280 | 41 | 13 | 34 | 114 | 40 | 71 | 67 |
| -63.24338213 | -68.83797673 | -103 | 42 | 196 | 6159 | 6 | -206 | 212 | -21 | -177 | -8 | -177 | 600 | 79 | 28 | 35 | 236 | 85 | 224 | 85 |
| 34.05391517 | -68.83797673 | -123 | 43 | 179 | 6964 | -5 | -243 | 238 | -144 | -22 | -22 | -214 | 777 | 95 | 31 | 32 | 271 | 94 | 94 | 182 |
| 43.78364489 | -68.83797673 | -216 | 54 | 217 | 6869 | -78 | -325 | 247 | -224 | -105 | -105 | -294 | 383 | 55 | 15 | 35 | 154 | 46 | 46 | 107 |
| 48.64850976 | -68.83797673 | -242 | 57 | 218 | 7084 | -88 | -347 | 260 | -263 | -123 | -123 | -317 | 356 | 44 | 15 | 27 | 124 | 53 | 53 | 106 |
| 53.51337462 | -68.83797673 | -317 | 62 | 228 | 7607 | -154 | -424 | 270 | -345 | -210 | -189 | -394 | 153 | 16 | 10 | 13 | 45 | 32 | 34 | 43 |
| 58.37823949 | -68.83797673 | -300 | 50 | 205 | 7027 | -149 | -394 | 245 | -179 | -279 | -179 | -370 | 134 | 14 | 8 | 15 | 41 | 26 | 41 | 36 |
| 63.24310435 | -68.83797673 | -259 | 44 | 192 | 6607 | -117 | -345 | 228 | -144 | -242 | -144 | -324 | 111 | 12 | 6 | 19 | 35 | 21 | 35 | 28 |
| 68.10796922 | -68.83797673 | -197 | 41 | 172 | 7232 | -49 | -288 | 239 | -71 | -227 | -71 | -269 | 271 | 35 | 15 | 28 | 101 | 49 | 101 | 66 |
| 72.97283408 | -68.83797673 | -136 | 39 | 161 | 7748 | 3 | -239 | 242 | -210 | -10 | -10 | -217 | 198 | 25 | 10 | 23 | 67 | 31 | 31 | 59 |
| 82.70256381 | -68.83797673 | -228 | 47 | 221 | 6128 | -97 | -312 | 215 | -272 | -153 | -122 | -291 | 183 | 22 | 9 | 27 | 63 | 29 | 30 | 52 |
| 87.56742868 | -68.83797673 | -262 | 49 | 225 | 6179 | -127 | -343 | 216 | -318 | -188 | -154 | -322 | 233 | 29 | 10 | 30 | 85 | 32 | 35 | 73 |
| 92.43229354 | -68.83797673 | -294 | 50 | 222 | 6445 | -151 | -374 | 223 | -350 | -216 | -180 | -352 | 162 | 20 | 6 | 31 | 59 | 22 | 22 | 54 |
| 97.29715841 | -68.83797673 | -312 | 51 | 211 | 7080 | -158 | -400 | 243 | -373 | -226 | -187 | -373 | 222 | 30 | 10 | 34 | 86 | 30 | 35 | 82 |
| 102.1620233 | -68.83797673 | -309 | 55 | 212 | 7380 | -150 | -408 | 258 | -372 | -218 | -180 | -377 | 149 | 21 | 7 | 32 | 59 | 21 | 25 | 42 |
| 107.0268881 | -68.83797673 | -302 | 54 | 213 | 7122 | -152 | -402 | 251 | -365 | -214 | -178 | -370 | 171 | 24 | 9 | 29 | 66 | 30 | 32 | 45 |
| 111.891753 | -68.83797673 | -262 | 48 | 215 | 6453 | -125 | -349 | 223 | -319 | -173 | -149 | -322 | 235 | 32 | 11 | 28 | 89 | 39 | 42 | 72 |
| 116.7566179 | -68.83797673 | -249 | 49 | 226 | 6178 | -117 | -334 | 217 | -305 | -164 | -141 | -308 | 320 | 41 | 14 | 26 | 115 | 52 | 59 | 96 |
| 121.6214827 | -68.83797673 | -265 | 51 | 231 | 6203 | -131 | -350 | 220 | -322 | -242 | -157 | -325 | 259 | 31 | 12 | 24 | 88 | 43 | 52 | 76 |
| 126.4863476 | -68.83797673 | -290 | 51 | 219 | 6651 | -146 | -380 | 233 | -347 | -264 | -174 | -353 | 326 | 38 | 13 | 23 | 108 | 48 | 73 | 87 |
| 131.3512125 | -68.83797673 | -302 | 57 | 223 | 7230 | -146 | -401 | 256 | -365 | -271 | -176 | -371 | 313 | 34 | 13 | 22 | 102 | 46 | 72 | 80 |
| 136.2160773 | -68.83797673 | -331 | 56 | 213 | 7498 | -169 | -430 | 261 | -398 | -296 | -200 | -401 | 251 | 27 | 10 | 23 | 80 | 34 | 63 | 71 |
| 141.0809422 | -68.83797673 | -288 | 44 | 191 | 6807 | -143 | -375 | 232 | -206 | -258 | -168 | -352 | 326 | 37 | 14 | 25 | 107 | 45 | 90 | 82 |
| 145.9458071 | -68.83797673 | -226 | 48 | 194 | 7207 | -75 | -323 | 249 | -137 | -193 | -101 | -297 | 541 | 68 | 20 | 28 | 191 | 68 | 147 | 127 |
| 150.8106719 | -68.83797673 | -187 | 53 | 203 | 7553 | -37 | -297 | 260 | -87 | -159 | -59 | -266 | 652 | 84 | 26 | 28 | 236 | 87 | 181 | 141 |
| -63.24338213 | -66.48662538 | -76 | 44 | 228 | 5484 | 15 | -179 | 194 | -3 | -132 | 4 | -148 | 725 | 89 | 36 | 27 | 258 | 119 | 227 | 120 |
| 53.51337462 | -66.48662538 | -205 | 52 | 219 | 6693 | -69 | -308 | 238 | -216 | -116 | -96 | -280 | 507 | 54 | 32 | 15 | 159 | 109 | 112 | 132 |
| 87.56742868 | -66.48662538 | -125 | 56 | 247 | 6439 | 4 | -224 | 228 | -187 | -17 | -17 | -194 | 458 | 53 | 17 | 30 | 157 | 59 | 59 | 137 |
| 97.29715841 | -66.48662538 | -143 | 56 | 260 | 5925 | -12 | -227 | 215 | -194 | -71 | -39 | -199 | 640 | 86 | 21 | 39 | 254 | 67 | 75 | 224 |
| 102.1620233 | -66.48662538 | -182 | 50 | 267 | 5175 | -69 | -256 | 187 | -228 | -115 | -92 | -231 | 279 | 36 | 11 | 33 | 107 | 35 | 39 | 94 |
| 107.0268881 | -66.48662538 | -121 | 40 | 220 | 5541 | -11 | -194 | 183 | -172 | -47 | -27 | -174 | 500 | 69 | 27 | 30 | 195 | 83 | 85 | 179 |
| 111.891753 | -66.48662538 | -170 | 48 | 245 | 5728 | -55 | -248 | 194 | -224 | -72 | -72 | -224 | 371 | 48 | 20 | 25 | 134 | 64 | 64 | 134 |
| 126.4863476 | -66.48662538 | -116 | 44 | 230 | 5616 | -5 | -194 | 189 | -168 | -103 | -22 | -172 | 1016 | 113 | 45 | 22 | 329 | 160 | 214 | 316 |
| 131.3512125 | -66.48662538 | -153 | 47 | 247 | 5356 | -44 | -233 | 189 | -203 | -85 | -63 | -208 | 778 | 85 | 34 | 24 | 254 | 120 | 151 | 238 |
| 136.2160773 | -66.48662538 | -150 | 44 | 227 | 5611 | -36 | -230 | 194 | -203 | -134 | -54 | -206 | 998 | 110 | 41 | 24 | 324 | 142 | 231 | 287 |
| -58.37851727 | -64.13527402 | -59 | 41 | 235 | 4830 | 17 | -157 | 174 | 3 | -117 | 7 | -126 | 530 | 59 | 32 | 21 | 175 | 99 | 163 | 102 |
| -68.108247 | -54.72986862 | 20 | 46 | 355 | 2959 | 86 | -43 | 129 | 59 | 15 | 60 | -21 | 643 | 63 | 42 | 12 | 187 | 135 | 173 | 137 |
| -72.97311186 | -52.37851727 | 45 | 36 | 320 | 2666 | 102 | -12 | 113 | 57 | 15 | 81 | 7 | 2006 | 200 | 133 | 12 | 596 | 421 | 505 | 439 |
| -72.97311186 | -50.02716592 | -23 | 53 | 337 | 3559 | 58 | -99 | 157 | -37 | 25 | 25 | -74 | 598 | 80 | 29 | 32 | 234 | 96 | 96 | 186 |
| -68.108247 | -50.02716592 | 93 | 51 | 323 | 3782 | 177 | 18 | 159 | 143 | 107 | 144 | 40 | 232 | 27 | 13 | 23 | 75 | 41 | 70 | 62 |
| -72.97311186 | -47.67581456 | 56 | 56 | 342 | 3461 | 139 | -24 | 163 | 43 | 104 | 104 | 6 | 895 | 104 | 43 | 26 | 306 | 154 | 154 | 270 |
| -68.108247 | -47.67581456 | 114 | 88 | 377 | 5230 | 238 | 4 | 234 | 53 | 159 | 185 | 41 | 175 | 21 | 12 | 21 | 61 | 36 | 44 | 52 |
| -72.97311186 | -45.32446321 | 65 | 52 | 346 | 3074 | 145 | -6 | 151 | 54 | 97 | 108 | 21 | 2624 | 309 | 161 | 21 | 840 | 505 | 516 | 800 |
| -68.108247 | -45.32446321 | 114 | 92 | 383 | 5258 | 241 | 1 | 240 | 85 | 105 | 185 | 40 | 204 | 28 | 10 | 29 | 75 | 34 | 43 | 67 |
| 170.2701314 | -45.32446321 | 95 | 63 | 367 | 3646 | 182 | 10 | 171 | 142 | 68 | 145 | 41 | 444 | 48 | 27 | 18 | 143 | 83 | 122 | 94 |
| -68.108247 | -42.97311186 | 92 | 98 | 382 | 5617 | 230 | -27 | 257 | 61 | 117 | 169 | 15 | 211 | 24 | 11 | 18 | 67 | 40 | 46 | 59 |
| 145.9458071 | -42.97311186 | 116 | 47 | 381 | 2425 | 186 | 64 | 122 | 86 | 152 | 152 | 84 | 2282 | 247 | 103 | 23 | 737 | 349 | 349 | 687 |
| -72.97311186 | -40.62176051 | 120 | 69 | 391 | 3452 | 220 | 43 | 177 | 85 | 170 | 170 | 75 | 1367 | 205 | 49 | 51 | 608 | 157 | 157 | 583 |
| -68.108247 | -40.62176051 | 104 | 101 | 374 | 6091 | 250 | -21 | 270 | 159 | 31 | 189 | 23 | 179 | 23 | 10 | 22 | 63 | 32 | 43 | 41 |
| -63.24338213 | -40.62176051 | 147 | 80 | 360 | 5192 | 262 | 39 | 223 | 196 | 76 | 219 | 76 | 380 | 49 | 19 | 27 | 142 | 65 | 107 | 65 |
| -72.97311186 | -38.27040916 | 122 | 84 | 424 | 3640 | 236 | 38 | 198 | 86 | 176 | 176 | 76 | 1077 | 227 | 19 | 77 | 575 | 62 | 62 | 534 |
| -68.108247 | -38.27040916 | 146 | 112 | 382 | 6431 | 295 | 0 | 295 | 201 | 69 | 234 | 56 | 217 | 29 | 12 | 26 | 72 | 37 | 52 | 45 |
| -63.24338213 | -38.27040916 | 148 | 91 | 368 | 5648 | 274 | 25 | 249 | 197 | 80 | 226 | 69 | 583 | 75 | 21 | 37 | 209 | 66 | 177 | 74 |
| -58.37851727 | -38.27040916 | 142 | 63 | 339 | 4344 | 239 | 53 | 186 | 189 | 88 | 202 | 84 | 791 | 84 | 49 | 17 | 240 | 148 | 221 | 155 |
| 145.9458071 | -38.27040916 | 128 | 56 | 366 | 3134 | 217 | 64 | 153 | 100 | 175 | 175 | 86 | 1248 | 139 | 50 | 27 | 407 | 163 | 163 | 362 |
| 175.1349962 | -38.27040916 | 142 | 63 | 384 | 3248 | 229 | 65 | 164 | 99 | 190 | 190 | 99 | 1421 | 159 | 82 | 20 | 467 | 255 | 255 | 467 |
| -68.108247 | -35.91905781 | 135 | 108 | 389 | 5881 | 271 | -6 | 277 | 185 | 53 | 215 | 53 | 440 | 68 | 9 | 59 | 194 | 33 | 193 | 33 |
| -63.24338213 | -35.91905781 | 157 | 90 | 366 | 5444 | 277 | 30 | 247 | 205 | 78 | 231 | 78 | 842 | 121 | 20 | 50 | 334 | 67 | 303 | 67 |
| -58.37851727 | -35.91905781 | 158 | 86 | 388 | 4885 | 268 | 45 | 223 | 224 | 89 | 224 | 89 | 925 | 101 | 49 | 25 | 291 | 155 | 291 | 155 |
| 141.0809422 | -35.91905781 | 151 | 93 | 407 | 4493 | 283 | 55 | 229 | 96 | 198 | 216 | 91 | 408 | 51 | 19 | 33 | 148 | 57 | 58 | 133 |
| 145.9458071 | -35.91905781 | 161 | 93 | 373 | 5393 | 299 | 48 | 251 | 103 | 238 | 238 | 88 | 468 | 49 | 30 | 15 | 141 | 96 | 96 | 135 |
| -68.108247 | -33.56770646 | 164 | 98 | 381 | 5523 | 284 | 27 | 257 | 231 | 96 | 236 | 83 | 240 | 43 | 6 | 64 | 125 | 19 | 113 | 21 |
| -63.24338213 | -33.56770646 | 174 | 88 | 370 | 5133 | 287 | 50 | 237 | 218 | 98 | 243 | 98 | 868 | 130 | 16 | 59 | 359 | 54 | 354 | 54 |
| -58.37851727 | -33.56770646 | 175 | 88 | 395 | 4880 | 284 | 60 | 223 | 222 | 106 | 242 | 106 | 1139 | 134 | 59 | 25 | 401 | 193 | 325 | 193 |
| -53.5136524 | -33.56770646 | 176 | 56 | 332 | 3990 | 258 | 90 | 169 | 122 | 199 | 230 | 122 | 1135 | 115 | 78 | 11 | 333 | 246 | 318 | 333 |
| 19.45932057 | -33.56770646 | 128 | 108 | 464 | 4094 | 247 | 13 | 233 | 74 | 183 | 183 | 71 | 432 | 58 | 16 | 42 | 169 | 50 | 50 | 168 |
| 24.32418544 | -33.56770646 | 152 | 112 | 498 | 3605 | 260 | 36 | 224 | 191 | 100 | 201 | 100 | 404 | 45 | 22 | 18 | 125 | 74 | 102 | 74 |
| 116.7566179 | -33.56770646 | 156 | 91 | 425 | 4012 | 278 | 64 | 215 | 105 | 207 | 213 | 105 | 539 | 95 | 13 | 63 | 273 | 49 | 54 | 273 |
| 121.6214827 | -33.56770646 | 169 | 67 | 412 | 3162 | 255 | 92 | 163 | 125 | 207 | 212 | 125 | 484 | 60 | 19 | 37 | 180 | 61 | 71 | 180 |
| 136.2160773 | -33.56770646 | 165 | 83 | 415 | 3890 | 274 | 75 | 200 | 114 | 219 | 219 | 111 | 368 | 49 | 14 | 40 | 145 | 48 | 48 | 141 |
| 141.0809422 | -33.56770646 | 179 | 103 | 398 | 5373 | 317 | 59 | 258 | 169 | 230 | 253 | 104 | 260 | 27 | 14 | 15 | 79 | 50 | 63 | 68 |
| 145.9458071 | -33.56770646 | 176 | 100 | 375 | 5940 | 316 | 48 | 268 | 255 | 231 | 257 | 94 | 382 | 41 | 27 | 15 | 111 | 85 | 95 | 86 |
| 150.8106719 | -33.56770646 | 178 | 72 | 387 | 4009 | 269 | 84 | 185 | 231 | 125 | 231 | 120 | 858 | 113 | 37 | 33 | 329 | 133 | 329 | 143 |
| -68.108247 | -31.21635511 | 166 | 102 | 387 | 5417 | 285 | 21 | 264 | 224 | 135 | 233 | 83 | 80 | 17 | 2 | 77 | 51 | 6 | 44 | 12 |
| -63.24338213 | -31.21635511 | 186 | 90 | 381 | 4976 | 297 | 60 | 236 | 251 | 125 | 251 | 111 | 803 | 132 | 12 | 66 | 378 | 38 | 378 | 45 |
| -58.37851727 | -31.21635511 | 188 | 88 | 400 | 4712 | 293 | 74 | 219 | 201 | 131 | 252 | 121 | 1179 | 148 | 49 | 30 | 417 | 153 | 314 | 177 |
| -53.5136524 | -31.21635511 | 177 | 84 | 422 | 4104 | 274 | 75 | 199 | 119 | 215 | 231 | 119 | 1354 | 145 | 94 | 12 | 400 | 288 | 377 | 400 |
| 19.45932057 | -31.21635511 | 185 | 128 | 473 | 4873 | 324 | 54 | 271 | 131 | 245 | 250 | 117 | 240 | 33 | 8 | 35 | 92 | 28 | 32 | 87 |
| 24.32418544 | -31.21635511 | 148 | 130 | 461 | 5196 | 283 | 2 | 281 | 195 | 117 | 213 | 69 | 348 | 63 | 10 | 62 | 177 | 32 | 130 | 33 |
| 29.1890503 | -31.21635511 | 168 | 89 | 482 | 3016 | 248 | 63 | 185 | 202 | 120 | 207 | 120 | 1001 | 135 | 28 | 48 | 391 | 90 | 389 | 90 |
| 116.7566179 | -31.21635511 | 186 | 102 | 403 | 5152 | 325 | 73 | 252 | 120 | 228 | 259 | 120 | 378 | 73 | 9 | 71 | 213 | 30 | 39 | 213 |
| 121.6214827 | -31.21635511 | 190 | 105 | 417 | 5128 | 321 | 68 | 252 | 257 | 181 | 257 | 117 | 254 | 30 | 14 | 22 | 85 | 42 | 68 | 70 |
| 126.4863476 | -31.21635511 | 187 | 110 | 459 | 4334 | 307 | 68 | 239 | 137 | 207 | 243 | 122 | 204 | 22 | 14 | 15 | 61 | 42 | 58 | 50 |
| 131.3512125 | -31.21635511 | 179 | 73 | 412 | 3483 | 269 | 92 | 177 | 169 | 221 | 226 | 128 | 237 | 27 | 13 | 22 | 75 | 39 | 41 | 71 |
| 136.2160773 | -31.21635511 | 192 | 105 | 396 | 5495 | 329 | 65 | 264 | 128 | 207 | 266 | 113 | 186 | 19 | 11 | 14 | 57 | 37 | 49 | 49 |
| 141.0809422 | -31.21635511 | 202 | 104 | 375 | 6127 | 340 | 62 | 278 | 283 | 160 | 283 | 113 | 179 | 20 | 12 | 16 | 56 | 38 | 56 | 43 |
| 145.9458071 | -31.21635511 | 193 | 101 | 374 | 6123 | 327 | 56 | 271 | 274 | 149 | 274 | 106 | 337 | 35 | 22 | 14 | 102 | 71 | 96 | 77 |
| 150.8106719 | -31.21635511 | 165 | 91 | 397 | 4928 | 273 | 44 | 229 | 228 | 92 | 228 | 92 | 688 | 105 | 36 | 35 | 295 | 114 | 295 | 114 |
| -68.108247 | -28.86500375 | 156 | 107 | 446 | 4378 | 262 | 21 | 241 | 201 | 102 | 207 | 87 | 93 | 24 | 1 | 95 | 66 | 4 | 61 | 5 |
| -63.24338213 | -28.86500375 | 206 | 89 | 383 | 4813 | 311 | 79 | 233 | 267 | 149 | 267 | 131 | 743 | 128 | 9 | 71 | 371 | 27 | 371 | 32 |
| -58.37851727 | -28.86500375 | 204 | 88 | 415 | 4394 | 304 | 92 | 212 | 217 | 152 | 263 | 141 | 1468 | 187 | 54 | 37 | 547 | 165 | 426 | 182 |
| -53.5136524 | -28.86500375 | 189 | 83 | 432 | 3826 | 279 | 88 | 192 | 182 | 227 | 237 | 133 | 1599 | 158 | 109 | 11 | 468 | 347 | 383 | 407 |
| 19.45932057 | -28.86500375 | 204 | 138 | 465 | 5615 | 346 | 50 | 296 | 259 | 132 | 274 | 121 | 125 | 27 | 2 | 74 | 78 | 9 | 62 | 13 |
| 24.32418544 | -28.86500375 | 187 | 133 | 439 | 5759 | 324 | 22 | 302 | 233 | 97 | 258 | 97 | 364 | 64 | 5 | 71 | 186 | 16 | 157 | 16 |
| 29.1890503 | -28.86500375 | 129 | 118 | 493 | 4110 | 230 | -9 | 239 | 176 | 62 | 176 | 62 | 1160 | 216 | 9 | 75 | 618 | 30 | 618 | 30 |
| 116.7566179 | -28.86500375 | 207 | 112 | 399 | 5933 | 353 | 72 | 280 | 146 | 226 | 289 | 126 | 303 | 51 | 7 | 55 | 145 | 23 | 73 | 137 |
| 121.6214827 | -28.86500375 | 215 | 112 | 401 | 6053 | 350 | 70 | 280 | 267 | 210 | 295 | 126 | 221 | 31 | 9 | 38 | 87 | 27 | 53 | 60 |
| 126.4863476 | -28.86500375 | 206 | 116 | 426 | 5491 | 339 | 66 | 273 | 275 | 205 | 276 | 124 | 245 | 29 | 13 | 26 | 82 | 39 | 76 | 47 |
| 131.3512125 | -28.86500375 | 194 | 110 | 411 | 5514 | 327 | 59 | 268 | 245 | 112 | 265 | 112 | 221 | 24 | 13 | 18 | 68 | 41 | 65 | 41 |
| 136.2160773 | -28.86500375 | 215 | 110 | 385 | 6247 | 354 | 69 | 285 | 296 | 138 | 297 | 124 | 152 | 20 | 8 | 32 | 60 | 27 | 55 | 30 |
| 141.0809422 | -28.86500375 | 220 | 109 | 376 | 6492 | 357 | 67 | 290 | 303 | 138 | 305 | 124 | 178 | 26 | 8 | 34 | 75 | 29 | 62 | 38 |
| 145.9458071 | -28.86500375 | 212 | 108 | 383 | 6188 | 345 | 62 | 283 | 291 | 170 | 291 | 120 | 354 | 47 | 15 | 30 | 126 | 49 | 126 | 68 |
| 150.8106719 | -28.86500375 | 192 | 100 | 418 | 4993 | 303 | 62 | 241 | 254 | 135 | 254 | 117 | 625 | 79 | 32 | 30 | 228 | 112 | 228 | 119 |
| -68.108247 | -26.5136524 | 14 | 129 | 518 | 4045 | 129 | -119 | 248 | 58 | -15 | 65 | -47 | 37 | 9 | 1 | 91 | 27 | 3 | 22 | 7 |

| | | | | | | | | | | | | | | | | | | | | | |
|---|---|---|---|---|---|---|---|---|---|---|---|---|---|---|---|---|---|---|---|---|---|
| -63.24338213 | -26.5136524 | 217 | 89 | 405 | 4316 | 314 | 96 | 219 | 270 | 168 | 270 | 148 | 809 | 155 | 6 | 81 | 436 | 18 | 436 | 22 |
| -58.37851727 | -26.5136524 | 221 | 87 | 428 | 4076 | 315 | 112 | 203 | 234 | 173 | 274 | 161 | 1368 | 166 | 48 | 36 | 483 | 144 | 436 | 166 |
| -53.5136524 | -26.5136524 | 187 | 80 | 457 | 3395 | 268 | 91 | 176 | 185 | 148 | 228 | 136 | 2361 | 270 | 151 | 15 | 751 | 459 | 605 | 496 |
| -48.64878754 | -26.5136524 | 219 | 56 | 395 | 2885 | 287 | 144 | 142 | 258 | 191 | 258 | 180 | 1987 | 261 | 102 | 30 | 753 | 330 | 753 | 372 |
| 19.45932057 | -26.5136524 | 209 | 134 | 449 | 5657 | 346 | 47 | 299 | 258 | 137 | 277 | 122 | 183 | 47 | 0 | 107 | 137 | 1 | 80 | 3 |
| 24.32418544 | -26.5136524 | 182 | 126 | 456 | 5152 | 300 | 25 | 276 | 242 | 120 | 242 | 99 | 486 | 94 | 3 | 78 | 255 | 11 | 255 | 13 |
| 29.1890503 | -26.5136524 | 157 | 122 | 509 | 3896 | 256 | 17 | 239 | 199 | 91 | 200 | 91 | 648 | 110 | 5 | 74 | 326 | 17 | 278 | 17 |
| 116.7566179 | -26.5136524 | 230 | 115 | 397 | 6257 | 371 | 82 | 289 | 162 | 217 | 313 | 141 | 222 | 35 | 5 | 52 | 92 | 19 | 85 | 91 |
| 121.6214827 | -26.5136524 | 236 | 115 | 389 | 6413 | 374 | 77 | 296 | 311 | 193 | 320 | 139 | 230 | 42 | 5 | 57 | 117 | 17 | 88 | 48 |
| 126.4863476 | -26.5136524 | 222 | 118 | 406 | 6191 | 355 | 65 | 290 | 145 | 227 | 299 | 128 | 187 | 23 | 5 | 32 | 62 | 23 | 61 | 53 |
| 131.3512125 | -26.5136524 | 202 | 116 | 400 | 6267 | 337 | 47 | 291 | 280 | 107 | 280 | 107 | 290 | 50 | 13 | 46 | 140 | 43 | 103 | 43 |
| 136.2160773 | -26.5136524 | 237 | 114 | 385 | 6607 | 374 | 77 | 297 | 319 | 197 | 322 | 138 | 180 | 27 | 8 | 38 | 80 | 26 | 68 | 32 |
| 141.0809422 | -26.5136524 | 234 | 111 | 380 | 6443 | 369 | 76 | 293 | 316 | 194 | 316 | 137 | 213 | 44 | 7 | 59 | 106 | 23 | 106 | 35 |
| 145.9458071 | -26.5136524 | 215 | 109 | 394 | 5879 | 341 | 65 | 276 | 284 | 139 | 288 | 126 | 499 | 80 | 19 | 46 | 222 | 64 | 218 | 80 |
| 150.8106719 | -26.5136524 | 201 | 95 | 427 | 4494 | 303 | 79 | 224 | 256 | 171 | 256 | 132 | 673 | 97 | 28 | 41 | 269 | 90 | 269 | 111 |
| -68.108247 | -24.16230105 | 7 | 145 | 588 | 3273 | 124 | -122 | 246 | 48 | 30 | 48 | -43 | 52 | 13 | 0 | 78 | 29 | 2 | 29 | 16 |
| -63.24338213 | -24.16230105 | 232 | 86 | 409 | 4072 | 329 | 119 | 210 | 280 | 187 | 280 | 168 | 648 | 127 | 4 | 82 | 368 | 14 | 368 | 16 |
| -58.37851727 | -24.16230105 | 234 | 85 | 449 | 3640 | 319 | 128 | 190 | 269 | 191 | 279 | 178 | 1243 | 163 | 28 | 44 | 451 | 90 | 418 | 134 |
| -53.5136524 | -24.16230105 | 224 | 78 | 479 | 2963 | 295 | 133 | 162 | 257 | 190 | 257 | 177 | 1445 | 157 | 67 | 25 | 464 | 211 | 464 | 234 |
| -48.64878754 | -24.16230105 | 181 | 78 | 509 | 2601 | 249 | 96 | 152 | 210 | 149 | 213 | 143 | 1548 | 219 | 66 | 38 | 621 | 204 | 597 | 228 |
| 14.59445571 | -24.16230105 | 202 | 79 | 565 | 2025 | 267 | 127 | 141 | 228 | 193 | 228 | 173 | 16 | 4 | 0 | 94 | 11 | 0 | 7 | 3 |
| 19.45932057 | -24.16230105 | 213 | 128 | 452 | 5146 | 338 | 55 | 283 | 250 | 130 | 272 | 130 | 249 | 60 | 0 | 99 | 170 | 2 | 110 | 2 |
| 24.32418544 | -24.16230105 | 203 | 122 | 467 | 4800 | 309 | 48 | 261 | 250 | 144 | 253 | 123 | 418 | 81 | 1 | 82 | 229 | 5 | 226 | 7 |
| 29.1890503 | -24.16230105 | 161 | 117 | 514 | 3665 | 253 | 26 | 227 | 200 | 99 | 201 | 99 | 499 | 94 | 2 | 82 | 277 | 9 | 232 | 9 |
| 34.05391517 | -24.16230105 | 237 | 87 | 508 | 2898 | 311 | 140 | 171 | 271 | 199 | 271 | 191 | 696 | 128 | 20 | 65 | 366 | 60 | 366 | 64 |
| 43.78364489 | -24.16230105 | 248 | 66 | 494 | 2414 | 306 | 171 | 134 | 275 | 216 | 277 | 211 | 456 | 106 | 7 | 90 | 286 | 23 | 264 | 33 |
| 116.7566179 | -24.16230105 | 247 | 114 | 405 | 5963 | 379 | 97 | 282 | 321 | 239 | 323 | 159 | 245 | 57 | 2 | 77 | 151 | 6 | 83 | 67 |
| 121.6214827 | -24.16230105 | 249 | 115 | 390 | 6265 | 382 | 87 | 295 | 316 | 258 | 327 | 152 | 294 | 65 | 3 | 72 | 169 | 11 | 109 | 50 |
| 126.4863476 | -24.16230105 | 250 | 115 | 394 | 6293 | 378 | 86 | 292 | 319 | 214 | 325 | 151 | 243 | 44 | 3 | 56 | 123 | 20 | 102 | 45 |
| 131.3512125 | -24.16230105 | 222 | 117 | 401 | 6247 | 352 | 60 | 292 | 295 | 144 | 299 | 126 | 289 | 46 | 5 | 50 | 122 | 25 | 98 | 48 |
| 136.2160773 | -24.16230105 | 242 | 113 | 381 | 6471 | 378 | 81 | 297 | 317 | 207 | 323 | 143 | 229 | 39 | 4 | 52 | 110 | 16 | 90 | 32 |
| 141.0809422 | -24.16230105 | 248 | 111 | 390 | 6035 | 376 | 91 | 286 | 317 | 212 | 321 | 155 | 235 | 50 | 6 | 69 | 136 | 19 | 122 | 32 |
| 145.9458071 | -24.16230105 | 223 | 104 | 412 | 5036 | 337 | 85 | 252 | 283 | 194 | 283 | 144 | 447 | 73 | 15 | 50 | 203 | 47 | 203 | 58 |
| 150.8106719 | -24.16230105 | 218 | 72 | 409 | 3627 | 296 | 120 | 176 | 261 | 162 | 261 | 162 | 890 | 156 | 26 | 63 | 447 | 82 | 442 | 82 |
| -68.108247 | -21.8109497 | 17 | 136 | 636 | 2565 | 120 | -94 | 214 | 49 | -12 | 49 | -22 | 46 | 26 | 0 | 190 | 53 | 0 | 53 | 1 |
| -63.24338213 | -21.8109497 | 226 | 83 | 441 | 3394 | 310 | 123 | 187 | 264 | 188 | 264 | 171 | 954 | 177 | 5 | 83 | 505 | 19 | 505 | 22 |
| -58.37851727 | -21.8109497 | 249 | 81 | 476 | 3010 | 323 | 153 | 170 | 283 | 201 | 283 | 201 | 1150 | 163 | 26 | 48 | 481 | 96 | 481 | 96 |
| -53.5136524 | -21.8109497 | 238 | 84 | 532 | 2377 | 304 | 147 | 158 | 261 | 198 | 262 | 198 | 1617 | 241 | 33 | 45 | 668 | 127 | 530 | 127 |
| -48.64878754 | -21.8109497 | 232 | 84 | 553 | 2058 | 296 | 144 | 152 | 249 | 210 | 251 | 196 | 1123 | 186 | 21 | 63 | 548 | 69 | 526 | 86 |
| -43.78392267 | -21.8109497 | 211 | 78 | 522 | 2077 | 280 | 130 | 150 | 233 | 179 | 238 | 179 | 1276 | 231 | 15 | 72 | 674 | 50 | 581 | 50 |
| 14.59445571 | -21.8109497 | 217 | 120 | 643 | 1967 | 302 | 116 | 186 | 243 | 199 | 243 | 187 | 40 | 17 | 0 | 145 | 42 | 0 | 42 | 0 |
| 19.45932057 | -21.8109497 | 197 | 123 | 473 | 4470 | 307 | 46 | 261 | 236 | 122 | 245 | 122 | 356 | 80 | 0 | 97 | 239 | 1 | 128 | 1 |
| 24.32418544 | -21.8109497 | 221 | 119 | 469 | 4412 | 325 | 72 | 253 | 263 | 166 | 263 | 145 | 341 | 78 | 0 | 91 | 211 | 2 | 211 | 3 |
| 29.1890503 | -21.8109497 | 223 | 112 | 498 | 3750 | 317 | 91 | 225 | 261 | 177 | 261 | 159 | 379 | 77 | 0 | 83 | 217 | 2 | 217 | 6 |
| 34.05391517 | -21.8109497 | 237 | 92 | 533 | 2783 | 310 | 137 | 173 | 269 | 201 | 269 | 192 | 715 | 161 | 4 | 97 | 441 | 16 | 441 | 18 |
| 43.78364489 | -21.8109497 | 254 | 75 | 513 | 2228 | 314 | 167 | 147 | 275 | 223 | 276 | 217 | 581 | 223 | 1 | 139 | 537 | 3 | 206 | 4 |
| 116.7566179 | -21.8109497 | 256 | 113 | 449 | 4779 | 373 | 121 | 252 | 309 | 254 | 313 | 183 | 392 | 100 | 1 | 100 | 288 | 5 | 233 | 51 |
| 121.6214827 | -21.8109497 | 265 | 114 | 419 | 5412 | 387 | 114 | 273 | 318 | 232 | 327 | 179 | 276 | 70 | 2 | 86 | 191 | 8 | 129 | 36 |
| 126.4863476 | -21.8109497 | 263 | 112 | 407 | 5665 | 381 | 107 | 275 | 317 | 233 | 326 | 172 | 321 | 64 | 7 | 73 | 186 | 23 | 159 | 43 |
| 131.3512125 | -21.8109497 | 244 | 113 | 409 | 5639 | 363 | 88 | 275 | 303 | 219 | 309 | 154 | 373 | 77 | 6 | 79 | 223 | 18 | 207 | 27 |
| 136.2160773 | -21.8109497 | 247 | 112 | 402 | 5717 | 370 | 92 | 279 | 308 | 174 | 314 | 156 | 278 | 63 | 3 | 85 | 187 | 14 | 157 | 21 |
| 141.0809422 | -21.8109497 | 247 | 109 | 413 | 5177 | 363 | 100 | 263 | 299 | 219 | 306 | 164 | 386 | 106 | 4 | 97 | 266 | 12 | 263 | 23 |
| 145.9458071 | -21.8109497 | 236 | 103 | 436 | 4346 | 343 | 107 | 236 | 279 | 211 | 285 | 166 | 501 | 99 | 9 | 72 | 291 | 30 | 261 | 51 |
| -68.108247 | -19.45959835 | 74 | 140 | 608 | 2724 | 183 | -48 | 231 | 99 | 31 | 104 | 31 | 180 | 66 | 0 | 148 | 178 | 1 | 39 | 1 |
| -63.24338213 | -19.45959835 | 215 | 78 | 493 | 2595 | 286 | 128 | 158 | 244 | 186 | 244 | 172 | 989 | 177 | 18 | 70 | 510 | 56 | 510 | 65 |
| -58.37851727 | -19.45959835 | 258 | 79 | 505 | 2383 | 325 | 169 | 156 | 282 | 218 | 283 | 218 | 1012 | 159 | 18 | 54 | 458 | 68 | 407 | 68 |
| -53.5136524 | -19.45959835 | 240 | 83 | 554 | 1673 | 305 | 154 | 150 | 251 | 209 | 257 | 209 | 1340 | 225 | 15 | 61 | 651 | 61 | 379 | 61 |
| -48.64878754 | -19.45959835 | 230 | 82 | 562 | 1569 | 295 | 149 | 146 | 239 | 202 | 245 | 202 | 1440 | 265 | 12 | 77 | 784 | 41 | 299 | 41 |
| -43.78392267 | -19.45959835 | 218 | 79 | 550 | 1834 | 281 | 137 | 144 | 234 | 194 | 239 | 186 | 1426 | 306 | 10 | 88 | 860 | 31 | 623 | 34 |
| 14.59445571 | -19.45959835 | 213 | 135 | 596 | 2665 | 313 | 87 | 226 | 232 | 165 | 237 | 165 | 252 | 71 | 0 | 120 | 201 | 0 | 54 | 0 |
| 19.45932057 | -19.45959835 | 214 | 117 | 466 | 3718 | 323 | 73 | 250 | 237 | 169 | 254 | 148 | 529 | 135 | 0 | 108 | 393 | 0 | 111 | 1 |
| 24.32418544 | -19.45959835 | 227 | 112 | 461 | 3714 | 334 | 90 | 244 | 255 | 182 | 267 | 161 | 480 | 119 | 0 | 105 | 330 | 0 | 122 | 1 |
| 29.1890503 | -19.45959835 | 197 | 112 | 508 | 3221 | 295 | 74 | 221 | 224 | 158 | 229 | 140 | 612 | 137 | 1 | 101 | 405 | 3 | 206 | 5 |
| 34.05391517 | -19.45959835 | 236 | 90 | 517 | 2646 | 313 | 139 | 174 | 262 | 224 | 264 | 192 | 1075 | 215 | 10 | 90 | 638 | 42 | 622 | 47 |
| 48.64850976 | -19.45959835 | 226 | 52 | 445 | 2361 | 280 | 164 | 116 | 254 | 201 | 254 | 192 | 2208 | 346 | 72 | 54 | 1000 | 246 | 1000 | 320 |
| 121.6214827 | -19.45959835 | 275 | 94 | 469 | 3570 | 358 | 158 | 200 | 307 | 250 | 311 | 216 | 400 | 94 | 1 | 105 | 280 | 4 | 221 | 27 |
| 126.4863476 | -19.45959835 | 273 | 109 | 436 | 4520 | 382 | 132 | 251 | 307 | 255 | 323 | 198 | 483 | 127 | 2 | 109 | 372 | 10 | 169 | 22 |
| 131.3512125 | -19.45959835 | 264 | 109 | 429 | 4884 | 370 | 116 | 254 | 307 | 202 | 315 | 183 | 450 | 109 | 3 | 100 | 321 | 13 | 172 | 20 |
| 136.2160773 | -19.45959835 | 262 | 110 | 434 | 4761 | 370 | 117 | 253 | 305 | 200 | 313 | 183 | 371 | 91 | 2 | 96 | 261 | 9 | 152 | 14 |
| 141.0809422 | -19.45959835 | 268 | 103 | 458 | 3884 | 363 | 139 | 224 | 298 | 215 | 310 | 202 | 512 | 144 | 2 | 113 | 400 | 7 | 193 | 15 |
| 145.9458071 | -19.45959835 | 233 | 91 | 457 | 3446 | 323 | 125 | 198 | 272 | 211 | 272 | 177 | 578 | 131 | 8 | 85 | 346 | 25 | 346 | 35 |
| -68.108247 | -17.108247 | 69 | 104 | 571 | 2162 | 146 | -35 | 181 | 91 | 41 | 91 | 33 | 493 | 135 | 3 | 99 | 342 | 12 | 175 | 13 |
| -63.24338213 | -17.108247 | 247 | 68 | 492 | 1930 | 311 | 172 | 138 | 263 | 228 | 268 | 214 | 1532 | 233 | 50 | 49 | 689 | 151 | 413 | 173 |
| -58.37851727 | -17.108247 | 259 | 76 | 504 | 1854 | 325 | 173 | 152 | 270 | 226 | 279 | 226 | 1107 | 194 | 14 | 68 | 554 | 50 | 325 | 50 |
| -53.5136524 | -17.108247 | 231 | 79 | 551 | 1193 | 296 | 153 | 142 | 234 | 208 | 245 | 208 | 1434 | 256 | 9 | 74 | 760 | 35 | 277 | 35 |
| -48.64878754 | -17.108247 | 232 | 82 | 574 | 1243 | 300 | 157 | 143 | 235 | 209 | 249 | 209 | 1361 | 252 | 6 | 79 | 726 | 21 | 306 | 21 |
| -43.78392267 | -17.108247 | 234 | 78 | 569 | 1541 | 297 | 159 | 138 | 241 | 206 | 249 | 206 | 903 | 206 | 4 | 92 | 578 | 12 | 339 | 12 |
| 14.59445571 | -17.108247 | 227 | 128 | 535 | 2644 | 338 | 99 | 239 | 232 | 179 | 256 | 179 | 460 | 114 | 0 | 114 | 338 | 0 | 105 | 0 |
| 19.45932057 | -17.108247 | 219 | 109 | 461 | 3010 | 329 | 93 | 236 | 229 | 166 | 259 | 166 | 593 | 149 | 0 | 108 | 415 | 0 | 41 | 0 |
| 24.32418544 | -17.108247 | 225 | 105 | 457 | 3070 | 332 | 102 | 230 | 238 | 191 | 263 | 170 | 696 | 171 | 0 | 113 | 494 | 0 | 176 | 1 |
| 29.1890503 | -17.108247 | 215 | 104 | 470 | 2961 | 318 | 98 | 220 | 231 | 182 | 255 | 165 | 777 | 194 | 1 | 113 | 563 | 3 | 169 | 4 |
| 34.05391517 | -17.108247 | 239 | 88 | 495 | 2472 | 323 | 146 | 177 | 256 | 232 | 268 | 197 | 640 | 158 | 3 | 110 | 452 | 10 | 130 | 12 |
| 48.64850976 | -17.108247 | 189 | 57 | 490 | 2130 | 244 | 127 | 117 | 212 | 181 | 212 | 157 | 2435 | 404 | 73 | 56 | 1126 | 239 | 1126 | 334 |
| 126.4863476 | -17.108247 | 271 | 102 | 482 | 3045 | 365 | 154 | 211 | 285 | 233 | 308 | 218 | 694 | 205 | 2 | 117 | 563 | 7 | 129 | 10 |
| 131.3512125 | -17.108247 | 276 | 104 | 463 | 3698 | 369 | 145 | 224 | 300 | 228 | 315 | 212 | 557 | 136 | 1 | 106 | 399 | 8 | 119 | 16 |
| 136.2160773 | -17.108247 | 260 | 110 | 496 | 3270 | 358 | 136 | 222 | 282 | 217 | 297 | 204 | 730 | 192 | 0 | 113 | 561 | 4 | 280 | 12 |
| 141.0809422 | -17.108247 | 269 | 77 | 467 | 2611 | 338 | 174 | 165 | 286 | 256 | 299 | 225 | 965 | 269 | 1 | 126 | 801 | 4 | 333 | 14 |
| -72.97311186 | -14.75689565 | 44 | 98 | 616 | 1356 | 114 | -45 | 159 | 55 | 21 | 56 | 21 | 606 | 137 | 4 | 98 | 403 | 16 | 144 | 16 |
| -68.108247 | -14.75689565 | 162 | 62 | 572 | 1323 | 206 | 98 | 108 | 173 | 139 | 175 | 139 | 1874 | 283 | 40 | 57 | 837 | 133 | 700 | 133 |
| -63.24338213 | -14.75689565 | 230 | 66 | 513 | 1317 | 286 | 158 | 129 | 239 | 206 | 244 | 206 | 1708 | 266 | 32 | 58 | 777 | 111 | 330 | 111 |
| -58.37851727 | -14.75689565 | 239 | 73 | 548 | 1196 | 297 | 163 | 134 | 246 | 216 | 252 | 216 | 1466 | 238 | 10 | 68 | 694 | 38 | 310 | 38 |
| -53.5136524 | -14.75689565 | 238 | 74 | 540 | 880 | 306 | 169 | 137 | 237 | 224 | 252 | 224 | 2501 | 474 | 10 | 84 | 1422 | 31 | 154 | 31 |
| -48.64878754 | -14.75689565 | 252 | 80 | 567 | 1011 | 326 | 186 | 140 | 247 | 238 | 270 | 238 | 1431 | 267 | 4 | 82 | 783 | 14 | 312 | 14 |
| -43.78392267 | -14.75689565 | 256 | 83 | 584 | 1235 | 325 | 183 | 142 | 259 | 233 | 271 | 233 | 872 | 214 | 0 | 98 | 599 | 2 | 149 | 2 |
| 14.59445571 | -14.75689565 | 213 | 118 | 547 | 1831 | 316 | 100 | 216 | 211 | 179 | 240 | 179 | 960 | 216 | 0 | 95 | 573 | 0 | 142 | 0 |
| 19.45932057 | -14.75689565 | 204 | 97 | 461 | 2318 | 310 | 99 | 211 | 207 | 167 | 244 | 167 | 1061 | 239 | 0 | 102 | 664 | 0 | 87 | 0 |
| 24.32418544 | -14.75689565 | 216 | 96 | 463 | 2483 | 319 | 112 | 208 | 218 | 175 | 257 | 175 | 968 | 225 | 0 | 109 | 659 | 0 | 72 | 0 |
| 29.1890503 | -14.75689565 | 239 | 95 | 442 | 2941 | 346 | 130 | 216 | 251 | 190 | 281 | 190 | 703 | 190 | 0 | 120 | 546 | 0 | 129 | 0 |
| 34.05391517 | -14.75689565 | 200 | 91 | 488 | 2485 | 291 | 104 | 187 | 215 | 157 | 233 | 157 | 1096 | 247 | 6 | 105 | 729 | 18 | 211 | 18 |
| 38.91878003 | -14.75689565 | 228 | 84 | 517 | 1915 | 309 | 146 | 162 | 242 | 220 | 252 | 196 | 1184 | 283 | 6 | 104 | 785 | 18 | 179 | 29 |
| 48.64850976 | -14.75689565 | 184 | 69 | 578 | 1395 | 242 | 122 | 120 | 195 | 169 | 198 | 162 | 1985 | 485 | 22 | 95 | 1323 | 75 | 781 | 109 |
| 126.4863476 | -14.75689565 | 266 | 86 | 515 | 1972 | 340 | 173 | 167 | 274 | 240 | 290 | 232 | 1125 | 275 | 0 | 106 | 785 | 7 | 219 | 17 |
| 131.3512125 | -14.75689565 | 271 | 95 | 491 | 2541 | 358 | 165 | 194 | 281 | 239 | 301 | 226 | 868 | 226 | 0 | 113 | 662 | 3 | 203 | 7 |
| -72.97311186 | -12.40554429 | 226 | 59 | 623 | 778 | 264 | 170 | 94 | 231 | 211 | 233 | 211 | 1212 | 182 | 28 | 53 | 536 | 96 | 385 | 96 |
| -68.108247 | -12.40554429 | 252 | 62 | 555 | 964 | 302 | 190 | 112 | 256 | 234 | 263 | 234 | 1973 | 287 | 39 | 57 | 860 | 125 | 590 | 125 |
| -63.24338213 | -12.40554429 | 253 | 62 | 530 | 794 | 307 | 190 | 118 | 254 | 239 | 264 | 239 | 1565 | 257 | 11 | 67 | 728 | 39 | 344 | 39 |
| -58.37851727 | -12.40554429 | 249 | 68 | 544 | 653 | 314 | 189 | 125 | 245 | 242 | 261 | 242 | 1829 | 355 | 4 | 82 | 1012 | 18 | 120 | 18 |
| -53.5136524 | -12.40554429 | 249 | 69 | 506 | 954 | 325 | 189 | 137 | 242 | 244 | 269 | 242 | 1838 | 300 | 2 | 73 | 888 | 12 | 135 | 864 |
| -48.64878754 | -12.40554429 | 264 | 71 | 561 | 1153 | 341 | 214 | 127 | 253 | 262 | 288 | 253 | 1516 | 272 | 3 | 85 | 808 | 11 | 63 | 808 |
| -43.78392267 | -12.40554429 | 256 | 85 | 596 | 1051 | 328 | 186 | 143 | 259 | 246 | 275 | 239 | 965 | 183 | 1 | 85 | 529 | 3 | 184 | 4 |
| -38.91905781 | -12.40554429 | 246 | 57 | 578 | 1338 | 296 | 198 | 98 | 245 | 244 | 261 | 225 | 1123 | 147 | 53 | 30 | 424 | 164 | 255 | 274 |
| 14.59445571 | -12.40554429 | 184 | 105 | 584 | 961 | 265 | 86 | 179 | 187 | 164 | 195 | 164 | 1267 | 318 | 0 | 92 | 750 | 0 | 276 | 0 |
| 19.45932057 | -12.40554429 | 200 | 91 | 460 | 1731 | 306 | 109 | 198 | 195 | 177 | 231 | 177 | 1153 | 224 | 0 | 87 | 607 | 0 | 44 | 0 |
| 24.32418544 | -12.40554429 | 207 | 90 | 466 | 1922 | 308 | 114 | 194 | 207 | 178 | 242 | 178 | 1405 | 303 | 0 | 97 | 816 | 0 | 158 | 0 |
| 29.1890503 | -12.40554429 | 212 | 91 | 464 | 2263 | 313 | 118 | 195 | 210 | 176 | 249 | 176 | 1421 | 339 | 0 | 112 | 987 | 0 | 58 | 0 |
| 34.05391517 | -12.40554429 | 230 | 62 | 441 | 2087 | 296 | 155 | 141 | 236 | 227 | 260 | 194 | 1537 | 380 | 2 | 101 | 988 | 8 | 228 | 30 |
| 38.91878003 | -12.40554429 | 238 | 83 | 547 | 1433 | 314 | 163 | 151 | 242 | 220 | 260 | 214 | 971 | 255 | 1 | 116 | 705 | 4 | 101 | 6 |
| 131.3512125 | -12.40554429 | 279 | 74 | 514 | 1553 | 346 | 202 | 144 | 281 | 251 | 299 | 251 | 1528 | 351 | 1 | 104 | 999 | 4 | 306 | 4 |
| -77.83797673 | -10.05419294 | 201 | 53 | 702 | 563 | 235 | 160 | 75 | 210 | 195 | 210 | 195 | 276 | 110 | 0 | 140 | 282 | 0 | 282 | 0 |
| -72.97311186 | -10.05419294 | 247 | 58 | 598 | 565 | 292 | 196 | 96 | 247 | 236 | 255 | 236 | 1558 | 231 | 41 | 49 | 654 | 131 | 359 | 131 |
| -68.108247 | -10.05419294 | 252 | 57 | 562 | 695 | 299 | 198 | 102 | 253 | 239 | 261 | 239 | 1631 | 245 | 33 | 55 | 701 | 100 | 373 | 100 |
| -63.24338213 | -10.05419294 | 251 | 57 | 544 | 488 | 306 | 200 | 106 | 249 | 246 | 259 | 246 | 2055 | 323 | 12 | 65 | 922 | 41 | 458 | 41 |
| -58.37851727 | -10.05419294 | 248 | 61 | 560 | 559 | 311 | 202 | 109 | 241 | 251 | 257 | 241 | 1941 | 304 | 8 | 69 | 904 | 35 | 205 | 904 |
| -53.5136524 | -10.05419294 | 244 | 64 | 508 | 1012 | 321 | 195 | 126 | 235 | 247 | 264 | 235 | 2571 | 421 | 4 | 74 | 1251 | 16 | 301 | 1193 |
| -48.64878754 | -10.05419294 | 264 | 68 | 591 | 1073 | 333 | 219 | 114 | 253 | 266 | 285 | 252 | 2082 | 329 | 5 | 75 | 982 | 18 | 150 | 969 |
| -43.78392267 | -10.05419294 | 262 | 79 | 587 | 1080 | 336 | 201 | 135 | 257 | 256 | 284 | 251 | 811 | 148 | 1 | 89 | 438 | 3 | 103 | 4 |
| -38.91905781 | -10.05419294 | 248 | 78 | 624 | 1499 | 312 | 187 | 126 | 262 | 235 | 263 | 225 | 450 | 78 | 11 | 52 | 210 | 38 | 144 | 82 |
| 14.59445571 | -10.05419294 | 241 | 88 | 597 | 824 | 303 | 156 | 148 | 248 | 226 | 249 | 226 | 776 | 177 | 0 | 93 | 515 | 0 | 229 | 0 |
| 19.45932057 | -10.05419294 | 203 | 80 | 524 | 1054 | 288 | 136 | 152 | 196 | 199 | 225 | 194 | 1349 | 242 | 0 | 75 | 645 | 1 | 151 | 505 |
| 24.32418544 | -10.05419294 | 215 | 84 | 500 | 1373 | 308 | 140 | 168 | 204 | 204 | 243 | 204 | 1481 | 292 | 0 | 95 | 864 | 0 | 21 | 864 |
| 29.1890503 | -10.05419294 | 214 | 85 | 496 | 1824 | 306 | 134 | 172 | 212 | 192 | 250 | 192 | 1141 | 230 | 0 | 94 | 625 | 0 | 118 | 0 |
| 34.05391517 | -10.05419294 | 240 | 85 | 474 | 2265 | 329 | 150 | 179 | 245 | 239 | 274 | 203 | 1037 | 246 | 2 | 98 | 645 | 7 | 123 | 26 |
| 38.91878003 | -10.05419294 | 242 | 74 | 596 | 1043 | 301 | 176 | 124 | 249 | 228 | 256 | 225 | 788 | 172 | 1 | 99 | 476 | 5 | 98 | 10 |
| -77.83797673 | -7.702841592 | 130 | 81 | 667 | 353 | 189 | 67 | 122 | 131 | 127 | 134 | 123 | 788 | 138 | 8 | 68 | 400 | 26 | 180 | 29 |
| -72.97311186 | -7.702841592 | 250 | 50 | 636 | 454 | 286 | 207 | 79 | 251 | 241 | 255 | 241 | 2180 | 295 | 53 | 44 | 856 | 188 | 647 | 188 |
| -68.108247 | -7.702841592 | 247 | 48 | 617 | 418 | 283 | 206 | 78 | 248 | 244 | 253 | 239 | 2289 | 318 | 40 | 54 | 931 | 136 | 536 | 140 |
| -63.24338213 | -7.702841592 | 255 | 50 | 588 | 322 | 299 | 215 | 84 | 253 | 252 | 260 | 252 | 2292 | 326 | 27 | 55 | 940 | 111 | 459 | 111 |
| -58.37851727 | -7.702841592 | 252 | 54 | 597 | 428 | 303 | 213 | 90 | 247 | 255 | 259 | 246 | 2502 | 389 | 17 | 63 | 1139 | 74 | 115 | 1095 |
| -53.5136524 | -7.702841592 | 249 | 57 | 526 | 911 | 320 | 211 | 109 | 240 | 256 | 266 | 240 | 1985 | 306 | 10 | 66 | 910 | 41 | 62 | 910 |
| -48.64878754 | -7.702841592 | 260 | 61 | 605 | 972 | 325 | 224 | 101 | 249 | 267 | 276 | 249 | 2144 | 331 | 9 | 67 | 985 | 38 | 214 | 985 |
| -43.78392267 | -7.702841592 | 278 | 69 | 576 | 1340 | 352 | 233 | 119 | 264 | 283 | 303 | 263 | 949 | 159 | 1 | 80 | 462 | 5 | 143 | 434 |
| -38.91905781 | -7.702841592 | 264 | 91 | 622 | 1348 | 340 | 194 | 146 | 266 | 259 | 283 | 243 | 802 | 193 | 3 | 94 | 533 | 17 | 179 | 51 |
| 14.59445571 | -7.702841592 | 225 | 77 | 560 | 653 | 283 | 145 | 137 | 230 | 213 | 232 | 213 | 1287 | 253 | 1 | 77 | 705 | 7 | 417 | 7 |
| 19.45932057 | -7.702841592 | 215 | 71 | 550 | 721 | 288 | 158 | 130 | 207 | 221 | 228 | 207 | 1324 | 201 | 2 | 67 | 575 | 10 | 58 | 575 |
| 24.32418544 | -7.702841592 | 224 | 75 | 570 | 1208 | 306 | 175 | 131 | 211 | 232 | 243 | 210 | 1465 | 238 | 2 | 71 | 687 | 7 | 44 | 650 |
| 29.1890503 | -7.702841592 | 217 | 83 | 520 | 1369 | 305 | 146 | 160 | 208 | 206 | 243 | 205 | 998 | 180 | 0 | 80 | 501 | 1 | 123 | 468 |
| 34.05391517 | -7.702841592 | 187 | 99 | 561 | 1501 | 278 | 103 | 176 | 184 | 165 | 215 | 165 | 1107 | 236 | 0 | 101 | 658 | 0 | 171 | 0 |
| 38.91878003 | -7.702841592 | 251 | 64 | 641 | 939 | 296 | 196 | 100 | 253 | 237 | 263 | 236 | 1209 | 311 | 16 | 83 | 790 | 51 | 267 | 71 |
| 111.891753 | -7.702841592 | 265 | 66 | 642 | 535 | 320 | 216 | 104 | 261 | 269 | 275 | 259 | 2138 | 335 | 30 | 63 | 974 | 106 | 216 | 936 |
| 126.4863476 | -7.702841592 | 223 | 15 | 389 | 645 | 242 | 204 | 38 | 225 | 216 | 231 | 212 | 1833 | 320 | 18 | 67 | 916 | 66 | 284 | 130 |
| 131.3512125 | -7.702841592 | 267 | 10 | 294 | 814 | 283 | 249 | 34 | 273 | 257 | 276 | 253 | 2024 | 355 | 12 | 69 | 940 | 44 | 545 | 140 |
| 141.0809422 | -7.702841592 | 258 | 48 | 596 | 612 | 302 | 222 | 80 | 258 | 248 | 268 | 248 | 2292 | 338 | 61 | 51 | 980 | 206 | 474 | 263 |
| 145.9458071 | -7.702841592 | 202 | 45 | 704 | 478 | 232 | 169 | 63 | 200 | 206 | 206 | 193 | 3248 | 411 | 199 | 25 | 1161 | 605 | 610 | 813 |
| -77.83797673 | -5.35149024 | 188 | 64 | 634 | 680 | 235 | 134 | 100 | 191 | 177 | 195 | 175 | 1923 | 220 | 118 | 18 | 590 | 354 | 504 | 396 |
| -72.97311186 | -5.35149024 | 252 | 46 | 665 | 373 | 286 | 216 | 69 | 254 | 247 | 256 | 244 | 2745 | 328 | 100 | 34 | 956 | 306 | 806 | 371 |
| -68.108247 | -5.35149024 | 249 | 45 | 657 | 363 | 283 | 215 | 68 | 250 | 246 | 254 | 242 | 2591 | 327 | 74 | 39 | 952 | 262 | 679 | 281 |
| -63.24338213 | -5.35149024 | 258 | 44 | 660 | 302 | 292 | 225 | 67 | 257 | 258 | 262 | 253 | 2401 | 315 | 81 | 43 | 926 | 246 | 484 | 258 |
| -58.37851727 | -5.35149024 | 261 | 52 | 659 | 577 | 305 | 226 | 79 | 255 | 266 | 270 | 254 | 2135 | 295 | 43 | 51 | 867 | 149 | 385 | 848 |
| -53.5136524 | -5.35149024 | 244 | 47 | 565 | 535 | 294 | 211 | 83 | 238 | 246 | 253 | 237 | 2115 | 356 | 28 | 68 | 1067 | 102 | 171 | 1050 |
| -48.64878754 | -5.35149024 | 266 | 55 | 592 | 663 | 322 | 229 | 93 | 258 | 275 | 277 | 257 | 1951 | 368 | 16 | 77 | 1055 | 48 | 110 | 932 |
| -43.78392267 | -5.35149024 | 268 | 59 | 580 | 1084 | 333 | 232 | 101 | 255 | 277 | 286 | 255 | 1381 | 297 | 6 | 87 | 820 | 21 | 40 | 820 |
| -38.91905781 | -5.35149024 | 279 | 88 | 695 | 843 | 349 | 222 | 127 | 269 | 289 | 291 | 269 | 897 | 215 | 3 | 99 | 629 | 10 | 17 | 629 |
| 14.59445571 | -5.35149024 | 239 | 60 | 537 | 673 | 285 | 173 | 111 | 246 | 229 | 246 | 226 | 1369 | 243 | 1 | 77 | 678 | 5 | 532 | 8 |
| 19.45932057 | -5.35149024 | 226 | 59 | 535 | 392 | 284 | 174 | 110 | 222 | 229 | 231 | 220 | 1625 | 228 | 13 | 53 | 656 | 45 | 304 | 610 |
| 24.32418544 | -5.35149024 | 235 | 68 | 571 | 870 | 308 | 188 | 120 | 228 | 247 | 248 | 224 | 1418 | 200 | 16 | 50 | 546 | 62 | 163 | 468 |
| 29.1890503 | -5.35149024 | 178 | 49 | 555 | 856 | 232 | 143 | 89 | 175 | 177 | 194 | 168 | 1332 | 240 | 1 | 73 | 665 | 6 | 104 | 473 |
| 34.05391517 | -5.35149024 | 231 | 97 | 572 | 1320 | 321 | 152 | 169 | 226 | 214 | 255 | 214 | 638 | 141 | 0 | 100 | 377 | 0 | 127 | 0 |
| 38.91878003 | -5.35149024 | 262 | 38 | 435 | 1361 | 307 | 220 | 86 | 262 | 280 | 280 | 242 | 1212 | 204 | 44 | 47 | 570 | 143 | 143 | 193 |
| 102.1620233 | -5.35149024 | 267 | 15 | 528 | 418 | 283 | 254 | 29 | 264 | 266 | 274 | 262 | 3594 | 497 | 161 | 38 | 1490 | 510 | 574 | 769 |
| 141.0809422 | -5.35149024 | 237 | 47 | 691 | 518 | 272 | 203 | 69 | 234 | 241 | 242 | 227 | 7123 | 715 | 448 | 12 | 2082 | 1414 | 1580 | 1930 |
| -77.83797673 | -3.000138889 | 246 | 64 | 629 | 691 | 297 | 195 | 102 | 245 | 244 | 252 | 233 | 1990 | 205 | 131 | 15 | 612 | 411 | 522 | 499 |
| -72.97311186 | -3.000138889 | 247 | 44 | 660 | 414 | 280 | 214 | 66 | 248 | 241 | 252 | 239 | 3003 | 321 | 198 | 16 | 963 | 610 | 666 | 668 |
| -68.108247 | -3.000138889 | 251 | 42 | 657 | 391 | 285 | 220 | 65 | 250 | 252 | 256 | 243 | 2802 | 284 | 153 | 20 | 843 | 466 | 622 | 617 |
| -63.24338213 | -3.000138889 | 256 | 42 | 662 | 340 | 291 | 227 | 64 | 255 | 254 | 262 | 251 | 2624 | 328 | 129 | 28 | 939 | 390 | 524 | 465 |
| -58.37851727 | -3.000138889 | 265 | 49 | 616 | 669 | 313 | 232 | 80 | 258 | 274 | 276 | 258 | 2243 | 341 | 80 | 47 | 947 | 241 | 274 | 833 |
| -53.5136524 | -3.000138889 | 250 | 44 | 628 | 515 | 289 | 219 | 69 | 245 | 255 | 258 | 245 | 2058 | 388 | 32 | 71 | 1085 | 107 | 136 | 841 |
| -48.64878754 | -3.000138889 | 266 | 53 | 648 | 658 | 314 | 232 | 82 | 258 | 273 | 276 | 258 | 1981 | 412 | 28 | 80 | 1132 | 86 | 120 | 1132 |
| -43.78392267 | -3.000138889 | 273 | 57 | 640 | 880 | 325 | 237 | 89 | 263 | 285 | 287 | 263 | 1764 | 346 | 19 | 83 | 1012 | 57 | 67 | 965 |
| 14.59445571 | -3.000138889 | 240 | 52 | 592 | 594 | 275 | 187 | 88 | 239 | 230 | 248 | 230 | 1465 | 225 | 6 | 59 | 621 | 29 | 542 | 29 |
| 19.45932057 | -3.000138889 | 241 | 47 | 604 | 343 | 281 | 203 | 78 | 236 | 242 | 245 | 236 | 1477 | 228 | 7 | 58 | 660 | 31 | 343 | 612 |
| 24.32418544 | -3.000138889 | 230 | 53 | 604 | 398 | 279 | 191 | 88 | 225 | 235 | 235 | 223 | 1793 | 222 | 39 | 39 | 663 | 125 | 125 | 608 |
| 29.1890503 | -3.000138889 | 242 | 74 | 648 | 971 | 309 | 195 | 114 | 236 | 243 | 262 | 232 | 929 | 133 | 7 | 53 | 390 | 32 | 127 | 306 |
| 34.05391517 | -3.000138889 | 222 | 87 | 665 | 977 | 296 | 165 | 131 | 218 | 218 | 242 | 211 | 904 | 159 | 3 | 77 | 449 | 14 | 86 | 346 |
| 38.91878003 | -3.000138889 | 257 | 74 | 593 | 1193 | 320 | 196 | 125 | 260 | 240 | 274 | 238 | 502 | 122 | 9 | 82 | 286 | 29 | 115 | 31 |
| 102.1620233 | -3.000138889 | 211 | 49 | 769 | 268 | 244 | 180 | 64 | 210 | 211 | 216 | 207 | 3864 | 501 | 179 | 31 | 1411 | 539 | 875 | 1198 |
| 111.891753 | -3.000138889 | 271 | 42 | 764 | 262 | 300 | 245 | 55 | 268 | 272 | 275 | 267 | 2962 | 374 | 129 | 33 | 1067 | 417 | 795 | 1010 |
| 121.6214827 | -3.000138889 | 228 | 43 | 608 | 500 | 267 | 196 | 71 | 229 | 233 | 234 | 219 | 2722 | 365 | 105 | 38 | 1037 | 323 | 417 | 633 |
| 136.2160773 | -3.000138889 | 264 | 48 | 768 | 261 | 296 | 234 | 62 | 261 | 268 | 268 | 260 | 4039 | 395 | 291 | 10 | 1128 | 882 | 882 | 1109 |
| 141.0809422 | -3.000138889 | 257 | 46 | 798 | 180 | 285 | 228 | 58 | 258 | 254 | 259 | 254 | 2559 | 332 | 139 | 32 | 943 | 419 | 752 | 419 |
| -77.83797673 | -0.648787537 | 172 | 62 | 639 | 511 | 219 | 122 | 97 | 173 | 174 | 177 | 162 | 3142 | 361 | 177 | 23 | 1075 | 550 | 733 | 886 |
| -72.97311186 | -0.648787537 | 242 | 40 | 657 | 567 | 272 | 211 | 61 | 240 | 248 | 248 | 232 | 3248 | 418 | 160 | 28 | 1220 | 522 | 522 | 960 |
| -68.108247 | -0.648787537 | 247 | 40 | 654 | 389 | 280 | 218 | 62 | 247 | 252 | 252 | 240 | 3350 | 350 | 220 | 15 | 1031 | 663 | 688 | 872 |
| -63.24338213 | -0.648787537 | 256 | 40 | 659 | 446 | 289 | 227 | 61 | 253 | 261 | 261 | 248 | 2340 | 326 | 127 | 33 | 933 | 407 | 458 | 688 |
| -58.37851727 | -0.648787537 | 253 | 48 | 640 | 499 | 296 | 221 | 75 | 250 | 258 | 262 | 247 | 1782 | 251 | 89 | 38 | 729 | 271 | 273 | 380 |
| -53.5136524 | -0.648787537 | 261 | 53 | 524 | 1124 | 325 | 224 | 101 | 254 | 282 | 282 | 250 | 2216 | 326 | 56 | 47 | 933 | 173 | 173 | 631 |
| -48.64878754 | -0.648787537 | 267 | 36 | 625 | 652 | 300 | 243 | 57 | 258 | 276 | 277 | 258 | 2911 | 551 | 24 | 77 | 1583 | 89 | 102 | 1583 |
| 9.729590841 | -0.648787537 | 253 | 39 | 540 | 715 | 286 | 213 | 72 | 252 | 242 | 263 | 242 | 1858 | 354 | 3 | 70 | 974 | 13 | 701 | 28 |
| 14.59445571 | -0.648787537 | 235 | 46 | 613 | 635 | 272 | 197 | 75 | 233 | 226 | 244 | 226 | 1588 | 228 | 14 | 50 | 618 | 76 | 535 | 76 |
| 19.45932057 | -0.648787537 | 242 | 44 | 686 | 407 | 275 | 211 | 64 | 237 | 240 | 248 | 237 | 1958 | 245 | 106 | 25 | 703 | 322 | 463 | 703 |
| 24.32418544 | -0.648787537 | 235 | 45 | 675 | 415 | 269 | 203 | 66 | 231 | 232 | 241 | 231 | 1782 | 202 | 102 | 23 | 606 | 311 | 412 | 403 |
| 29.1890503 | -0.648787537 | 178 | 69 | 818 | 325 | 222 | 137 | 85 | 177 | 180 | 181 | 173 | 1903 | 227 | 95 | 28 | 667 | 321 | 541 | 467 |
| 34.05391517 | -0.648787537 | 221 | 55 | 767 | 271 | 256 | 185 | 71 | 222 | 216 | 224 | 216 | 1165 | 218 | 38 | 52 | 579 | 128 | 241 | 128 |
| 38.91878003 | -0.648787537 | 256 | 83 | 666 | 904 | 318 | 194 | 124 | 254 | 245 | 269 | 242 | 486 | 143 | 1 | 112 | 321 | 4 | 128 | 5 |
| 102.1620233 | -0.648787537 | 257 | 51 | 763 | 351 | 293 | 227 | 67 | 253 | 260 | 263 | 252 | 2577 | 290 | 124 | 29 | 859 | 379 | 456 | 812 |
| 111.891753 | -0.648787537 | 232 | 52 | 773 | 227 | 266 | 198 | 67 | 229 | 232 | 236 | 229 | 4187 | 455 | 230 | 24 | 1341 | 708 | 953 | 1293 |
| 116.7566179 | -0.648787537 | 261 | 42 | 793 | 192 | 288 | 235 | 53 | 264 | 260 | 265 | 259 | 2269 | 248 | 125 | 23 | 734 | 380 | 647 | 681 |
| -77.83797673 | 1.702563814 | 245 | 47 | 763 | 298 | 279 | 217 | 62 | 248 | 247 | 248 | 240 | 4622 | 539 | 234 | 21 | 1581 | 757 | 1581 | 1167 |
| -72.97311186 | 1.702563814 | 242 | 44 | 593 | 737 | 284 | 210 | 74 | 234 | 251 | 253 | 230 | 2773 | 366 | 51 | 44 | 1089 | 217 | 321 | 1056 |
| -68.108247 | 1.702563814 | 242 | 37 | 670 | 359 | 271 | 216 | 56 | 239 | 243 | 247 | 235 | 3506 | 403 | 195 | 25 | 1204 | 631 | 696 | 1129 |
| -63.24338213 | 1.702563814 | 256 | 50 | 640 | 672 | 301 | 223 | 78 | 248 | 265 | 265 | 245 | 1853 | 286 | 88 | 43 | 782 | 266 | 266 | 718 |
| -58.37851727 | 1.702563814 | 238 | 51 | 627 | 615 | 287 | 205 | 82 | 232 | 249 | 249 | 231 | 2323 | 412 | 75 | 57 | 1154 | 251 | 257 | 923 |
| -53.5136524 | 1.702563814 | 254 | 58 | 525 | 1177 | 327 | 216 | 111 | 247 | 273 | 275 | 243 | 2075 | 331 | 34 | 55 | 941 | 110 | 138 | 676 |
| 9.729590841 | 1.702563814 | 256 | 36 | 576 | 715 | 288 | 225 | 63 | 251 | 245 | 265 | 245 | 2757 | 529 | 31 | 58 | 1364 | 163 | 847 | 163 |
| 14.59445571 | 1.702563814 | 233 | 47 | 633 | 637 | 275 | 200 | 75 | 239 | 234 | 242 | 225 | 1568 | 211 | 55 | 40 | 601 | 187 | 407 | 235 |
| 19.45932057 | 1.702563814 | 244 | 46 | 596 | 619 | 288 | 211 | 77 | 238 | 245 | 254 | 237 | 1689 | 205 | 58 | 32 | 576 | 192 | 359 | 554 |
| 24.32418544 | 1.702563814 | 233 | 46 | 630 | 554 | 275 | 202 | 73 | 229 | 234 | 242 | 227 | 1972 | 259 | 63 | 35 | 725 | 207 | 426 | 503 |
| 29.1890503 | 1.702563814 | 215 | 60 | 648 | 673 | 271 | 179 | 92 | 220 | 219 | 225 | 207 | 1813 | 207 | 74 | 28 | 571 | 242 | 390 | 514 |
| 34.05391517 | 1.702563814 | 237 | 86 | 603 | 1465 | 318 | 177 | 142 | 233 | 250 | 259 | 217 | 1080 | 146 | 22 | 43 | 425 | 77 | 197 | 379 |
| 38.91878003 | 1.702563814 | 265 | 83 | 690 | 792 | 329 | 209 | 120 | 269 | 261 | 277 | 253 | 471 | 147 | 2 | 111 | 335 | 6 | 97 | 198 |
| 43.78364489 | 1.702563814 | 274 | 68 | 560 | 1274 | 343 | 222 | 121 | 276 | 290 | 293 | 257 | 444 | 91 | 1 | 79 | 257 | 4 | 19 | 94 |
| 111.891753 | 1.702563814 | 254 | 56 | 764 | 357 | 294 | 221 | 73 | 251 | 257 | 260 | 249 | 3676 | 455 | 183 | 24 | 1258 | 573 | 693 | 1247 |
| 116.7566179 | 1.702563814 | 253 | 52 | 836 | 236 | 284 | 222 | 62 | 253 | 249 | 256 | 249 | 2759 | 296 | 167 | 17 | 859 | 524 | 793 | 524 |
| -72.97311186 | 4.053915165 | 261 | 59 | 621 | 770 | 314 | 220 | 94 | 252 | 267 | 272 | 248 | 3012 | 428 | 27 | 55 | 1264 | 123 | 143 | 1180 |
| -68.108247 | 4.053915165 | 243 | 40 | 718 | 324 | 273 | 216 | 56 | 237 | 245 | 247 | 237 | 3186 | 495 | 84 | 54 | 1477 | 282 | 609 | 1477 |
| -63.24338213 | 4.053915165 | 227 | 59 | 690 | 407 | 270 | 185 | 85 | 222 | 228 | 232 | 219 | 3582 | 473 | 183 | 31 | 1321 | 551 | 705 | 1220 |
| -58.37851727 | 4.053915165 | 254 | 49 | 656 | 475 | 295 | 220 | 75 | 250 | 261 | 262 | 248 | 1818 | 340 | 46 | 62 | 973 | 166 | 178 | 850 |
| -53.5136524 | 4.053915165 | 256 | 58 | 592 | 828 | 319 | 221 | 98 | 253 | 271 | 271 | 247 | 2558 | 378 | 73 | 42 | 1034 | 219 | 219 | 741 |
| 9.729590841 | 4.053915165 | 258 | 34 | 551 | 848 | 293 | 232 | 61 | 245 | 263 | 269 | 245 | 3486 | 645 | 27 | 74 | 1925 | 95 | 376 | 1925 |
| 14.59445571 | 4.053915165 | 231 | 59 | 542 | 779 | 291 | 181 | 110 | 223 | 231 | 243 | 222 | 1589 | 269 | 22 | 56 | 779 | 76 | 262 | 479 |
| 19.45932057 | 4.053915165 | 250 | 62 | 530 | 949 | 315 | 198 | 117 | 242 | 250 | 266 | 240 | 1673 | 228 | 24 | 48 | 659 | 91 | 295 | 574 |
| 24.32418544 | 4.053915165 | 237 | 57 | 542 | 771 | 298 | 192 | 106 | 230 | 239 | 250 | 228 | 1570 | 232 | 25 | 49 | 657 | 80 | 241 | 503 |
| 29.1890503 | 4.053915165 | 236 | 69 | 544 | 1331 | 315 | 188 | 127 | 219 | 247 | 257 | 219 | 1775 | 244 | 19 | 55 | 691 | 61 | 95 | 691 |
| 34.05391517 | 4.053915165 | 220 | 86 | 697 | 819 | 287 | 164 | 123 | 228 | 218 | 232 | 207 | 854 | 118 | 16 | 43 | 317 | 64 | 200 | 305 |
| 38.91878003 | 4.053915165 | 207 | 88 | 609 | 1235 | 289 | 144 | 145 | 202 | 193 | 228 | 193 | 731 | 174 | 19 | 82 | 434 | 63 | 79 | 73 |
| 43.78364489 | 4.053915165 | 262 | 90 | 627 | 1070 | 343 | 199 | 144 | 261 | 275 | 282 | 250 | 361 | 103 | 1 | 124 | 239 | 5 | 29 | 15 |
| 97.29715841 | 4.053915165 | 186 | 58 | 783 | 360 | 225 | 151 | 74 | 183 | 184 | 192 | 182 | 3602 | 420 | 199 | 22 | 1181 | 640 | 997 | 1022 |
| 102.1620233 | 4.053915165 | 262 | 63 | 699 | 627 | 306 | 215 | 91 | 254 | 267 | 270 | 252 | 2473 | 317 | 137 | 29 | 944 | 421 | 486 | 735 |
| 116.7566179 | 4.053915165 | 258 | 58 | 823 | 285 | 294 | 223 | 71 | 261 | 254 | 261 | 253 | 2725 | 313 | 137 | 24 | 883 | 415 | 883 | 459 |
| -72.97311186 | 6.405266517 | 162 | 71 | 751 | 259 | 207 | 112 | 95 | 163 | 158 | 165 | 158 | 1657 | 223 | 65 | 37 | 613 | 220 | 588 | 220 |
| -68.108247 | 6.405266517 | 262 | 52 | 517 | 1213 | 325 | 226 | 100 | 247 | 271 | 283 | 247 | 2316 | 424 | 4 | 78 | 1216 | 37 | 77 | 1216 |
| -63.24338213 | 6.405266517 | 250 | 61 | 637 | 663 | 303 | 208 | 96 | 243 | 257 | 262 | 242 | 2684 | 505 | 28 | 79 | 1485 | 85 | 218 | 1400 |
| -58.37851727 | 6.405266517 | 252 | 41 | 638 | 459 | 287 | 222 | 64 | 252 | 248 | 260 | 245 | 2454 | 349 | 113 | 36 | 999 | 343 | 420 | 438 |
| -9.729868619 | 6.405266517 | 251 | 46 | 592 | 778 | 293 | 216 | 77 | 241 | 255 | 260 | 239 | 2801 | 505 | 28 | 66 | 1441 | 111 | 301 | 1224 |
| -4.865003754 | 6.405266517 | 258 | 54 | 519 | 1053 | 315 | 211 | 104 | 257 | 259 | 273 | 243 | 999 | 169 | 12 | 54 | 470 | 47 | 235 | 186 |
| -0.000138889 | 6.405266517 | 268 | 58 | 576 | 1136 | 326 | 226 | 100 | 265 | 272 | 284 | 251 | 1224 | 186 | 20 | 50 | 523 | 71 | 269 | 272 |
| 4.864725976 | 6.405266517 | 264 | 42 | 518 | 1081 | 313 | 231 | 82 | 251 | 270 | 278 | 247 | 1900 | 309 | 13 | 64 | 889 | 49 | 270 | 733 |
| 9.729590841 | 6.405266517 | 257 | 69 | 457 | 1665 | 350 | 199 | 151 | 237 | 267 | 284 | 236 | 1896 | 343 | 12 | 76 | 977 | 36 | 197 | 894 |
| 14.59445571 | 6.405266517 | 230 | 81 | 517 | 1349 | 322 | 165 | 157 | 215 | 232 | 254 | 214 | 1330 | 227 | 4 | 76 | 680 | 12 | 117 | 662 |
| 19.45932057 | 6.405266517 | 248 | 75 | 494 | 1349 | 330 | 179 | 151 | 234 | 244 | 271 | 234 | 1503 | 268 | 5 | 78 | 785 | 18 | 117 | 785 |
| 24.32418544 | 6.405266517 | 246 | 78 | 538 | 1338 | 333 | 188 | 145 | 232 | 252 | 268 | 230 | 1571 | 265 | 6 | 75 | 773 | 24 | 129 | 764 |
| 29.1890503 | 6.405266517 | 262 | 77 | 517 | 1809 | 355 | 206 | 149 | 241 | 269 | 292 | 241 | 971 | 171 | 1 | 81 | 503 | 3 | 52 | 503 |
| 34.05391517 | 6.405266517 | 262 | 76 | 597 | 1004 | 333 | 205 | 128 | 255 | 271 | 280 | 250 | 1068 | 159 | 19 | 54 | 472 | 63 | 134 | 129 |
| 38.91878003 | 6.405266517 | 173 | 80 | 548 | 1132 | 256 | 110 | 146 | 160 | 177 | 193 | 160 | 1432 | 214 | 18 | 58 | 625 | 74 | 202 | 625 |
| 43.78364489 | 6.405266517 | 274 | 88 | 643 | 994 | 348 | 211 | 137 | 269 | 276 | 289 | 257 | 293 | 82 | 0 | 126 | 232 | 2 | 35 | 133 |
| 48.64850976 | 6.405266517 | 282 | 78 | 676 | 936 | 340 | 224 | 116 | 290 | 289 | 293 | 269 | 159 | 49 | 0 | 122 | 124 | 0 | 55 | 10 |
| 116.7566179 | 6.405266517 | 245 | 43 | 693 | 553 | 278 | 216 | 62 | 237 | 249 | 253 | 237 | 2680 | 413 | 144 | 37 | 1190 | 449 | 600 | 1190 |
| -82.70284159 | 8.756617868 | 202 | 40 | 593 | 529 | 242 | 174 | 68 | 196 | 204 | 211 | 195 | 3505 | 572 | 58 | 59 | 1629 | 185 | 426 | 1402 |
| -77.83797673 | 8.756617868 | 252 | 35 | 772 | 250 | 275 | 229 | 45 | 251 | 252 | 255 | 248 | 3451 | 445 | 41 | 57 | 1327 | 142 | 913 | 1247 |
| -72.97311186 | 8.756617868 | 268 | 64 | 630 | 706 | 323 | 221 | 102 | 269 | 260 | 278 | 258 | 2528 | 346 | 63 | 41 | 981 | 231 | 672 | 286 |
| -68.108247 | 8.756617868 | 276 | 62 | 532 | 1329 | 348 | 232 | 116 | 261 | 288 | 298 | 260 | 1341 | 243 | 3 | 79 | 697 | 13 | 25 | 637 |
| -63.24338213 | 8.756617868 | 263 | 66 | 605 | 757 | 323 | 214 | 109 | 258 | 264 | 274 | 252 | 1110 | 200 | 13 | 68 | 577 | 43 | 69 | 130 |
| -9.729868619 | 8.756617868 | 238 | 68 | 467 | 1252 | 319 | 172 | 146 | 225 | 236 | 260 | 222 | 2473 | 408 | 14 | 66 | 1211 | 68 | 222 | 1124 |
| -4.865003754 | 8.756617868 | 263 | 74 | 460 | 1650 | 353 | 193 | 160 | 244 | 261 | 291 | 243 | 1109 | 195 | 7 | 68 | 562 | 27 | 128 | 484 |
| -0.000138889 | 8.756617868 | 273 | 76 | 486 | 1674 | 362 | 205 | 157 | 252 | 274 | 301 | 252 | 1224 | 239 | 4 | 77 | 665 | 18 | 123 | 552 |
| 4.864725976 | 8.756617868 | 268 | 75 | 469 | 1771 | 358 | 198 | 160 | 247 | 268 | 298 | 246 | 997 | 199 | 6 | 80 | 542 | 18 | 82 | 439 |
| 9.729590841 | 8.756617868 | 277 | 83 | 484 | 1888 | 374 | 203 | 170 | 255 | 272 | 309 | 253 | 1044 | 196 | 0 | 89 | 578 | 1 | 165 | 560 |
| 14.59445571 | 8.756617868 | 274 | 95 | 471 | 2190 | 381 | 179 | 201 | 254 | 270 | 316 | 254 | 983 | 244 | 0 | 107 | 699 | 0 | 82 | 0 |
| 19.45932057 | 8.756617868 | 271 | 87 | 489 | 2007 | 370 | 192 | 178 | 247 | 266 | 306 | 247 | 990 | 226 | 0 | 100 | 658 | 0 | 96 | 658 |
| 24.32418544 | 8.756617868 | 240 | 90 | 515 | 1789 | 337 | 163 | 174 | 221 | 230 | 273 | 221 | 986 | 237 | 0 | 107 | 692 | 0 | 55 | 692 |
| 29.1890503 | 8.756617868 | 271 | 84 | 491 | 1997 | 370 | 199 | 171 | 248 | 285 | 306 | 248 | 898 | 200 | 0 | 99 | 577 | 1 | 63 | 577 |
| 34.05391517 | 8.756617868 | 259 | 74 | 527 | 1470 | 342 | 202 | 140 | 244 | 257 | 285 | 244 | 935 | 175 | 3 | 81 | 513 | 11 | 52 | 513 |
| 38.91878003 | 8.756617868 | 185 | 94 | 609 | 1199 | 262 | 108 | 154 | 174 | 174 | 204 | 174 | 884 | 217 | 5 | 95 | 649 | 17 | 183 | 649 |
| 43.78364489 | 8.756617868 | 226 | 101 | 641 | 945 | 299 | 142 | 157 | 232 | 209 | 235 | 209 | 394 | 104 | 3 | 90 | 246 | 10 | 123 | 10 |
| 48.64850976 | 8.756617868 | 269 | 102 | 635 | 1549 | 340 | 179 | 161 | 281 | 245 | 287 | 245 | 133 | 48 | 0 | 134 | 118 | 1 | 33 | 1 |
| 77.83769895 | 8.756617868 | 282 | 54 | 517 | 1225 | 333 | 227 | 105 | 272 | 286 | 301 | 263 | 676 | 166 | 12 | 92 | 494 | 50 | 115 | 141 |
| -72.97311186 | 11.10796922 | 259 | 54 | 642 | 683 | 300 | 215 | 85 | 260 | 250 | 268 | 249 | 968 | 205 | 8 | 73 | 555 | 27 | 219 | 65 |
| -14.59473348 | 11.10796922 | 273 | 75 | 519 | 1118 | 350 | 205 | 145 | 257 | 276 | 289 | 257 | 2169 | 606 | 0 | 119 | 1693 | 0 | 4 | 1693 |
| -9.729868619 | 11.10796922 | 269 | 94 | 467 | 2375 | 375 | 174 | 201 | 248 | 273 | 312 | 248 | 1213 | 296 | 1 | 102 | 809 | 3 | 83 | 809 |
| -4.865003754 | 11.10796922 | 268 | 95 | 477 | 2063 | 370 | 172 | 198 | 249 | 247 | 305 | 247 | 847 | 227 | 1 | 104 | 626 | 5 | 88 | 5 |
| -0.000138889 | 11.10796922 | 281 | 97 | 504 | 2068 | 381 | 190 | 192 | 259 | 262 | 317 | 259 | 864 | 232 | 0 | 106 | 634 | 1 | 88 | 634 |
| 4.864725976 | 11.10796922 | 278 | 94 | 473 | 2242 | 384 | 185 | 198 | 256 | 258 | 319 | 256 | 1060 | 269 | 0 | 110 | 750 | 0 | 70 | 750 |
| 9.729590841 | 11.10796922 | 264 | 105 | 453 | 2692 | 379 | 148 | 231 | 251 | 242 | 306 | 223 | 1024 | 323 | 0 | 131 | 927 | 0 | 158 | 0 |
| 14.59445571 | 11.10796922 | 280 | 105 | 445 | 2758 | 398 | 162 | 236 | 264 | 258 | 325 | 241 | 680 | 223 | 0 | 130 | 614 | 0 | 101 | 0 |
| 19.45932057 | 11.10796922 | 269 | 100 | 466 | 2308 | 379 | 165 | 214 | 250 | 261 | 309 | 244 | 885 | 281 | 0 | 123 | 773 | 0 | 47 | 0 |
| 24.32418544 | 11.10796922 | 264 | 99 | 470 | 2263 | 377 | 166 | 211 | 249 | 253 | 303 | 238 | 597 | 169 | 0 | 125 | 503 | 0 | 62 | 0 |
| 29.1890503 | 11.10796922 | 268 | 96 | 489 | 2004 | 376 | 179 | 197 | 249 | 263 | 305 | 249 | 658 | 167 | 0 | 109 | 466 | 0 | 25 | 466 |
| 34.05391517 | 11.10796922 | 259 | 88 | 514 | 1887 | 358 | 186 | 172 | 238 | 265 | 293 | 238 | 659 | 159 | 0 | 104 | 454 | 0 | 33 | 454 |
| 38.91878003 | 11.10796922 | 187 | 96 | 573 | 1412 | 270 | 103 | 167 | 180 | 169 | 210 | 169 | 973 | 279 | 12 | 105 | 802 | 38 | 201 | 43 |
| 48.64850976 | 11.10796922 | 291 | 77 | 450 | 3216 | 370 | 198 | 172 | 324 | 326 | 326 | 243 | 62 | 10 | 1 | 55 | 27 | 6 | 15 | 9 |
| 77.83769895 | 11.10796922 | 285 | 81 | 489 | 1973 | 371 | 204 | 166 | 282 | 270 | 317 | 256 | 614 | 145 | 8 | 78 | 367 | 25 | 175 | 63 |
| 107.0268881 | 11.10796922 | 272 | 60 | 508 | 876 | 328 | 210 | 118 | 269 | 264 | 286 | 259 | 2252 | 391 | 12 | 79 | 1168 | 40 | 203 | 60 |
| -87.56770646 | 13.45932057 | 287 | 70 | 635 | 757 | 346 | 236 | 111 | 284 | 278 | 299 | 278 | 1546 | 361 | 1 | 95 | 930 | 6 | 81 | 51 |
| -14.59473348 | 13.45932057 | 290 | 104 | 529 | 1931 | 390 | 194 | 196 | 279 | 313 | 321 | 262 | 741 | 214 | 0 | 130 | 608 | 0 | 18 | 1 |
| -9.729868619 | 13.45932057 | 287 | 105 | 468 | 2820 | 403 | 178 | 225 | 267 | 271 | 336 | 251 | 1000 | 299 | 0 | 126 | 841 | 0 | 76 | 0 |
| -4.865003754 | 13.45932057 | 289 | 108 | 477 | 2714 | 398 | 171 | 227 | 276 | 266 | 332 | 246 | 676 | 216 | 0 | 128 | 600 | 0 | 86 | 0 |
| -0.000138889 | 13.45932057 | 288 | 111 | 467 | 3039 | 402 | 164 | 238 | 281 | 255 | 334 | 237 | 571 | 183 | 0 | 128 | 510 | 0 | 81 | 0 |
| 4.864725976 | 13.45932057 | 290 | 110 | 467 | 2888 | 406 | 171 | 235 | 279 | 262 | 335 | 244 | 540 | 179 | 0 | 135 | 508 | 0 | 73 | 0 |
| 9.729590841 | 13.45932057 | 277 | 115 | 447 | 3736 | 396 | 140 | 257 | 279 | 232 | 329 | 214 | 395 | 144 | 0 | 153 | 421 | 0 | 36 | 0 |
| 14.59445571 | 13.45932057 | 287 | 112 | 449 | 3569 | 403 | 154 | 249 | 286 | 243 | 336 | 227 | 290 | 133 | 0 | 167 | 345 | 0 | 15 | 0 |
| 19.45932057 | 13.45932057 | 285 | 115 | 474 | 2823 | 402 | 159 | 242 | 273 | 255 | 328 | 241 | 368 | 160 | 0 | 158 | 417 | 0 | 26 | 0 |
| 24.32418544 | 13.45932057 | 216 | 122 | 491 | 2879 | 329 | 80 | 249 | 217 | 181 | 257 | 167 | 592 | 242 | 0 | 163 | 683 | 0 | 35 | 0 |
| 29.1890503 | 13.45932057 | 267 | 116 | 478 | 3024 | 379 | 137 | 242 | 269 | 225 | 307 | 214 | 277 | 107 | 0 | 152 | 297 | 0 | 13 | 0 |
| 34.05391517 | 13.45932057 | 286 | 106 | 515 | 2023 | 395 | 189 | 206 | 269 | 270 | 321 | 262 | 509 | 169 | 0 | 139 | 480 | 0 | 28 | 0 |
| 38.91878003 | 13.45932057 | 212 | 107 | 569 | 1709 | 307 | 120 | 188 | 204 | 199 | 242 | 189 | 813 | 240 | 3 | 122 | 713 | 9 | 126 | 10 |
| 43.78364489 | 13.45932057 | 267 | 94 | 513 | 2649 | 357 | 173 | 184 | 290 | 227 | 302 | 227 | 435 | 104 | 3 | 84 | 270 | 12 | 173 | 12 |
| 77.83769895 | 13.45932057 | 243 | 89 | 476 | 2231 | 340 | 153 | 187 | 238 | 230 | 278 | 211 | 677 | 159 | 2 | 86 | 448 | 8 | 164 | 79 |
| 102.1620233 | 13.45932057 | 272 | 57 | 499 | 1068 | 321 | 206 | 115 | 273 | 254 | 285 | 254 | 1634 | 306 | 6 | 76 | 867 | 26 | 492 | 26 |
| 107.0268881 | 13.45932057 | 269 | 58 | 458 | 1131 | 329 | 203 | 126 | 268 | 249 | 287 | 249 | 2198 | 473 | 2 | 95 | 1380 | 9 | 198 | 9 |
| -92.43257132 | 15.81067192 | 221 | 77 | 578 | 1027 | 289 | 156 | 133 | 223 | 211 | 236 | 204 | 1997 | 427 | 10 | 93 | 1205 | 32 | 379 | 36 |
| -87.56770646 | 15.81067192 | 263 | 53 | 551 | 1414 | 311 | 214 | 97 | 255 | 267 | 279 | 241 | 2592 | 365 | 76 | 42 | 1084 | 259 | 491 | 769 |
| -14.59473348 | 15.81067192 | 283 | 118 | 522 | 2712 | 391 | 164 | 227 | 295 | 293 | 315 | 234 | 355 | 121 | 0 | 144 | 323 | 0 | 40 | 3 |
| -9.729868619 | 15.81067192 | 302 | 112 | 443 | 3673 | 420 | 169 | 252 | 306 | 239 | 352 | 239 | 293 | 110 | 0 | 141 | 290 | 1 | 59 | 1 |
| -4.865003754 | 15.81067192 | 290 | 115 | 449 | 3946 | 408 | 152 | 255 | 304 | 240 | 341 | 222 | 248 | 93 | 0 | 147 | 254 | 0 | 50 | 0 |
| -0.000138889 | 15.81067192 | 306 | 118 | 457 | 3800 | 423 | 165 | 258 | 314 | 255 | 354 | 240 | 231 | 93 | 0 | 153 | 251 | 0 | 50 | 0 |
| 4.864725976 | 15.81067192 | 285 | 122 | 457 | 4046 | 405 | 139 | 267 | 302 | 232 | 334 | 215 | 252 | 93 | 0 | 150 | 265 | 0 | 68 | 0 |
| 9.729590841 | 15.81067192 | 280 | 124 | 437 | 4851 | 408 | 126 | 283 | 307 | 215 | 337 | 198 | 97 | 43 | 0 | 172 | 119 | 0 | 17 | 0 |
| 14.59445571 | 15.81067192 | 278 | 122 | 440 | 4774 | 407 | 128 | 278 | 309 | 214 | 335 | 199 | 69 | 37 | 0 | 187 | 93 | 0 | 10 | 0 |
| 19.45932057 | 15.81067192 | 288 | 121 | 458 | 4074 | 408 | 144 | 264 | 304 | 232 | 338 | 219 | 76 | 38 | 0 | 173 | 94 | 0 | 18 | 0 |
| 24.32418544 | 15.81067192 | 258 | 125 | 443 | 5009 | 384 | 101 | 283 | 297 | 192 | 313 | 176 | 84 | 46 | 0 | 194 | 116 | 0 | 7 | 0 |
| 29.1890503 | 15.81067192 | 274 | 121 | 438 | 4975 | 396 | 120 | 276 | 313 | 204 | 328 | 193 | 80 | 32 | 0 | 173 | 94 | 0 | 9 | 0 |
| 34.05391517 | 15.81067192 | 295 | 117 | 498 | 3175 | 403 | 167 | 236 | 304 | 247 | 335 | 241 | 170 | 67 | 0 | 154 | 185 | 0 | 33 | 0 |
| 38.91878003 | 15.81067192 | 266 | 104 | 532 | 2871 | 368 | 173 | 195 | 289 | 302 | 302 | 223 | 324 | 54 | 12 | 47 | 152 | 45 | 45 | 90 |
| 43.78364489 | 15.81067192 | 202 | 110 | 550 | 2551 | 303 | 103 | 199 | 225 | 190 | 234 | 163 | 193 | 59 | 1 | 111 | 145 | 4 | 53 | 4 |
| 48.64850976 | 15.81067192 | 263 | 127 | 470 | 4623 | 393 | 122 | 271 | 316 | 205 | 320 | 194 | 72 | 16 | 0 | 92 | 37 | 1 | 33 | 3 |
| 77.83769895 | 15.81067192 | 284 | 93 | 452 | 2755 | 389 | 184 | 205 | 280 | 267 | 331 | 247 | 674 | 149 | 1 | 98 | 434 | 5 | 106 | 28 |
| 102.1620233 | 15.81067192 | 271 | 65 | 431 | 1972 | 335 | 184 | 151 | 276 | 236 | 293 | 236 | 1176 | 260 | 4 | 80 | 685 | 12 | 396 | 12 |
| 107.0268881 | 15.81067192 | 219 | 57 | 445 | 2142 | 272 | 144 | 128 | 224 | 192 | 241 | 181 | 1882 | 318 | 36 | 61 | 934 | 116 | 523 | 147 |
| -102.1623011 | 18.16202327 | 269 | 50 | 563 | 881 | 310 | 222 | 88 | 275 | 265 | 280 | 257 | 768 | 186 | 2 | 108 | 529 | 8 | 269 | 41 |
| -97.29743619 | 18.16202327 | 213 | 90 | 538 | 1820 | 295 | 128 | 167 | 226 | 191 | 237 | 183 | 620 | 131 | 7 | 90 | 351 | 22 | 125 | 25 |
| -92.43257132 | 18.16202327 | 265 | 60 | 474 | 1981 | 328 | 202 | 126 | 268 | 272 | 287 | 234 | 2037 | 335 | 52 | 54 | 983 | 161 | 250 | 375 |
| -77.83797673 | 18.16202327 | 257 | 23 | 423 | 1031 | 283 | 229 | 54 | 264 | 242 | 270 | 242 | 1903 | 271 | 63 | 44 | 771 | 195 | 580 | 195 |
| -72.97311186 | 18.16202327 | 275 | 31 | 469 | 1061 | 307 | 242 | 65 | 283 | 263 | 290 | 261 | 1250 | 186 | 35 | 46 | 517 | 133 | 375 | 155 |
| -14.59473348 | 18.16202327 | 291 | 125 | 500 | 3730 | 408 | 159 | 249 | 323 | 317 | 333 | 228 | 128 | 44 | 0 | 142 | 124 | 1 | 4 | 6 |
| -9.729868619 | 18.16202327 | 293 | 117 | 425 | 4979 | 418 | 144 | 274 | 336 | 233 | 350 | 211 | 94 | 37 | 0 | 145 | 92 | 0 | 16 | 0 |
| -4.865003754 | 18.16202327 | 288 | 120 | 425 | 5253 | 415 | 133 | 282 | 334 | 291 | 346 | 202 | 82 | 33 | 0 | 143 | 81 | 0 | 13 | 5 |
| -0.000138889 | 18.16202327 | 299 | 122 | 437 | 4981 | 422 | 142 | 280 | 337 | 232 | 358 | 217 | 135 | 52 | 0 | 154 | 145 | 0 | 30 | 0 |
| 4.864725976 | 18.16202327 | 293 | 128 | 435 | 5391 | 421 | 127 | 293 | 338 | 219 | 353 | 204 | 84 | 37 | 0 | 168 | 101 | 0 | 16 | 0 |
| 9.729590841 | 18.16202327 | 267 | 127 | 414 | 5930 | 406 | 100 | 306 | 319 | 188 | 333 | 172 | 30 | 17 | 0 | 189 | 40 | 0 | 8 | 0 |
| 14.59445571 | 18.16202327 | 268 | 133 | 430 | 5730 | 408 | 99 | 310 | 319 | 191 | 331 | 176 | 19 | 10 | 0 | 183 | 25 | 0 | 3 | 0 |
| 19.45932057 | 18.16202327 | 284 | 126 | 431 | 5524 | 419 | 126 | 293 | 332 | 209 | 345 | 196 | 10 | 6 | 0 | 201 | 14 | 0 | 2 | 0 |
| 24.32418544 | 18.16202327 | 242 | 132 | 433 | 6046 | 380 | 74 | 305 | 307 | 161 | 307 | 148 | 16 | 9 | 0 | 204 | 23 | 0 | 23 | 0 |
| 29.1890503 | 18.16202327 | 273 | 131 | 427 | 6129 | 412 | 105 | 307 | 339 | 189 | 339 | 179 | 22 | 11 | 0 | 203 | 31 | 0 | 31 | 0 |
| 34.05391517 | 18.16202327 | 305 | 124 | 457 | 4702 | 428 | 156 | 272 | 345 | 235 | 356 | 231 | 48 | 24 | 0 | 175 | 61 | 0 | 4 | 0 |
| 43.78364489 | 18.16202327 | 202 | 122 | 476 | 4382 | 325 | 68 | 257 | 196 | 211 | 257 | 139 | 127 | 34 | 0 | 96 | 94 | 6 | 18 | 18 |
| 48.64850976 | 18.16202327 | 293 | 128 | 408 | 6097 | 441 | 128 | 313 | 295 | 218 | 367 | 199 | 55 | 21 | 0 | 143 | 57 | 0 | 3 | 2 |
| 53.51337462 | 18.16202327 | 276 | 131 | 457 | 5075 | 416 | 128 | 288 | 278 | 255 | 335 | 197 | 42 | 11 | 0 | 114 | 31 | 0 | 3 | 2 |
| 77.83769895 | 18.16202327 | 275 | 91 | 392 | 3338 | 396 | 165 | 231 | 267 | 233 | 335 | 232 | 1049 | 266 | 4 | 114 | 752 | 13 | 59 | 27 |
| 82.70256381 | 18.16202327 | 227 | 76 | 405 | 2605 | 319 | 131 | 188 | 234 | 186 | 265 | 186 | 1134 | 216 | 4 | 87 | 626 | 14 | 221 | 14 |
| 97.29715841 | 18.16202327 | 229 | 70 | 354 | 2108 | 325 | 127 | 198 | 234 | 207 | 259 | 193 | 2450 | 561 | 6 | 102 | 1617 | 18 | 128 | 22 |
| 102.1620233 | 18.16202327 | 257 | 66 | 424 | 2326 | 316 | 160 | 156 | 274 | 215 | 278 | 215 | 1744 | 327 | 4 | 90 | 955 | 18 | 821 | 18 |
| 121.6214827 | 18.16202327 | 269 | 35 | 388 | 1775 | 312 | 223 | 90 | 264 | 272 | 289 | 240 | 1801 | 377 | 32 | 68 | 1067 | 104 | 357 | 324 |
| -102.1623011 | 20.51337462 | 188 | 112 | 529 | 2260 | 295 | 82 | 212 | 197 | 182 | 220 | 152 | 679 | 176 | 5 | 109 | 486 | 16 | 67 | 33 |
| -97.29743619 | 20.51337462 | 249 | 58 | 424 | 2564 | 308 | 171 | 137 | 270 | 229 | 277 | 209 | 1333 | 281 | 42 | 59 | 700 | 135 | 366 | 187 |
| -87.56770646 | 20.51337462 | 255 | 56 | 516 | 1611 | 308 | 200 | 109 | 268 | 243 | 273 | 231 | 1181 | 190 | 36 | 53 | 528 | 109 | 419 | 172 |
| -14.59473348 | 20.51337462 | 287 | 129 | 474 | 4763 | 418 | 146 | 272 | 340 | 272 | 343 | 214 | 76 | 26 | 1 | 133 | 75 | 3 | 48 | 6 |
| -9.729868619 | 20.51337462 | 275 | 130 | 417 | 6063 | 422 | 110 | 312 | 337 | 298 | 349 | 182 | 41 | 11 | 1 | 97 | 32 | 3 | 22 | 7 |
| -4.865003754 | 20.51337462 | 290 | 137 | 419 | 6504 | 441 | 113 | 328 | 364 | 321 | 367 | 189 | 37 | 12 | 0 | 110 | 27 | 1 | 8 | 7 |
| -0.000138889 | 20.51337462 | 284 | 132 | 413 | 6317 | 431 | 110 | 321 | 351 | 208 | 357 | 186 | 64 | 29 | 0 | 156 | 67 | 0 | 22 | 2 |
| 4.864725976 | 20.51337462 | 273 | 142 | 437 | 6181 | 422 | 98 | 325 | 337 | 191 | 343 | 177 | 18 | 6 | 0 | 135 | 17 | 0 | 11 | 0 |
| 9.729590841 | 20.51337462 | 258 | 137 | 415 | 6615 | 411 | 80 | 330 | 325 | 171 | 330 | 154 | 11 | 5 | 0 | 157 | 12 | 0 | 6 | 0 |
| 14.59445571 | 20.51337462 | 274 | 145 | 425 | 6549 | 428 | 87 | 341 | 338 | 188 | 347 | 172 | 12 | 6 | 0 | 173 | 15 | 0 | 7 | 0 |
| 19.45932057 | 20.51337462 | 247 | 142 | 432 | 6488 | 396 | 66 | 329 | 298 | 162 | 318 | 148 | 3 | 2 | 0 | 238 | 4 | 0 | 0 | 0 |
| 24.32418544 | 20.51337462 | 233 | 147 | 443 | 6551 | 383 | 51 | 332 | 306 | 147 | 306 | 132 | 3 | 2 | 0 | 238 | 5 | 0 | 5 | 0 |
| 29.1890503 | 20.51337462 | 259 | 140 | 427 | 6591 | 410 | 82 | 328 | 331 | 169 | 333 | 159 | 4 | 3 | 0 | 255 | 6 | 0 | 5 | 0 |
| 34.05391517 | 20.51337462 | 294 | 139 | 453 | 5744 | 435 | 127 | 308 | 354 | 211 | 354 | 206 | 7 | 4 | 0 | 204 | 10 | 0 | 10 | 0 |
| 43.78364489 | 20.51337462 | 278 | 128 | 423 | 5901 | 423 | 120 | 303 | 274 | 350 | 352 | 190 | 46 | 18 | 0 | 143 | 49 | 0 | 2 | 10 |
| 48.64850976 | 20.51337462 | 292 | 134 | 391 | 7003 | 457 | 114 | 343 | 290 | 375 | 378 | 185 | 49 | 18 | 0 | 138 | 49 | 0 | 0 | 7 |
| 53.51337462 | 20.51337462 | 296 | 133 | 413 | 6407 | 451 | 130 | 322 | 244 | 359 | 374 | 200 | 63 | 22 | 0 | 134 | 61 | 1 | 9 | 9 |
| 58.37823949 | 20.51337462 | 283 | 81 | 438 | 3202 | 378 | 194 | 184 | 290 | 293 | 329 | 235 | 53 | 11 | 0 | 78 | 32 | 3 | 7 | 9 |
| 72.97283408 | 20.51337462 | 273 | 65 | 400 | 2100 | 348 | 184 | 163 | 281 | 244 | 303 | 238 | 2107 | 874 | 0 | 155 | 2052 | 1 | 11 | 3 |
| 77.83769895 | 20.51337462 | 272 | 96 | 361 | 4157 | 411 | 145 | 267 | 286 | 318 | 345 | 215 | 1014 | 298 | 9 | 119 | 765 | 33 | 36 | 33 |
| 82.70256381 | 20.51337462 | 263 | 90 | 356 | 3910 | 394 | 142 | 252 | 266 | 211 | 328 | 206 | 1629 | 471 | 2 | 127 | 1383 | 15 | 82 | 20 |
| 97.29715841 | 20.51337462 | 214 | 75 | 343 | 2847 | 316 | 96 | 219 | 228 | 177 | 246 | 163 | 1662 | 355 | 6 | 91 | 997 | 20 | 445 | 24 |
| 102.1620233 | 20.51337462 | 196 | 62 | 417 | 2846 | 248 | 100 | 148 | 224 | 161 | 226 | 148 | 1606 | 366 | 20 | 90 | 1036 | 64 | 989 | 64 |
| -102.1623011 | 22.86472598 | 181 | 125 | 531 | 3136 | 291 | 57 | 235 | 209 | 164 | 219 | 131 | 317 | 62 | 3 | 83 | 180 | 11 | 116 | 41 |
| -82.70284159 | 22.86472598 | 249 | 34 | 359 | 2048 | 292 | 197 | 95 | 266 | 220 | 275 | 220 | 1620 | 265 | 41 | 56 | 710 | 161 | 560 | 206 |
| -14.59473348 | 22.86472598 | 246 | 133 | 487 | 4387 | 390 | 117 | 273 | 276 | 212 | 308 | 183 | 51 | 17 | 1 | 126 | 45 | 3 | 15 | 9 |
| -9.729868619 | 22.86472598 | 273 | 127 | 383 | 6566 | 441 | 110 | 332 | 348 | 355 | 366 | 177 | 31 | 11 | 0 | 110 | 25 | 1 | 6 | 7 |
| -4.865003754 | 22.86472598 | 280 | 138 | 384 | 7382 | 456 | 97 | 359 | 358 | 372 | 376 | 171 | 22 | 4 | 0 | 66 | 9 | 1 | 6 | 6 |
| -0.000138889 | 22.86472598 | 275 | 145 | 400 | 7419 | 448 | 85 | 363 | 349 | 307 | 366 | 164 | 20 | 6 | 0 | 102 | 16 | 1 | 1 | 4 |
| 4.864725976 | 22.86472598 | 251 | 131 | 402 | 6602 | 399 | 72 | 327 | 312 | 173 | 330 | 152 | 42 | 8 | 0 | 71 | 22 | 1 | 11 | 2 |
| 9.729590841 | 22.86472598 | 249 | 144 | 411 | 7054 | 410 | 59 | 351 | 324 | 273 | 329 | 140 | 7 | 3 | 0 | 148 | 6 | 0 | 1 | 2 |
| 14.59445571 | 22.86472598 | 246 | 145 | 412 | 7214 | 408 | 55 | 353 | 298 | 151 | 325 | 135 | 1 | 1 | 0 | 332 | 2 | 0 | 0 | 0 |
| 19.45932057 | 22.86472598 | 198 | 146 | 440 | 6506 | 351 | 21 | 331 | 245 | 112 | 270 | 100 | 2 | 1 | 0 | 224 | 3 | 0 | 0 | 0 |
| 34.05391517 | 22.86472598 | 271 | 156 | 455 | 6593 | 430 | 86 | 344 | 303 | 180 | 346 | 172 | 4 | 2 | 0 | 187 | 4 | 0 | 0 | 0 |
| 38.91878003 | 22.86472598 | 294 | 57 | 344 | 3861 | 370 | 205 | 165 | 292 | 336 | 340 | 237 | 46 | 15 | 0 | 118 | 31 | 0 | 0 | 19 |
| 43.78364489 | 22.86472598 | 259 | 126 | 391 | 6765 | 414 | 92 | 322 | 245 | 341 | 345 | 160 | 114 | 38 | 0 | 124 | 106 | 0 | 0 | 29 |
| 48.64850976 | 22.86472598 | 282 | 133 | 373 | 7709 | 457 | 100 | 357 | 272 | 375 | 377 | 168 | 66 | 20 | 0 | 117 | 56 | 0 | 0 | 23 |
| 53.51337462 | 22.86472598 | 284 | 134 | 420 | 6394 | 445 | 126 | 318 | 225 | 344 | 364 | 191 | 54 | 20 | 0 | 144 | 50 | 0 | 0 | 17 |
| 58.37823949 | 22.86472598 | 272 | 120 | 437 | 5253 | 413 | 137 | 276 | 228 | 286 | 338 | 195 | 103 | 22 | 0 | 90 | 65 | 4 | 27 | 13 |
| 72.97283408 | 22.86472598 | 276 | 110 | 406 | 3941 | 404 | 134 | 270 | 284 | 264 | 330 | 211 | 851 | 341 | 1 | 154 | 791 | 3 | 7 | 5 |
| 77.83769895 | 22.86472598 | 268 | 103 | 353 | 4706 | 418 | 125 | 293 | 269 | 310 | 344 | 197 | 1123 | 361 | 2 | 142 | 1083 | 10 | 14 | 37 |
| 82.70256381 | 22.86472598 | 248 | 98 | 344 | 4628 | 392 | 108 | 284 | 261 | 187 | 317 | 175 | 1582 | 476 | 6 | 129 | 1414 | 23 | 42 | 58 |
| 87.56742868 | 22.86472598 | 260 | 86 | 345 | 4234 | 364 | 114 | 250 | 287 | 202 | 307 | 185 | 1413 | 308 | 4 | 95 | 922 | 29 | 257 | 30 |
| 92.43229354 | 22.86472598 | 253 | 74 | 360 | 3133 | 331 | 125 | 206 | 277 | 195 | 281 | 195 | 3130 | 609 | 13 | 87 | 1825 | 48 | 821 | 48 |
| 97.29715841 | 22.86472598 | 223 | 76 | 350 | 3303 | 312 | 96 | 216 | 251 | 212 | 252 | 165 | 1160 | 229 | 5 | 82 | 645 | 16 | 479 | 27 |
| 102.1620233 | 22.86472598 | 184 | 68 | 403 | 3543 | 249 | 80 | 169 | 220 | 138 | 222 | 126 | 1964 | 382 | 27 | 78 | 1104 | 83 | 1079 | 95 |
| 107.0268881 | 22.86472598 | 201 | 64 | 306 | 5161 | 287 | 77 | 211 | 261 | 122 | 261 | 122 | 1803 | 322 | 33 | 74 | 960 | 115 | 960 | 115 |
| 111.891753 | 22.86472598 | 215 | 67 | 293 | 5438 | 307 | 79 | 228 | 264 | 159 | 279 | 133 | 1411 | 205 | 32 | 55 | 611 | 107 | 561 | 128 |
| -111.8920308 | 25.21607733 | 241 | 92 | 456 | 3814 | 340 | 138 | 202 | 295 | 254 | 295 | 191 | 133 | 28 | 0 | 87 | 82 | 0 | 82 | 57 |
| -107.0271659 | 25.21607733 | 205 | 110 | 488 | 3785 | 306 | 82 | 224 | 246 | 220 | 248 | 150 | 1038 | 269 | 5 | 102 | 770 | 17 | 523 | 158 |
| -102.1623011 | 25.21607733 | 223 | 124 | 473 | 4332 | 342 | 80 | 262 | 268 | 198 | 275 | 156 | 206 | 37 | 4 | 68 | 107 | 13 | 68 | 28 |
| -14.59473348 | 25.21607733 | 217 | 81 | 492 | 2471 | 302 | 138 | 164 | 199 | 223 | 252 | 183 | 47 | 13 | 0 | 91 | 31 | 1 | 9 | 27 |
| -9.729868619 | 25.21607733 | 254 | 137 | 402 | 6379 | 436 | 96 | 340 | 284 | 291 | 352 | 163 | 32 | 8 | 0 | 96 | 23 | 1 | 11 | 6 |
| -4.865003754 | 25.21607733 | 272 | 135 | 358 | 7897 | 462 | 86 | 375 | 253 | 373 | 380 | 157 | 31 | 5 | 0 | 70 | 15 | 0 | 4 | 5 |
| -0.000138889 | 25.21607733 | 268 | 139 | 365 | 8211 | 457 | 75 | 382 | 167 | 297 | 374 | 148 | 12 | 2 | 0 | 71 | 6 | 1 | 2 | 5 |
| 4.864725976 | 25.21607733 | 245 | 132 | 369 | 7684 | 412 | 53 | 358 | 184 | 146 | 340 | 132 | 55 | 12 | 1 | 70 | 26 | 5 | 10 | 8 |
| 9.729590841 | 25.21607733 | 228 | 126 | 370 | 7400 | 385 | 43 | 342 | 170 | 314 | 314 | 116 | 15 | 3 | 0 | 58 | 7 | 1 | 1 | 6 |
| 14.59445571 | 25.21607733 | 239 | 129 | 381 | 7373 | 397 | 58 | 339 | 127 | 322 | 322 | 127 | 4 | 2 | 0 | 187 | 4 | 0 | 0 | 4 |
| 19.45932057 | 25.21607733 | 231 | 130 | 404 | 6846 | 382 | 60 | 323 | 228 | 301 | 308 | 128 | 3 | 1 | 0 | 173 | 3 | 0 | 0 | 2 |
| 24.32418544 | 25.21607733 | 229 | 145 | 421 | 7025 | 390 | 46 | 344 | 173 | 136 | 310 | 124 | 1 | 1 | 0 | 332 | 2 | 0 | 0 | 0 |
| 34.05391517 | 25.21607733 | 250 | 143 | 432 | 6563 | 407 | 77 | 330 | 279 | 161 | 328 | 153 | 4 | 3 | 0 | 255 | 6 | 0 | 0 | 0 |
| 38.91878003 | 25.21607733 | 285 | 135 | 404 | 6943 | 442 | 108 | 333 | 211 | 362 | 366 | 183 | 45 | 9 | 0 | 84 | 25 | 0 | 0 | 23 |
| 43.78364489 | 25.21607733 | 257 | 128 | 366 | 7709 | 425 | 76 | 349 | 238 | 348 | 353 | 145 | 218 | 64 | 0 | 105 | 172 | 0 | 0 | 68 |
| 48.64850976 | 25.21607733 | 265 | 133 | 365 | 8134 | 446 | 82 | 364 | 194 | 364 | 365 | 147 | 82 | 18 | 0 | 101 | 46 | 0 | 0 | 46 |
| 68.10796922 | 25.21607733 | 267 | 119 | 413 | 5045 | 395 | 106 | 289 | 302 | 254 | 326 | 181 | 215 | 72 | 0 | 138 | 208 | 3 | 30 | 7 |
| 72.97283408 | 25.21607733 | 261 | 115 | 377 | 5222 | 399 | 94 | 305 | 286 | 288 | 328 | 175 | 461 | 160 | 1 | 147 | 460 | 5 | 90 | 7 |
| 77.83769895 | 25.21607733 | 252 | 106 | 331 | 5740 | 407 | 86 | 321 | 275 | 285 | 331 | 160 | 905 | 298 | 2 | 141 | 879 | 9 | 153 | 39 |
| 82.70256381 | 25.21607733 | 258 | 99 | 321 | 5690 | 402 | 94 | 307 | 291 | 186 | 328 | 165 | 1065 | 340 | 3 | 138 | 1002 | 14 | 183 | 50 |
| 87.56742868 | 25.21607733 | 243 | 84 | 341 | 4548 | 343 | 96 | 246 | 281 | 185 | 287 | 163 | 1203 | 292 | 1 | 104 | 800 | 11 | 446 | 33 |
| 92.43229354 | 25.21607733 | 233 | 68 | 355 | 3755 | 296 | 106 | 190 | 265 | 167 | 272 | 167 | 3746 | 836 | 4 | 89 | 2245 | 26 | 1756 | 26 |
| 97.29715841 | 25.21607733 | 247 | 76 | 372 | 3669 | 321 | 117 | 204 | 281 | 184 | 283 | 184 | 2161 | 513 | 9 | 104 | 1526 | 28 | 1319 | 28 |
| 102.1620233 | 25.21607733 | 167 | 83 | 375 | 4395 | 260 | 39 | 222 | 214 | 114 | 214 | 96 | 780 | 180 | 10 | 89 | 510 | 31 | 372 | 32 |
| 107.0268881 | 25.21607733 | 167 | 67 | 281 | 5929 | 268 | 29 | 239 | 224 | 76 | 239 | 76 | 1386 | 256 | 21 | 74 | 743 | 65 | 621 | 65 |
| 111.891753 | 25.21607733 | 153 | 66 | 245 | 6917 | 275 | 5 | 270 | 181 | 84 | 241 | 53 | 1729 | 280 | 55 | 51 | 806 | 179 | 455 | 215 |
| 116.7566179 | 25.21607733 | 171 | 75 | 292 | 6012 | 285 | 28 | 257 | 223 | 109 | 247 | 83 | 2010 | 374 | 39 | 66 | 1087 | 121 | 652 | 159 |
| 121.6214827 | 25.21607733 | 217 | 27 | 157 | 4986 | 300 | 128 | 172 | 244 | 279 | 283 | 147 | 3363 | 456 | 148 | 29 | 1288 | 518 | 630 | 804 |
| -107.0271659 | 27.56742868 | 139 | 125 | 445 | 4865 | 278 | -2 | 280 | 190 | 166 | 199 | 68 | 636 | 180 | 6 | 116 | 514 | 24 | 420 | 56 |
| -102.1623011 | 27.56742868 | 168 | 115 | 420 | 5341 | 291 | 17 | 274 | 213 | 133 | 230 | 87 | 402 | 67 | 9 | 59 | 184 | 30 | 161 | 47 |
| -9.729868619 | 27.56742868 | 219 | 138 | 431 | 5510 | 394 | 75 | 319 | 160 | 288 | 306 | 141 | 58 | 9 | 0 | 66 | 25 | 1 | 14 | 23 |
| -4.865003754 | 27.56742868 | 259 | 129 | 336 | 8211 | 456 | 72 | 383 | 289 | 367 | 376 | 142 | 28 | 5 | 0 | 69 | 13 | 0 | 2 | 9 |
| -0.000138889 | 27.56742868 | 259 | 129 | 335 | 8583 | 453 | 69 | 385 | 160 | 336 | 374 | 136 | 17 | 3 | 0 | 84 | 9 | 0 | 0 | 9 |
| 4.864725976 | 27.56742868 | 245 | 125 | 337 | 8448 | 428 | 57 | 370 | 148 | 353 | 353 | 124 | 24 | 7 | 0 | 94 | 16 | 1 | 1 | 13 |
| 9.729590841 | 27.56742868 | 240 | 123 | 349 | 8078 | 407 | 54 | 353 | 146 | 334 | 337 | 120 | 25 | 5 | 0 | 69 | 13 | 0 | 2 | 9 |
| 14.59445571 | 27.56742868 | 239 | 122 | 370 | 7405 | 396 | 65 | 331 | 273 | 324 | 324 | 129 | 13 | 3 | 0 | 80 | 7 | 0 | 0 | 5 |
| 19.45932057 | 27.56742868 | 225 | 121 | 396 | 6610 | 373 | 67 | 306 | 151 | 300 | 302 | 128 | 13 | 4 | 0 | 110 | 9 | 0 | 0 | 8 |
| 24.32418544 | 27.56742868 | 222 | 134 | 406 | 6944 | 381 | 51 | 330 | 165 | 305 | 305 | 121 | 8 | 3 | 0 | 127 | 7 | 0 | 0 | 3 |
| 29.1890503 | 27.56742868 | 221 | 138 | 422 | 6605 | 381 | 55 | 327 | 165 | 133 | 302 | 125 | 1 | 1 | 0 | 332 | 2 | 0 | 0 | 0 |
| 38.91878003 | 27.56742868 | 228 | 131 | 375 | 7543 | 397 | 48 | 349 | 119 | 316 | 323 | 119 | 69 | 12 | 0 | 79 | 33 | 0 | 0 | 33 |
| 43.78364489 | 27.56742868 | 244 | 130 | 348 | 8668 | 426 | 53 | 373 | 223 | 346 | 351 | 119 | 158 | 31 | 0 | 88 | 80 | 0 | 0 | 58 |
| 48.64850976 | 27.56742868 | 275 | 116 | 330 | 8381 | 450 | 99 | 352 | 156 | 379 | 380 | 156 | 96 | 23 | 0 | 100 | 62 | 0 | 0 | 62 |
| 53.51337462 | 27.56742868 | 260 | 134 | 372 | 7902 | 436 | 76 | 359 | 147 | 341 | 359 | 147 | 213 | 54 | 1 | 116 | 156 | 3 | 11 | 156 |
| 58.37823949 | 27.56742868 | 278 | 138 | 379 | 7938 | 456 | 92 | 365 | 172 | 325 | 377 | 162 | 133 | 38 | 0 | 131 | 114 | 2 | 2 | 90 |
| 63.24310435 | 27.56742868 | 256 | 136 | 375 | 7857 | 431 | 68 | 362 | 199 | 316 | 355 | 139 | 68 | 14 | 0 | 79 | 41 | 2 | 10 | 30 |
| 68.10796922 | 27.56742868 | 270 | 140 | 383 | 7409 | 439 | 75 | 365 | 358 | 234 | 358 | 156 | 118 | 52 | 1 | 144 | 110 | 3 | 15 | 11 |
| 72.97283408 | 27.56742868 | 272 | 117 | 350 | 6563 | 422 | 89 | 333 | 316 | 192 | 353 | 164 | 264 | 90 | 1 | 135 | 246 | 4 | 61 | 7 |
| 77.83769895 | 27.56742868 | 256 | 108 | 328 | 6401 | 410 | 80 | 329 | 298 | 176 | 338 | 152 | 697 | 240 | 3 | 141 | 685 | 13 | 71 | 37 |
| 82.70256381 | 27.56742868 | 244 | 101 | 341 | 5697 | 369 | 73 | 296 | 291 | 174 | 308 | 150 | 1588 | 484 | 7 | 122 | 1268 | 33 | 526 | 49 |
| 87.56742868 | 27.56742868 | 32 | 72 | 315 | 5485 | 128 | -99 | 227 | 101 | -19 | 104 | -41 | 3563 | 758 | 20 | 90 | 2136 | 79 | 2124 | 84 |
| 92.43229354 | 27.56742868 | 71 | 67 | 302 | 5450 | 163 | -58 | 221 | 138 | -6 | 139 | -6 | 1724 | 333 | 14 | 77 | 951 | 56 | 901 | 56 |
| 97.29715841 | 27.56742868 | 177 | 71 | 333 | 4934 | 259 | 45 | 214 | 236 | 103 | 236 | 103 | 4492 | 1106 | 19 | 105 | 3106 | 58 | 3106 | 58 |
| 102.1620233 | 27.56742868 | 136 | 80 | 357 | 4948 | 229 | 4 | 225 | 195 | 59 | 195 | 59 | 1186 | 277 | 4 | 102 | 797 | 13 | 717 | 13 |
| 107.0268881 | 27.56742868 | 158 | 67 | 245 | 6879 | 285 | 13 | 271 | 219 | 65 | 249 | 57 | 1010 | 181 | 21 | 62 | 506 | 66 | 376 | 71 |
| 111.891753 | 27.56742868 | 179 | 68 | 218 | 8228 | 328 | 16 | 311 | 211 | 98 | 290 | 63 | 1345 | 191 | 44 | 42 | 562 | 162 | 345 | 178 |
| 116.7566179 | 27.56742868 | 190 | 72 | 236 | 7792 | 333 | 30 | 303 | 257 | 109 | 293 | 77 | 1499 | 296 | 38 | 67 | 850 | 121 | 353 | 150 |
| -111.8920308 | 29.91878003 | 216 | 120 | 408 | 6135 | 358 | 65 | 293 | 294 | 231 | 294 | 131 | 198 | 42 | 2 | 81 | 122 | 6 | 122 | 60 |
| -107.0271659 | 29.91878003 | 173 | 129 | 402 | 6186 | 331 | 11 | 320 | 241 | 126 | 252 | 83 | 337 | 80 | 4 | 102 | 238 | 15 | 177 | 37 |
| -102.1623011 | 29.91878003 | 212 | 116 | 374 | 6806 | 353 | 42 | 311 | 272 | 111 | 295 | 111 | 329 | 50 | 12 | 44 | 138 | 36 | 96 | 36 |
| -97.29743619 | 29.91878003 | 204 | 93 | 329 | 6476 | 338 | 57 | 281 | 230 | 289 | 289 | 110 | 904 | 115 | 46 | 28 | 305 | 150 | 150 | 187 |
| -92.43257132 | 29.91878003 | 204 | 61 | 272 | 5787 | 304 | 80 | 224 | 274 | 154 | 277 | 122 | 1465 | 171 | 98 | 18 | 478 | 303 | 467 | 363 |
| -82.70284159 | 29.91878003 | 203 | 79 | 342 | 5313 | 305 | 75 | 230 | 269 | 185 | 269 | 127 | 1359 | 194 | 56 | 40 | 576 | 185 | 576 | 273 |
| -9.729868619 | 29.91878003 | 215 | 86 | 383 | 4164 | 332 | 109 | 223 | 159 | 268 | 282 | 159 | 187 | 34 | 0 | 79 | 98 | 1 | 2 | 98 |
| -4.865003754 | 29.91878003 | 248 | 123 | 323 | 8299 | 442 | 62 | 380 | 153 | 358 | 365 | 130 | 64 | 14 | 0 | 80 | 39 | 1 | 2 | 24 |
| -0.000138889 | 29.91878003 | 234 | 122 | 319 | 8716 | 431 | 49 | 382 | 113 | 350 | 354 | 113 | 32 | 6 | 0 | 76 | 17 | 0 | 0 | 17 |
| 4.864725976 | 29.91878003 | 239 | 121 | 321 | 8731 | 431 | 54 | 377 | 139 | 355 | 355 | 116 | 31 | 8 | 0 | 93 | 21 | 0 | 0 | 18 |
| 9.729590841 | 29.91878003 | 229 | 120 | 340 | 8118 | 405 | 51 | 353 | 113 | 331 | 331 | 113 | 29 | 5 | 0 | 76 | 15 | 0 | 0 | 15 |
| 14.59445571 | 29.91878003 | 214 | 123 | 397 | 6628 | 369 | 60 | 309 | 250 | 294 | 294 | 118 | 54 | 9 | 0 | 64 | 24 | 0 | 0 | 23 |
| 19.45932057 | 29.91878003 | 219 | 104 | 411 | 5473 | 344 | 92 | 252 | 141 | 280 | 284 | 141 | 95 | 26 | 0 | 111 | 76 | 0 | 0 | 76 |
| 24.32418544 | 29.91878003 | 220 | 119 | 407 | 6150 | 370 | 78 | 292 | 134 | 294 | 298 | 134 | 19 | 5 | 0 | 114 | 14 | 0 | 0 | 14 |
| 29.1890503 | 29.91878003 | 213 | 119 | 431 | 5527 | 354 | 78 | 276 | 135 | 261 | 283 | 135 | 29 | 8 | 0 | 113 | 21 | 0 | 0 | 21 |
| 34.05391517 | 29.91878003 | 203 | 126 | 430 | 5767 | 351 | 58 | 293 | 119 | 271 | 275 | 119 | 30 | 7 | 0 | 97 | 20 | 0 | 0 | 20 |
| 38.91878003 | 29.91878003 | 228 | 130 | 358 | 8190 | 405 | 41 | 364 | 109 | 324 | 331 | 109 | 48 | 9 | 0 | 79 | 26 | 0 | 0 | 26 |
| 43.78364489 | 29.91878003 | 241 | 128 | 334 | 9134 | 432 | 47 | 384 | 110 | 350 | 355 | 110 | 72 | 14 | 0 | 88 | 38 | 0 | 0 | 38 |
| 53.51337462 | 29.91878003 | 168 | 133 | 348 | 8619 | 364 | -18 | 382 | 49 | 249 | 283 | 49 | 248 | 55 | 0 | 108 | 160 | 0 | 2 | 160 |
| 58.37823949 | 29.91878003 | 282 | 129 | 346 | 8459 | 462 | 88 | 374 | 215 | 351 | 390 | 157 | 90 | 22 | 0 | 113 | 64 | 0 | 2 | 50 |
| 63.24310435 | 29.91878003 | 229 | 137 | 350 | 8670 | 422 | 31 | 390 | 117 | 296 | 343 | 102 | 66 | 18 | 0 | 111 | 46 | 0 | 4 | 42 |
| 68.10796922 | 29.91878003 | 246 | 125 | 343 | 8032 | 419 | 55 | 364 | 328 | 270 | 341 | 126 | 198 | 46 | 2 | 76 | 118 | 12 | 101 | 46 |
| 72.97283408 | 29.91878003 | 259 | 123 | 349 | 7590 | 415 | 62 | 353 | 331 | 226 | 347 | 140 | 204 | 68 | 2 | 114 | 162 | 6 | 44 | 15 |
| 77.83769895 | 29.91878003 | 235 | 112 | 348 | 6266 | 382 | 61 | 321 | 288 | 160 | 311 | 136 | 1116 | 349 | 5 | 127 | 1023 | 31 | 175 | 103 |
| 82.70256381 | 29.91878003 | -42 | 99 | 314 | 7483 | 95 | -221 | 316 | 61 | -68 | 61 | -144 | 503 | 114 | 7 | 83 | 342 | 37 | 342 | 65 |
| 87.56742868 | 29.91878003 | -3 | 104 | 333 | 7010 | 133 | -181 | 313 | 88 | -98 | 88 | -100 | 177 | 68 | 0 | 157 | 195 | 0 | 195 | 0 |
| 92.43229354 | 29.91878003 | 19 | 83 | 306 | 6304 | 141 | -130 | 271 | 103 | -60 | 103 | -67 | 268 | 70 | 1 | 114 | 206 | 3 | 206 | 3 |
| 97.29715841 | 29.91878003 | 27 | 77 | 282 | 6603 | 148 | -126 | 274 | 114 | -65 | 114 | -65 | 291 | 66 | 1 | 96 | 198 | 4 | 198 | 4 |
| 102.1620233 | 29.91878003 | 106 | 73 | 301 | 5745 | 211 | -32 | 243 | 178 | 22 | 180 | 22 | 874 | 155 | 5 | 79 | 459 | 15 | 422 | 15 |
| 107.0268881 | 29.91878003 | 183 | 69 | 248 | 7101 | 317 | 39 | 278 | 245 | 90 | 281 | 82 | 1146 | 180 | 18 | 61 | 525 | 56 | 427 | 62 |
| 111.891753 | 29.91878003 | 177 | 76 | 239 | 8213 | 322 | 5 | 317 | 251 | 59 | 286 | 59 | 1179 | 177 | 30 | 50 | 511 | 94 | 425 | 94 |
| 116.7566179 | 29.91878003 | 175 | 74 | 233 | 8407 | 323 | 5 | 318 | 248 | 89 | 288 | 55 | 1310 | 240 | 36 | 56 | 655 | 124 | 376 | 136 |
| 121.6214827 | 29.91878003 | 176 | 50 | 181 | 7742 | 310 | 35 | 275 | 232 | 109 | 284 | 71 | 1375 | 194 | 47 | 41 | 526 | 158 | 451 | 210 |
| -116.7568956 | 32.27013138 | 156 | 102 | 455 | 4166 | 279 | 56 | 223 | 117 | 204 | 218 | 105 | 335 | 67 | 2 | 87 | 197 | 6 | 12 | 165 |
| -111.8920308 | 32.27013138 | 226 | 126 | 384 | 7336 | 390 | 61 | 329 | 320 | 249 | 322 | 126 | 264 | 58 | 2 | 74 | 165 | 11 | 104 | 61 |
| -107.0271659 | 32.27013138 | 180 | 120 | 350 | 7632 | 350 | 9 | 342 | 270 | 167 | 279 | 71 | 245 | 54 | 6 | 77 | 156 | 20 | 111 | 36 |
| -102.1623011 | 32.27013138 | 177 | 120 | 357 | 7665 | 335 | 0 | 336 | 248 | 66 | 275 | 66 | 414 | 66 | 14 | 53 | 179 | 42 | 147 | 42 |
| -97.29743619 | 32.27013138 | 190 | 98 | 308 | 7543 | 344 | 27 | 317 | 216 | 293 | 293 | 83 | 890 | 106 | 46 | 21 | 290 | 154 | 154 | 195 |
| -92.43257132 | 32.27013138 | 187 | 90 | 315 | 7020 | 323 | 38 | 285 | 218 | 208 | 278 | 87 | 1372 | 141 | 77 | 19 | 415 | 250 | 268 | 403 |
| -87.56770646 | 32.27013138 | 184 | 91 | 327 | 6841 | 315 | 36 | 279 | 131 | 205 | 271 | 86 | 1411 | 173 | 70 | 24 | 481 | 225 | 324 | 412 |
| -82.70284159 | 32.27013138 | 180 | 87 | 322 | 6512 | 308 | 36 | 272 | 265 | 150 | 265 | 89 | 1214 | 131 | 67 | 20 | 390 | 203 | 390 | 312 |
| -4.865003754 | 32.27013138 | 152 | 117 | 344 | 7141 | 334 | -5 | 339 | 126 | 258 | 258 | 57 | 202 | 28 | 4 | 38 | 77 | 16 | 16 | 52 |
| -0.000138889 | 32.27013138 | 199 | 117 | 318 | 8404 | 395 | 26 | 370 | 171 | 314 | 320 | 86 | 109 | 15 | 1 | 47 | 40 | 6 | 7 | 34 |
| 4.864725976 | 32.27013138 | 226 | 113 | 313 | 8406 | 415 | 54 | 361 | 132 | 344 | 344 | 112 | 60 | 9 | 0 | 63 | 27 | 0 | 0 | 25 |
| 9.729590841 | 32.27013138 | 210 | 117 | 353 | 7329 | 383 | 51 | 332 | 147 | 309 | 309 | 108 | 69 | 14 | 0 | 73 | 34 | 0 | 0 | 30 |
| 14.59445571 | 32.27013138 | 210 | 88 | 363 | 5403 | 335 | 94 | 242 | 137 | 282 | 282 | 137 | 216 | 47 | 0 | 89 | 139 | 0 | 0 | 139 |
| 38.91878003 | 32.27013138 | 186 | 126 | 349 | 8338 | 372 | 9 | 362 | 112 | 289 | 295 | 70 | 111 | 18 | 0 | 77 | 54 | 0 | 0 | 45 |
| 43.78364489 | 32.27013138 | 244 | 126 | 317 | 9623 | 445 | 47 | 397 | 108 | 363 | 368 | 108 | 102 | 24 | 0 | 92 | 62 | 0 | 0 | 62 |
| 48.64850976 | 32.27013138 | 250 | 137 | 326 | 10010 | 471 | 52 | 419 | 113 | 377 | 381 | 113 | 401 | 92 | 0 | 104 | 272 | 0 | 0 | 272 |
| 53.51337462 | 32.27013138 | 141 | 129 | 326 | 9030 | 340 | -57 | 397 | 65 | 226 | 262 | 14 | 162 | 34 | 0 | 97 | 96 | 0 | 2 | 89 |
| 58.37823949 | 32.27013138 | 221 | 131 | 338 | 8947 | 414 | 26 | 389 | 146 | 338 | 338 | 92 | 93 | 23 | 0 | 114 | 66 | 0 | 0 | 52 |
| 63.24310435 | 32.27013138 | 222 | 137 | 330 | 9647 | 434 | 20 | 415 | 141 | 350 | 350 | 86 | 131 | 35 | 0 | 119 | 99 | 0 | 0 | 79 |
| 68.10796922 | 32.27013138 | 130 | 130 | 328 | 9249 | 324 | -72 | 396 | 48 | 220 | 249 | 1 | 247 | 54 | 0 | 86 | 156 | 10 | 45 | 108 |
| 72.97283408 | 32.27013138 | 241 | 111 | 322 | 7484 | 400 | 55 | 345 | 312 | 209 | 332 | 126 | 516 | 137 | 7 | 98 | 407 | 27 | 81 | 59 |
| 77.83769895 | 32.27013138 | -51 | 93 | 298 | 7650 | 97 | -216 | 313 | -121 | -83 | 59 | -151 | 639 | 86 | 19 | 38 | 245 | 67 | 202 | 186 |
| 82.70256381 | 32.27013138 | -35 | 122 | 338 | 8156 | 139 | -223 | 361 | 75 | -141 | 75 | -144 | 135 | 50 | 1 | 145 | 142 | 3 | 142 | 3 |
| 87.56742868 | 32.27013138 | -30 | 117 | 343 | 7756 | 131 | -210 | 342 | 72 | -92 | 73 | -136 | 234 | 78 | 1 | 137 | 227 | 3 | 172 | 5 |
| 92.43229354 | 32.27013138 | -25 | 108 | 325 | 7424 | 127 | -205 | 332 | 71 | -110 | 75 | -130 | 585 | 152 | 3 | 109 | 420 | 10 | 382 | 11 |
| 97.29715841 | 32.27013138 | -12 | 99 | 321 | 6970 | 133 | -175 | 308 | 81 | -91 | 84 | -107 | 689 | 145 | 3 | 99 | 429 | 11 | 411 | 11 |
| 102.1620233 | 32.27013138 | 3 | 93 | 328 | 6379 | 131 | -151 | 282 | 86 | -85 | 89 | -85 | 1087 | 206 | 4 | 85 | 568 | 14 | 487 | 14 |
| 107.0268881 | 32.27013138 | 116 | 77 | 267 | 7322 | 252 | -34 | 287 | 210 | 11 | 215 | 11 | 1515 | 315 | 12 | 79 | 827 | 38 | 795 | 38 |
| 111.891753 | 32.27013138 | 164 | 85 | 258 | 8484 | 314 | -16 | 330 | 273 | 42 | 273 | 42 | 871 | 142 | 18 | 57 | 425 | 58 | 377 | 58 |
| 116.7566179 | 32.27013138 | 159 | 83 | 245 | 8901 | 314 | -24 | 338 | 273 | 31 | 276 | 31 | 1010 | 190 | 21 | 57 | 529 | 81 | 450 | 81 |
| 131.3512125 | 32.27013138 | 135 | 61 | 234 | 6945 | 261 | 0 | 261 | 218 | 41 | 233 | 41 | 3349 | 592 | 74 | 57 | 1566 | 262 | 1217 | 262 |
| -116.7568956 | 34.62148273 | 159 | 132 | 386 | 7252 | 346 | 4 | 342 | 63 | 217 | 263 | 63 | 213 | 33 | 2 | 54 | 93 | 9 | 47 | 93 |
| -111.8920308 | 34.62148273 | 182 | 123 | 353 | 8065 | 364 | 15 | 349 | 289 | 259 | 291 | 74 | 319 | 48 | 8 | 41 | 132 | 24 | 80 | 95 |
| -107.0271659 | 34.62148273 | 143 | 122 | 343 | 8372 | 324 | -31 | 355 | 250 | 44 | 255 | 27 | 232 | 49 | 8 | 69 | 138 | 26 | 94 | 28 |
| -102.1623011 | 34.62148273 | 147 | 121 | 346 | 8136 | 319 | -30 | 349 | 226 | 32 | 254 | 32 | 457 | 81 | 12 | 60 | 219 | 39 | 197 | 39 |
| -97.29743619 | 34.62148273 | 172 | 97 | 286 | 8360 | 338 | 0 | 339 | 201 | 66 | 286 | 53 | 994 | 147 | 43 | 36 | 389 | 141 | 205 | 141 |
| -92.43257132 | 34.62148273 | 168 | 91 | 284 | 8023 | 324 | 4 | 320 | 200 | 274 | 274 | 53 | 1375 | 140 | 82 | 16 | 420 | 262 | 262 | 335 |
| -87.56770646 | 34.62148273 | 156 | 84 | 285 | 7477 | 299 | 5 | 294 | 96 | 178 | 255 | 51 | 1565 | 174 | 80 | 22 | 508 | 259 | 330 | 496 |
| -82.70284159 | 34.62148273 | 161 | 88 | 300 | 7294 | 305 | 11 | 294 | 101 | 180 | 257 | 59 | 1369 | 149 | 93 | 13 | 424 | 290 | 342 | 379 |
| -77.83797673 | 34.62148273 | 179 | 67 | 263 | 6498 | 298 | 44 | 253 | 265 | 156 | 265 | 90 | 1354 | 186 | 76 | 31 | 518 | 232 | 518 | 295 |
| -4.865003754 | 34.62148273 | 156 | 93 | 365 | 5180 | 297 | 43 | 254 | 107 | 236 | 236 | 90 | 936 | 179 | 2 | 74 | 485 | 10 | 10 | 453 |
| -0.000138889 | 34.62148273 | 146 | 113 | 350 | 6910 | 329 | 7 | 322 | 87 | 242 | 252 | 59 | 257 | 31 | 4 | 38 | 87 | 20 | 24 | 77 |
| 4.864725976 | 34.62148273 | 212 | 108 | 304 | 8341 | 407 | 50 | 357 | 238 | 328 | 335 | 104 | 121 | 19 | 3 | 41 | 45 | 12 | 15 | 28 |
| 9.729590841 | 34.62148273 | 196 | 101 | 348 | 6385 | 353 | 63 | 290 | 229 | 278 | 289 | 112 | 250 | 45 | 4 | 55 | 124 | 15 | 22 | 79 |
| 38.91878003 | 34.62148273 | 200 | 120 | 329 | 8586 | 391 | 27 | 364 | 124 | 308 | 313 | 81 | 118 | 21 | 0 | 81 | 59 | 0 | 0 | 53 |
| 43.78364489 | 34.62148273 | 236 | 125 | 303 | 10149 | 450 | 39 | 412 | 146 | 367 | 371 | 97 | 217 | 37 | 0 | 88 | 109 | 0 | 0 | 107 |
| 48.64850976 | 34.62148273 | 90 | 123 | 300 | 9588 | 310 | -98 | 408 | 0 | 192 | 227 | -38 | 337 | 61 | 1 | 75 | 166 | 3 | 4 | 132 |
| 53.51337462 | 34.62148273 | 214 | 121 | 295 | 9886 | 421 | 10 | 411 | 127 | 308 | 342 | 74 | 91 | 18 | 0 | 88 | 51 | 0 | 3 | 46 |
| 58.37823949 | 34.62148273 | 190 | 126 | 321 | 9293 | 387 | -5 | 392 | 108 | 305 | 311 | 58 | 176 | 39 | 0 | 98 | 114 | 0 | 1 | 86 |
| 63.24310435 | 34.62148273 | 59 | 131 | 363 | 7840 | 252 | -110 | 362 | -10 | 166 | 166 | -48 | 571 | 151 | 0 | 108 | 421 | 0 | 0 | 287 |
| 68.10796922 | 34.62148273 | 15 | 119 | 310 | 8645 | 206 | -177 | 383 | -6 | 132 | 132 | -106 | 196 | 43 | 1 | 86 | 124 | 4 | 4 | 77 |
| 72.97283408 | 34.62148273 | 132 | 98 | 324 | 7123 | 276 | -27 | 303 | 70 | 102 | 222 | 31 | 1462 | 213 | 40 | 46 | 583 | 140 | 491 | 312 |
| 77.83769895 | 34.62148273 | -116 | 101 | 284 | 8509 | 61 | -292 | 353 | 8 | -155 | 8 | -231 | 164 | 26 | 4 | 46 | 75 | 16 | 75 | 32 |
| 82.70256381 | 34.62148273 | -86 | 121 | 348 | 7638 | 84 | -263 | 347 | 15 | -166 | 21 | -190 | 285 | 83 | 1 | 117 | 219 | 5 | 219 | 8 |
| 87.56742868 | 34.62148273 | -65 | 121 | 342 | 8059 | 108 | -247 | 355 | 46 | -156 | 46 | -175 | 321 | 96 | 1 | 125 | 274 | 7 | 274 | 9 |
| 92.43229354 | 34.62148273 | -42 | 121 | 346 | 7659 | 121 | -229 | 349 | 56 | -132 | 61 | -151 | 170 | 47 | 1 | 112 | 124 | 3 | 123 | 3 |
| 97.29715841 | 34.62148273 | -39 | 123 | 336 | 7799 | 127 | -237 | 364 | 60 | -128 | 65 | -152 | 344 | 76 | 4 | 90 | 213 | 12 | 206 | 14 |
| 102.1620233 | 34.62148273 | 17 | 111 | 336 | 7090 | 165 | -164 | 330 | 111 | -87 | 111 | -87 | 396 | 82 | 1 | 90 | 234 | 5 | 234 | 5 |
| 107.0268881 | 34.62148273 | 119 | 95 | 281 | 8432 | 281 | -57 | 338 | 228 | -4 | 228 | -4 | 712 | 135 | 4 | 76 | 368 | 14 | 368 | 14 |
| 111.891753 | 34.62148273 | 137 | 94 | 267 | 9170 | 301 | -52 | 353 | 255 | 3 | 255 | 3 | 639 | 152 | 6 | 80 | 380 | 18 | 380 | 18 |
| 116.7566179 | 34.62148273 | 146 | 91 | 255 | 9426 | 308 | -48 | 355 | 267 | 9 | 267 | 9 | 772 | 214 | 14 | 90 | 526 | 42 | 526 | 42 |
| 126.4863476 | 34.62148273 | 145 | 39 | 143 | 7889 | 278 | 9 | 269 | 237 | 39 | 258 | 38 | 1328 | 233 | 30 | 63 | 654 | 96 | 649 | 139 |
| 136.2160773 | 34.62148273 | 140 | 68 | 233 | 7773 | 287 | -4 | 290 | 232 | 38 | 251 | 35 | 2016 | 285 | 64 | 40 | 743 | 206 | 574 | 249 |
| -121.6217605 | 36.97283408 | 129 | 99 | 474 | 3598 | 244 | 34 | 210 | 78 | 173 | 177 | 78 | 545 | 118 | 1 | 93 | 334 | 5 | 5 | 334 |
| -116.7568956 | 36.97283408 | 169 | 118 | 315 | 8854 | 368 | -6 | 374 | 66 | 246 | 293 | 50 | 103 | 15 | 3 | 38 | 43 | 11 | 26 | 35 |
| -111.8920308 | 36.97283408 | 125 | 117 | 311 | 9060 | 326 | -51 | 377 | 249 | 209 | 249 | 3 | 225 | 27 | 7 | 28 | 75 | 25 | 75 | 59 |
| -107.0271659 | 36.97283408 | 85 | 124 | 332 | 8346 | 278 | -96 | 374 | 198 | 163 | 199 | -29 | 453 | 56 | 19 | 23 | 154 | 66 | 103 | 127 |
| -102.1623011 | 36.97283408 | 134 | 124 | 328 | 8900 | 325 | -52 | 377 | 253 | 25 | 253 | 11 | 406 | 63 | 11 | 58 | 185 | 33 | 177 | 33 |
| -97.29743619 | 36.97283408 | 148 | 102 | 273 | 9304 | 333 | -41 | 374 | 232 | 17 | 275 | 17 | 886 | 122 | 25 | 42 | 359 | 85 | 278 | 85 |
| -92.43257132 | 36.97283408 | 137 | 89 | 262 | 8561 | 307 | -33 | 340 | 170 | 26 | 252 | 16 | 1051 | 122 | 53 | 21 | 349 | 177 | 249 | 180 |
| -87.56770646 | 36.97283408 | 144 | 86 | 264 | 8396 | 305 | -22 | 327 | 178 | 169 | 254 | 25 | 1357 | 137 | 84 | 15 | 393 | 255 | 300 | 345 |
| -82.70284159 | 36.97283408 | 120 | 84 | 284 | 7501 | 266 | -30 | 296 | 218 | 141 | 219 | 16 | 1267 | 134 | 72 | 16 | 379 | 224 | 346 | 327 |
| -77.83797673 | 36.97283408 | 150 | 88 | 283 | 7786 | 305 | -5 | 310 | 253 | 71 | 254 | 44 | 1130 | 111 | 80 | 11 | 333 | 244 | 315 | 274 |
| -4.865003754 | 36.97283408 | 165 | 101 | 357 | 5868 | 320 | 37 | 284 | 106 | 253 | 253 | 89 | 649 | 107 | 2 | 66 | 318 | 14 | 14 | 277 |
| 9.729590841 | 36.97283408 | 187 | 64 | 294 | 5332 | 310 | 91 | 219 | 124 | 255 | 267 | 121 | 630 | 91 | 3 | 60 | 273 | 21 | 21 | 245 |
| 14.59445571 | 36.97283408 | 177 | 44 | 220 | 5333 | 285 | 87 | 198 | 136 | 244 | 257 | 110 | 442 | 76 | 3 | 66 | 200 | 11 | 27 | 144 |
| 29.1890503 | 36.97283408 | 139 | 100 | 315 | 7711 | 314 | -3 | 317 | 44 | 252 | 252 | 42 | 662 | 135 | 5 | 83 | 395 | 17 | 17 | 334 |
| 34.05391517 | 36.97283408 | 73 | 96 | 305 | 7730 | 244 | -72 | 316 | -26 | 186 | 186 | -26 | 527 | 100 | 3 | 73 | 288 | 13 | 13 | 288 |
| 38.91878003 | 36.97283408 | 189 | 116 | 300 | 9527 | 400 | 14 | 386 | 63 | 322 | 322 | 63 | 365 | 65 | 0 | 81 | 194 | 0 | 0 | 194 |
| 43.78364489 | 36.97283408 | 144 | 111 | 287 | 9508 | 354 | -34 | 388 | 108 | 277 | 277 | 18 | 721 | 123 | 2 | 77 | 359 | 7 | 7 | 278 |
| 48.64850976 | 36.97283408 | 108 | 107 | 302 | 8331 | 284 | -70 | 354 | 137 | 217 | 217 | -9 | 332 | 52 | 10 | 48 | 153 | 31 | 31 | 77 |
| 53.51337462 | 36.97283408 | 192 | 71 | 258 | 7238 | 330 | 56 | 274 | 176 | 280 | 290 | 93 | 463 | 64 | 18 | 40 | 187 | 56 | 58 | 121 |
| 58.37823949 | 36.97283408 | 65 | 120 | 328 | 8451 | 259 | -107 | 367 | 43 | 180 | 180 | -49 | 380 | 75 | 2 | 71 | 197 | 11 | 11 | 134 |
| 63.24310435 | 36.97283408 | 171 | 121 | 310 | 9475 | 378 | -14 | 391 | 92 | 292 | 301 | 41 | 216 | 49 | 0 | 91 | 130 | 0 | 1 | 102 |
| 68.10796922 | 36.97283408 | 194 | 110 | 287 | 9508 | 395 | 10 | 385 | 114 | 322 | 322 | 63 | 177 | 45 | 0 | 97 | 118 | 0 | 0 | 73 |
| 72.97283408 | 36.97283408 | -59 | 96 | 278 | 8578 | 127 | -220 | 347 | -47 | -103 | 69 | -170 | 115 | 21 | 4 | 59 | 59 | 14 | 19 | 23 |
| 77.83769895 | 36.97283408 | -2 | 108 | 310 | 7942 | 152 | -195 | 348 | 71 | -87 | 102 | -121 | 130 | 26 | 1 | 79 | 68 | 3 | 57 | 5 |
| 82.70256381 | 36.97283408 | 114 | 113 | 294 | 9370 | 291 | -93 | 384 | 228 | 64 | 230 | -29 | 56 | 15 | 0 | 99 | 41 | 1 | 29 | 3 |
| 87.56742868 | 36.97283408 | -66 | 124 | 339 | 8164 | 107 | -258 | 365 | 39 | -113 | 46 | -182 | 261 | 80 | 1 | 122 | 216 | 5 | 206 | 16 |
| 92.43229354 | 36.97283408 | 23 | 129 | 330 | 8961 | 207 | -186 | 392 | 138 | -82 | 141 | -106 | 71 | 19 | 0 | 115 | 56 | 1 | 47 | 2 |
| 97.29715841 | 36.97283408 | 38 | 126 | 304 | 9647 | 229 | -184 | 413 | 158 | -15 | 164 | -104 | 91 | 21 | 0 | 103 | 63 | 2 | 55 | 5 |
| 102.1620233 | 36.97283408 | -8 | 118 | 308 | 8524 | 171 | -212 | 382 | 104 | -132 | 104 | -132 | 581 | 129 | 2 | 95 | 382 | 6 | 333 | 6 |
| 107.0268881 | 36.97283408 | 90 | 109 | 268 | 10028 | 281 | -127 | 408 | 214 | -59 | 220 | -59 | 321 | 77 | 1 | 94 | 226 | 5 | 180 | 5 |
| 111.891753 | 36.97283408 | 108 | 100 | 257 | 10059 | 289 | -101 | 390 | 231 | -41 | 238 | -41 | 434 | 107 | 4 | 89 | 281 | 12 | 263 | 12 |
| 116.7566179 | 36.97283408 | 139 | 97 | 253 | 10226 | 314 | -69 | 383 | 265 | -11 | 269 | -11 | 634 | 196 | 4 | 109 | 488 | 14 | 470 | 14 |
| 121.6214827 | 36.97283408 | 117 | 33 | 116 | 8755 | 254 | -29 | 283 | 238 | -2 | 238 | -2 | 842 | 203 | 12 | 92 | 601 | 40 | 601 | 40 |
| -121.6217605 | 39.32418544 | 181 | 123 | 385 | 6663 | 366 | 46 | 320 | 93 | 275 | 275 | 93 | 608 | 125 | 1 | 85 | 358 | 5 | 5 | 358 |
| -116.7568956 | 39.32418544 | 86 | 116 | 309 | 8774 | 296 | -81 | 377 | 100 | 217 | 217 | -30 | 288 | 35 | 19 | 22 | 101 | 57 | 57 | 63 |
| -111.8920308 | 39.32418544 | 106 | 117 | 296 | 9407 | 323 | -71 | 394 | 33 | 235 | 242 | -20 | 292 | 29 | 15 | 18 | 85 | 47 | 62 | 76 |
| -107.0271659 | 39.32418544 | 58 | 121 | 330 | 8182 | 255 | -112 | 367 | 32 | 129 | 174 | -51 | 326 | 31 | 20 | 10 | 91 | 67 | 79 | 81 |
| -102.1623011 | 39.32418544 | 108 | 122 | 316 | 9250 | 310 | -74 | 385 | 233 | -17 | 234 | -17 | 430 | 73 | 8 | 68 | 215 | 24 | 197 | 24 |
| -97.29743619 | 39.32418544 | 127 | 101 | 258 | 9961 | 322 | -67 | 390 | 219 | -14 | 260 | -14 | 702 | 103 | 16 | 51 | 306 | 53 | 246 | 53 |
| -92.43257132 | 39.32418544 | 124 | 91 | 247 | 9478 | 305 | -62 | 367 | 214 | -11 | 248 | -11 | 1051 | 127 | 44 | 33 | 379 | 138 | 323 | 138 |
| -87.56770646 | 39.32418544 | 128 | 87 | 245 | 9322 | 304 | -52 | 356 | 166 | 2 | 249 | -4 | 1013 | 113 | 56 | 19 | 324 | 181 | 306 | 209 |
| -82.70284159 | 39.32418544 | 124 | 84 | 254 | 8587 | 289 | -41 | 330 | 236 | 151 | 237 | 5 | 946 | 101 | 59 | 16 | 289 | 188 | 269 | 216 |
| -77.83797673 | 39.32418544 | 129 | 84 | 258 | 8419 | 296 | -32 | 328 | 241 | 15 | 241 | 15 | 934 | 90 | 59 | 13 | 267 | 181 | 267 | 194 |
| -4.865003754 | 39.32418544 | 165 | 96 | 313 | 6963 | 339 | 33 | 307 | 91 | 270 | 270 | 76 | 536 | 71 | 6 | 51 | 210 | 20 | 20 | 209 |
| 29.1890503 | 39.32418544 | 119 | 97 | 323 | 7166 | 284 | -15 | 299 | 52 | 222 | 222 | 26 | 751 | 138 | 16 | 58 | 366 | 55 | 55 | 288 |
| 34.05391517 | 39.32418544 | 109 | 104 | 309 | 8211 | 292 | -45 | 338 | 134 | 225 | 225 | 0 | 361 | 45 | 6 | 43 | 127 | 21 | 21 | 125 |
| 38.91878003 | 39.32418544 | 13 | 94 | 268 | 8875 | 201 | -150 | 351 | -71 | 142 | 142 | -105 | 824 | 115 | 4 | 60 | 330 | 15 | 15 | 277 |
| 43.78364489 | 39.32418544 | 28 | 103 | 280 | 8756 | 211 | -155 | 366 | 51 | 150 | 150 | -91 | 512 | 71 | 11 | 46 | 213 | 35 | 35 | 122 |
| 48.64850976 | 39.32418544 | 170 | 81 | 258 | 7984 | 335 | 22 | 314 | 202 | 281 | 281 | 64 | 426 | 69 | 6 | 47 | 173 | 26 | 26 | 120 |
| 53.51337462 | 39.32418544 | 172 | 63 | 200 | 8930 | 336 | 23 | 313 | 86 | 269 | 300 | 57 | 120 | 18 | 2 | 57 | 49 | 7 | 9 | 37 |
| 58.37823949 | 39.32418544 | 174 | 113 | 267 | 10797 | 395 | -28 | 422 | 159 | 315 | 320 | 26 | 114 | 20 | 1 | 66 | 59 | 4 | 7 | 39 |
| 63.24310435 | 39.32418544 | 166 | 121 | 294 | 10377 | 383 | -30 | 413 | 157 | 299 | 307 | 23 | 124 | 26 | 0 | 85 | 74 | 1 | 3 | 48 |
| 68.10796922 | 39.32418544 | 75 | 96 | 300 | 7651 | 238 | -82 | 320 | 62 | 147 | 177 | -31 | 348 | 60 | 6 | 58 | 172 | 20 | 25 | 91 |
| 72.97283408 | 39.32418544 | -160 | 99 | 295 | 7743 | 7 | -328 | 335 | -133 | -85 | -50 | -268 | 383 | 61 | 15 | 39 | 164 | 50 | 77 | 69 |
| 77.83769895 | 39.32418544 | 127 | 117 | 267 | 11182 | 330 | -110 | 440 | 259 | 70 | 261 | -47 | 47 | 9 | 1 | 81 | 26 | 3 | 25 | 3 |
| 82.70256381 | 39.32418544 | 125 | 132 | 268 | 12491 | 346 | -145 | 491 | 249 | 56 | 273 | -72 | 31 | 10 | 0 | 123 | 26 | 0 | 24 | 2 |
| 87.56742868 | 39.32418544 | 140 | 125 | 263 | 12235 | 364 | -114 | 478 | 292 | -12 | 292 | -48 | 12 | 4 | 0 | 115 | 10 | 0 | 10 | 1 |
| 92.43229354 | 39.32418544 | 70 | 130 | 302 | 10386 | 278 | -154 | 432 | 205 | -58 | 205 | -81 | 11 | 4 | 0 | 144 | 10 | 0 | 10 | 0 |
| 97.29715841 | 39.32418544 | -73 | 124 | 315 | 9047 | 113 | -280 | 393 | 46 | -174 | 46 | -202 | 291 | 78 | 3 | 102 | 209 | 10 | 198 | 11 |
| 102.1620233 | 39.32418544 | 76 | 122 | 266 | 11505 | 295 | -161 | 456 | 221 | -69 | 227 | -90 | 117 | 29 | 1 | 103 | 85 | 3 | 76 | 3 |
| 107.0268881 | 39.32418544 | 81 | 120 | 262 | 11546 | 298 | -162 | 460 | 223 | -89 | 231 | -89 | 210 | 62 | 1 | 109 | 174 | 3 | 122 | 3 |
| 111.891753 | 39.32418544 | 63 | 105 | 243 | 10945 | 262 | -167 | 430 | 197 | -101 | 204 | -101 | 467 | 115 | 3 | 95 | 321 | 11 | 293 | 11 |
| 116.7566179 | 39.32418544 | 129 | 101 | 247 | 11075 | 315 | -93 | 408 | 263 | -33 | 267 | -33 | 549 | 178 | 3 | 127 | 498 | 9 | 423 | 9 |
| 126.4863476 | 39.32418544 | 68 | 85 | 212 | 10698 | 247 | -156 | 403 | 210 | -75 | 210 | -91 | 1359 | 386 | 20 | 102 | 1038 | 60 | 1038 | 65 |
| 141.0809422 | 39.32418544 | 119 | 66 | 213 | 8420 | 278 | -31 | 309 | 240 | 5 | 240 | 5 | 1278 | 176 | 54 | 40 | 524 | 164 | 524 | 164 |
| -121.6217605 | 41.67553679 | 66 | 120 | 359 | 7188 | 265 | -69 | 333 | -13 | 176 | 176 | -24 | 397 | 47 | 10 | 36 | 138 | 42 | 42 | 133 |
| -116.7568956 | 41.67553679 | 89 | 112 | 297 | 8804 | 306 | -73 | 379 | -9 | 223 | 223 | -27 | 313 | 37 | 10 | 33 | 107 | 34 | 34 | 101 |
| -111.8920308 | 41.67553679 | 102 | 114 | 289 | 9335 | 326 | -70 | 396 | 123 | 239 | 239 | -25 | 351 | 38 | 13 | 25 | 113 | 47 | 47 | 102 |
| -107.0271659 | 41.67553679 | 63 | 113 | 295 | 9172 | 278 | -106 | 384 | 82 | -57 | 195 | -57 | 234 | 33 | 11 | 33 | 93 | 33 | 61 | 33 |
| -102.1623011 | 41.67553679 | 94 | 115 | 302 | 9326 | 301 | -81 | 381 | 178 | -28 | 226 | -28 | 462 | 83 | 8 | 67 | 235 | 24 | 169 | 24 |
| -97.29743619 | 41.67553679 | 98 | 100 | 250 | 10405 | 300 | -99 | 399 | 197 | -47 | 235 | -47 | 764 | 128 | 15 | 58 | 369 | 50 | 318 | 50 |
| -92.43257132 | 41.67553679 | 97 | 91 | 231 | 10368 | 289 | -103 | 392 | 198 | -44 | 231 | -52 | 875 | 117 | 27 | 44 | 348 | 83 | 319 | 85 |
| -87.56770646 | 41.67553679 | 100 | 74 | 215 | 9296 | 272 | -74 | 346 | 182 | -26 | 225 | -29 | 899 | 100 | 42 | 26 | 294 | 130 | 281 | 147 |
| -82.70284159 | 41.67553679 | 103 | 68 | 208 | 9151 | 271 | -59 | 330 | 184 | -22 | 227 | -22 | 861 | 91 | 44 | 20 | 266 | 139 | 252 | 139 |
| -77.83797673 | 41.67553679 | 78 | 79 | 240 | 8601 | 244 | -83 | 328 | 157 | -40 | 193 | -40 | 1073 | 116 | 61 | 18 | 339 | 190 | 302 | 190 |
| -72.97311186 | 41.67553679 | 104 | 76 | 231 | 8594 | 271 | -59 | 330 | 28 | -13 | 221 | -13 | 1134 | 106 | 76 | 8 | 304 | 235 | 294 | 235 |
| -4.865003754 | 41.67553679 | 124 | 92 | 335 | 6128 | 283 | 9 | 273 | 95 | 216 | 216 | 46 | 459 | 52 | 15 | 29 | 147 | 48 | 48 | 137 |
| -0.000138889 | 41.67553679 | 153 | 95 | 337 | 6310 | 310 | 27 | 283 | 204 | 247 | 247 | 72 | 445 | 50 | 21 | 23 | 144 | 81 | 81 | 96 |
| 14.59445571 | 41.67553679 | 125 | 71 | 286 | 6101 | 260 | 12 | 248 | 69 | 206 | 215 | 47 | 832 | 115 | 39 | 31 | 325 | 125 | 129 | 200 |
| 24.32418544 | 41.67553679 | 80 | 89 | 290 | 7583 | 240 | -68 | 308 | 3 | 156 | 186 | -21 | 647 | 75 | 38 | 19 | 210 | 116 | 125 | 179 |
| 34.05391517 | 41.67553679 | 75 | 88 | 304 | 7081 | 227 | -60 | 288 | 7 | 166 | 176 | -18 | 783 | 97 | 36 | 29 | 286 | 124 | 124 | 208 |
| 43.78364489 | 41.67553679 | 51 | 89 | 280 | 8002 | 212 | -106 | 318 | 118 | -56 | 161 | -58 | 655 | 93 | 35 | 34 | 276 | 107 | 187 | 111 |
| 48.64850976 | 41.67553679 | 141 | 62 | 216 | 7988 | 290 | 1 | 289 | 174 | 253 | 253 | 35 | 375 | 46 | 20 | 26 | 137 | 66 | 66 | 76 |
| 53.51337462 | 41.67553679 | 154 | 104 | 249 | 11041 | 379 | -38 | 417 | 130 | 256 | 307 | 11 | 120 | 20 | 3 | 47 | 52 | 11 | 12 | 28 |
| 58.37823949 | 41.67553679 | 142 | 113 | 255 | 11690 | 376 | -67 | 443 | 129 | 292 | 299 | -17 | 111 | 20 | 2 | 63 | 58 | 6 | 8 | 31 |
| 63.24310435 | 41.67553679 | 147 | 115 | 264 | 11411 | 375 | -61 | 435 | 136 | 241 | 300 | -10 | 109 | 20 | 1 | 69 | 58 | 3 | 7 | 32 |
| 68.10796922 | 41.67553679 | 149 | 118 | 282 | 10659 | 370 | -49 | 419 | 137 | 241 | 292 | 2 | 233 | 37 | 1 | 67 | 109 | 5 | 17 | 89 |
| 72.97283408 | 41.67553679 | 92 | 94 | 276 | 8520 | 257 | -83 | 340 | 131 | 176 | 203 | -29 | 384 | 55 | 12 | 38 | 156 | 43 | 71 | 76 |
| 77.83769895 | 41.67553679 | -29 | 97 | 282 | 8128 | 128 | -216 | 344 | 71 | -149 | 76 | -149 | 166 | 29 | 3 | 71 | 85 | 10 | 85 | 10 |
| 82.70256381 | 41.67553679 | 108 | 108 | 248 | 11299 | 308 | -127 | 436 | 244 | -26 | 244 | -65 | 63 | 13 | 1 | 74 | 35 | 4 | 35 | 5 |
| 87.56742868 | 41.67553679 | 88 | 120 | 261 | 11667 | 305 | -153 | 458 | 233 | -60 | 234 | -87 | 87 | 22 | 0 | 100 | 60 | 1 | 59 | 3 |
| 92.43229354 | 41.67553679 | 94 | 128 | 262 | 12399 | 327 | -162 | 489 | 222 | 12 | 253 | -92 | 20 | 6 | 0 | 102 | 14 | 1 | 14 | 1 |
| 97.29715841 | 41.67553679 | 56 | 121 | 262 | 11779 | 281 | -182 | 463 | 211 | -92 | 211 | -114 | 75 | 23 | 1 | 115 | 61 | 3 | 61 | 3 |
| 102.1620233 | 41.67553679 | 98 | 123 | 244 | 13213 | 343 | -163 | 506 | 263 | -66 | 271 | -95 | 40 | 11 | 0 | 115 | 33 | 0 | 27 | 0 |
| 107.0268881 | 41.67553679 | 55 | 119 | 245 | 12607 | 292 | -195 | 487 | 213 | -127 | 221 | -127 | 153 | 44 | 1 | 110 | 127 | 3 | 97 | 3 |
| 111.891753 | 41.67553679 | 40 | 108 | 231 | 12284 | 264 | -203 | 467 | 192 | -139 | 199 | -139 | 307 | 79 | 3 | 102 | 233 | 9 | 191 | 9 |
| 116.7566179 | 41.67553679 | 52 | 96 | 220 | 11832 | 256 | -181 | 437 | 198 | -122 | 202 | -122 | 391 | 116 | 1 | 117 | 314 | 3 | 295 | 3 |
| 121.6214827 | 41.67553679 | 85 | 94 | 215 | 12021 | 284 | -151 | 435 | 236 | -71 | 236 | -92 | 565 | 152 | 3 | 104 | 416 | 9 | 416 | 10 |
| 126.4863476 | 41.67553679 | 58 | 85 | 190 | 12392 | 253 | -194 | 448 | 215 | -109 | 215 | -128 | 668 | 165 | 8 | 92 | 461 | 24 | 461 | 28 |
| -121.6217605 | 44.02688814 | 40 | 105 | 365 | 6176 | 213 | -76 | 289 | -38 | 135 | 135 | -38 | 591 | 97 | 13 | 62 | 291 | 41 | 41 | 291 |
| -116.7568956 | 44.02688814 | 120 | 112 | 300 | 8685 | 335 | -38 | 373 | 3 | 251 | 251 | 3 | 333 | 50 | 6 | 51 | 149 | 24 | 24 | 149 |
| -111.8920308 | 44.02688814 | 69 | 110 | 290 | 8937 | 284 | -97 | 380 | 138 | 200 | 200 | -52 | 361 | 46 | 20 | 26 | 137 | 64 | 64 | 98 |
| -107.0271659 | 44.02688814 | 11 | 121 | 315 | 8953 | 227 | -158 | 385 | 86 | -97 | 142 | -107 | 420 | 71 | 15 | 51 | 211 | 47 | 89 | 52 |
| -102.1623011 | 44.02688814 | 93 | 112 | 279 | 10073 | 308 | -94 | 402 | 181 | -41 | 236 | -41 | 421 | 82 | 8 | 70 | 237 | 25 | 150 | 25 |
| -97.29743619 | 44.02688814 | 72 | 97 | 231 | 11164 | 283 | -137 | 420 | 179 | -85 | 218 | -85 | 666 | 112 | 14 | 56 | 307 | 46 | 256 | 46 |
| -92.43257132 | 44.02688814 | 78 | 89 | 219 | 10876 | 278 | -128 | 406 | 218 | -69 | 218 | -77 | 818 | 122 | 21 | 52 | 344 | 65 | 344 | 75 |
| -87.56770646 | 44.02688814 | 80 | 56 | 176 | 9055 | 240 | -80 | 320 | 209 | -41 | 209 | -41 | 769 | 85 | 33 | 28 | 254 | 106 | 254 | 125 |
| -82.70284159 | 44.02688814 | 78 | 47 | 160 | 8628 | 227 | -66 | 293 | 178 | -14 | 202 | -39 | 827 | 87 | 50 | 15 | 252 | 156 | 226 | 159 |
| -77.83797673 | 44.02688814 | 78 | 66 | 195 | 9380 | 246 | -94 | 341 | -1 | -53 | 206 | -53 | 865 | 86 | 58 | 12 | 250 | 182 | 212 | 182 |
| -72.97311186 | 44.02688814 | 56 | 79 | 221 | 9414 | 230 | -128 | 359 | 179 | -76 | 179 | -76 | 1228 | 130 | 71 | 16 | 381 | 223 | 381 | 223 |
| -0.000138889 | 44.02688814 | 135 | 82 | 343 | 5440 | 266 | 27 | 239 | 77 | 206 | 213 | 64 | 837 | 85 | 49 | 15 | 250 | 161 | 165 | 243 |
| 4.864725976 | 44.02688814 | 152 | 73 | 285 | 6216 | 293 | 37 | 256 | 175 | 236 | 243 | 72 | 630 | 87 | 23 | 36 | 256 | 83 | 103 | 138 |
| 19.45932057 | 44.02688814 | 94 | 88 | 293 | 7403 | 250 | -50 | 300 | 166 | -3 | 194 | -6 | 930 | 99 | 56 | 16 | 288 | 175 | 241 | 205 |
| 24.32418544 | 44.02688814 | 122 | 86 | 266 | 8431 | 287 | -37 | 324 | 205 | 11 | 234 | 5 | 543 | 65 | 34 | 23 | 190 | 103 | 153 | 111 |
| 43.78364489 | 44.02688814 | 113 | 90 | 264 | 8874 | 287 | -53 | 339 | 196 | -6 | 232 | -6 | 517 | 81 | 23 | 39 | 227 | 71 | 157 | 71 |
| 48.64850976 | 44.02688814 | 128 | 17 | 67 | 8614 | 261 | 2 | 260 | 222 | 242 | 252 | 17 | 189 | 22 | 12 | 18 | 58 | 39 | 40 | 40 |
| 53.51337462 | 44.02688814 | 112 | 102 | 240 | 11559 | 344 | -82 | 426 | 167 | 214 | 268 | -40 | 170 | 27 | 7 | 40 | 73 | 21 | 35 | 35 |
| 58.37823949 | 44.02688814 | 111 | 84 | 193 | 12268 | 336 | -98 | 434 | 85 | 271 | 274 | -54 | 111 | 18 | 3 | 47 | 51 | 11 | 17 | 22 |
| 63.24310435 | 44.02688814 | 120 | 110 | 240 | 12450 | 359 | -99 | 458 | 109 | 224 | 285 | -50 | 107 | 16 | 3 | 47 | 46 | 9 | 13 | 31 |
| 68.10796922 | 44.02688814 | 79 | 113 | 259 | 11502 | 306 | -129 | 435 | 69 | 177 | 231 | -79 | 174 | 27 | 4 | 50 | 76 | 12 | 24 | 47 |
| 72.97283408 | 44.02688814 | 117 | 114 | 252 | 11892 | 344 | -108 | 452 | 106 | 268 | 271 | -50 | 262 | 32 | 7 | 40 | 94 | 24 | 37 | 72 |
| 77.83769895 | 44.02688814 | 78 | 116 | 280 | 10267 | 284 | -130 | 414 | 129 | -61 | 211 | -68 | 284 | 45 | 11 | 50 | 131 | 33 | 62 | 38 |
| 82.70256381 | 44.02688814 | -17 | 117 | 279 | 10005 | 181 | -237 | 419 | 80 | -152 | 110 | -163 | 207 | 29 | 8 | 41 | 86 | 25 | 71 | 30 |
| 87.56742868 | 44.02688814 | 95 | 110 | 243 | 11958 | 308 | -145 | 454 | 216 | -66 | 244 | -79 | 189 | 25 | 7 | 35 | 72 | 23 | 67 | 24 |
| 92.43229354 | 44.02688814 | 14 | 120 | 244 | 12743 | 245 | -247 | 491 | 175 | -155 | 175 | -175 | 153 | 34 | 4 | 74 | 95 | 13 | 95 | 17 |
| 97.29715841 | 44.02688814 | 71 | 122 | 247 | 12952 | 314 | -182 | 496 | 242 | -91 | 242 | -114 | 29 | 10 | 0 | 115 | 24 | 1 | 24 | 3 |
| 102.1620233 | 44.02688814 | 59 | 119 | 237 | 13260 | 304 | -199 | 503 | 224 | -132 | 233 | -132 | 141 | 35 | 2 | 93 | 99 | 6 | 88 | 6 |
| 107.0268881 | 44.02688814 | 42 | 117 | 230 | 13512 | 288 | -220 | 508 | 212 | -153 | 218 | -153 | 116 | 31 | 1 | 110 | 93 | 3 | 80 | 3 |
| 111.891753 | 44.02688814 | 42 | 113 | 216 | 14144 | 294 | -230 | 524 | 219 | -141 | 224 | -164 | 133 | 36 | 1 | 109 | 106 | 3 | 91 | 3 |
| 116.7566179 | 44.02688814 | 20 | 102 | 206 | 13629 | 255 | -237 | 492 | 189 | -162 | 194 | -177 | 346 | 104 | 3 | 113 | 266 | 9 | 263 | 9 |
| 121.6214827 | 44.02688814 | 75 | 97 | 207 | 13227 | 294 | -174 | 467 | 240 | -119 | 240 | -119 | 349 | 111 | 1 | 118 | 284 | 4 | 284 | 4 |
| 126.4863476 | 44.02688814 | 57 | 87 | 180 | 13661 | 272 | -209 | 481 | 226 | -148 | 226 | -148 | 642 | 184 | 5 | 104 | 470 | 16 | 470 | 16 |
| 131.3512125 | 44.02688814 | 55 | 85 | 191 | 12433 | 254 | -189 | 443 | 215 | -132 | 215 | -132 | 617 | 127 | 7 | 77 | 350 | 26 | 350 | 26 |
| -121.6217605 | 46.37823949 | 44 | 92 | 350 | 5930 | 204 | -60 | 264 | -32 | 132 | 132 | -32 | 2052 | 363 | 24 | 70 | 1065 | 96 | 96 | 1065 |
| -116.7568956 | 46.37823949 | 108 | 103 | 316 | 7603 | 299 | -27 | 326 | 125 | 224 | 224 | 8 | 521 | 59 | 23 | 27 | 170 | 69 | 69 | 146 |
| -111.8920308 | 46.37823949 | 17 | 110 | 315 | 8042 | 216 | -134 | 350 | 81 | -78 | 137 | -88 | 477 | 89 | 16 | 53 | 247 | 53 | 142 | 61 |
| -107.0271659 | 46.37823949 | 91 | 111 | 270 | 10176 | 315 | -96 | 410 | 176 | -35 | 239 | -45 | 315 | 60 | 9 | 58 | 171 | 29 | 84 | 34 |
| -102.1623011 | 46.37823949 | 68 | 103 | 247 | 10800 | 288 | -130 | 418 | 166 | -80 | 217 | -80 | 425 | 86 | 10 | 68 | 238 | 30 | 152 | 30 |
| -97.29743619 | 46.37823949 | 63 | 95 | 217 | 11807 | 280 | -159 | 439 | 214 | -96 | 215 | -104 | 568 | 101 | 13 | 62 | 277 | 43 | 226 | 43 |
| -92.43257132 | 46.37823949 | 56 | 93 | 220 | 11201 | 261 | -161 | 422 | 199 | -96 | 199 | -105 | 827 | 120 | 23 | 48 | 356 | 76 | 356 | 89 |
| -87.56770646 | 46.37823949 | 48 | 68 | 206 | 9201 | 213 | -118 | 331 | 146 | -80 | 173 | -80 | 799 | 92 | 36 | 25 | 268 | 116 | 247 | 116 |
| -82.70284159 | 46.37823949 | 42 | 70 | 195 | 9876 | 216 | -146 | 362 | 86 | -98 | 174 | -98 | 1013 | 113 | 57 | 18 | 328 | 184 | 263 | 184 |
| -77.83797673 | 46.37823949 | 42 | 86 | 211 | 10773 | 236 | -169 | 405 | 143 | -66 | 179 | -111 | 890 | 97 | 50 | 22 | 286 | 150 | 275 | 156 |
| -72.97311186 | 46.37823949 | 61 | 81 | 200 | 10854 | 253 | -151 | 404 | 163 | -92 | 200 | -92 | 866 | 84 | 48 | 15 | 250 | 156 | 242 | 156 |
| -68.108247 | 46.37823949 | 47 | 78 | 211 | 9786 | 225 | -147 | 371 | 175 | -50 | 175 | -91 | 1062 | 106 | 62 | 15 | 316 | 202 | 316 | 205 |
| -63.24338213 | 46.37823949 | 57 | 44 | 147 | 8615 | 202 | -98 | 300 | -2 | 163 | 180 | -65 | 1159 | 125 | 79 | 16 | 368 | 237 | 249 | 281 |
| -0.000138889 | 46.37823949 | 119 | 77 | 336 | 5333 | 246 | 18 | 228 | 64 | 190 | 196 | 51 | 841 | 100 | 55 | 20 | 291 | 166 | 167 | 260 |
| 4.864725976 | 46.37823949 | 124 | 75 | 294 | 6234 | 265 | 9 | 255 | 95 | 71 | 212 | 42 | 811 | 86 | 54 | 15 | 255 | 162 | 201 | 199 |
| 9.729590841 | 46.37823949 | -29 | 80 | 300 | 6372 | 111 | -154 | 265 | 61 | -109 | 61 | -113 | 1270 | 154 | 54 | 34 | 456 | 171 | 456 | 190 |
| 14.59445571 | 46.37823949 | 32 | 84 | 289 | 7085 | 182 | -108 | 290 | 129 | -59 | 129 | -64 | 1583 | 181 | 82 | 24 | 535 | 252 | 535 | 268 |
| 19.45932057 | 46.37823949 | 110 | 87 | 277 | 8006 | 273 | -39 | 312 | 189 | 42 | 215 | -1 | 557 | 75 | 32 | 26 | 206 | 97 | 159 | 115 |
| 24.32418544 | 46.37823949 | 100 | 80 | 264 | 7926 | 255 | -47 | 302 | 202 | -4 | 204 | -9 | 542 | 86 | 19 | 49 | 250 | 60 | 190 | 73 |
| 29.1890503 | 46.37823949 | 100 | 80 | 250 | 8666 | 268 | -53 | 321 | 186 | 18 | 216 | -15 | 544 | 72 | 28 | 29 | 202 | 96 | 166 | 110 |
| 34.05391517 | 46.37823949 | 109 | 84 | 252 | 8898 | 289 | -46 | 334 | 196 | 21 | 231 | -9 | 401 | 47 | 26 | 19 | 135 | 81 | 105 | 88 |
| 38.91878003 | 46.37823949 | 110 | 81 | 237 | 9279 | 292 | -50 | 342 | 200 | 20 | 237 | -13 | 588 | 65 | 37 | 19 | 185 | 118 | 149 | 133 |
| 43.78364489 | 46.37823949 | 96 | 88 | 232 | 10305 | 302 | -79 | 381 | 197 | -4 | 236 | -42 | 414 | 53 | 23 | 23 | 149 | 75 | 108 | 81 |
| 48.64850976 | 46.37823949 | 109 | 78 | 200 | 11025 | 318 | -74 | 392 | 221 | -39 | 257 | -39 | 185 | 20 | 10 | 16 | 58 | 33 | 56 | 33 |
| 53.51337462 | 46.37823949 | 107 | 92 | 206 | 12584 | 343 | -104 | 446 | 236 | 217 | 275 | -64 | 156 | 18 | 7 | 28 | 53 | 25 | 48 | 29 |
| 58.37823949 | 46.37823949 | 84 | 95 | 208 | 12975 | 326 | -129 | 455 | 152 | 196 | 256 | -92 | 149 | 18 | 6 | 28 | 52 | 24 | 39 | 27 |
| 63.24310435 | 46.37823949 | 90 | 105 | 225 | 13124 | 334 | -135 | 468 | 76 | 254 | 261 | -90 | 135 | 18 | 4 | 32 | 50 | 16 | 23 | 31 |
| 68.10796922 | 46.37823949 | 85 | 109 | 234 | 12948 | 326 | -137 | 463 | 157 | 196 | 253 | -91 | 137 | 18 | 4 | 33 | 53 | 15 | 28 | 38 |
| 72.97283408 | 46.37823949 | 70 | 100 | 216 | 13042 | 304 | -158 | 462 | 53 | 183 | 237 | -109 | 162 | 20 | 4 | 29 | 54 | 21 | 49 | 39 |
| 77.83769895 | 46.37823949 | 85 | 109 | 226 | 13299 | 323 | -160 | 483 | 253 | 200 | 253 | -101 | 163 | 20 | 8 | 27 | 56 | 25 | 56 | 37 |
| 82.70256381 | 46.37823949 | 76 | 107 | 236 | 12294 | 299 | -154 | 453 | 23 | 186 | 230 | -99 | 278 | 32 | 12 | 26 | 88 | 41 | 84 | 62 |
| 87.56742868 | 46.37823949 | 78 | 120 | 224 | 14647 | 330 | -206 | 536 | 259 | -121 | 259 | -136 | 79 | 14 | 2 | 46 | 36 | 8 | 36 | 14 |
| 92.43229354 | 46.37823949 | -55 | 115 | 248 | 12129 | 169 | -295 | 464 | 102 | -213 | 102 | -227 | 119 | 36 | 1 | 109 | 92 | 3 | 92 | 4 |
| 97.29715841 | 46.37823949 | 0 | 115 | 238 | 12888 | 235 | -252 | 486 | 169 | -164 | 169 | -181 | 96 | 27 | 1 | 103 | 69 | 3 | 69 | 3 |
| 102.1620233 | 46.37823949 | -39 | 118 | 247 | 12091 | 187 | -292 | 479 | 119 | -211 | 119 | -211 | 351 | 108 | 2 | 116 | 285 | 7 | 285 | 7 |
| 107.0268881 | 46.37823949 | 8 | 111 | 220 | 13538 | 253 | -252 | 505 | 177 | -186 | 185 | -186 | 213 | 65 | 1 | 127 | 187 | 4 | 167 | 4 |
| 111.891753 | 46.37823949 | 19 | 109 | 201 | 14765 | 271 | -270 | 541 | 207 | -176 | 207 | -197 | 216 | 66 | 2 | 114 | 170 | 6 | 170 | 6 |
| 116.7566179 | 46.37823949 | 18 | 101 | 196 | 14497 | 265 | -250 | 515 | 199 | -173 | 203 | -190 | 232 | 67 | 2 | 112 | 177 | 6 | 173 | 7 |
| 121.6214827 | 46.37823949 | 36 | 95 | 195 | 13660 | 261 | -224 | 485 | 208 | -144 | 208 | -164 | 458 | 154 | 2 | 123 | 391 | 6 | 391 | 6 |
| 126.4863476 | 46.37823949 | 40 | 89 | 172 | 14946 | 274 | -248 | 522 | 225 | -160 | 225 | -186 | 523 | 157 | 3 | 111 | 396 | 11 | 396 | 11 |
| 131.3512125 | 46.37823949 | 24 | 84 | 173 | 14022 | 249 | -239 | 488 | 199 | -162 | 199 | -186 | 534 | 126 | 4 | 98 | 370 | 12 | 370 | 14 |
| 136.2160773 | 46.37823949 | -6 | 81 | 172 | 13398 | 211 | -258 | 468 | 167 | -185 | 167 | -205 | 954 | 180 | 15 | 68 | 511 | 47 | 511 | 60 |
| -121.6217605 | 48.72959084 | 105 | 84 | 323 | 6095 | 258 | -1 | 260 | 33 | 196 | 196 | 25 | 1424 | 217 | 37 | 54 | 644 | 111 | 111 | 627 |
| -116.7568956 | 48.72959084 | 38 | 97 | 297 | 7962 | 226 | -102 | 328 | -56 | 154 | 154 | -68 | 808 | 100 | 38 | 31 | 299 | 120 | 120 | 290 |
| -111.8920308 | 48.72959084 | 68 | 104 | 284 | 8951 | 268 | -100 | 368 | 144 | -42 | 196 | -53 | 335 | 78 | 9 | 74 | 210 | 28 | 118 | 30 |
| -107.0271659 | 48.72959084 | 62 | 106 | 252 | 10745 | 280 | -139 | 419 | 158 | -56 | 210 | -87 | 287 | 59 | 9 | 67 | 164 | 29 | 98 | 34 |
| -102.1623011 | 48.72959084 | 43 | 99 | 229 | 11566 | 261 | -171 | 431 | 152 | -82 | 194 | -120 | 424 | 81 | 11 | 64 | 218 | 34 | 164 | 41 |
| -97.29743619 | 48.72959084 | 42 | 92 | 205 | 12346 | 261 | -189 | 450 | 198 | -125 | 199 | -134 | 494 | 85 | 12 | 60 | 235 | 39 | 201 | 45 |
| -92.43257132 | 48.72959084 | 36 | 92 | 209 | 11823 | 249 | -192 | 441 | 186 | -125 | 187 | -136 | 705 | 102 | 22 | 48 | 305 | 75 | 278 | 94 |
| -87.56770646 | 48.72959084 | 28 | 44 | 145 | 8493 | 173 | -133 | 306 | 133 | -59 | 144 | -95 | 785 | 86 | 40 | 21 | 252 | 128 | 230 | 140 |
| -82.70284159 | 48.72959084 | 22 | 89 | 207 | 11547 | 227 | -204 | 431 | 130 | -142 | 168 | -142 | 777 | 88 | 41 | 24 | 259 | 131 | 251 | 153 |
| -77.83797673 | 48.72959084 | 17 | 88 | 201 | 11699 | 224 | -211 | 435 | 125 | -150 | 166 | -150 | 931 | 114 | 39 | 33 | 335 | 133 | 334 | 133 |
| -72.97311186 | 48.72959084 | 11 | 83 | 198 | 11244 | 208 | -210 | 417 | 154 | -99 | 154 | -150 | 955 | 120 | 55 | 24 | 330 | 173 | 330 | 175 |
| -58.37851727 | 48.72959084 | 33 | 60 | 199 | 8135 | 182 | -120 | 302 | -26 | -9 | 149 | -83 | 1113 | 116 | 60 | 18 | 343 | 197 | 276 | 287 |
| -0.000138889 | 48.72959084 | 108 | 64 | 319 | 4859 | 219 | 17 | 202 | 60 | 86 | 178 | 46 | 713 | 78 | 47 | 16 | 228 | 149 | 157 | 173 |
| 4.864725976 | 48.72959084 | 108 | 72 | 302 | 5893 | 241 | 2 | 240 | 48 | 89 | 192 | 31 | 893 | 95 | 58 | 11 | 268 | 183 | 228 | 253 |
| 9.729590841 | 48.72959084 | 94 | 74 | 284 | 6601 | 233 | -28 | 261 | 157 | 37 | 185 | 7 | 990 | 105 | 70 | 12 | 298 | 216 | 276 | 240 |
| 14.59445571 | 48.72959084 | 79 | 76 | 275 | 7082 | 223 | -53 | 276 | 170 | -14 | 174 | -17 | 716 | 100 | 38 | 38 | 296 | 116 | 278 | 124 |
| 19.45932057 | 48.72959084 | 44 | 75 | 263 | 7437 | 194 | -93 | 286 | 118 | -53 | 144 | -57 | 1034 | 131 | 60 | 24 | 363 | 184 | 315 | 202 |
| 24.32418544 | 48.72959084 | 68 | 76 | 256 | 7938 | 219 | -77 | 297 | 170 | -39 | 171 | -40 | 897 | 123 | 38 | 40 | 368 | 124 | 317 | 132 |
| 29.1890503 | 48.72959084 | 87 | 75 | 240 | 8721 | 253 | -61 | 314 | 200 | 5 | 201 | -29 | 568 | 81 | 32 | 34 | 241 | 100 | 195 | 108 |
| 34.05391517 | 48.72959084 | 85 | 80 | 235 | 9365 | 263 | -75 | 338 | 207 | -6 | 207 | -41 | 551 | 70 | 32 | 23 | 192 | 100 | 167 | 113 |
| 38.91878003 | 48.72959084 | 89 | 82 | 229 | 9910 | 278 | -79 | 357 | 218 | -8 | 218 | -45 | 499 | 59 | 31 | 21 | 175 | 96 | 175 | 107 |
| 43.78364489 | 48.72959084 | 86 | 86 | 219 | 10961 | 295 | -99 | 394 | -23 | 63 | 230 | -62 | 429 | 48 | 29 | 17 | 137 | 87 | 119 | 98 |
| 48.64850976 | 48.72959084 | 86 | 91 | 211 | 12027 | 314 | -116 | 430 | 29 | -77 | 246 | -77 | 219 | 24 | 12 | 18 | 67 | 40 | 62 | 40 |
| 53.51337462 | 48.72959084 | 83 | 94 | 207 | 12920 | 324 | -133 | 457 | -46 | 194 | 254 | -93 | 184 | 23 | 10 | 27 | 68 | 32 | 40 | 44 |
| 58.37823949 | 48.72959084 | 59 | 94 | 205 | 13090 | 299 | -158 | 457 | 3 | 172 | 230 | -119 | 272 | 28 | 16 | 17 | 84 | 51 | 64 | 65 |
| 63.24310435 | 48.72959084 | 68 | 101 | 211 | 13750 | 315 | -164 | 479 | 10 | -120 | 247 | -120 | 157 | 18 | 9 | 20 | 50 | 29 | 41 | 29 |
| 68.10796922 | 48.72959084 | 44 | 101 | 215 | 13267 | 282 | -185 | 467 | 119 | 158 | 215 | -138 | 188 | 23 | 9 | 27 | 64 | 29 | 60 | 36 |
| 72.97283408 | 48.72959084 | 35 | 100 | 229 | 12294 | 259 | -180 | 439 | 194 | -84 | 194 | -132 | 279 | 40 | 16 | 31 | 108 | 48 | 108 | 50 |
| 77.83769895 | 48.72959084 | 23 | 105 | 234 | 12363 | 248 | -200 | 448 | 183 | -146 | 183 | -146 | 212 | 39 | 8 | 52 | 106 | 25 | 106 | 25 |
| 82.70256381 | 48.72959084 | 45 | 106 | 230 | 12809 | 274 | -187 | 461 | 208 | -74 | 208 | -132 | 281 | 33 | 14 | 25 | 94 | 47 | 94 | 73 |
| 87.56742868 | 48.72959084 | -43 | 101 | 243 | 11029 | 164 | -251 | 415 | 101 | -140 | 101 | -196 | 361 | 66 | 11 | 56 | 183 | 34 | 183 | 47 |
| 92.43229354 | 48.72959084 | 2 | 114 | 218 | 14326 | 241 | -282 | 523 | 181 | -191 | 181 | -208 | 111 | 37 | 1 | 120 | 94 | 3 | 94 | 4 |
| 97.29715841 | 48.72959084 | -29 | 107 | 226 | 12577 | 198 | -273 | 470 | 136 | -192 | 136 | -204 | 184 | 58 | 2 | 111 | 146 | 7 | 146 | 10 |
| 102.1620233 | 48.72959084 | -9 | 105 | 224 | 12347 | 210 | -261 | 471 | 142 | -189 | 152 | -189 | 352 | 114 | 1 | 127 | 294 | 4 | 292 | 4 |
| 107.0268881 | 48.72959084 | -19 | 101 | 202 | 13577 | 215 | -284 | 499 | 150 | -196 | 157 | -216 | 358 | 96 | 3 | 110 | 268 | 10 | 258 | 12 |
| 111.891753 | 48.72959084 | -2 | 103 | 198 | 14152 | 239 | -279 | 518 | 172 | -187 | 181 | -208 | 355 | 109 | 2 | 118 | 285 | 6 | 273 | 7 |
| 116.7566179 | 48.72959084 | 10 | 101 | 185 | 15529 | 268 | -280 | 548 | 200 | -196 | 208 | -215 | 280 | 93 | 1 | 132 | 253 | 3 | 228 | 4 |
| 121.6214827 | 48.72959084 | -16 | 87 | 175 | 14319 | 219 | -280 | 499 | 161 | -202 | 167 | -224 | 505 | 133 | 3 | 109 | 383 | 9 | 346 | 11 |
| 126.4863476 | 48.72959084 | 14 | 87 | 164 | 15397 | 256 | -275 | 532 | 207 | -187 | 207 | -214 | 503 | 147 | 3 | 111 | 381 | 9 | 381 | 11 |
| 131.3512125 | 48.72959084 | 5 | 84 | 161 | 15229 | 245 | -275 | 520 | 192 | -193 | 195 | -222 | 637 | 138 | 7 | 82 | 395 | 23 | 326 | 32 |
| 136.2160773 | 48.72959084 | 22 | 78 | 155 | 14735 | 256 | -246 | 503 | 205 | -174 | 205 | -196 | 709 | 148 | 13 | 73 | 427 | 41 | 344 | 52 |
| -126.4866254 | 51.08094219 | 74 | 73 | 319 | 5591 | 207 | -24 | 230 | 45 | 146 | 156 | 2 | 1508 | 201 | 65 | 38 | 599 | 213 | 227 | 521 |
| -121.6217605 | 51.08094219 | 57 | 93 | 288 | 7929 | 233 | -90 | 323 | 160 | -2 | 168 | -52 | 284 | 39 | 13 | 33 | 109 | 43 | 91 | 65 |
| -116.7568956 | 51.08094219 | 24 | 84 | 259 | 8441 | 200 | -126 | 326 | -94 | 10 | 141 | -94 | 604 | 80 | 29 | 31 | 232 | 91 | 138 | 232 |
| -111.8920308 | 51.08094219 | 52 | 103 | 269 | 9970 | 249 | -135 | 384 | 146 | -76 | 182 | -87 | 340 | 64 | 14 | 56 | 167 | 44 | 131 | 47 |
| -107.0271659 | 51.08094219 | 37 | 100 | 235 | 11434 | 250 | -174 | 424 | 180 | -114 | 184 | -124 | 344 | 66 | 11 | 60 | 178 | 38 | 132 | 49 |
| -102.1623011 | 51.08094219 | 20 | 92 | 210 | 12024 | 235 | -201 | 436 | 136 | -141 | 173 | -150 | 489 | 85 | 17 | 52 | 223 | 58 | 199 | 75 |
| -97.29743619 | 51.08094219 | 25 | 89 | 195 | 12667 | 245 | -210 | 454 | 144 | -148 | 186 | -157 | 534 | 82 | 18 | 47 | 219 | 60 | 217 | 73 |
| -92.43257132 | 51.08094219 | 9 | 88 | 195 | 12454 | 226 | -227 | 453 | 166 | -162 | 168 | -173 | 608 | 90 | 22 | 47 | 262 | 71 | 254 | 83 |
| -87.56770646 | 51.08094219 | 2 | 89 | 196 | 12417 | 220 | -234 | 454 | 160 | -171 | 161 | -177 | 759 | 101 | 29 | 38 | 292 | 96 | 287 | 118 |
| -82.70284159 | 51.08094219 | 6 | 92 | 198 | 12606 | 232 | -230 | 463 | 167 | -115 | 167 | -173 | 721 | 95 | 32 | 34 | 265 | 104 | 265 | 132 |
| -77.83797673 | 51.08094219 | -2 | 87 | 192 | 12358 | 215 | -237 | 452 | 113 | -127 | 153 | -182 | 821 | 108 | 30 | 40 | 311 | 102 | 298 | 104 |
| -72.97311186 | 51.08094219 | -5 | 79 | 180 | 12051 | 201 | -237 | 437 | 146 | -39 | 146 | -180 | 1042 | 136 | 53 | 33 | 398 | 168 | 382 | 179 |
| -68.108247 | 51.08094219 | -11 | 76 | 184 | 11311 | 187 | -228 | 416 | 132 | -173 | 135 | -174 | 1072 | 124 | 62 | 19 | 349 | 194 | 320 | 222 |
| -63.24338213 | 51.08094219 | 1 | 71 | 182 | 10631 | 185 | -205 | 390 | 141 | -153 | 141 | -153 | 894 | 94 | 42 | 23 | 280 | 140 | 280 | 140 |
| -0.000138889 | 51.08094219 | 99 | 53 | 299 | 4342 | 197 | 20 | 176 | 88 | 155 | 163 | 44 | 863 | 95 | 56 | 20 | 281 | 174 | 188 | 204 |
| 4.864725976 | 51.08094219 | 106 | 67 | 305 | 5400 | 227 | 6 | 221 | 182 | 86 | 182 | 34 | 781 | 74 | 47 | 13 | 220 | 154 | 220 | 175 |
| 9.729590841 | 51.08094219 | 82 | 68 | 279 | 6239 | 215 | -30 | 245 | 164 | 0 | 169 | 0 | 797 | 83 | 49 | 15 | 247 | 159 | 232 | 159 |
| 14.59445571 | 51.08094219 | 82 | 71 | 271 | 6891 | 223 | -40 | 263 | 176 | -8 | 176 | -9 | 643 | 71 | 39 | 19 | 210 | 123 | 210 | 146 |
| 19.45932057 | 51.08094219 | 84 | 72 | 263 | 7298 | 231 | -43 | 275 | 179 | -13 | 182 | -13 | 632 | 91 | 31 | 37 | 262 | 96 | 243 | 110 |
| 24.32418544 | 51.08094219 | 81 | 73 | 251 | 7950 | 236 | -55 | 290 | 185 | 4 | 185 | -26 | 575 | 79 | 28 | 34 | 231 | 87 | 201 | 93 |
| 29.1890503 | 51.08094219 | 75 | 75 | 239 | 8676 | 241 | -71 | 312 | 189 | -9 | 189 | -41 | 642 | 96 | 33 | 37 | 279 | 101 | 279 | 108 |
| 34.05391517 | 51.08094219 | 72 | 74 | 224 | 9392 | 247 | -85 | 332 | 195 | -54 | 195 | -54 | 620 | 86 | 33 | 30 | 248 | 107 | 248 | 107 |
| 38.91878003 | 51.08094219 | 72 | 77 | 217 | 10067 | 259 | -96 | 356 | 204 | -28 | 204 | -63 | 597 | 66 | 34 | 20 | 192 | 105 | 192 | 118 |
| 43.78364489 | 51.08094219 | 67 | 80 | 207 | 10976 | 270 | -117 | 387 | 210 | -46 | 210 | -82 | 495 | 53 | 28 | 20 | 159 | 86 | 159 | 97 |
| 48.64850976 | 51.08094219 | 68 | 85 | 201 | 12064 | 289 | -134 | 423 | 226 | -55 | 226 | -96 | 329 | 39 | 18 | 23 | 115 | 55 | 115 | 65 |
| 53.51337462 | 51.08094219 | 59 | 89 | 201 | 12604 | 289 | -152 | 441 | 223 | -68 | 223 | -114 | 264 | 29 | 15 | 19 | 86 | 45 | 86 | 49 |
| 58.37823949 | 51.08094219 | 49 | 90 | 204 | 12805 | 278 | -165 | 443 | 215 | -125 | 215 | -125 | 322 | 39 | 18 | 25 | 116 | 56 | 116 | 56 |
| 63.24310435 | 51.08094219 | 39 | 96 | 206 | 13468 | 277 | -189 | 466 | 213 | -96 | 213 | -145 | 265 | 32 | 14 | 26 | 94 | 42 | 94 | 47 |
| 68.10796922 | 51.08094219 | 36 | 95 | 206 | 13344 | 270 | -191 | 462 | 208 | -98 | 208 | -146 | 262 | 37 | 14 | 29 | 103 | 43 | 103 | 48 |
| 72.97283408 | 51.08094219 | 17 | 95 | 217 | 12540 | 239 | -198 | 437 | 178 | -106 | 178 | -154 | 363 | 61 | 15 | 43 | 164 | 48 | 164 | 59 |
| 77.83769895 | 51.08094219 | 44 | 100 | 212 | 13404 | 278 | -193 | 471 | 216 | -86 | 216 | -142 | 238 | 39 | 10 | 44 | 110 | 31 | 110 | 34 |
| 82.70256381 | 51.08094219 | 30 | 96 | 227 | 11813 | 245 | -180 | 425 | 185 | -128 | 185 | -130 | 544 | 65 | 27 | 27 | 172 | 84 | 172 | 99 |
| 87.56742868 | 51.08094219 | -32 | 100 | 248 | 10494 | 164 | -240 | 404 | 107 | -169 | 107 | -181 | 495 | 103 | 9 | 79 | 292 | 27 | 292 | 32 |
| 92.43229354 | 51.08094219 | -96 | 101 | 217 | 11643 | 125 | -341 | 466 | 68 | -266 | 68 | -281 | 367 | 85 | 7 | 81 | 230 | 22 | 230 | 33 |
| 97.29715841 | 51.08094219 | -46 | 95 | 208 | 12483 | 173 | -284 | 457 | 118 | -151 | 118 | -222 | 369 | 95 | 6 | 90 | 243 | 19 | 243 | 36 |
| 102.1620233 | 51.08094219 | -78 | 100 | 217 | 12461 | 136 | -325 | 461 | 79 | -238 | 84 | -257 | 561 | 151 | 8 | 105 | 409 | 25 | 399 | 30 |
| 107.0268881 | 51.08094219 | 14 | 97 | 187 | 14280 | 259 | -260 | 520 | 189 | -101 | 200 | -193 | 211 | 60 | 2 | 112 | 162 | 6 | 157 | 10 |
| 111.891753 | 51.08094219 | 0 | 90 | 184 | 13608 | 236 | -252 | 488 | 170 | -177 | 180 | -196 | 321 | 94 | 3 | 113 | 248 | 9 | 240 | 10 |
| 116.7566179 | 51.08094219 | -30 | 97 | 177 | 15376 | 222 | -326 | 548 | 158 | -230 | 167 | -256 | 357 | 100 | 3 | 109 | 270 | 11 | 254 | 12 |
| 121.6214827 | 51.08094219 | -33 | 90 | 166 | 15683 | 218 | -325 | 544 | 159 | -236 | 167 | -263 | 488 | 133 | 4 | 108 | 363 | 12 | 347 | 15 |
| 126.4863476 | 51.08094219 | -6 | 87 | 163 | 15678 | 244 | -288 | 532 | 184 | -207 | 193 | -234 | 511 | 130 | 3 | 102 | 360 | 10 | 345 | 15 |
| 131.3512125 | 51.08094219 | -24 | 88 | 153 | 16672 | 236 | -337 | 572 | 177 | -238 | 183 | -276 | 678 | 151 | 5 | 93 | 449 | 16 | 407 | 22 |
| 136.2160773 | 51.08094219 | -4 | 84 | 159 | 15426 | 242 | -287 | 529 | 187 | -206 | 188 | -232 | 672 | 116 | 17 | 59 | 333 | 52 | 280 | 61 |
| -126.4866254 | 53.43229354 | 0 | 78 | 292 | 6606 | 146 | -120 | 266 | -74 | -21 | 93 | -88 | 865 | 109 | 38 | 34 | 322 | 125 | 185 | 294 |
| -121.6217605 | 53.43229354 | 29 | 86 | 269 | 8233 | 200 | -121 | 321 | -86 | 15 | 140 | -86 | 866 | 95 | 46 | 19 | 277 | 150 | 218 | 277 |
| -116.7568956 | 53.43229354 | 33 | 96 | 276 | 8870 | 214 | -133 | 347 | 147 | -75 | 150 | -91 | 630 | 117 | 20 | 62 | 334 | 71 | 299 | 88 |
| -111.8920308 | 53.43229354 | 24 | 94 | 236 | 10875 | 222 | -174 | 396 | 160 | -117 | 161 | -129 | 445 | 84 | 14 | 65 | 245 | 49 | 216 | 63 |
| -107.0271659 | 53.43229354 | 19 | 91 | 213 | 11903 | 228 | -198 | 427 | 167 | -138 | 168 | -151 | 401 | 69 | 15 | 56 | 203 | 49 | 175 | 58 |
| -102.1623011 | 53.43229354 | 1 | 90 | 202 | 12429 | 220 | -226 | 445 | 158 | -165 | 159 | -178 | 425 | 70 | 15 | 52 | 194 | 45 | 168 | 48 |
| -97.29743619 | 53.43229354 | 1 | 79 | 168 | 13544 | 230 | -238 | 469 | 175 | -183 | 175 | -192 | 465 | 72 | 17 | 50 | 210 | 52 | 210 | 60 |
| -92.43257132 | 53.43229354 | -7 | 88 | 186 | 13287 | 225 | -249 | 474 | 162 | -190 | 164 | -199 | 544 | 81 | 18 | 48 | 235 | 59 | 233 | 73 |
| -87.56770646 | 53.43229354 | -15 | 88 | 184 | 13266 | 221 | -254 | 476 | 157 | -200 | 157 | -203 | 602 | 90 | 18 | 50 | 270 | 60 | 270 | 75 |
| -82.70284159 | 53.43229354 | -17 | 83 | 184 | 12593 | 204 | -246 | 450 | 145 | -146 | 145 | -200 | 562 | 88 | 20 | 50 | 255 | 64 | 255 | 65 |
| -77.83797673 | 53.43229354 | -20 | 78 | 175 | 12436 | 192 | -249 | 442 | 99 | -207 | 135 | -207 | 685 | 104 | 22 | 47 | 294 | 75 | 251 | 75 |
| -72.97311186 | 53.43229354 | -25 | 74 | 167 | 12488 | 186 | -255 | 442 | 93 | -208 | 133 | -208 | 756 | 112 | 25 | 46 | 329 | 89 | 313 | 89 |
| -68.108247 | 53.43229354 | -34 | 68 | 157 | 12374 | 175 | -260 | 435 | 123 | -213 | 126 | -213 | 796 | 104 | 39 | 30 | 286 | 127 | 282 | 127 |
| -63.24338213 | 53.43229354 | -15 | 71 | 168 | 11919 | 189 | -235 | 424 | 134 | -187 | 141 | -187 | 833 | 97 | 49 | 21 | 272 | 156 | 267 | 156 |
| -58.37851727 | 53.43229354 | -13 | 72 | 190 | 10332 | 171 | -207 | 378 | 125 | -163 | 125 | -163 | 978 | 110 | 62 | 18 | 314 | 199 | 314 | 199 |
| -9.729868619 | 53.43229354 | 105 | 26 | 240 | 2911 | 162 | 55 | 107 | 80 | 103 | 148 | 69 | 1616 | 181 | 90 | 25 | 537 | 278 | 350 | 445 |
| -0.000138889 | 53.43229354 | 100 | 43 | 264 | 4213 | 190 | 28 | 162 | 89 | 47 | 163 | 47 | 584 | 57 | 39 | 9 | 163 | 129 | 151 | 129 |
| 9.729590841 | 53.43229354 | 92 | 64 | 276 | 5940 | 219 | -12 | 231 | 176 | 70 | 176 | 14 | 741 | 78 | 42 | 17 | 230 | 142 | 230 | 146 |
| 14.59445571 | 53.43229354 | 90 | 64 | 256 | 6652 | 226 | -22 | 249 | 178 | 3 | 183 | 3 | 512 | 62 | 28 | 23 | 180 | 91 | 170 | 91 |
| 19.45932057 | 53.43229354 | 79 | 68 | 254 | 7235 | 225 | -44 | 269 | 174 | -17 | 177 | -17 | 577 | 82 | 25 | 38 | 240 | 81 | 220 | 81 |
| 24.32418544 | 53.43229354 | 74 | 69 | 238 | 7962 | 230 | -61 | 290 | 179 | -33 | 179 | -33 | 596 | 79 | 30 | 29 | 230 | 95 | 201 | 95 |
| 29.1890503 | 53.43229354 | 66 | 71 | 226 | 8739 | 233 | -81 | 314 | 182 | -52 | 182 | -52 | 612 | 83 | 31 | 34 | 247 | 96 | 247 | 96 |
| 34.05391517 | 53.43229354 | 61 | 71 | 215 | 9268 | 233 | -96 | 330 | 183 | -65 | 183 | -65 | 632 | 83 | 30 | 30 | 243 | 99 | 243 | 99 |
| 38.91878003 | 53.43229354 | 56 | 73 | 208 | 10007 | 241 | -113 | 354 | 188 | -45 | 188 | -80 | 557 | 77 | 27 | 31 | 219 | 83 | 219 | 92 |
| 43.78364489 | 53.43229354 | 53 | 76 | 203 | 10615 | 248 | -126 | 374 | 192 | -55 | 192 | -91 | 543 | 64 | 30 | 22 | 184 | 90 | 184 | 97 |
| 48.64850976 | 53.43229354 | 53 | 80 | 200 | 11405 | 260 | -140 | 400 | 202 | -63 | 202 | -103 | 553 | 62 | 29 | 24 | 185 | 87 | 185 | 98 |
| 53.51337462 | 53.43229354 | 40 | 81 | 197 | 11850 | 254 | -158 | 411 | 194 | -78 | 194 | -122 | 526 | 60 | 27 | 25 | 166 | 83 | 166 | 97 |
| 58.37823949 | 53.43229354 | 13 | 83 | 203 | 11768 | 221 | -186 | 408 | 166 | -98 | 166 | -147 | 504 | 73 | 21 | 40 | 210 | 63 | 210 | 68 |
| 63.24310435 | 53.43229354 | 34 | 91 | 204 | 12953 | 259 | -186 | 445 | 201 | -92 | 201 | -142 | 328 | 55 | 12 | 44 | 154 | 38 | 154 | 46 |
| 68.10796922 | 53.43229354 | 24 | 87 | 198 | 12884 | 246 | -194 | 440 | 190 | -101 | 190 | -151 | 357 | 67 | 12 | 52 | 178 | 36 | 178 | 40 |
| 72.97283408 | 53.43229354 | 29 | 91 | 198 | 13457 | 262 | -200 | 462 | 203 | -101 | 203 | -155 | 296 | 39 | 12 | 34 | 110 | 37 | 110 | 43 |
| 77.83769895 | 53.43229354 | 23 | 92 | 195 | 13530 | 259 | -210 | 469 | 199 | -107 | 199 | -162 | 286 | 48 | 10 | 50 | 132 | 30 | 132 | 35 |
| 82.70256381 | 53.43229354 | 28 | 89 | 196 | 12945 | 254 | -199 | 453 | 186 | -92 | 197 | -150 | 379 | 60 | 15 | 43 | 164 | 46 | 163 | 64 |
| 87.56742868 | 53.43229354 | 14 | 88 | 208 | 11589 | 225 | -195 | 420 | 160 | -83 | 170 | -147 | 940 | 105 | 44 | 27 | 304 | 132 | 284 | 178 |
| 92.43229354 | 53.43229354 | 19 | 90 | 197 | 12570 | 240 | -215 | 455 | 175 | -82 | 185 | -163 | 520 | 82 | 15 | 52 | 238 | 47 | 230 | 70 |
| 97.29715841 | 53.43229354 | -78 | 86 | 199 | 11809 | 135 | -297 | 432 | 73 | -231 | 81 | -246 | 618 | 141 | 6 | 91 | 399 | 20 | 382 | 26 |
| 102.1620233 | 53.43229354 | 6 | 93 | 196 | 13062 | 232 | -242 | 474 | 168 | -98 | 178 | -184 | 405 | 83 | 10 | 74 | 243 | 30 | 221 | 45 |
| 107.0268881 | 53.43229354 | -49 | 69 | 159 | 12767 | 162 | -275 | 437 | 116 | -159 | 116 | -230 | 313 | 85 | 3 | 111 | 249 | 9 | 249 | 14 |
| 111.891753 | 53.43229354 | -46 | 93 | 181 | 14349 | 201 | -311 | 512 | 135 | -232 | 145 | -251 | 406 | 126 | 2 | 119 | 322 | 6 | 320 | 8 |
| 116.7566179 | 53.43229354 | -45 | 99 | 174 | 16254 | 215 | -355 | 570 | 151 | -249 | 162 | -288 | 375 | 109 | 2 | 112 | 287 | 8 | 279 | 8 |
| 121.6214827 | 53.43229354 | -26 | 93 | 167 | 16450 | 238 | -321 | 559 | 173 | -232 | 185 | -266 | 463 | 117 | 4 | 100 | 321 | 13 | 314 | 18 |
| 126.4863476 | 53.43229354 | -17 | 87 | 160 | 16322 | 244 | -296 | 541 | 182 | -218 | 193 | -251 | 480 | 119 | 4 | 94 | 315 | 13 | 310 | 18 |
| 131.3512125 | 53.43229354 | -45 | 89 | 152 | 17316 | 228 | -361 | 588 | 163 | -261 | 171 | -308 | 572 | 122 | 5 | 82 | 359 | 17 | 303 | 26 |
| 136.2160773 | 53.43229354 | -19 | 81 | 159 | 15061 | 227 | -280 | 507 | 171 | -212 | 171 | -238 | 620 | 106 | 12 | 62 | 302 | 40 | 302 | 53 |
| -160.5406794 | 55.78364489 | 16 | 34 | 202 | 4772 | 107 | -61 | 169 | 75 | -41 | 89 | -41 | 865 | 120 | 37 | 35 | 347 | 121 | 292 | 121 |
| -155.6758146 | 55.78364489 | 48 | 15 | 121 | 3874 | 118 | -7 | 126 | 71 | 18 | 109 | 5 | 932 | 121 | 50 | 30 | 361 | 155 | 253 | 169 |
| -131.3514902 | 55.78364489 | 32 | 53 | 245 | 5875 | 151 | -66 | 216 | 52 | 109 | 115 | -44 | 3495 | 516 | 148 | 38 | 1376 | 450 | 617 | 1138 |
| -126.4866254 | 55.78364489 | 27 | 74 | 242 | 8123 | 188 | -117 | 305 | -86 | 10 | 136 | -86 | 484 | 54 | 18 | 25 | 161 | 64 | 121 | 161 |
| -121.6217605 | 55.78364489 | 14 | 86 | 241 | 9602 | 196 | -161 | 358 | 137 | -104 | 138 | -122 | 458 | 74 | 22 | 45 | 217 | 69 | 172 | 75 |
| -116.7568956 | 55.78364489 | 22 | 92 | 237 | 10610 | 214 | -177 | 391 | 155 | -66 | 155 | -130 | 506 | 87 | 20 | 51 | 241 | 64 | 241 | 99 |
| -111.8920308 | 55.78364489 | 12 | 91 | 222 | 11301 | 215 | -193 | 409 | 154 | -135 | 154 | -149 | 526 | 102 | 18 | 65 | 288 | 59 | 288 | 70 |
| -107.0271659 | 55.78364489 | 8 | 91 | 201 | 12610 | 230 | -220 | 451 | 168 | -99 | 168 | -174 | 458 | 80 | 19 | 51 | 223 | 57 | 223 | 68 |
| -102.1623011 | 55.78364489 | -8 | 89 | 189 | 13415 | 227 | -246 | 473 | 164 | -124 | 164 | -200 | 479 | 79 | 18 | 51 | 224 | 56 | 224 | 69 |
| -97.29743619 | 55.78364489 | -19 | 89 | 182 | 13835 | 225 | -262 | 487 | 160 | -205 | 160 | -216 | 549 | 88 | 16 | 52 | 254 | 54 | 254 | 68 |
| -92.43257132 | 55.78364489 | -30 | 87 | 178 | 13841 | 216 | -272 | 488 | 152 | -220 | 152 | -225 | 401 | 58 | 14 | 49 | 172 | 45 | 172 | 57 |
| -87.56770646 | 55.78364489 | -48 | 68 | 158 | 12711 | 157 | -272 | 430 | 89 | -234 | 116 | -234 | 499 | 75 | 16 | 48 | 219 | 53 | 215 | 53 |
| -72.97311186 | 55.78364489 | -45 | 70 | 156 | 12839 | 173 | -273 | 446 | 81 | -234 | 121 | -234 | 650 | 98 | 22 | 47 | 288 | 76 | 271 | 76 |
| -68.108247 | 55.78364489 | -43 | 69 | 160 | 12388 | 169 | -266 | 434 | 78 | -175 | 119 | -224 | 740 | 94 | 37 | 35 | 282 | 117 | 280 | 118 |
| -63.24338213 | 55.78364489 | -44 | 69 | 170 | 11461 | 156 | -250 | 406 | 109 | -12 | 109 | -210 | 834 | 114 | 44 | 29 | 312 | 144 | 312 | 160 |
| -4.865003754 | 55.78364489 | 95 | 43 | 293 | 3746 | 177 | 29 | 148 | 111 | 100 | 150 | 49 | 1575 | 178 | 80 | 28 | 520 | 242 | 348 | 410 |
| 9.729590841 | 55.78364489 | 82 | 41 | 202 | 5739 | 192 | -12 | 204 | 66 | 51 | 164 | 7 | 753 | 83 | 43 | 21 | 247 | 138 | 198 | 154 |
| 24.32418544 | 55.78364489 | 71 | 65 | 225 | 7885 | 227 | -65 | 292 | 176 | -35 | 176 | -35 | 592 | 76 | 28 | 28 | 214 | 91 | 210 | 91 |
| 29.1890503 | 55.78364489 | 60 | 66 | 210 | 8592 | 224 | -89 | 312 | 175 | -27 | 175 | -57 | 659 | 85 | 34 | 31 | 250 | 104 | 250 | 107 |
| 34.05391517 | 55.78364489 | 52 | 66 | 201 | 9144 | 224 | -104 | 328 | 175 | -38 | 175 | -72 | 644 | 86 | 31 | 34 | 251 | 94 | 251 | 102 |
| 38.91878003 | 55.78364489 | 54 | 68 | 197 | 9823 | 237 | -110 | 347 | 187 | -41 | 187 | -79 | 604 | 87 | 30 | 32 | 241 | 90 | 241 | 98 |
| 43.78364489 | 55.78364489 | 49 | 70 | 193 | 10393 | 241 | -123 | 364 | 189 | -52 | 189 | -91 | 482 | 65 | 20 | 33 | 186 | 64 | 186 | 78 |
| 48.64850976 | 55.78364489 | 47 | 74 | 192 | 11058 | 250 | -137 | 386 | 195 | -61 | 195 | -103 | 501 | 64 | 23 | 33 | 189 | 72 | 189 | 82 |
| 53.51337462 | 55.78364489 | 41 | 77 | 193 | 11529 | 251 | -149 | 400 | 183 | -69 | 195 | -115 | 483 | 61 | 20 | 33 | 181 | 64 | 175 | 80 |
| 58.37823949 | 55.78364489 | 29 | 77 | 193 | 11462 | 235 | -164 | 399 | 181 | -74 | 181 | -126 | 519 | 86 | 19 | 48 | 238 | 57 | 238 | 77 |
| 63.24310435 | 55.78364489 | 27 | 83 | 195 | 12308 | 243 | -182 | 425 | 188 | -85 | 188 | -141 | 387 | 68 | 12 | 54 | 191 | 37 | 191 | 56 |
| 68.10796922 | 55.78364489 | 20 | 83 | 190 | 12675 | 239 | -197 | 436 | 185 | -96 | 185 | -154 | 385 | 68 | 12 | 54 | 189 | 38 | 189 | 55 |
| 72.97283408 | 55.78364489 | 14 | 84 | 187 | 12976 | 240 | -209 | 449 | 184 | -104 | 184 | -165 | 394 | 68 | 13 | 52 | 185 | 39 | 185 | 59 |
| 77.83769895 | 55.78364489 | 11 | 83 | 183 | 12965 | 236 | -214 | 450 | 182 | -102 | 182 | -169 | 393 | 65 | 13 | 47 | 178 | 41 | 178 | 64 |
| 82.70256381 | 55.78364489 | 11 | 79 | 178 | 12745 | 233 | -209 | 443 | 167 | -99 | 180 | -167 | 415 | 64 | 13 | 47 | 188 | 40 | 173 | 65 |
| 87.56742868 | 55.78364489 | 13 | 76 | 181 | 11995 | 230 | -193 | 423 | 164 | -87 | 176 | -153 | 491 | 68 | 17 | 42 | 202 | 53 | 197 | 76 |
| 92.43229354 | 55.78364489 | 7 | 80 | 182 | 12249 | 228 | -210 | 437 | 161 | -94 | 173 | -165 | 443 | 68 | 15 | 47 | 198 | 45 | 188 | 65 |
| 97.29715841 | 55.78364489 | 13 | 83 | 182 | 12662 | 239 | -217 | 455 | 172 | -89 | 184 | -168 | 426 | 77 | 15 | 52 | 207 | 45 | 195 | 63 |
| 102.1620233 | 55.78364489 | -16 | 84 | 174 | 13602 | 224 | -260 | 484 | 156 | -124 | 167 | -211 | 409 | 73 | 11 | 56 | 214 | 36 | 183 | 67 |
| 107.0268881 | 55.78364489 | -31 | 82 | 165 | 14116 | 216 | -282 | 498 | 148 | -146 | 159 | -233 | 442 | 82 | 11 | 66 | 246 | 35 | 220 | 62 |
| 111.891753 | 55.78364489 | -63 | 82 | 161 | 14844 | 193 | -319 | 512 | 129 | -252 | 138 | -271 | 455 | 105 | 6 | 91 | 297 | 20 | 281 | 28 |
| 116.7566179 | 55.78364489 | -71 | 88 | 163 | 15875 | 194 | -347 | 541 | 129 | -267 | 139 | -298 | 344 | 91 | 2 | 106 | 245 | 7 | 244 | 10 |
| 121.6214827 | 55.78364489 | -41 | 90 | 161 | 16785 | 232 | -327 | 559 | 167 | -245 | 179 | -281 | 366 | 85 | 4 | 93 | 241 | 13 | 230 | 17 |
| 126.4863476 | 55.78364489 | -59 | 82 | 158 | 15996 | 205 | -318 | 523 | 141 | -253 | 154 | -280 | 653 | 133 | 7 | 84 | 399 | 21 | 366 | 27 |
| 131.3512125 | 55.78364489 | -104 | 83 | 155 | 16153 | 163 | -371 | 534 | 98 | -302 | 108 | -331 | 559 | 115 | 6 | 82 | 331 | 20 | 316 | 27 |
| 136.2160773 | 55.78364489 | -51 | 61 | 135 | 14133 | 170 | -278 | 448 | 132 | -231 | 132 | -250 | 685 | 145 | 10 | 78 | 393 | 33 | 393 | 45 |
| 155.6755368 | 55.78364489 | 9 | 41 | 143 | 8568 | 149 | -140 | 289 | 60 | -79 | 128 | -109 | 689 | 106 | 20 | 51 | 289 | 64 | 285 | 69 |
| 160.5404017 | 55.78364489 | -70 | 58 | 164 | 9985 | 118 | -236 | 353 | -201 | -50 | 71 | -203 | 827 | 97 | 43 | 24 | 285 | 139 | 226 | 261 |
| -155.6758146 | 58.13499625 | 21 | 55 | 230 | 6652 | 148 | -91 | 239 | 94 | -58 | 116 | -60 | 634 | 93 | 31 | 40 | 279 | 97 | 248 | 118 |
| -136.2163551 | 58.13499625 | 40 | 39 | 199 | 5651 | 141 | -55 | 195 | 64 | 90 | 119 | -32 | 3230 | 581 | 88 | 50 | 1568 | 327 | 520 | 901 |
| -131.3514902 | 58.13499625 | 17 | 60 | 212 | 7956 | 168 | -115 | 283 | 37 | -4 | 124 | -89 | 324 | 54 | 10 | 47 | 146 | 32 | 73 | 110 |
| -126.4866254 | 58.13499625 | -24 | 74 | 221 | 9285 | 147 | -186 | 333 | 97 | -39 | 97 | -153 | 503 | 80 | 15 | 42 | 220 | 50 | 220 | 122 |
| -121.6217605 | 58.13499625 | 15 | 90 | 211 | 12166 | 225 | -200 | 425 | 168 | -80 | 168 | -159 | 384 | 70 | 19 | 52 | 197 | 58 | 197 | 64 |
| -116.7568956 | 58.13499625 | 8 | 91 | 209 | 12459 | 222 | -213 | 435 | 165 | -153 | 165 | -169 | 425 | 75 | 17 | 53 | 211 | 54 | 211 | 62 |
| -111.8920308 | 58.13499625 | 8 | 89 | 196 | 12922 | 233 | -223 | 456 | 173 | -14 | 173 | -177 | 450 | 79 | 19 | 55 | 225 | 58 | 225 | 66 |
| -107.0271659 | 58.13499625 | -20 | 85 | 183 | 13156 | 211 | -254 | 465 | 150 | -195 | 150 | -208 | 443 | 79 | 15 | 56 | 218 | 49 | 218 | 60 |
| -102.1623011 | 58.13499625 | -34 | 83 | 175 | 13707 | 206 | -270 | 476 | 144 | -216 | 144 | -227 | 444 | 71 | 15 | 48 | 198 | 53 | 198 | 70 |
| -97.29743619 | 58.13499625 | -54 | 83 | 168 | 14270 | 199 | -293 | 492 | 136 | -242 | 136 | -248 | 457 | 73 | 15 | 54 | 218 | 46 | 218 | 50 |
| -72.97311186 | 58.13499625 | -61 | 68 | 152 | 12857 | 162 | -287 | 449 | 108 | -249 | 108 | -249 | 491 | 72 | 20 | 43 | 206 | 63 | 206 | 63 |
| -68.108247 | 58.13499625 | -42 | 64 | 155 | 11889 | 160 | -252 | 412 | 113 | -98 | 113 | -218 | 489 | 62 | 24 | 31 | 179 | 75 | 179 | 90 |
| -63.24338213 | 58.13499625 | -76 | 64 | 164 | 11157 | 115 | -276 | 391 | 74 | -53 | 74 | -240 | 862 | 115 | 45 | 32 | 331 | 138 | 331 | 152 |
| -4.865003754 | 58.13499625 | 41 | 30 | 248 | 3284 | 106 | -14 | 121 | 34 | 41 | 90 | 1 | 2088 | 244 | 108 | 29 | 721 | 335 | 412 | 556 |
| 14.59445571 | 58.13499625 | 61 | 67 | 250 | 6822 | 208 | -59 | 268 | 156 | -11 | 156 | -28 | 613 | 75 | 29 | 28 | 211 | 95 | 211 | 99 |
| 29.1890503 | 58.13499625 | 54 | 65 | 206 | 8577 | 222 | -95 | 317 | 166 | -33 | 172 | -63 | 734 | 85 | 38 | 25 | 254 | 122 | 239 | 124 |
| 34.05391517 | 58.13499625 | 45 | 65 | 196 | 9215 | 222 | -110 | 332 | 172 | -45 | 172 | -80 | 647 | 87 | 30 | 33 | 243 | 94 | 243 | 98 |
| 38.91878003 | 58.13499625 | 42 | 66 | 191 | 9829 | 228 | -119 | 347 | 167 | -52 | 177 | -91 | 670 | 86 | 32 | 31 | 248 | 98 | 246 | 112 |
| 43.78364489 | 58.13499625 | 38 | 68 | 186 | 10333 | 232 | -133 | 365 | 180 | -58 | 180 | -101 | 618 | 84 | 26 | 36 | 242 | 79 | 242 | 91 |
| 48.64850976 | 58.13499625 | 34 | 69 | 182 | 10796 | 236 | -145 | 381 | 181 | -65 | 181 | -112 | 627 | 89 | 29 | 34 | 245 | 88 | 245 | 99 |
| 53.51337462 | 58.13499625 | 24 | 70 | 181 | 11050 | 229 | -160 | 389 | 175 | -74 | 175 | -125 | 590 | 79 | 27 | 34 | 223 | 81 | 223 | 90 |
| 58.37823949 | 58.13499625 | 10 | 70 | 177 | 11156 | 214 | -178 | 393 | 148 | -82 | 162 | -143 | 784 | 98 | 35 | 31 | 282 | 109 | 278 | 162 |
| 63.24310435 | 58.13499625 | 23 | 74 | 180 | 11795 | 235 | -178 | 413 | 167 | -74 | 182 | -140 | 478 | 79 | 15 | 49 | 215 | 48 | 213 | 73 |
| 68.10796922 | 58.13499625 | 13 | 78 | 180 | 12426 | 233 | -200 | 433 | 165 | -92 | 179 | -159 | 461 | 74 | 15 | 49 | 212 | 47 | 202 | 71 |
| 72.97283408 | 58.13499625 | 6 | 74 | 172 | 12406 | 224 | -207 | 431 | 158 | -96 | 172 | -168 | 473 | 72 | 14 | 51 | 214 | 42 | 207 | 70 |
| 77.83769895 | 58.13499625 | -1 | 77 | 172 | 12857 | 226 | -222 | 448 | 158 | -105 | 172 | -183 | 550 | 82 | 17 | 45 | 240 | 53 | 220 | 90 |
| 82.70256381 | 58.13499625 | 0 | 76 | 167 | 12918 | 229 | -223 | 452 | 161 | -103 | 176 | -182 | 461 | 70 | 15 | 47 | 205 | 45 | 186 | 71 |
| 87.56742868 | 58.13499625 | -4 | 74 | 164 | 12892 | 228 | -224 | 452 | 159 | -105 | 173 | -185 | 512 | 75 | 18 | 41 | 217 | 56 | 189 | 87 |
| 92.43229354 | 58.13499625 | -2 | 75 | 160 | 13398 | 239 | -232 | 471 | 168 | -107 | 182 | -191 | 437 | 62 | 15 | 40 | 181 | 46 | 163 | 82 |
| 97.29715841 | 58.13499625 | -7 | 78 | 157 | 14195 | 241 | -255 | 496 | 171 | -116 | 185 | -211 | 364 | 50 | 14 | 39 | 148 | 42 | 135 | 64 |
| 102.1620233 | 58.13499625 | -18 | 77 | 153 | 14549 | 236 | -269 | 504 | 166 | -129 | 180 | -225 | 346 | 58 | 11 | 50 | 164 | 34 | 155 | 53 |
| 107.0268881 | 58.13499625 | -32 | 77 | 148 | 15068 | 227 | -294 | 521 | 171 | -145 | 171 | -249 | 326 | 67 | 8 | 66 | 177 | 25 | 177 | 43 |
| 111.891753 | 58.13499625 | -60 | 74 | 144 | 14848 | 197 | -315 | 512 | 129 | -172 | 141 | -273 | 604 | 92 | 18 | 48 | 270 | 60 | 255 | 91 |
| 116.7566179 | 58.13499625 | -65 | 74 | 144 | 15367 | 198 | -320 | 518 | 143 | -181 | 143 | -283 | 362 | 92 | 4 | 100 | 252 | 12 | 252 | 17 |
| 121.6214827 | 58.13499625 | -33 | 78 | 142 | 16466 | 243 | -303 | 546 | 174 | -156 | 188 | -265 | 309 | 61 | 6 | 76 | 173 | 18 | 172 | 32 |
| 126.4863476 | 58.13499625 | -84 | 76 | 141 | 16443 | 190 | -348 | 538 | 122 | -288 | 136 | -313 | 547 | 93 | 13 | 68 | 279 | 43 | 276 | 55 |
| 131.3512125 | 58.13499625 | -80 | 79 | 138 | 17701 | 208 | -364 | 573 | 153 | -212 | 153 | -330 | 354 | 69 | 7 | 71 | 192 | 23 | 192 | 41 |
| 136.2160773 | 58.13499625 | -81 | 77 | 135 | 17913 | 207 | -366 | 573 | 136 | -210 | 148 | -337 | 478 | 78 | 10 | 56 | 227 | 38 | 198 | 68 |
| 160.5404017 | 58.13499625 | -63 | 60 | 155 | 11062 | 135 | -252 | 387 | -12 | -170 | 87 | -214 | 771 | 99 | 31 | 38 | 286 | 99 | 279 | 155 |
| -160.5406794 | 60.4863476 | -10 | 65 | 197 | 9437 | 161 | -167 | 328 | 115 | -123 | 118 | -131 | 589 | 119 | 23 | 58 | 315 | 72 | 208 | 83 |
| -155.6758146 | 60.4863476 | 10 | 64 | 208 | 8875 | 176 | -134 | 310 | 129 | -94 | 132 | -101 | 589 | 113 | 21 | 57 | 317 | 69 | 236 | 88 |
| -150.8109497 | 60.4863476 | 37 | 54 | 205 | 7546 | 181 | -83 | 264 | 113 | 13 | 142 | -57 | 418 | 70 | 15 | 47 | 195 | 49 | 147 | 85 |
| -141.08122 | 60.4863476 | -94 | 44 | 208 | 6063 | 19 | -192 | 210 | -78 | -32 | -12 | -167 | 2252 | 321 | 103 | 35 | 925 | 337 | 508 | 580 |
| -136.2163551 | 60.4863476 | -23 | 65 | 198 | 9159 | 150 | -177 | 327 | -1 | -45 | 102 | -145 | 433 | 54 | 13 | 33 | 161 | 50 | 106 | 117 |
| -131.3514902 | 60.4863476 | -33 | 74 | 202 | 10257 | 154 | -210 | 365 | -7 | 18 | 104 | -176 | 458 | 49 | 18 | 25 | 146 | 68 | 131 | 128 |
| -126.4866254 | 60.4863476 | -38 | 78 | 190 | 12027 | 168 | -245 | 414 | 117 | -50 | 117 | -208 | 570 | 87 | 28 | 37 | 245 | 86 | 245 | 118 |
| -121.6217605 | 60.4863476 | -30 | 86 | 183 | 13993 | 204 | -266 | 470 | 148 | -203 | 148 | -228 | 412 | 66 | 17 | 49 | 192 | 55 | 192 | 62 |
| -116.7568956 | 60.4863476 | -16 | 85 | 183 | 13715 | 220 | -246 | 466 | 156 | -142 | 161 | -206 | 354 | 47 | 17 | 38 | 139 | 51 | 135 | 62 |
| -111.8920308 | 60.4863476 | -17 | 83 | 173 | 13924 | 223 | -256 | 479 | 163 | -50 | 163 | -213 | 347 | 55 | 13 | 44 | 143 | 44 | 143 | 58 |
| -107.0271659 | 60.4863476 | -47 | 79 | 165 | 13998 | 194 | -286 | 481 | 134 | -234 | 136 | -242 | 359 | 55 | 12 | 49 | 164 | 41 | 156 | 53 |
| -102.1623011 | 60.4863476 | -64 | 78 | 159 | 14225 | 188 | -300 | 487 | 127 | -252 | 127 | -255 | 379 | 64 | 11 | 55 | 176 | 38 | 176 | 51 |
| -97.29743619 | 60.4863476 | -79 | 75 | 152 | 14591 | 184 | -310 | 495 | 123 | -270 | 123 | -270 | 354 | 58 | 12 | 56 | 171 | 37 | 171 | 37 |
| -72.97311186 | 60.4863476 | -84 | 65 | 145 | 12906 | 143 | -303 | 446 | 88 | -269 | 88 | -269 | 363 | 60 | 13 | 51 | 170 | 40 | 170 | 40 |
| -68.108247 | 60.4863476 | -71 | 29 | 102 | 8925 | 46 | -238 | 284 | 32 | -127 | 32 | -214 | 432 | 55 | 18 | 32 | 158 | 58 | 158 | 88 |
| -43.78392267 | 60.4863476 | -19 | 44 | 216 | 5841 | 85 | -121 | 206 | -42 | 56 | 57 | -93 | 2044 | 201 | 121 | 16 | 598 | 373 | 451 | 537 |
| 4.864725976 | 60.4863476 | 81 | 29 | 186 | 4518 | 166 | 9 | 157 | 100 | 86 | 149 | 25 | 1742 | 208 | 83 | 30 | 615 | 264 | 400 | 414 |
| 9.729590841 | 60.4863476 | 2 | 64 | 239 | 7100 | 152 | -118 | 269 | 101 | -64 | 103 | -87 | 855 | 101 | 39 | 32 | 299 | 121 | 286 | 132 |
| 14.59445571 | 60.4863476 | 46 | 68 | 233 | 7705 | 206 | -87 | 293 | 151 | -29 | 154 | -54 | 616 | 77 | 27 | 31 | 229 | 91 | 217 | 92 |
| 24.32418544 | 60.4863476 | 52 | 49 | 171 | 7906 | 203 | -84 | 286 | 163 | 74 | 163 | -56 | 653 | 80 | 34 | 31 | 233 | 103 | 233 | 116 |
| 29.1890503 | 60.4863476 | 48 | 53 | 170 | 8654 | 211 | -98 | 309 | 165 | -45 | 168 | -72 | 591 | 76 | 29 | 31 | 216 | 93 | 177 | 95 |
| 34.05391517 | 60.4863476 | 41 | 63 | 188 | 9333 | 222 | -113 | 335 | 163 | -51 | 172 | -85 | 654 | 77 | 33 | 27 | 226 | 109 | 204 | 110 |
| 38.91878003 | 60.4863476 | 31 | 63 | 181 | 9807 | 217 | -134 | 351 | 156 | -102 | 168 | -102 | 607 | 73 | 26 | 31 | 219 | 86 | 209 | 86 |
| 43.78364489 | 60.4863476 | 25 | 64 | 175 | 10120 | 216 | -148 | 364 | 166 | -66 | 166 | -112 | 611 | 79 | 27 | 32 | 226 | 81 | 226 | 115 |
| 48.64850976 | 60.4863476 | 23 | 68 | 177 | 10692 | 228 | -158 | 386 | 173 | -70 | 173 | -121 | 624 | 77 | 29 | 30 | 222 | 87 | 222 | 122 |
| 53.51337462 | 60.4863476 | 18 | 70 | 176 | 11085 | 230 | -167 | 397 | 173 | -75 | 173 | -131 | 557 | 70 | 25 | 30 | 202 | 79 | 202 | 109 |
| 58.37823949 | 60.4863476 | -10 | 70 | 169 | 11703 | 205 | -211 | 417 | 134 | -102 | 151 | -174 | 751 | 94 | 30 | 33 | 280 | 98 | 265 | 136 |
| 63.24310435 | 60.4863476 | 6 | 74 | 171 | 12291 | 230 | -204 | 435 | 158 | -90 | 175 | -164 | 462 | 72 | 16 | 48 | 210 | 50 | 197 | 71 |
| 68.10796922 | 60.4863476 | -2 | 72 | 162 | 12690 | 227 | -216 | 443 | 158 | -104 | 173 | -176 | 526 | 74 | 20 | 43 | 222 | 60 | 209 | 88 |
| 72.97283408 | 60.4863476 | -8 | 71 | 161 | 12634 | 221 | -220 | 441 | 153 | -109 | 168 | -181 | 567 | 83 | 20 | 46 | 247 | 62 | 236 | 90 |
| 77.83769895 | 60.4863476 | -11 | 74 | 161 | 13249 | 229 | -233 | 463 | 159 | -116 | 174 | -193 | 502 | 79 | 15 | 51 | 227 | 46 | 208 | 68 |
| 82.70256381 | 60.4863476 | -20 | 74 | 156 | 13515 | 223 | -249 | 472 | 153 | -124 | 168 | -208 | 589 | 81 | 21 | 41 | 241 | 67 | 227 | 94 |
| 87.56742868 | 60.4863476 | -22 | 73 | 152 | 13704 | 225 | -255 | 480 | 155 | -126 | 168 | -214 | 529 | 71 | 19 | 39 | 211 | 58 | 199 | 94 |
| 92.43229354 | 60.4863476 | -37 | 71 | 142 | 14432 | 217 | -283 | 500 | 147 | -224 | 161 | -242 | 519 | 79 | 18 | 44 | 219 | 60 | 180 | 79 |
| 97.29715841 | 60.4863476 | -45 | 75 | 140 | 15492 | 219 | -314 | 533 | 149 | -155 | 163 | -271 | 543 | 83 | 20 | 41 | 231 | 60 | 189 | 93 |
| 102.1620233 | 60.4863476 | -40 | 76 | 142 | 15707 | 231 | -305 | 537 | 158 | -153 | 173 | -265 | 403 | 60 | 13 | 44 | 173 | 41 | 155 | 65 |
| 107.0268881 | 60.4863476 | -43 | 76 | 141 | 16030 | 232 | -310 | 542 | 174 | -160 | 174 | -271 | 323 | 58 | 8 | 58 | 159 | 28 | 159 | 42 |
| 111.891753 | 60.4863476 | -52 | 74 | 134 | 16472 | 226 | -324 | 550 | 155 | -172 | 170 | -287 | 378 | 54 | 13 | 41 | 158 | 41 | 145 | 70 |
| 116.7566179 | 60.4863476 | -37 | 75 | 138 | 16634 | 243 | -303 | 546 | 171 | -163 | 186 | -270 | 285 | 56 | 7 | 64 | 150 | 22 | 149 | 45 |
| 121.6214827 | 60.4863476 | -61 | 78 | 136 | 17542 | 231 | -344 | 575 | 173 | -195 | 173 | -307 | 359 | 62 | 10 | 56 | 172 | 32 | 172 | 59 |
| 126.4863476 | 60.4863476 | -77 | 79 | 136 | 18025 | 220 | -363 | 583 | 146 | -218 | 162 | -327 | 358 | 62 | 8 | 67 | 186 | 28 | 182 | 41 |
| 131.3512125 | 60.4863476 | -73 | 81 | 126 | 20226 | 246 | -396 | 642 | 170 | -225 | 188 | -359 | 250 | 46 | 6 | 64 | 133 | 20 | 118 | 34 |
| 136.2160773 | 60.4863476 | -87 | 78 | 120 | 20765 | 233 | -416 | 649 | 159 | -238 | 176 | -384 | 324 | 54 | 9 | 59 | 157 | 28 | 138 | 36 |
| 141.0809422 | 60.4863476 | -66 | 71 | 143 | 15678 | 196 | -301 | 497 | 135 | -190 | 141 | -276 | 361 | 66 | 7 | 68 | 190 | 23 | 174 | 36 |
| 145.9458071 | 60.4863476 | -64 | 66 | 138 | 15225 | 187 | -292 | 479 | 134 | -187 | 135 | -268 | 478 | 77 | 12 | 58 | 225 | 37 | 203 | 58 |
| 150.8106719 | 60.4863476 | -94 | 62 | 132 | 14791 | 154 | -314 | 468 | 100 | -219 | 104 | -289 | 351 | 58 | 7 | 59 | 170 | 24 | 156 | 41 |
| 165.4052665 | 60.4863476 | -39 | 47 | 140 | 10191 | 133 | -201 | 334 | 104 | -139 | 104 | -170 | 461 | 73 | 21 | 44 | 209 | 63 | 209 | 85 |
| 170.2701314 | 60.4863476 | -19 | 43 | 149 | 8781 | 132 | -156 | 289 | 107 | -8 | 107 | -128 | 498 | 70 | 25 | 33 | 204 | 75 | 204 | 113 |
| -160.5406794 | 62.83769895 | -13 | 65 | 178 | 10797 | 179 | -187 | 366 | 126 | -51 | 136 | -151 | 398 | 82 | 17 | 62 | 219 | 55 | 143 | 60 |
| -155.6758146 | 62.83769895 | 2 | 68 | 181 | 11224 | 201 | -176 | 377 | 144 | -132 | 157 | -143 | 411 | 75 | 16 | 54 | 213 | 52 | 167 | 69 |
| -150.8109497 | 62.83769895 | -60 | 59 | 170 | 10462 | 121 | -227 | 348 | 74 | -76 | 82 | -198 | 1020 | 191 | 36 | 56 | 516 | 116 | 350 | 197 |
| -145.9460848 | 62.83769895 | -38 | 64 | 176 | 10711 | 152 | -210 | 362 | 99 | -52 | 107 | -180 | 638 | 92 | 26 | 42 | 260 | 86 | 251 | 115 |
| -141.08122 | 62.83769895 | -22 | 76 | 181 | 12538 | 190 | -230 | 419 | 142 | -31 | 142 | -198 | 333 | 80 | 9 | 82 | 223 | 27 | 223 | 33 |
| -136.2163551 | 62.83769895 | -11 | 75 | 185 | 11899 | 200 | -207 | 407 | 150 | -30 | 150 | -172 | 306 | 63 | 8 | 62 | 164 | 24 | 164 | 49 |
| -131.3514902 | 62.83769895 | -61 | 68 | 178 | 11320 | 141 | -244 | 385 | 94 | -86 | 94 | -210 | 487 | 78 | 17 | 42 | 210 | 58 | 210 | 100 |
| -126.4866254 | 62.83769895 | -75 | 71 | 174 | 12308 | 141 | -267 | 408 | 92 | -183 | 92 | -235 | 542 | 108 | 20 | 67 | 296 | 60 | 296 | 70 |
| -121.6217605 | 62.83769895 | -36 | 82 | 164 | 15183 | 219 | -284 | 502 | 152 | -170 | 162 | -246 | 340 | 50 | 14 | 44 | 146 | 46 | 143 | 57 |
| -116.7568956 | 62.83769895 | -42 | 78 | 157 | 14776 | 213 | -280 | 493 | 147 | -87 | 154 | -242 | 265 | 44 | 9 | 44 | 123 | 33 | 91 | 52 |
| -111.8920308 | 62.83769895 | -62 | 77 | 154 | 14762 | 197 | -302 | 499 | 132 | -212 | 135 | -257 | 303 | 44 | 13 | 45 | 130 | 39 | 120 | 47 |
| -107.0271659 | 62.83769895 | -79 | 74 | 150 | 14523 | 179 | -313 | 492 | 118 | -269 | 118 | -269 | 296 | 51 | 10 | 53 | 145 | 31 | 145 | 31 |
| -102.1623011 | 62.83769895 | -90 | 73 | 146 | 14853 | 178 | -323 | 500 | 116 | -282 | 116 | -282 | 265 | 47 | 5 | 64 | 137 | 18 | 137 | 18 |
| -97.29743619 | 62.83769895 | -100 | 71 | 141 | 15113 | 172 | -330 | 503 | 112 | -296 | 112 | -296 | 267 | 47 | 7 | 63 | 138 | 22 | 138 | 22 |
| -92.43257132 | 62.83769895 | -100 | 61 | 129 | 14660 | 147 | -328 | 475 | 102 | -297 | 102 | -297 | 304 | 57 | 8 | 57 | 154 | 26 | 154 | 26 |
| -82.70284159 | 62.83769895 | -86 | 28 | 83 | 11299 | 74 | -270 | 344 | 42 | -252 | 59 | -252 | 313 | 46 | 11 | 49 | 134 | 35 | 130 | 35 |
| -68.108247 | 62.83769895 | -113 | 58 | 141 | 12132 | 90 | -320 | 410 | 48 | -250 | 48 | -288 | 485 | 73 | 24 | 43 | 214 | 72 | 214 | 74 |
| -48.64878754 | 62.83769895 | -112 | 61 | 211 | 8156 | 32 | -254 | 287 | -39 | -217 | -2 | -217 | 771 | 88 | 45 | 24 | 260 | 137 | 224 | 137 |
| -43.78392267 | 62.83769895 | -141 | 49 | 203 | 6881 | -9 | -249 | 240 | -221 | -113 | -41 | -223 | 954 | 98 | 60 | 16 | 285 | 192 | 200 | 264 |
| 9.729590841 | 62.83769895 | 27 | 56 | 232 | 6543 | 161 | -80 | 241 | 118 | 46 | 120 | -54 | 681 | 76 | 32 | 22 | 209 | 101 | 208 | 181 |
| 14.59445571 | 62.83769895 | 32 | 64 | 223 | 7697 | 190 | -97 | 287 | 141 | -40 | 141 | -66 | 593 | 86 | 27 | 36 | 236 | 85 | 236 | 88 |
| 24.32418544 | 62.83769895 | 30 | 66 | 203 | 8665 | 205 | -118 | 323 | 146 | -58 | 153 | -87 | 615 | 83 | 26 | 35 | 243 | 88 | 217 | 89 |
| 29.1890503 | 62.83769895 | 31 | 63 | 188 | 9173 | 211 | -125 | 335 | 153 | -61 | 160 | -95 | 649 | 80 | 34 | 28 | 232 | 110 | 206 | 111 |
| 34.05391517 | 62.83769895 | 25 | 61 | 180 | 9470 | 208 | -131 | 339 | 150 | -104 | 158 | -104 | 612 | 85 | 30 | 32 | 233 | 96 | 186 | 96 |
| 38.91878003 | 62.83769895 | 16 | 63 | 177 | 9859 | 209 | -150 | 359 | 144 | -75 | 156 | -117 | 595 | 82 | 25 | 35 | 235 | 79 | 204 | 83 |
| 43.78364489 | 62.83769895 | 14 | 64 | 170 | 10367 | 216 | -163 | 379 | 147 | -78 | 162 | -125 | 585 | 72 | 28 | 28 | 209 | 85 | 188 | 124 |
| 48.64850976 | 62.83769895 | 4 | 66 | 169 | 10721 | 211 | -180 | 391 | 140 | -88 | 156 | -142 | 598 | 71 | 28 | 29 | 210 | 86 | 198 | 118 |
| 53.51337462 | 62.83769895 | -5 | 68 | 169 | 11050 | 208 | -194 | 401 | 135 | -96 | 152 | -156 | 572 | 73 | 25 | 34 | 217 | 79 | 204 | 105 |
| 58.37823949 | 62.83769895 | -23 | 69 | 159 | 12082 | 201 | -231 | 432 | 128 | -118 | 145 | -192 | 761 | 86 | 38 | 26 | 251 | 124 | 220 | 159 |
| 63.24310435 | 62.83769895 | -15 | 71 | 159 | 12597 | 216 | -229 | 446 | 144 | -114 | 161 | -189 | 503 | 70 | 19 | 41 | 208 | 58 | 199 | 82 |
| 68.10796922 | 62.83769895 | -23 | 69 | 154 | 12813 | 212 | -236 | 448 | 142 | -126 | 157 | -197 | 560 | 80 | 19 | 47 | 238 | 63 | 232 | 83 |
| 72.97283408 | 62.83769895 | -32 | 69 | 152 | 13133 | 210 | -245 | 455 | 142 | -141 | 155 | -205 | 559 | 80 | 20 | 44 | 237 | 66 | 226 | 89 |
| 77.83769895 | 62.83769895 | -35 | 70 | 149 | 13518 | 213 | -254 | 467 | 145 | -146 | 157 | -214 | 513 | 71 | 17 | 46 | 213 | 55 | 206 | 75 |
| 82.70256381 | 62.83769895 | -37 | 69 | 146 | 13641 | 213 | -257 | 470 | 144 | -144 | 157 | -220 | 601 | 84 | 23 | 39 | 244 | 75 | 216 | 103 |
| 87.56742868 | 62.83769895 | -32 | 68 | 140 | 14254 | 223 | -264 | 487 | 8 | -142 | 168 | -226 | 603 | 73 | 27 | 32 | 213 | 83 | 180 | 125 |
| 92.43229354 | 62.83769895 | -62 | 67 | 130 | 15205 | 203 | -311 | 514 | 135 | -177 | 148 | -273 | 703 | 90 | 28 | 34 | 252 | 90 | 215 | 129 |
| 97.29715841 | 62.83769895 | -69 | 70 | 127 | 16452 | 209 | -342 | 551 | 138 | -187 | 151 | -303 | 438 | 69 | 16 | 43 | 190 | 50 | 152 | 74 |
| 102.1620233 | 62.83769895 | -71 | 75 | 130 | 16996 | 216 | -357 | 574 | 142 | -192 | 157 | -315 | 350 | 60 | 11 | 50 | 165 | 33 | 133 | 55 |
| 107.0268881 | 62.83769895 | -69 | 74 | 131 | 16987 | 220 | -347 | 567 | 145 | -286 | 160 | -310 | 385 | 66 | 10 | 54 | 180 | 36 | 153 | 52 |
| 111.891753 | 62.83769895 | -68 | 73 | 129 | 17315 | 225 | -342 | 567 | 151 | -286 | 166 | -308 | 308 | 49 | 9 | 52 | 143 | 31 | 132 | 43 |
| 116.7566179 | 62.83769895 | -70 | 76 | 129 | 18121 | 231 | -356 | 587 | 172 | -209 | 172 | -322 | 278 | 46 | 8 | 54 | 130 | 24 | 130 | 43 |
| 121.6214827 | 62.83769895 | -78 | 75 | 127 | 18470 | 226 | -366 | 592 | 168 | -220 | 168 | -333 | 282 | 52 | 8 | 58 | 141 | 26 | 141 | 36 |
| 126.4863476 | 62.83769895 | -87 | 76 | 121 | 19835 | 233 | -393 | 626 | 174 | -335 | 174 | -363 | 246 | 38 | 6 | 57 | 111 | 20 | 111 | 28 |
| 131.3512125 | 62.83769895 | -100 | 78 | 116 | 21590 | 238 | -432 | 670 | 161 | -375 | 179 | -401 | 284 | 50 | 7 | 65 | 144 | 21 | 141 | 26 |
| 136.2160773 | 62.83769895 | -100 | 79 | 120 | 21423 | 231 | -428 | 660 | 156 | -373 | 175 | -398 | 382 | 68 | 8 | 69 | 204 | 24 | 190 | 28 |
| 141.0809422 | 62.83769895 | -144 | 63 | 118 | 17391 | 141 | -389 | 531 | 79 | -295 | 92 | -365 | 290 | 69 | 5 | 90 | 191 | 16 | 184 | 25 |
| 145.9458071 | 62.83769895 | -112 | 67 | 121 | 17962 | 182 | -376 | 558 | 129 | -147 | 129 | -349 | 313 | 68 | 7 | 71 | 174 | 23 | 174 | 41 |
| 150.8106719 | 62.83769895 | -96 | 66 | 115 | 18462 | 206 | -365 | 571 | 134 | -133 | 152 | -341 | 388 | 71 | 10 | 56 | 193 | 34 | 185 | 67 |
| 155.6755368 | 62.83769895 | -101 | 61 | 117 | 16734 | 178 | -346 | 524 | 112 | -147 | 126 | -319 | 316 | 52 | 9 | 51 | 152 | 29 | 135 | 63 |
| 160.5404017 | 62.83769895 | -78 | 60 | 133 | 14260 | 166 | -288 | 453 | 110 | -126 | 116 | -261 | 290 | 44 | 11 | 44 | 131 | 33 | 113 | 66 |
| 165.4052665 | 62.83769895 | -62 | 61 | 137 | 13816 | 177 | -266 | 443 | 122 | -18 | 128 | -236 | 433 | 60 | 19 | 32 | 171 | 57 | 135 | 116 |
| 170.2701314 | 62.83769895 | -71 | 59 | 139 | 13059 | 159 | -267 | 427 | 107 | -197 | 112 | -234 | 328 | 62 | 11 | 61 | 182 | 34 | 150 | 48 |
| 175.1349962 | 62.83769895 | -57 | 59 | 149 | 12007 | 156 | -240 | 396 | 108 | -120 | 108 | -207 | 437 | 78 | 19 | 46 | 215 | 59 | 215 | 90 |
| -175.135274 | 65.1890503 | -64 | 41 | 129 | 10096 | 104 | -217 | 322 | 75 | -134 | 75 | -194 | 422 | 59 | 20 | 35 | 160 | 64 | 160 | 88 |
| -165.4055443 | 65.1890503 | -42 | 54 | 153 | 10983 | 144 | -210 | 355 | 106 | -159 | 106 | -178 | 332 | 70 | 15 | 61 | 190 | 45 | 190 | 49 |
| -160.5406794 | 65.1890503 | -43 | 68 | 165 | 12569 | 175 | -239 | 414 | 120 | -91 | 133 | -200 | 306 | 79 | 10 | 82 | 210 | 34 | 127 | 40 |
| -155.6758146 | 65.1890503 | -29 | 68 | 161 | 12899 | 193 | -231 | 424 | 130 | 42 | 149 | -197 | 348 | 64 | 13 | 53 | 183 | 39 | 145 | 64 |
| -150.8109497 | 65.1890503 | -55 | 68 | 159 | 13352 | 172 | -260 | 432 | 109 | -79 | 127 | -229 | 442 | 93 | 12 | 72 | 268 | 41 | 220 | 68 |
| -145.9460848 | 65.1890503 | -50 | 74 | 165 | 13771 | 179 | -267 | 446 | 118 | -217 | 135 | -233 | 353 | 68 | 12 | 62 | 197 | 40 | 172 | 54 |
| -141.08122 | 65.1890503 | -45 | 75 | 173 | 13301 | 182 | -253 | 435 | 120 | -68 | 136 | -221 | 305 | 62 | 7 | 73 | 176 | 23 | 170 | 37 |
| -136.2163551 | 65.1890503 | -72 | 70 | 169 | 12578 | 149 | -263 | 412 | 91 | -108 | 103 | -230 | 489 | 80 | 16 | 47 | 222 | 51 | 213 | 94 |
| -131.3514902 | 65.1890503 | -56 | 69 | 168 | 12532 | 167 | -242 | 409 | 120 | -204 | 120 | -211 | 419 | 93 | 15 | 69 | 248 | 47 | 248 | 51 |
| -126.4866254 | 65.1890503 | -33 | 78 | 160 | 15089 | 225 | -261 | 486 | 159 | -77 | 171 | -227 | 295 | 48 | 11 | 52 | 143 | 35 | 132 | 48 |
| -121.6217605 | 65.1890503 | -53 | 60 | 130 | 14772 | 190 | -276 | 466 | 141 | -201 | 143 | -243 | 238 | 35 | 10 | 42 | 98 | 30 | 75 | 43 |
| -116.7568956 | 65.1890503 | -70 | 71 | 147 | 14723 | 190 | -296 | 485 | 125 | -133 | 131 | -258 | 276 | 48 | 8 | 52 | 139 | 31 | 106 | 33 |
| -111.8920308 | 65.1890503 | -92 | 68 | 143 | 14369 | 167 | -310 | 477 | 105 | -273 | 105 | -273 | 317 | 58 | 10 | 56 | 161 | 32 | 123 | 32 |
| -107.0271659 | 65.1890503 | -105 | 69 | 142 | 14687 | 163 | -325 | 488 | 101 | -290 | 101 | -290 | 241 | 50 | 6 | 67 | 139 | 18 | 139 | 18 |
| -102.1623011 | 65.1890503 | -113 | 70 | 137 | 15326 | 165 | -344 | 509 | 103 | -309 | 103 | -309 | 192 | 43 | 3 | 78 | 121 | 10 | 121 | 10 |
| -97.29743619 | 65.1890503 | -119 | 68 | 132 | 15544 | 160 | -354 | 514 | 99 | -321 | 99 | -321 | 214 | 38 | 5 | 69 | 113 | 16 | 113 | 16 |
| -92.43257132 | 65.1890503 | -123 | 65 | 129 | 15151 | 150 | -354 | 504 | 90 | -322 | 90 | -322 | 236 | 48 | 5 | 68 | 133 | 16 | 133 | 16 |
| -87.56770646 | 65.1890503 | -110 | 55 | 121 | 13931 | 132 | -324 | 456 | 37 | -266 | 83 | -297 | 371 | 55 | 16 | 37 | 151 | 52 | 147 | 54 |
| -72.97311186 | 65.1890503 | -105 | 60 | 130 | 13704 | 137 | -328 | 464 | 81 | -295 | 81 | -295 | 302 | 56 | 10 | 55 | 150 | 31 | 150 | 31 |
| -68.108247 | 65.1890503 | -101 | 51 | 125 | 11894 | 109 | -295 | 404 | 61 | -236 | 61 | -267 | 417 | 59 | 18 | 36 | 167 | 58 | 167 | 66 |
| -48.64878754 | 65.1890503 | -117 | 51 | 196 | 7550 | 22 | -240 | 262 | -49 | -209 | -9 | -209 | 458 | 59 | 22 | 33 | 173 | 68 | 146 | 68 |
| -43.78392267 | 65.1890503 | -182 | 53 | 199 | 7619 | -34 | -300 | 266 | -120 | -154 | -69 | -271 | 464 | 50 | 28 | 17 | 140 | 86 | 111 | 125 |
| -19.45959835 | 65.1890503 | 3 | 50 | 255 | 5144 | 109 | -86 | 195 | 53 | 13 | 79 | -55 | 457 | 50 | 24 | 18 | 144 | 76 | 142 | 106 |
| -14.59473348 | 65.1890503 | 39 | 44 | 256 | 4508 | 130 | -42 | 172 | 50 | 78 | 107 | -12 | 718 | 87 | 32 | 30 | 245 | 102 | 144 | 240 |
| 14.59445571 | 65.1890503 | -5 | 52 | 192 | 7515 | 143 | -128 | 271 | 100 | 15 | 100 | -102 | 746 | 85 | 33 | 25 | 229 | 105 | 229 | 181 |
| 19.45932057 | 65.1890503 | 6 | 67 | 205 | 8967 | 184 | -143 | 327 | 126 | -78 | 134 | -110 | 669 | 95 | 34 | 34 | 263 | 103 | 252 | 114 |
| 29.1890503 | 65.1890503 | 12 | 61 | 182 | 9340 | 194 | -142 | 336 | 136 | -79 | 145 | -114 | 583 | 72 | 29 | 29 | 214 | 93 | 196 | 96 |
| 34.05391517 | 65.1890503 | 16 | 57 | 173 | 9147 | 195 | -132 | 327 | 140 | -75 | 146 | -107 | 506 | 74 | 20 | 41 | 211 | 64 | 184 | 66 |
| 43.78364489 | 65.1890503 | 5 | 63 | 173 | 9914 | 202 | -163 | 365 | 136 | -128 | 146 | -128 | 577 | 73 | 26 | 34 | 215 | 84 | 202 | 84 |
| 48.64850976 | 65.1890503 | -9 | 64 | 166 | 10435 | 196 | -188 | 384 | 127 | -150 | 140 | -150 | 577 | 70 | 25 | 30 | 203 | 84 | 185 | 84 |
| 53.51337462 | 65.1890503 | -9 | 65 | 162 | 10996 | 204 | -195 | 400 | 134 | -103 | 148 | -157 | 545 | 70 | 27 | 29 | 201 | 83 | 176 | 88 |
| 58.37823949 | 65.1890503 | -29 | 64 | 151 | 11919 | 196 | -228 | 424 | 77 | -129 | 140 | -190 | 714 | 86 | 38 | 28 | 253 | 118 | 208 | 122 |
| 63.24310435 | 65.1890503 | -42 | 67 | 149 | 12809 | 195 | -253 | 448 | 126 | -211 | 139 | -213 | 535 | 77 | 21 | 44 | 231 | 67 | 214 | 77 |
| 68.10796922 | 65.1890503 | -43 | 67 | 150 | 12903 | 198 | -249 | 447 | 130 | -211 | 141 | -211 | 518 | 72 | 20 | 43 | 216 | 66 | 207 | 66 |
| 72.97283408 | 65.1890503 | -49 | 66 | 146 | 13260 | 198 | -257 | 455 | 132 | -220 | 140 | -220 | 497 | 73 | 18 | 44 | 207 | 61 | 203 | 61 |
| 77.83769895 | 65.1890503 | -54 | 66 | 139 | 13886 | 203 | -271 | 474 | 137 | -231 | 144 | -231 | 490 | 67 | 20 | 40 | 199 | 66 | 184 | 66 |
| 82.70256381 | 65.1890503 | -54 | 65 | 136 | 14187 | 206 | -274 | 480 | 140 | -231 | 149 | -236 | 496 | 66 | 20 | 37 | 190 | 68 | 173 | 89 |
| 87.56742868 | 65.1890503 | -53 | 63 | 128 | 14765 | 212 | -281 | 493 | 85 | -236 | 156 | -245 | 528 | 73 | 20 | 38 | 209 | 70 | 164 | 100 |
| 92.43229354 | 65.1890503 | -80 | 62 | 118 | 16014 | 194 | -333 | 527 | 130 | -209 | 140 | -295 | 581 | 80 | 23 | 40 | 238 | 72 | 224 | 118 |
| 97.29715841 | 65.1890503 | -96 | 64 | 114 | 17215 | 191 | -369 | 560 | 127 | -234 | 136 | -331 | 344 | 56 | 11 | 50 | 160 | 35 | 139 | 52 |
| 102.1620233 | 65.1890503 | -98 | 69 | 120 | 17454 | 191 | -383 | 573 | 124 | -234 | 135 | -341 | 420 | 71 | 12 | 60 | 211 | 36 | 199 | 54 |
| 107.0268881 | 65.1890503 | -93 | 71 | 124 | 17321 | 197 | -377 | 574 | 126 | -319 | 139 | -336 | 337 | 67 | 7 | 69 | 183 | 23 | 154 | 31 |
| 111.891753 | 65.1890503 | -95 | 71 | 122 | 17892 | 205 | -379 | 584 | 132 | -324 | 146 | -341 | 283 | 49 | 6 | 64 | 147 | 21 | 139 | 29 |
| 116.7566179 | 65.1890503 | -94 | 74 | 123 | 18578 | 216 | -382 | 598 | 157 | -240 | 157 | -347 | 215 | 45 | 5 | 74 | 124 | 16 | 124 | 23 |
| 121.6214827 | 65.1890503 | -104 | 73 | 116 | 19901 | 219 | -409 | 628 | 160 | -352 | 160 | -378 | 275 | 49 | 8 | 57 | 133 | 26 | 133 | 33 |
| 126.4863476 | 65.1890503 | -112 | 70 | 116 | 19292 | 204 | -394 | 598 | 133 | -348 | 146 | -368 | 366 | 56 | 10 | 57 | 157 | 30 | 138 | 35 |
| 131.3512125 | 65.1890503 | -135 | 69 | 118 | 19061 | 175 | -412 | 587 | 120 | -372 | 120 | -384 | 338 | 79 | 5 | 92 | 223 | 16 | 223 | 19 |
| 136.2160773 | 65.1890503 | -134 | 69 | 115 | 19815 | 179 | -424 | 603 | 111 | -384 | 128 | -394 | 238 | 50 | 6 | 79 | 142 | 18 | 139 | 20 |
| 141.0809422 | 65.1890503 | -152 | 65 | 113 | 18796 | 148 | -427 | 575 | 84 | -308 | 100 | -399 | 224 | 57 | 4 | 93 | 151 | 13 | 150 | 17 |
| 145.9458071 | 65.1890503 | -174 | 66 | 124 | 17111 | 112 | -418 | 529 | 45 | -320 | 60 | -392 | 248 | 61 | 5 | 88 | 161 | 17 | 158 | 26 |
| 150.8106719 | 65.1890503 | -94 | 68 | 116 | 18975 | 217 | -373 | 590 | 141 | -137 | 162 | -344 | 298 | 52 | 10 | 50 | 140 | 32 | 137 | 63 |
| 155.6755368 | 65.1890503 | -96 | 68 | 114 | 18782 | 208 | -383 | 591 | 154 | -136 | 154 | -351 | 232 | 45 | 7 | 53 | 113 | 23 | 113 | 51 |
| 160.5404017 | 65.1890503 | -92 | 63 | 117 | 17196 | 194 | -346 | 539 | 126 | -145 | 144 | -315 | 257 | 50 | 8 | 59 | 136 | 25 | 124 | 50 |
| 165.4052665 | 65.1890503 | -121 | 59 | 122 | 15288 | 141 | -342 | 482 | 80 | -178 | 93 | -312 | 271 | 40 | 11 | 42 | 116 | 34 | 110 | 63 |
| 170.2701314 | 65.1890503 | -76 | 62 | 131 | 14929 | 184 | -291 | 475 | 124 | -146 | 137 | -257 | 396 | 61 | 18 | 39 | 176 | 56 | 141 | 77 |
| 175.1349962 | 65.1890503 | -66 | 63 | 144 | 13446 | 174 | -265 | 439 | 117 | -33 | 124 | -230 | 358 | 52 | 14 | 40 | 156 | 46 | 127 | 91 |
| -160.5406794 | 67.54040165 | -87 | 67 | 160 | 13028 | 147 | -274 | 420 | 88 | -46 | 102 | -239 | 386 | 103 | 10 | 85 | 272 | 35 | 153 | 49 |
| -155.6758146 | 67.54040165 | -115 | 66 | 158 | 12999 | 117 | -302 | 418 | 56 | -61 | 73 | -269 | 472 | 112 | 13 | 73 | 305 | 44 | 200 | 78 |
| -150.8109497 | 67.54040165 | -86 | 64 | 145 | 13796 | 156 | -282 | 438 | 93 | -132 | 113 | -250 | 385 | 71 | 11 | 57 | 197 | 35 | 155 | 70 |
| -145.9460848 | 67.54040165 | -62 | 66 | 151 | 13804 | 181 | -259 | 440 | 116 | -107 | 137 | -227 | 170 | 31 | 6 | 54 | 86 | 20 | 67 | 32 |
| -141.08122 | 67.54040165 | -82 | 70 | 154 | 14099 | 165 | -286 | 451 | 100 | -130 | 118 | -254 | 268 | 54 | 7 | 70 | 153 | 22 | 129 | 33 |
| -136.2163551 | 67.54040165 | -127 | 71 | 161 | 13681 | 117 | -326 | 443 | 52 | -292 | 64 | -296 | 417 | 64 | 18 | 48 | 191 | 54 | 172 | 58 |
| -131.3514902 | 67.54040165 | -68 | 74 | 150 | 15345 | 201 | -291 | 492 | 132 | -129 | 146 | -257 | 236 | 43 | 10 | 54 | 121 | 30 | 108 | 40 |
| -126.4866254 | 67.54040165 | -81 | 72 | 148 | 15044 | 188 | -297 | 485 | 118 | -152 | 131 | -266 | 201 | 36 | 9 | 51 | 103 | 27 | 81 | 31 |
| -121.6217605 | 67.54040165 | -104 | 67 | 137 | 15120 | 166 | -323 | 489 | 103 | -291 | 108 | -291 | 202 | 40 | 6 | 62 | 113 | 20 | 85 | 20 |
| -116.7568956 | 67.54040165 | -103 | 70 | 144 | 14854 | 168 | -320 | 488 | 101 | -288 | 103 | -288 | 225 | 41 | 7 | 62 | 122 | 21 | 99 | 21 |
| -111.8920308 | 67.54040165 | -111 | 48 | 119 | 13175 | 104 | -301 | 406 | 70 | -100 | 70 | -275 | 211 | 45 | 7 | 69 | 123 | 22 | 123 | 24 |
| -107.0271659 | 67.54040165 | -132 | 67 | 131 | 15670 | 145 | -368 | 513 | 90 | -329 | 90 | -329 | 185 | 40 | 5 | 71 | 112 | 17 | 112 | 17 |
| -102.1623011 | 67.54040165 | -131 | 64 | 125 | 15690 | 143 | -366 | 509 | 89 | -331 | 89 | -331 | 152 | 37 | 3 | 86 | 104 | 9 | 104 | 9 |
| -97.29743619 | 67.54040165 | -132 | 65 | 124 | 15871 | 148 | -376 | 524 | 88 | -341 | 88 | -341 | 174 | 33 | 4 | 72 | 99 | 12 | 99 | 12 |
| -92.43257132 | 67.54040165 | -145 | 63 | 123 | 15455 | 131 | -383 | 514 | 70 | -349 | 70 | -349 | 229 | 50 | 5 | 74 | 138 | 15 | 138 | 15 |
| -82.70284159 | 67.54040165 | -134 | 50 | 111 | 13892 | 101 | -346 | 447 | 55 | -320 | 55 | -320 | 244 | 47 | 6 | 61 | 131 | 20 | 131 | 20 |
| -68.108247 | 67.54040165 | -125 | 49 | 121 | 12245 | 92 | -313 | 404 | 47 | -286 | 47 | -286 | 296 | 51 | 6 | 60 | 144 | 20 | 144 | 20 |
| -53.5136524 | 67.54040165 | -45 | 41 | 140 | 8339 | 98 | -191 | 289 | 38 | -160 | 68 | -160 | 301 | 43 | 13 | 43 | 127 | 39 | 120 | 39 |
| -48.64878754 | 67.54040165 | -113 | 51 | 193 | 7752 | 30 | -237 | 267 | -43 | -157 | -2 | -205 | 249 | 34 | 12 | 35 | 96 | 39 | 75 | 40 |
| -43.78392267 | 67.54040165 | -190 | 58 | 205 | 8194 | -31 | -315 | 284 | -124 | -282 | -68 | -283 | 276 | 32 | 17 | 24 | 96 | 52 | 78 | 55 |
| -38.91905781 | 67.54040165 | -187 | 55 | 212 | 7374 | -36 | -297 | 261 | -131 | -101 | -73 | -267 | 358 | 37 | 20 | 16 | 106 | 68 | 78 | 95 |
| -34.05419294 | 67.54040165 | -97 | 43 | 185 | 6892 | 35 | -197 | 231 | -35 | -79 | 5 | -175 | 745 | 95 | 35 | 26 | 255 | 150 | 150 | 188 |
| 14.59445571 | 67.54040165 | 54 | 26 | 149 | 5269 | 152 | -25 | 176 | 72 | 97 | 134 | -9 | 1205 | 157 | 69 | 27 | 455 | 211 | 260 | 283 |
| 19.45932057 | 67.54040165 | -11 | 56 | 180 | 8740 | 159 | -150 | 309 | 110 | -95 | 114 | -121 | 527 | 88 | 24 | 44 | 236 | 73 | 226 | 79 |
| 24.32418544 | 67.54040165 | 4 | 61 | 180 | 9479 | 188 | -149 | 337 | 132 | -85 | 140 | -119 | 581 | 86 | 30 | 34 | 228 | 94 | 225 | 108 |
| 29.1890503 | 67.54040165 | 0 | 59 | 178 | 9214 | 180 | -150 | 331 | 132 | -30 | 132 | -122 | 532 | 69 | 28 | 32 | 197 | 85 | 197 | 90 |
| 34.05391517 | 67.54040165 | -1 | 57 | 180 | 8875 | 174 | -146 | 320 | 120 | -90 | 125 | -120 | 488 | 68 | 22 | 38 | 194 | 70 | 189 | 71 |
| 38.91878003 | 67.54040165 | -10 | 52 | 181 | 7935 | 147 | -142 | 289 | 100 | -91 | 100 | -117 | 484 | 62 | 22 | 35 | 183 | 70 | 163 | 72 |
| 48.64850976 | 67.54040165 | -21 | 50 | 154 | 9059 | 147 | -175 | 322 | 74 | -112 | 104 | -146 | 442 | 56 | 22 | 35 | 167 | 68 | 155 | 80 |
| 53.51337462 | 67.54040165 | -28 | 60 | 160 | 10405 | 175 | -201 | 375 | 114 | -129 | 118 | -168 | 438 | 61 | 21 | 34 | 170 | 65 | 137 | 68 |
| 58.37823949 | 67.54040165 | -42 | 61 | 155 | 11115 | 175 | -220 | 395 | 110 | -95 | 116 | -189 | 392 | 57 | 19 | 36 | 158 | 57 | 123 | 69 |
| 63.24310435 | 67.54040165 | -57 | 60 | 145 | 11807 | 172 | -241 | 413 | 63 | -113 | 112 | -211 | 572 | 65 | 29 | 22 | 191 | 95 | 164 | 123 |
| 68.10796922 | 67.54040165 | -62 | 63 | 144 | 12621 | 176 | -260 | 436 | 114 | -227 | 115 | -227 | 439 | 67 | 18 | 46 | 194 | 58 | 169 | 58 |
| 77.83769895 | 67.54040165 | -75 | 65 | 139 | 13940 | 184 | -284 | 469 | 123 | -250 | 123 | -250 | 393 | 60 | 18 | 40 | 169 | 58 | 140 | 58 |
| 82.70256381 | 67.54040165 | -77 | 65 | 135 | 14333 | 188 | -290 | 478 | 126 | -254 | 127 | -254 | 434 | 57 | 18 | 38 | 166 | 60 | 156 | 60 |
| 87.56742868 | 67.54040165 | -75 | 62 | 125 | 15044 | 198 | -297 | 495 | 76 | -260 | 138 | -262 | 542 | 67 | 26 | 32 | 193 | 85 | 165 | 108 |
| 92.43229354 | 67.54040165 | -111 | 56 | 111 | 15492 | 162 | -342 | 504 | 103 | -308 | 103 | -310 | 522 | 79 | 19 | 46 | 229 | 64 | 229 | 84 |
| 97.29715841 | 67.54040165 | -117 | 59 | 113 | 16175 | 160 | -362 | 522 | 104 | -325 | 104 | -327 | 332 | 64 | 8 | 70 | 186 | 26 | 186 | 33 |
| 102.1620233 | 67.54040165 | -89 | 62 | 118 | 16250 | 185 | -338 | 523 | 127 | -297 | 132 | -302 | 264 | 51 | 7 | 66 | 145 | 23 | 126 | 30 |
| 107.0268881 | 67.54040165 | -95 | 62 | 118 | 16364 | 181 | -347 | 529 | 120 | -304 | 129 | -311 | 290 | 54 | 7 | 70 | 159 | 22 | 149 | 26 |
| 111.891753 | 67.54040165 | -111 | 63 | 112 | 17402 | 181 | -377 | 559 | 127 | -332 | 127 | -341 | 327 | 58 | 10 | 59 | 162 | 31 | 162 | 38 |
| 116.7566179 | 67.54040165 | -112 | 64 | 111 | 18110 | 189 | -387 | 577 | 135 | -341 | 135 | -353 | 279 | 48 | 8 | 63 | 140 | 25 | 140 | 30 |
| 121.6214827 | 67.54040165 | -116 | 65 | 109 | 19092 | 195 | -406 | 601 | 125 | -356 | 140 | -375 | 263 | 42 | 8 | 55 | 120 | 25 | 118 | 30 |
| 126.4863476 | 67.54040165 | -138 | 62 | 113 | 17856 | 161 | -393 | 554 | 97 | -355 | 107 | -367 | 421 | 70 | 11 | 61 | 205 | 33 | 186 | 37 |
| 131.3512125 | 67.54040165 | -131 | 70 | 119 | 18979 | 179 | -409 | 588 | 125 | -292 | 125 | -378 | 198 | 42 | 6 | 72 | 116 | 20 | 116 | 23 |
| 136.2160773 | 67.54040165 | -117 | 67 | 112 | 19719 | 196 | -403 | 599 | 145 | -284 | 145 | -376 | 210 | 45 | 5 | 75 | 122 | 16 | 122 | 21 |
| 141.0809422 | 67.54040165 | -120 | 64 | 116 | 18305 | 180 | -378 | 557 | 114 | -166 | 129 | -351 | 250 | 53 | 6 | 73 | 149 | 20 | 139 | 34 |
| 145.9458071 | 67.54040165 | -113 | 64 | 113 | 18476 | 192 | -374 | 566 | 124 | -281 | 140 | -346 | 281 | 55 | 8 | 63 | 149 | 27 | 148 | 45 |
| 150.8106719 | 67.54040165 | -117 | 63 | 111 | 18211 | 186 | -379 | 566 | 115 | -176 | 132 | -349 | 207 | 33 | 6 | 48 | 97 | 20 | 89 | 43 |
| 155.6755368 | 67.54040165 | -103 | 63 | 113 | 17672 | 189 | -366 | 555 | 118 | -155 | 136 | -335 | 240 | 44 | 8 | 54 | 118 | 25 | 114 | 46 |
| 160.5404017 | 67.54040165 | -122 | 64 | 118 | 17143 | 163 | -378 | 541 | 93 | -174 | 112 | -346 | 238 | 44 | 8 | 55 | 116 | 26 | 114 | 45 |
| 165.4052665 | 67.54040165 | -101 | 61 | 122 | 15873 | 165 | -334 | 499 | 102 | -159 | 120 | -302 | 253 | 46 | 7 | 51 | 120 | 24 | 118 | 56 |
| 170.2701314 | 67.54040165 | -129 | 60 | 129 | 14682 | 127 | -338 | 465 | 66 | -197 | 80 | -310 | 274 | 51 | 9 | 60 | 151 | 29 | 128 | 43 |
| 175.1349962 | 67.54040165 | -109 | 59 | 133 | 13948 | 136 | -309 | 445 | 81 | -249 | 91 | -279 | 261 | 57 | 8 | 68 | 151 | 28 | 141 | 45 |
| -160.5406794 | 69.891753 | -102 | 56 | 131 | 13476 | 132 | -297 | 429 | 80 | -265 | 86 | -265 | 171 | 61 | 3 | 123 | 164 | 10 | 93 | 10 |
| -155.6758146 | 69.891753 | -108 | 65 | 136 | 14683 | 156 | -323 | 479 | 90 | -76 | 100 | -287 | 152 | 36 | 4 | 80 | 103 | 14 | 75 | 17 |
| -150.8109497 | 69.891753 | -98 | 57 | 133 | 13489 | 137 | -289 | 426 | 83 | -250 | 89 | -261 | 139 | 33 | 3 | 81 | 95 | 11 | 71 | 16 |
| -145.9460848 | 69.891753 | -90 | 58 | 139 | 13378 | 139 | -280 | 419 | 88 | -237 | 97 | -251 | 149 | 35 | 4 | 81 | 100 | 12 | 75 | 15 |
| -131.3514902 | 69.891753 | -98 | 34 | 97 | 12176 | 88 | -266 | 354 | 66 | -238 | 66 | -245 | 137 | 26 | 6 | 56 | 71 | 19 | 71 | 21 |
| -116.7568956 | 69.891753 | -112 | 46 | 115 | 12855 | 105 | -296 | 401 | 64 | -271 | 64 | -271 | 160 | 30 | 5 | 65 | 87 | 15 | 87 | 15 |
| -111.8920308 | 69.891753 | -140 | 66 | 130 | 15523 | 141 | -368 | 509 | 79 | -332 | 79 | -332 | 114 | 25 | 2 | 89 | 75 | 6 | 62 | 6 |
| -107.0271659 | 69.891753 | -142 | 63 | 123 | 15640 | 139 | -371 | 510 | 79 | -338 | 79 | -338 | 127 | 27 | 3 | 76 | 78 | 9 | 78 | 9 |
| -102.1623011 | 69.891753 | -139 | 45 | 104 | 13868 | 90 | -341 | 431 | 49 | -317 | 49 | -317 | 113 | 26 | 3 | 78 | 71 | 9 | 71 | 9 |
| -97.29743619 | 69.891753 | -136 | 36 | 91 | 13054 | 71 | -323 | 394 | 38 | -304 | 38 | -304 | 127 | 24 | 4 | 64 | 66 | 12 | 66 | 12 |
| -92.43257132 | 69.891753 | -148 | 47 | 103 | 14459 | 90 | -365 | 455 | 47 | -340 | 47 | -340 | 172 | 32 | 6 | 58 | 87 | 18 | 87 | 18 |
| -72.97311186 | 69.891753 | -172 | 49 | 116 | 13030 | 53 | -364 | 418 | 12 | -338 | 12 | -338 | 235 | 51 | 5 | 76 | 141 | 15 | 141 | 15 |
| -68.108247 | 69.891753 | -120 | 31 | 90 | 11656 | 50 | -295 | 345 | -43 | -264 | 32 | -275 | 230 | 38 | 7 | 52 | 108 | 22 | 80 | 26 |
| -53.5136524 | 69.891753 | -85 | 47 | 157 | 8667 | 78 | -223 | 301 | -4 | -192 | 41 | -192 | 293 | 45 | 12 | 45 | 118 | 38 | 113 | 40 |
| -48.64878754 | 69.891753 | -124 | 48 | 183 | 7795 | 19 | -242 | 261 | -56 | -168 | -9 | -212 | 294 | 39 | 16 | 30 | 107 | 49 | 82 | 51 |
| -43.78392267 | 69.891753 | -202 | 56 | 198 | 8327 | -42 | -323 | 281 | -137 | -243 | -76 | -293 | 295 | 33 | 18 | 20 | 96 | 57 | 77 | 58 |
| -38.91905781 | 69.891753 | -242 | 61 | 203 | 8660 | -70 | -369 | 299 | -177 | -336 | -109 | -336 | 189 | 22 | 10 | 24 | 66 | 34 | 55 | 34 |
| -34.05419294 | 69.891753 | -217 | 54 | 191 | 8231 | -51 | -336 | 285 | -99 | -292 | -89 | -307 | 213 | 34 | 10 | 37 | 84 | 36 | 59 | 47 |
| -29.18932808 | 69.891753 | -154 | 53 | 183 | 8572 | 17 | -272 | 289 | -31 | -127 | -20 | -246 | 379 | 52 | 21 | 26 | 134 | 65 | 92 | 93 |
| -24.32446321 | 69.891753 | -137 | 38 | 168 | 6907 | -14 | -240 | 226 | -119 | -67 | -39 | -217 | 569 | 66 | 30 | 24 | 189 | 90 | 110 | 156 |
| 19.45932057 | 69.891753 | 35 | 33 | 162 | 5915 | 144 | -59 | 203 | 54 | 85 | 122 | -39 | 783 | 102 | 39 | 29 | 281 | 123 | 169 | 220 |
| 24.32418544 | 69.891753 | -10 | 50 | 181 | 7777 | 144 | -134 | 279 | 101 | -44 | 102 | -108 | 529 | 65 | 27 | 29 | 195 | 84 | 175 | 106 |
| 29.1890503 | 69.891753 | -12 | 57 | 188 | 8438 | 153 | -152 | 306 | 105 | -43 | 107 | -122 | 503 | 69 | 25 | 36 | 205 | 76 | 190 | 91 |
| 68.10796922 | 69.891753 | -77 | 57 | 145 | 11487 | 133 | -259 | 392 | 52 | -148 | 86 | -227 | 325 | 47 | 17 | 39 | 139 | 51 | 129 | 57 |
| 77.83769895 | 69.891753 | -94 | 63 | 138 | 13588 | 156 | -298 | 454 | 55 | -230 | 102 | -263 | 344 | 43 | 19 | 28 | 128 | 58 | 122 | 65 |
| 82.70256381 | 69.891753 | -92 | 63 | 135 | 14083 | 167 | -297 | 464 | 113 | -80 | 113 | -264 | 379 | 46 | 19 | 27 | 133 | 59 | 133 | 80 |
| 87.56742868 | 69.891753 | -89 | 60 | 128 | 14412 | 176 | -292 | 469 | 64 | -71 | 121 | -263 | 513 | 63 | 29 | 27 | 183 | 88 | 180 | 95 |
| 92.43229354 | 69.891753 | -135 | 56 | 124 | 14037 | 122 | -334 | 456 | 70 | -306 | 70 | -306 | 394 | 64 | 13 | 55 | 190 | 42 | 190 | 42 |
| 97.29715841 | 69.891753 | -142 | 61 | 128 | 14692 | 121 | -357 | 479 | 68 | -279 | 68 | -324 | 448 | 77 | 16 | 59 | 229 | 48 | 229 | 50 |
| 102.1620233 | 69.891753 | -124 | 62 | 117 | 16402 | 159 | -368 | 527 | 102 | -332 | 104 | -332 | 286 | 52 | 10 | 60 | 149 | 30 | 131 | 30 |
| 107.0268881 | 69.891753 | -135 | 59 | 111 | 16562 | 150 | -378 | 528 | 92 | -342 | 97 | -342 | 340 | 63 | 10 | 67 | 187 | 30 | 173 | 30 |
| 111.891753 | 69.891753 | -123 | 60 | 107 | 17552 | 175 | -383 | 558 | 122 | -342 | 122 | -345 | 254 | 46 | 8 | 57 | 126 | 26 | 126 | 32 |
| 116.7566179 | 69.891753 | -116 | 58 | 107 | 17314 | 173 | -374 | 548 | 122 | -334 | 122 | -340 | 236 | 36 | 8 | 50 | 103 | 26 | 103 | 33 |
| 121.6214827 | 69.891753 | -126 | 59 | 106 | 17931 | 168 | -394 | 561 | 108 | -352 | 117 | -360 | 236 | 35 | 9 | 49 | 101 | 28 | 97 | 33 |
| 126.4863476 | 69.891753 | -123 | 57 | 104 | 17646 | 167 | -377 | 544 | 111 | -341 | 118 | -347 | 387 | 57 | 16 | 51 | 166 | 48 | 144 | 52 |
| 131.3512125 | 69.891753 | -138 | 61 | 115 | 17307 | 151 | -384 | 535 | 93 | -198 | 99 | -353 | 317 | 65 | 10 | 74 | 191 | 30 | 176 | 40 |
| 136.2160773 | 69.891753 | -141 | 63 | 113 | 17996 | 155 | -401 | 556 | 103 | -306 | 103 | -370 | 289 | 62 | 7 | 72 | 165 | 24 | 165 | 34 |
| 141.0809422 | 69.891753 | -142 | 62 | 116 | 17393 | 147 | -389 | 535 | 88 | -304 | 96 | -357 | 241 | 54 | 6 | 74 | 144 | 21 | 138 | 27 |
| 145.9458071 | 69.891753 | -129 | 59 | 110 | 17201 | 157 | -374 | 532 | 97 | -194 | 106 | -345 | 240 | 42 | 8 | 56 | 118 | 24 | 114 | 33 |
| 150.8106719 | 69.891753 | -140 | 59 | 116 | 16343 | 139 | -371 | 510 | 77 | -213 | 86 | -343 | 207 | 35 | 6 | 54 | 101 | 19 | 94 | 33 |
| 155.6755368 | 69.891753 | -123 | 59 | 120 | 15553 | 144 | -347 | 491 | 81 | -193 | 89 | -318 | 150 | 28 | 5 | 55 | 74 | 15 | 72 | 25 |
| -121.6217605 | 72.24310435 | -154 | 61 | 121 | 15787 | 116 | -389 | 506 | 61 | -337 | 68 | -350 | 132 | 27 | 3 | 75 | 79 | 10 | 63 | 11 |
| -116.7568956 | 72.24310435 | -157 | 64 | 121 | 16388 | 127 | -404 | 531 | 68 | -363 | 75 | -363 | 144 | 31 | 4 | 80 | 93 | 12 | 76 | 12 |
| -111.8920308 | 72.24310435 | -165 | 58 | 117 | 15315 | 104 | -390 | 494 | 48 | -355 | 54 | -355 | 157 | 37 | 3 | 87 | 106 | 9 | 81 | 9 |
| -107.0271659 | 72.24310435 | -161 | 56 | 114 | 15391 | 105 | -390 | 495 | 52 | -354 | 56 | -354 | 104 | 28 | 2 | 89 | 72 | 6 | 44 | 6 |
| -97.29743619 | 72.24310435 | -168 | 50 | 105 | 15300 | 85 | -398 | 482 | 43 | -362 | 43 | -362 | 127 | 27 | 3 | 71 | 75 | 10 | 75 | 10 |
| -87.56770646 | 72.24310435 | -169 | 55 | 104 | 16564 | 104 | -425 | 529 | 58 | -388 | 58 | -388 | 159 | 32 | 4 | 71 | 91 | 12 | 91 | 12 |
| -82.70284159 | 72.24310435 | -175 | 56 | 102 | 17061 | 110 | -436 | 546 | 57 | -405 | 58 | -405 | 263 | 49 | 5 | 67 | 145 | 16 | 122 | 16 |
| -77.83797673 | 72.24310435 | -159 | 44 | 95 | 14701 | 86 | -374 | 460 | 44 | -350 | 44 | -350 | 209 | 39 | 5 | 65 | 116 | 16 | 116 | 16 |
| -53.5136524 | 72.24310435 | -124 | 48 | 159 | 9339 | 52 | -254 | 306 | -44 | -224 | 18 | -228 | 219 | 32 | 9 | 42 | 86 | 29 | 73 | 34 |
| -48.64878754 | 72.24310435 | -190 | 47 | 170 | 8397 | -33 | -308 | 275 | -124 | -236 | -62 | -279 | 238 | 28 | 13 | 26 | 80 | 39 | 74 | 40 |
| -43.78392267 | 72.24310435 | -239 | 54 | 181 | 9031 | -67 | -364 | 297 | -113 | -334 | -100 | -334 | 201 | 23 | 11 | 25 | 67 | 35 | 61 | 35 |
| -38.91905781 | 72.24310435 | -267 | 57 | 181 | 9536 | -84 | -399 | 315 | -133 | -361 | -119 | -368 | 124 | 16 | 7 | 29 | 45 | 21 | 40 | 22 |
| -34.05419294 | 72.24310435 | -255 | 57 | 180 | 9466 | -70 | -388 | 319 | -122 | -299 | -107 | -356 | 100 | 17 | 5 | 47 | 46 | 15 | 37 | 17 |
| -29.18932808 | 72.24310435 | -174 | 49 | 162 | 8935 | 3 | -297 | 299 | -45 | -146 | -33 | -271 | 126 | 16 | 6 | 26 | 44 | 19 | 34 | 32 |
| -24.32446321 | 72.24310435 | -111 | 43 | 135 | 10084 | 66 | -253 | 319 | 35 | -89 | 37 | -225 | 219 | 26 | 9 | 28 | 71 | 29 | 50 | 65 |
| 53.51337462 | 72.24310435 | -71 | 47 | 160 | 8510 | 84 | -210 | 294 | 27 | -133 | 50 | -181 | 326 | 48 | 14 | 45 | 142 | 44 | 137 | 52 |
| 77.83769895 | 72.24310435 | -99 | 44 | 121 | 11594 | 89 | -273 | 362 | 39 | -94 | 65 | -244 | 306 | 37 | 17 | 22 | 107 | 53 | 95 | 73 |
| 82.70256381 | 72.24310435 | -110 | 52 | 124 | 13062 | 115 | -305 | 421 | 37 | -102 | 77 | -272 | 335 | 38 | 15 | 28 | 113 | 45 | 105 | 95 |
| 87.56742868 | 72.24310435 | -111 | 57 | 126 | 13965 | 141 | -313 | 454 | 92 | -97 | 92 | -280 | 331 | 41 | 17 | 26 | 119 | 51 | 119 | 82 |
| 92.43229354 | 72.24310435 | -111 | 56 | 120 | 14545 | 150 | -317 | 467 | 100 | -92 | 100 | -287 | 322 | 38 | 17 | 23 | 111 | 53 | 111 | 69 |
| 97.29715841 | 72.24310435 | -118 | 57 | 118 | 15208 | 151 | -333 | 484 | 100 | -304 | 100 | -304 | 272 | 38 | 15 | 33 | 107 | 46 | 107 | 46 |
| 102.1620233 | 72.24310435 | -121 | 57 | 116 | 15609 | 152 | -344 | 496 | 100 | -314 | 100 | -314 | 268 | 37 | 12 | 41 | 110 | 36 | 110 | 36 |
| 107.0268881 | 72.24310435 | -133 | 56 | 114 | 15725 | 139 | -356 | 495 | 88 | -327 | 88 | -327 | 231 | 39 | 9 | 50 | 109 | 29 | 104 | 29 |
| 111.891753 | 72.24310435 | -137 | 59 | 113 | 16684 | 147 | -378 | 525 | 93 | -346 | 94 | -346 | 177 | 32 | 6 | 59 | 90 | 18 | 86 | 18 |
| 116.7566179 | 72.24310435 | -136 | 58 | 112 | 16665 | 145 | -375 | 520 | 92 | -345 | 93 | -345 | 192 | 30 | 8 | 51 | 90 | 25 | 84 | 25 |
| 121.6214827 | 72.24310435 | -138 | 56 | 110 | 16503 | 136 | -375 | 510 | 32 | -211 | 86 | -345 | 234 | 31 | 12 | 35 | 92 | 37 | 87 | 39 |
| 126.4863476 | 72.24310435 | -141 | 58 | 112 | 16830 | 134 | -384 | 518 | 87 | -223 | 87 | -353 | 291 | 46 | 13 | 44 | 134 | 40 | 134 | 49 |
| 141.0809422 | 72.24310435 | -136 | 50 | 109 | 15055 | 107 | -354 | 462 | 64 | -295 | 64 | -326 | 150 | 30 | 5 | 67 | 88 | 15 | 88 | 17 |
| 145.9458071 | 72.24310435 | -132 | 42 | 103 | 13735 | 82 | -329 | 411 | 48 | -202 | 48 | -306 | 189 | 31 | 7 | 52 | 91 | 21 | 91 | 26 |
| -111.8920308 | 74.59445571 | -153 | 33 | 89 | 12352 | 46 | -324 | 371 | 14 | -304 | 21 | -304 | 105 | 23 | 3 | 71 | 63 | 9 | 44 | 9 |
| -97.29743619 | 74.59445571 | -145 | 27 | 75 | 12023 | 43 | -311 | 354 | 15 | -296 | 20 | -296 | 143 | 28 | 4 | 62 | 75 | 13 | 49 | 13 |
| -87.56770646 | 74.59445571 | -187 | 40 | 89 | 14637 | 45 | -402 | 447 | 11 | -378 | 12 | -378 | 194 | 36 | 6 | 62 | 102 | 19 | 76 | 19 |
| -82.70284159 | 74.59445571 | -166 | 34 | 81 | 14215 | 51 | -372 | 423 | 24 | -351 | 25 | -351 | 270 | 52 | 6 | 62 | 148 | 20 | 112 | 20 |
| -53.5136524 | 74.59445571 | -157 | 40 | 145 | 8603 | 3 | -272 | 275 | -89 | -248 | -23 | -248 | 244 | 38 | 10 | 46 | 108 | 31 | 86 | 31 |
| -48.64878754 | 74.59445571 | -223 | 46 | 157 | 9250 | -50 | -345 | 295 | -92 | -318 | -78 | -318 | 168 | 21 | 8 | 37 | 63 | 24 | 61 | 25 |
| -43.78392267 | 74.59445571 | -260 | 52 | 163 | 9989 | -73 | -394 | 321 | -120 | -361 | -104 | -364 | 120 | 17 | 5 | 42 | 49 | 15 | 45 | 16 |
| -38.91905781 | 74.59445571 | -271 | 55 | 164 | 10395 | -74 | -411 | 336 | -126 | -380 | -108 | -380 | 68 | 10 | 3 | 49 | 30 | 9 | 29 | 9 |
| -34.05419294 | 74.59445571 | -266 | 53 | 160 | 10276 | -69 | -403 | 334 | -122 | -315 | -103 | -374 | 72 | 12 | 3 | 55 | 35 | 9 | 32 | 11 |
| -29.18932808 | 74.59445571 | -229 | 49 | 156 | 9584 | -43 | -359 | 316 | -94 | -196 | -77 | -331 | 108 | 15 | 5 | 34 | 44 | 16 | 36 | 23 |
| -24.32446321 | 74.59445571 | -159 | 42 | 141 | 9252 | 13 | -286 | 299 | -25 | -136 | -16 | -261 | 138 | 18 | 7 | 29 | 51 | 21 | 37 | 39 |
| -19.45959835 | 74.59445571 | -97 | 28 | 114 | 8564 | 40 | -210 | 249 | -186 | -11 | 26 | -190 | 265 | 32 | 9 | 32 | 92 | 28 | 80 | 84 |
| 58.37823949 | 74.59445571 | -92 | 29 | 119 | 7965 | 21 | -220 | 242 | 3 | -95 | 13 | -197 | 251 | 38 | 11 | 37 | 106 | 33 | 87 | 53 |
| 87.56742868 | 74.59445571 | -123 | 44 | 114 | 12257 | 79 | -303 | 381 | 50 | -108 | 50 | -274 | 274 | 34 | 12 | 27 | 97 | 38 | 97 | 77 |
| 92.43229354 | 74.59445571 | -138 | 52 | 119 | 13784 | 99 | -342 | 441 | 59 | -118 | 59 | -308 | 270 | 36 | 13 | 32 | 107 | 39 | 107 | 62 |
| 97.29715841 | 74.59445571 | -145 | 53 | 115 | 14531 | 102 | -361 | 462 | 60 | -327 | 60 | -327 | 253 | 39 | 13 | 41 | 113 | 39 | 113 | 39 |
| 102.1620233 | 74.59445571 | -137 | 53 | 113 | 14722 | 112 | -354 | 465 | 70 | -322 | 70 | -322 | 260 | 36 | 13 | 37 | 107 | 39 | 107 | 39 |
| 107.0268881 | 74.59445571 | -141 | 50 | 111 | 14434 | 102 | -348 | 450 | 62 | -320 | 62 | -320 | 200 | 29 | 9 | 41 | 85 | 29 | 85 | 29 |
| -116.7568956 | 76.94580706 | -155 | 33 | 87 | 12478 | 45 | -332 | 377 | 15 | -294 | 22 | -308 | 107 | 22 | 3 | 66 | 60 | 10 | 41 | 11 |
| -87.56770646 | 76.94580706 | -200 | 37 | 84 | 15021 | 35 | -408 | 443 | 6 | -382 | 11 | -382 | 133 | 26 | 4 | 67 | 74 | 13 | 55 | 13 |
| -82.70284159 | 76.94580706 | -210 | 41 | 86 | 16087 | 43 | -434 | 477 | 10 | -407 | 17 | -407 | 212 | 48 | 4 | 73 | 134 | 14 | 98 | 14 |
| -68.108247 | 76.94580706 | -173 | 39 | 116 | 11154 | 9 | -331 | 340 | -17 | -239 | -12 | -306 | 118 | 22 | 5 | 52 | 61 | 15 | 42 | 18 |
| -63.24338213 | 76.94580706 | -179 | 35 | 116 | 9539 | -10 | -310 | 299 | -40 | -286 | -34 | -288 | 300 | 53 | 11 | 55 | 148 | 34 | 105 | 37 |
| -58.37851727 | 76.94580706 | -193 | 39 | 124 | 9951 | -11 | -322 | 311 | -48 | -299 | -37 | -299 | 321 | 53 | 11 | 55 | 155 | 35 | 125 | 40 |
| -53.5136524 | 76.94580706 | -224 | 41 | 133 | 9888 | -41 | -353 | 311 | -83 | -328 | -68 | -328 | 168 | 28 | 6 | 54 | 78 | 19 | 74 | 21 |
| -48.64878754 | 76.94580706 | -253 | 46 | 140 | 10318 | -63 | -387 | 325 | -91 | -361 | -91 | -361 | 103 | 18 | 3 | 58 | 50 | 10 | 50 | 10 |
| -43.78392267 | 76.94580706 | -270 | 50 | 144 | 10831 | -68 | -412 | 343 | -99 | -383 | -99 | -383 | 59 | 11 | 2 | 65 | 31 | 6 | 31 | 6 |
| -38.91905781 | 76.94580706 | -267 | 51 | 144 | 10983 | -60 | -412 | 352 | -114 | -321 | -92 | -383 | 49 | 8 | 2 | 54 | 24 | 6 | 22 | 8 |
| -34.05419294 | 76.94580706 | -251 | 49 | 139 | 10908 | -45 | -395 | 349 | -100 | -307 | -77 | -367 | 69 | 11 | 3 | 50 | 33 | 9 | 29 | 12 |
| -29.18932808 | 76.94580706 | -221 | 43 | 131 | 10264 | -27 | -355 | 328 | -76 | -276 | -57 | -330 | 95 | 15 | 5 | 44 | 43 | 15 | 38 | 18 |
| -24.32446321 | 76.94580706 | -147 | 40 | 124 | 10316 | 42 | -284 | 326 | 1 | -35 | 13 | -261 | 111 | 14 | 6 | 28 | 41 | 18 | 32 | 30 |
| -19.45959835 | 76.94580706 | -136 | 35 | 116 | 10124 | 30 | -274 | 304 | -240 | -29 | 10 | -249 | 157 | 17 | 5 | 29 | 50 | 17 | 34 | 49 |
| 68.10796922 | 76.94580706 | -101 | 27 | 106 | 8862 | 23 | -234 | 257 | 16 | -102 | 16 | -212 | 282 | 34 | 17 | 19 | 94 | 56 | 94 | 73 |
| 102.1620233 | 76.94580706 | -144 | 45 | 110 | 13295 | 68 | -342 | 410 | 38 | -294 | 38 | -311 | 236 | 34 | 11 | 40 | 101 | 39 | 101 | 42 |
| 107.0268881 | 76.94580706 | -138 | 31 | 92 | 11465 | 36 | -300 | 336 | 17 | -240 | 17 | -279 | 214 | 29 | 11 | 33 | 86 | 37 | 86 | 47 |
| -92.43257132 | 79.29715841 | -221 | 38 | 77 | 16923 | 49 | -442 | 491 | 20 | -401 | 27 | -414 | 107 | 26 | 2 | 90 | 74 | 6 | 54 | 7 |
| -87.56770646 | 79.29715841 | -193 | 38 | 81 | 15720 | 67 | -403 | 470 | 26 | -365 | 38 | -378 | 79 | 16 | 2 | 76 | 48 | 6 | 40 | 7 |
| -82.70284159 | 79.29715841 | -210 | 40 | 88 | 14928 | 40 | -411 | 450 | 0 | -385 | 10 | -385 | 108 | 21 | 3 | 66 | 61 | 10 | 50 | 10 |
| -77.83797673 | 79.29715841 | -181 | 31 | 78 | 13649 | 44 | -356 | 400 | 19 | -334 | 22 | -334 | 100 | 20 | 3 | 68 | 58 | 9 | 44 | 9 |
| -63.24338213 | 79.29715841 | -170 | 36 | 105 | 11221 | 24 | -319 | 343 | -7 | -297 | 1 | -297 | 104 | 21 | 3 | 73 | 59 | 9 | 53 | 9 |
| -58.37851727 | 79.29715841 | -197 | 39 | 119 | 10473 | -7 | -335 | 328 | -33 | -312 | -33 | -312 | 122 | 27 | 2 | 79 | 69 | 7 | 69 | 7 |
| -53.5136524 | 79.29715841 | -217 | 40 | 122 | 10481 | -25 | -353 | 328 | -51 | -330 | -51 | -330 | 108 | 24 | 2 | 74 | 62 | 7 | 62 | 7 |
| -48.64878754 | 79.29715841 | -238 | 41 | 122 | 10689 | -42 | -376 | 333 | -69 | -352 | -69 | -352 | 92 | 20 | 2 | 72 | 53 | 7 | 53 | 7 |
| -43.78392267 | 79.29715841 | -256 | 43 | 123 | 11129 | -52 | -400 | 348 | -79 | -374 | -79 | -374 | 65 | 14 | 2 | 75 | 38 | 6 | 38 | 6 |
| -38.91905781 | 79.29715841 | -255 | 45 | 125 | 11325 | -47 | -403 | 357 | -75 | -376 | -75 | -376 | 51 | 10 | 2 | 65 | 27 | 6 | 27 | 6 |
| -34.05419294 | 79.29715841 | -235 | 44 | 124 | 11174 | -27 | -382 | 355 | -56 | -199 | -56 | -355 | 65 | 11 | 3 | 47 | 29 | 10 | 29 | 12 |
| -29.18932808 | 79.29715841 | -195 | 40 | 117 | 10678 | 4 | -335 | 338 | -40 | -167 | -25 | -310 | 90 | 13 | 4 | 35 | 35 | 14 | 32 | 20 |
| -24.32446321 | 79.29715841 | -172 | 35 | 107 | 10561 | 16 | -311 | 327 | -19 | -59 | -9 | -289 | 113 | 13 | 4 | 28 | 37 | 16 | 32 | 29 |
| -19.45959835 | 79.29715841 | -141 | 29 | 91 | 10760 | 35 | -281 | 316 | -249 | -24 | 18 | -259 | 132 | 14 | 3 | 30 | 40 | 13 | 29 | 38 |
| 14.59445571 | 79.29715841 | -116 | 32 | 144 | 7024 | -2 | -222 | 219 | -16 | -111 | -16 | -197 | 336 | 37 | 18 | 20 | 108 | 55 | 108 | 85 |
| 97.29715841 | 79.29715841 | -160 | 32 | 97 | 11589 | 11 | -325 | 335 | -2 | -149 | -2 | -298 | 190 | 31 | 9 | 43 | 89 | 30 | 89 | 39 |
| -87.56770646 | 81.64850976 | -230 | 33 | 73 | 15886 | 21 | -435 | 456 | -2 | -403 | 2 | -408 | 93 | 21 | 2 | 90 | 61 | 6 | 52 | 7 |
| -82.70284159 | 81.64850976 | -258 | 35 | 79 | 15230 | -7 | -453 | 446 | -35 | -425 | -30 | -425 | 116 | 25 | 3 | 82 | 75 | 10 | 63 | 10 |
| -77.83797673 | 81.64850976 | -229 | 32 | 77 | 14085 | 10 | -409 | 419 | -21 | -379 | -14 | -384 | 78 | 18 | 2 | 86 | 53 | 6 | 43 | 7 |
| -72.97311186 | 81.64850976 | -199 | 38 | 88 | 14365 | 49 | -388 | 437 | 12 | -180 | 21 | -358 | 41 | 8 | 1 | 66 | 24 | 4 | 19 | 6 |
| -68.108247 | 81.64850976 | -214 | 41 | 93 | 14491 | 30 | -409 | 439 | -3 | -377 | 4 | -377 | 77 | 13 | 4 | 50 | 37 | 12 | 31 | 12 |
| -58.37851727 | 81.64850976 | -200 | 41 | 95 | 14205 | 39 | -392 | 431 | 5 | -362 | 13 | -362 | 131 | 28 | 3 | 79 | 74 | 9 | 70 | 9 |
| -53.5136524 | 81.64850976 | -221 | 40 | 101 | 13177 | 3 | -396 | 400 | -22 | -369 | -22 | -369 | 102 | 21 | 3 | 71 | 53 | 9 | 53 | 9 |
| -48.64878754 | 81.64850976 | -194 | 43 | 106 | 13264 | 36 | -370 | 406 | 9 | -341 | 9 | -341 | 155 | 31 | 4 | 66 | 81 | 13 | 81 | 13 |

| | | | | | | | | | | | | | | | | | | | | | |
|---|---|---|---|---|---|---|---|---|---|---|---|---|---|---|---|---|---|---|---|---|---|
| -43.78392267 | 81.64850976 | -193 | 36 | 96 | 12547 | 25 | -353 | 379 | 0 | -330 | 0 | -330 | 135 | 27 | 3 | 72 | 73 | 10 | 73 | 10 |
| -38.91905781 | 81.64850976 | -228 | 37 | 100 | 11949 | -17 | -383 | 367 | -55 | -359 | -43 | -359 | 103 | 18 | 3 | 64 | 52 | 10 | 50 | 10 |
| -34.05419294 | 81.64850976 | -210 | 38 | 95 | 13182 | 21 | -382 | 403 | -98 | -354 | -7 | -354 | 69 | 10 | 3 | 44 | 29 | 10 | 27 | 10 |
| -29.18932808 | 81.64850976 | -231 | 43 | 94 | 14978 | 23 | -430 | 453 | -14 | -197 | -4 | -396 | 81 | 11 | 3 | 36 | 31 | 12 | 28 | 14 |
| -24.32446321 | 81.64850976 | -198 | 37 | 83 | 14942 | 45 | -396 | 441 | 16 | -364 | 24 | -366 | 127 | 16 | 6 | 31 | 46 | 22 | 41 | 22 |
| -14.59473348 | 81.64850976 | -140 | 22 | 75 | 10040 | 20 | -267 | 287 | -46 | -33 | 8 | -250 | 238 | 32 | 11 | 29 | 86 | 35 | 57 | 51 |
| 63.24310435 | 81.64850976 | -138 | 25 | 95 | 9083 | -3 | -264 | 261 | -243 | -48 | -14 | -243 | 246 | 28 | 13 | 21 | 78 | 40 | 57 | 78 |

| country | birth | death | gdp |
|---|---|---|---|
| AFG | 33.314 | 8.034 | 1861.12433 |
| AGO | 44.998 | 13.424 | 6634.40227 |
| ALB | 13.632 | 7.463 | 11504.7482 |
| ARE | 10.592 | 1.641 | 70245.9325 |
| ARG | 17.359 | 7.55 | 20337.716 |
| ARM | 13.008 | 9.316 | 8709.54109 |
| ATG | 16.202 | 6.087 | 21415.708 |
| AUS | 12.7 | 6.6 | 46475.9278 |
| AUT | 9.8 | 9.6 | 49419.3301 |
| AZE | 17.2 | 5.7 | 17779.7737 |
| BDI | 43.497 | 11.066 | 796.994415 |
| BEL | 10.9 | 9.8 | 45608.4293 |
| BEN | 35.592 | 9.244 | 2115.7956 |
| BFA | 39.53 | 9.343 | 1651.23267 |
| BGD | 19.455 | 5.35 | 3335.33726 |
| BGR | 9.2 | 15.3 | 18248.8319 |
| BHR | 14.205 | 2.367 | 47333.848 |
| BHS | 15.044 | 6.234 | 23072.8147 |
| BIH | 8.818 | 10.69 | 11687.0708 |
| BLR | 12.5 | 12.6 | 18344.8161 |
| BLZ | 22.758 | 5.568 | 8583.1383 |
| BOL | 23.591 | 7.359 | 6954.30177 |
| BRA | 14.513 | 6.163 | 15615.3431 |
| BRN | 15.691 | 3.024 | 79429 |
| BTN | 17.404 | 6.218 | 8236.35647 |
| BWA | 24.57 | 7.536 | 16350.4743 |
| CAF | 33.435 | 14.088 | 666.958172 |
| CAN | 10.9 | 7.5 | 44204.9466 |
| CHE | 10.2 | 8.1 | 62499.6372 |
| CHL | 13.048 | 5.137 | 23579.4623 |
| CHN | 12.1 | 7.1 | 14448.2656 |
| CIV | 36.895 | 13.323 | 3461.60381 |
| CMR | 36.233 | 11.197 | 3184.81054 |
| COD | 41.527 | 10.096 | 799.076704 |
| COG | 36.028 | 8.308 | 5901.6826 |
| COL | 15.483 | 5.93 | 13825.9213 |
| COM | 33.433 | 7.38 | 1504.52598 |
| CPV | 21.108 | 5.344 | 6301.69607 |
| CRI | 14.524 | 4.889 | 15879.5968 |
| CYP | 11.206 | 6.897 | 31539.5268 |
| CZE | 10.5 | 10.5 | 33743.1909 |
| DEU | 9 | 11.3 | 47998.8627 |
| DJI | 24.721 | 8.551 | 3342.47733 |
| DMA | NA | NA | 10792.9615 |
| DNK | 10.2 | 9.2 | 48980.7917 |
| DOM | 20.506 | 6.087 | 14237.0579 |
| DZA | 23.692 | 5.114 | 14612.6804 |
| ECU | 20.487 | 5.119 | 11474.1372 |
| EGY | 27.298 | 6.066 | 10749.0946 |
| ESP | 9 | 9 | 34696.3362 |
| EST | 10.6 | 11.6 | 28946.7899 |
| ETH | 31.906 | 7.198 | 1632.34501 |
| FIN | 10.1 | 9.6 | 42275.2399 |
| FJI | 19.688 | 6.982 | 9323.15657 |
| FRA | 12 | 9 | 41178.1418 |
| GAB | 29.718 | 8.391 | 17926.4291 |
| GBR | 11.9 | 9.3 | 41767.2897 |
| GEO | 13.359 | 11.533 | 9609.32603 |
| GHA | 32.262 | 8.821 | 4184.04756 |
| GIN | 36.409 | 9.678 | 1260.68464 |
| GMB | 41.698 | 8.58 | 1691.00974 |
| GNB | 36.647 | 11.9 | 1516.48439 |
| GNQ | 34.626 | 10.594 | 29000.9177 |
| GRC | 8.5 | 11.2 | 26357.938 |
| GRD | 19.019 | 7.117 | 13558.7991 |
| GTM | 26.752 | 5.339 | 7764.77359 |
| GUY | 19.288 | 8.242 | 7520.03423 |
| HKG | 8.2 | 6.3 | 56951.9637 |
| HND | 20.881 | 5.015 | 4590.23694 |
| HRV | 8.9 | 12.9 | 22488.7462 |
| HTI | 24.578 | 8.629 | 1758.11173 |
| HUN | 9.4 | 13.4 | 26436.2078 |
| IDN | 19.579 | 7.167 | 11038.7916 |
| IND | 19.658 | 7.306 | 6126.52266 |
| IRL | 14.2 | 6.4 | 67974.1633 |
| IRN | 17.084 | 4.607 | 17046.4389 |
| IRQ | 34.233 | 5.112 | 15895.2266 |
| ISL | 12.5 | 6.6 | 47690.3816 |
| ISR | 21.3 | 5.3 | 36545.699 |
| ITA | 8 | 10.7 | 37255.17 |
| JAM | 17.23 | 6.914 | 8630.07386 |
| JOR | 26.512 | 3.839 | 9040.64198 |
| JPN | 7.9 | 10.2 | 40686.0244 |
| KAZ | 22.66 | 7.48 | 25044.878 |
| KEN | 34.084 | 8.025 | 3019.19363 |
| KGZ | 27.4 | 5.8 | 3447.17155 |
| KHM | 23.776 | 6.038 | 3503.97435 |
| KOR | 8.6 | 5.4 | 34421.5796 |
| KWT | 19.559 | 2.523 | 73817.0753 |
| LAO | 26.27 | 6.629 | 5785.93143 |
| LBN | 15.397 | 4.547 | 13934.5054 |
| LBR | 34.613 | 8.52 | 836.075571 |
| LCA | 14.933 | 7.257 | 11368.5283 |
| LKA | 15.601 | 6.844 | 11777.8718 |
| LSO | 28.461 | 14.693 | 2956.22201 |
| LTU | 10.8 | 14.1 | 28936.2736 |
| LUX | 10.7 | 7 | 104206.109 |
| LVA | 11.1 | 14.4 | 24919.4788 |
| MAC | 11.681 | 4.816 | 107024.284 |
| MAR | 20.38 | 5.682 | 7757.25318 |
| MDA | 10.566 | 11.363 | 5054.0422 |
| MDG | 34.223 | 6.52 | 1465.42355 |
| MDV | 20.804 | 3.761 | 12770.2713 |
| MEX | 18.47 | 4.814 | 17244.3658 |
| MKD | 11.257 | 9.492 | 14022.7729 |
| MLI | 42.908 | 10.118 | 2043.45923 |
| MLT | 10 | 8 | 35880.6148 |
| MMR | 17.49 | 8.263 | 5399.36051 |
| MNE | 11.441 | 10.035 | 16183.0827 |
| MNG | 23.429 | 6.087 | 12147.9422 |
| MOZ | 38.81 | 11.235 | 1190.60139 |
| MRT | 33.002 | 7.861 | 3834.74526 |
| MUS | 10.1 | 7.7 | 20085.1709 |
| MWI | 38.542 | 7.471 | 1159.17299 |
| MYS | 16.789 | 4.983 | 26606.3433 |
| NAM | 29.414 | 7.033 | 10554.1592 |
| NER | 49.211 | 8.819 | 955.58681 |
| NGA | 39.125 | 12.704 | 6037.69743 |
| NIC | 19.926 | 4.753 | 5282.03008 |
| NLD | 10 | 8.7 | 49546.9564 |
| NOR | 11.4 | 7.8 | 62053.2129 |
| NPL | 20.216 | 6.339 | 2449.82433 |
| NZL | 13.27 | 6.87 | 37948.9196 |
| OMN | 19.322 | 2.666 | 42737.1283 |
| PAK | 28.845 | 7.339 | 4998.75825 |
| PAN | 19.124 | 5.044 | 22012.538 |
| PER | 19.6 | 5.619 | 12529.2282 |
| PHL | 23.316 | 6.773 | 7319.57126 |
| PNG | 28.151 | 7.649 | NA |
| POL | 9.7 | 10.4 | 26855.7747 |
| PRT | 8.3 | 10.5 | 29687.7904 |
| PRY | 21.167 | 5.693 | 9198.50408 |
| PSE | 32.363 | 3.582 | 2865.80511 |
| QAT | 11.806 | 1.486 | 127500.756 |
| ROU | 9.3 | 13.2 | 22070.5305 |
| RUS | 13.3 | 13 | 23702.8445 |
| RWA | 31.163 | 6.833 | 1826.95569 |
| SAU | 19.688 | 3.418 | 54007.0362 |
| SDN | 32.645 | 7.662 | 4568.08615 |
| SEN | 37.522 | 5.99 | 2445.40838 |
| SGP | 9.7 | 4.8 | 86128.1725 |
| SLB | 29.298 | 5.7 | 2186.35699 |
| SLE | 35.445 | 13.268 | 1401.24809 |
| SLV | 17.175 | 6.833 | 8352.97116 |
| SOM | 43.373 | 11.807 | NA |
| SRB | 9.3 | 14.6 | 14111.9386 |
| STP | 33.532 | 6.814 | 3132.83955 |
| SUR | 17.946 | 7.461 | 15722.6476 |
| SVK | 10.3 | 9.9 | 29907.0594 |
| SVN | 10 | 9.6 | 31964.654 |
| SWE | 11.7 | 9.3 | 47823.298 |
| SWZ | 29.372 | 14.485 | 8575.05739 |
| SYC | 17 | 7.5 | 27177.1741 |
| SYR | 22.509 | 5.715 | NA |
| TCD | 44.792 | 13.799 | 2180.17983 |
| TGO | 35.068 | 8.699 | 1438.24836 |
| THA | 10.532 | 8.028 | 16222.9718 |
| TJK | 30.225 | 5.568 | 2811.51371 |
| TLS | 37.097 | 6.809 | 2290.34727 |
| TON | 24.331 | 5.985 | 5524.73679 |
| TTO | 13.979 | 9.507 | 33308.4225 |
| TUN | 17.94 | 6.593 | 11445.6928 |
| TUR | 16.489 | 5.735 | 24054.1778 |
| TZA | 38.535 | 6.682 | 2652.19566 |
| UGA | 42.528 | 9.324 | 1802.08621 |
| UKR | 10.7 | 14.9 | 7948.1421 |
| URY | 14.148 | 9.344 | 21115.1266 |
| USA | 12.4 | 8.2 | 56207.0368 |
| UZB | 23.5 | 4.9 | 6069.20979 |
| VCT | 15.776 | 7.127 | 11140.3499 |
| VEN | 19.268 | 5.567 | NA |
| VNM | 16.941 | 5.847 | 6034.25995 |
| VUT | 26.167 | 4.718 | 2988.47194 |
| WSM | 25.097 | 5.334 | 5918.61003 |
| XKX | 17.1 | 7 | 9685.99066 |
| YEM | 31.889 | 6.829 | 2811.99956 |
| ZAF | 20.415 | 12.409 | 13229.6246 |
| ZMB | 39.672 | 8.671 | 3861.98923 |
| ZWE | 34.521 | 9.086 | 2013.16712 |

```
#______________________________________________________________________________________________

#----  A copula-based measure for quantifying asymmetry in dependence and associations    --
--
#----                             Supporting Information 4                               --
--
#----                                                                                    --
--
#----                 Robert R. Junker, Florian Griessenberger, Wolfgang Trutschnig      --
--
#______________________________________________________________________________________________


library(qad)

#______________________________________________________________________________________________

#----                                Parabola; Fig. 1                                    --
--
#______________________________________________________________________________________________


# Function to generate the quadratic function
quadratic <- function(n,a){
  x <- seq(-1,1, length.out = n)
  y <- x^2 + runif(n, -a, a)
  A <- data.frame(x=x,y=y)
  A <- data.frame(x=(A$x-min(A$x))/(max(A$x)-min(A$x)) ,y=(A$y-min(A$y))/(max(A$y)-
min(A$y)))
  return(A)
}

rt <- quadratic(1000, 0.01) # quadratic(sample size, noise)

plot(rt$x, rt$y, col = rgb(255/255, 120/255, 180/255, 0.55), pch = 16, cex = 2)
plot(rank(rt$x)/length(rt$x), rank(rt$y)/length(rt$x), col = rgb(255/255, 120/255, 180/255,
0.55), pch = 16, cex = 2)

qr <- qad(rt$x,rt$y, print=TRUE, permutation = FALSE)
plot(qr, copula = TRUE, density = FALSE)


#______________________________________________________________________________________________

#----                 Dependence between two random variables; Fig. 2                    --
--
#______________________________________________________________________________________________


#Linear
linear <- function(n,a){
  x <- seq(0,1, length.out = n)
  y <- x + runif(n, -a, a)
  A <- data.frame(x=x,y=y)
  A <- data.frame(x=(A$x-min(A$x))/(max(A$x)-min(A$x)) ,y=(A$y-min(A$y))/(max(A$y)-
min(A$y)))
  return(A)
}

# x
xs <- function(n,a){
  x1 <- seq(0,1, length.out = n/2)
  y1 <- x1 + runif(n, -a, a)
  x2 <- seq(0,1, length.out = n/2)
  y2 <- 1-x2 + runif(n, -a, a)
  x <- c(x1, x2)
  y <- c(y1, y2)
  A <- data.frame(x=x,y=y)
  A <- data.frame(x=(A$x-min(A$x))/(max(A$x)-min(A$x)) ,y=(A$y-min(A$y))/(max(A$y)-
min(A$y)))
  return(A)
```

```
}

# two paralell lines
tpl <- function(n,a){

  x <- seq(0,1, length.out = n)
  y <- c(seq(0,1,length.out = n/2), seq(0,1,length.out = n/2))

  x <- x + runif(length(x), -a, a)
  y <- y + runif(length(y), -a, a)


  A <- data.frame(x=x,y=y)
  A <- data.frame(x=(A$x-min(A$x))/(max(A$x)-min(A$x)) ,y=(A$y-min(A$y))/(max(A$y)-
min(A$y)))
  return(A)
}

#two lines:
two.lines <- function(n=2000,a=0.05){
  x <- runif(n,0,1)
  #y <- rnorm(n,0,a)
  y <- runif(n,-a,a)
  A <- data.frame(x=x,y=y)
  A$case <- sample(c(0,1),n,replace=TRUE)
  A$u <- 0
  A$v <- 0

  phi <- c(pi/20,pi/4)
  Rot0 <- matrix(c(cos(phi[1]),sin(phi[1]),-sin(phi[1]),cos(phi[1])),nrow=2)
  Rot1 <- matrix(c(cos(phi[2]),sin(phi[2]),-sin(phi[2]),cos(phi[2])),nrow=2)

  for(i in 1:n){
    z <- Rot0 %*% c(A$x[i],A$y[i])
    if(A$case[i]==1){z <- Rot1 %*% c(A$x[i],A$y[i])}
    A[i,4:5] <- as.vector(z)
  }
  A <- subset(A,select = c(u,v))
  names(A) <- c("x","y")
  A <- data.frame(x=(A$x-min(A$x))/(max(A$x)-min(A$x)) ,y=(A$y-min(A$y))/(max(A$y)-
min(A$y)))
  return(A)
}

#Non-coexistence
non.coexistence <- function(n,a){
  A <- data.frame(x=runif(n),y=runif(n))
  A <- subset(A,A$x<=a | A$y<=a)
  A <- data.frame(x=(A$x-min(A$x))/(max(A$x)-min(A$x)) ,y=(A$y-min(A$y))/(max(A$y)-
min(A$y)))
  return(A)
}

# quadratic
quadratic <- function(n,a){
  x <- seq(-1,1, length.out = n)
  y <- x^2 + runif(n, -a, a)
  A <- data.frame(x=x,y=y)
  A <- data.frame(x=(A$x-min(A$x))/(max(A$x)-min(A$x)) ,y=(A$y-min(A$y))/(max(A$y)-
min(A$y)))
  return(A)
}

# sinus
sinus <- function(n,a){
  x <- seq(-8,8, length.out = n)
  y <- sin(x) + runif(n, -a, a)
  A <- data.frame(x=x,y=y)
  A <- data.frame(x=(A$x-min(A$x))/(max(A$x)-min(A$x)) ,y=(A$y-min(A$y))/(max(A$y)-
min(A$y)))
  return(A)
}
```

```
#Torus
torus <- function(n,a){
  r <- sqrt(runif(n,1-a,1+a))
  phi <- runif(n,0,2*pi)
  A <- data.frame(x=r*cos(phi),y=r*sin(phi))
  A <- data.frame(x=(A$x-min(A$x))/(max(A$x)-min(A$x)) ,y=(A$y-min(A$y))/(max(A$y)-
min(A$y)))
  return(A)
}

cbs <- function(n,a){

  #x1 <- seq(0,1, length.out = 16)
  #yp <- seq(0,1, length.out = 4)
  #y1 <- c(sample(yp, 4, replace = FALSE), sample(yp, 4, replace = FALSE), sample(yp, 4,
replace = FALSE), sample(yp, 4, replace = FALSE))

  x1 <- c(0.00000000, 0.06666667, 0.13333333, 0.20000000, 0.26666667, 0.33333333,
0.40000000, 0.46666667, 0.53333333, 0.60000000, 0.66666667, 0.73333333, 0.80000000,
0.86666667, 0.93333333, 1.00000000)
  y1 <- c(0.6666667, 0.0000000, 1.0000000, 0.3333333, 0.6666667, 0.3333333, 0.0000000,
1.0000000, 0.6666667, 0.3333333, 0.0000000, 1.0000000, 0.6666667, 0.3333333, 1.0000000,
0.0000000)

  x2 <- rep(x1, round(n/length(x1), 0))
  y2 <- rep(y1, round(n/length(y1), 0))
  x <- x2 + runif(length(x2), -a, a)
  y <- y2 + runif(length(y2), -a, a)


  A <- data.frame(x=x,y=y)
  A <- data.frame(x=(A$x-min(A$x))/(max(A$x)-min(A$x)) ,y=(A$y-min(A$y))/(max(A$y)-
min(A$y)))
  return(A)
}

### loop [example: Torus]
n = 1000
noise <- seq(0,1, length.out = 100)
simnum <- 10

sto_Pr <- rep(NA, (simnum * length(noise)))
sto_Sr <- rep(NA, (simnum * length(noise)))
sto_MIC <- rep(NA, (simnum * length(noise)))
sto_x1x2 <- rep(NA, (simnum * length(noise)))
sto_x2x1 <- rep(NA, (simnum * length(noise)))
sto_asy <- rep(NA, (simnum * length(noise)))


noi <- rep(NA, (simnum * length(noise)))

pos <- 0
for (i in 1:length(noise)){
  ai <- noise[i]
  for (j in 1:simnum){
    pos <- pos +1
    rt <- torus(n, ai)
    sto_Pr[pos] <- cor(rt$x,rt$y)
    sto_Sr[pos] <- cor(rt$x,rt$y, method = "spearman")
    mr <- mine(rt$x,rt$y)
    sto_MIC[pos] <- mr$MIC
    qr <- qad(rt$x,rt$y, print=FALSE)
    sto_x1x2[pos] <- qr$results$coef[1]
    sto_x2x1[pos] <- qr$results$coef[2]
    sto_asy[pos] <- qr$results$coef[4]

    noi[pos] <- ai

    print(pos)
    }
}
```

```
plot(noi, sto_x1x2)

#___________________________________________________________________________________________

#----                Relationship between qad and Pearson's r; Fig. 3                    --
--
#___________________________________________________________________________________________


source("http://janhove.github.io/RCode/plot_r.R")

rV <- seq(-0.999, 0.999, length.out = 200)

x1x2 <- matrix(nrow = length(rV), ncol = 16)
x2x1 <- matrix(nrow = length(rV), ncol = 16)
asy <- matrix(nrow = length(rV), ncol = 16)
MIC <- matrix(nrow = length(rV), ncol = 16)
rho <- matrix(nrow = length(rV), ncol = 16)
rf <- matrix(nrow = length(rV), ncol = 16)

for(i in 1:length(rV)){
  rd <- plot_r(r = rV[i], n = 500, showdata = TRUE)

  for(j in 1:16){
    data <- as.data.frame(rd$data[j])
    qr <- qad(data[,1], data[,2])
    x1x2[i, j] <- qr$results$coef[1]
    x2x1[i, j] <- qr$results$coef[2]
    asy[i, j] <- qr$results$coef[4]
    mr <- mine(data[,1], data[,2])
    MIC[i, j] <- mr$MIC
    rf[i, j] <- rV[i]
    sr <- cor.test(data[,1], data[,2], method = "spearman")
    rho[i, j] <- as.numeric(sr$estimate)
  }
}


#___________________________________________________________________________________________

#----                               Global climate; Fig. 4                               --
--
#___________________________________________________________________________________________


setwd("folder containing data") #
m <- read.table("Results_BioClim_Raster_Coords.txt", header = T) ## data are provided in SIX
m <- m[3:ncol(m)]

### quantification of asymmetric dependence (q) for all variable pairs

pwq <- pairwise.qad(m, permutation = FALSE)
heatmap.qad(pwq, select = "dependence", fontsize = 2) + theme(axis.text.x =
element_text(angle=90, hjust=0, vjust=1))
heatmap.qad(pwq, select = "asymmetry", fontsize = 2) + theme(axis.text.x =
element_text(angle=90, hjust=0, vjust=1))

### quantification of symmetric dependence (Pearson's r^2 and r) for all variable pairs

r2 <- as.matrix(cor(m, method="pearson")^2)
diag(r2) <- NA
r <- as.matrix(cor(m, method="pearson"))
diag(r) <- NA

### Association between annual mean temperature [°C] and annual Precipitation [mm]

plot((m$AMT_R/10), m$AP_R, col=rgb(16/255, 78/255, 139/255, 0.15), cex = 1.5, pch = 16, log
= "y", xlab = "Annual mean temperature [°C]",  ylab = "Annual Precipitation [mm]")
plot(rank((m$AMT_R/10)), rank(m$AP_R), col=rgb(16/255, 78/255, 139/255, 0.15), cex = 1.5,
pch = 16,  xlab = "Annual mean temperature [°C]", ylab = "Annual Precipitation [mm]")
```

```
### Asymmetric dependence between annual mean temperature [°C] and annual Precipitation [mm]
### empirical checkerboard copula is shown, which can be used as prediction tool

r <- qad((m$AMT_R/10), log(m$AP_R+1), print=TRUE, permutation = TRUE, nperm = 100)
plot(r, copula = FALSE, density = FALSE)


#_______________________________________________________________________________________________

#----                              Microbiome [Fig. 5]                                       --
--
#_______________________________________________________________________________________________


list.files()
m<-read.table("microbiome.txt", header=T)
m <- m[,2:ncol(m)]

### remove columns with too many non-unique values

unique.prop <- rep(NA, ncol(m))
prop.single.value <- rep(NA, ncol(m))

for(n in 1:ncol(m)){
  unique.prop[n] <- length(unique(m[,n])) / nrow(m)
  uV <- unique(m[,n])

  numV <- rep(NA, length(uV))
  for(i in 1:length(uV)){
    numV[i] <- length(m[,n] [m[,n] == uV[i]] )
  }
  prop.single.value[n] <- max(numV, na.rm = T)/nrow(m)
}

m <- m[,prop.single.value < 0.25]


### quantification of asymmetric dependence (q) for all variable pairs

pwq <- pairwise.qad(m)

### quantification of symmetric dependence (Pearson's r^2 and r) for all variable pairs

r2 <- as.matrix(cor(m, method="pearson")^2)
diag(r2) <- NA
r <- as.matrix(cor(m, method="pearson"))
diag(r) <- NA


### identification of key-species

AdepM <- matrix(NA, ncol(pwq$q), nrow(pwq$q))
colnames(AdepM) <- colnames(pwq$q)
AdepMean <- rep(NA, ncol(pwq$q))
AdepMedian <- rep(NA, ncol(pwq$q))
AdepQ2.5 <- rep(NA, ncol(pwq$q))
AdepQ97.5 <- rep(NA, ncol(pwq$q))
AdepMIN <- rep(NA, ncol(pwq$q))
AdepMAX <- rep(NA, ncol(pwq$q))
AdepSD <- rep(NA, ncol(pwq$q))
DepMean <- rep(NA, ncol(pwq$q))
DetMean <- rep(NA, ncol(pwq$q))
ColName <- rep(NA, ncol(pwq$q))

for(i in 1:ncol(pwq$q)){

  adt <-  as.numeric(pwq$q[i,]) - pwq$q[,i]
  AdepM[,i] <- adt
  AdepMean[i] <- mean(adt, na.rm = TRUE)
  AdepMedian[i] <- median(adt, na.rm = TRUE)
  AdepQ2.5[i] <- as.numeric(t.test(adt, mu = 0)$conf.int[1])
```

```
  AdepQ97.5[i] <- as.numeric(t.test(adt, mu = 0)$conf.int[2])
  AdepMIN[i] <- min(adt, na.rm = TRUE)
  AdepMAX[i] <- max(adt, na.rm = TRUE)
  AdepSD[i] <- sd(adt, na.rm = TRUE)
  DepMean[i] <- mean(as.numeric(pwq$q[,i]), na.rm=TRUE)
  DetMean[i] <- mean(as.numeric(pwq$q[i,]), na.rm=TRUE)
  ColName[i] <- colnames(pwq$q)[i]

}

Summary <- as.data.frame(cbind(AdepMean, AdepMedian, AdepQ2.5, AdepQ97.5, AdepMIN, AdepMAX,
AdepSD, DepMean, DetMean))
row.names(Summary) <- ColName

### Fig. 5a
SummaryO <- Summary[ order( Summary[,2]), ]  ## 1 = mean; 2 = median
AdepMO <- AdepM[, row.names(SummaryO)]

ii <- cut(SummaryO$AdepMedian, breaks = seq(min(SummaryO$AdepMedian),
max(SummaryO$AdepMedian), len = 500),
          include.lowest = TRUE)
colors_1 <- colorRampPalette(c(rgb(80/255, 235/255, 150/255, 1),rgb(64/255,161/255, 255/255,
1), col=rgb(255/255, 120/255, 180/255, 1)))(499)[ii]

boxplot(AdepMO, las = 2, cex.axis = 0.7, ylim=c(min(AdepMO, na.rm=TRUE), 0.5), col =
colors_1, range = 0, xaxt = "n")
xnames <- sub("OTU_", "", row.names(SummaryO))
axis(1,at=1:ncol(AdepMO),labels=xnames, las = 2, cex.axis = 0.65)

points(seq(-1,ncol(AdepMO)+2), rep(0, ncol(AdepMO)+4), type ="l", lty =1, col = "black", lwd
= 1.5)
points(seq(1,ncol(AdepMO),1), SummaryO$DepMean, type ="l", lty =1, col = rgb(80/255,
235/255, 150/255, 1), lwd = 3)
points(seq(1,ncol(AdepMO),1), SummaryO$DetMean, type ="l", lty =2, col=rgb(255/255, 120/255,
180/255, 1), lwd = 3)


#_______________________________________________________________________________________________

#----                    World development indicators  ; Fig. 6                           --
--
#_______________________________________________________________________________________________



setwd("folder containing data") #

m <- read.table("WDI_data.txt", header = T) ## data are provided in SIX
attach(m)
head(m)

### quantification of asymmetric dependence (q)
r <- qad(birth, death, permutation = TRUE, nperm = 100, DoParallel = TRUE, print=TRUE)
plot(r, copula = FALSE, margins = TRUE)

r <- qad(birth, log(gdp+1), permutation = TRUE, nperm = 100, DoParallel = TRUE, print=TRUE)
plot(r, copula = FALSE, margins = TRUE)

r <- qad(death, log(gdp+1), permutation = TRUE, nperm = 100, DoParallel = TRUE, print=TRUE)
plot(r, copula = FALSE, margins = TRUE)

mean_gdp <- mean(log(gdp+1), na.rm = TRUE)
r <- qad(birth[log(gdp+1) < mean_gdp], death[log(gdp+1) < mean_gdp], permutation = TRUE,
nperm = 100, DoParallel = TRUE, print=TRUE)
plot(r, copula = FALSE, margins = TRUE)

r <- qad(birth[log(gdp+1) > mean_gdp], death[log(gdp+1) > mean_gdp], permutation = TRUE,
nperm = 100, DoParallel = TRUE, print=TRUE)
plot(r, copula = FALSE, margins = TRUE)


### Fig. 4
```

```
ii <- cut(log(gdp+1),  include.lowest = TRUE, breaks = 300)
colT <- colorRampPalette(c(rgb(80/255, 235/255, 150/255, 1),rgb(64/255,161/255, 255/255, 1),
col=rgb(255/255, 120/255, 180/255, 1), rgb(247/255,110/255, 94/255, 1)))(300)[ii]
plot(birth, death, xlab = "Birth_rate_crude_per_1000_people",ylab =
"Death_rate_crude_per_1000_people", pch = 16,cex = 2, col= colT)
points(birth, death, xlab = "Birth_rate_crude_per_1000_people",ylab =
"Death_rate_crude_per_1000_people", pch = 1,cex = 2, col= "black")
text(birth, death+0.3, labels=country, cex= 0.7)

###linear model

m1 <- aov(death ~ birth*log(gdp+1))
summary(m1)
```